**Research Thesis | *Doctoral Program – Doctorate in Business Administration***

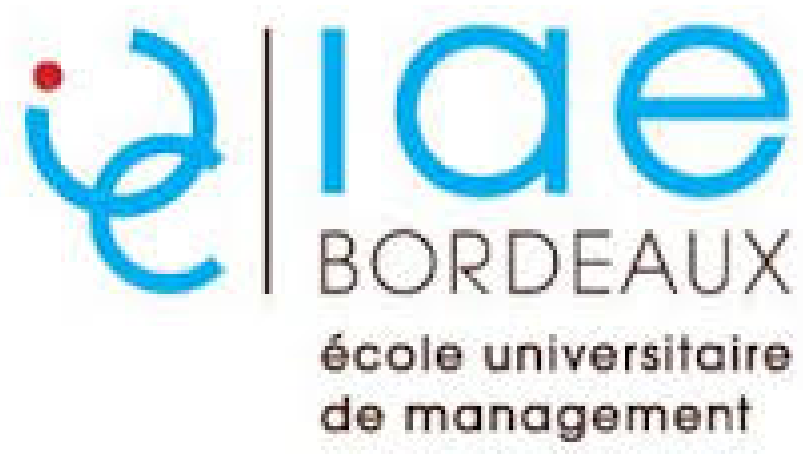

THESIS TITLE

# Designing for Ethical AI: HCI Feature Considerations to Improve Fairness and User Experience in AutoML use for Human Resources

Addressing bias and discrimination in models developed using AutoML with HCI features

**Sundaraparipurnan Narayanan**

**Doctorate in Business Administration**

**28 October 2025**

University of Bordeaux | IAE

<table>
<tr><td colspan="2">This Study by: Sundaraparipurnan Narayanan<br>Entitled: Designing for Ethical AI: HCI Feature Considerations to Improve Fairness and User Experience in AutoML use for Human Resources<br><br>has been approved as meeting the dissertation requirements for the Degree of Doctorate in Business Administration.</td></tr>
<tr><td>Date</td><td>Professor</td></tr>
<tr><td></td><td>Dr. Stéphane Trébucq, Professor of Management Sciences</td></tr>
<tr><td></td><td>Dr. Olivier Herrbach, Professor of Management and Dean of IAE Bordeaux</td></tr>
<tr><td></td><td>Dr. Pascal Barneto, University Professor and Director of the Doctorate in Business Administration</td></tr>
</table>

—--------—-----------------------------------------------------------------------------------------

**Dedication**

This thesis is dedicated to those who guided, supported, and inspired me throughout this journey.

To **Dr. Stéphane Trébucq**, whose insightful mentorship, patience, and encouragement were invaluable in shaping my academic growth and guiding this research with clarity and purpose.

To **Arva Shah** and **Sandeep Vishwakarma**, for their expertise, collaboration, and unwavering support with algorithms or code.

And to my **family**, whose love, understanding, and constant encouragement formed the foundation of my strength and exploration. This thesis is dedicated to their love and trust.

## Table of Contents

---

## Part II – METHODOLOGY & FRAMEWORK DEVELOPMENT

---

## Part III – EMPIRICAL EVALUATION

---

## Part IV – BUSINESS AND DESIGN IMPLICATIONS

---

## Part V – CONTRIBUTIONS AND FUTURE WORK

---

## Appendices

# Abstract

This thesis looks at a big problem: how fair are Automated Machine Learning (AutoML) tools when used for hiring, especially in human resources (HR)? It brings together ideas from regulations, business, and how people interact with computers (HCI). It argues that fairness isn't just about ethics anymore. It's now a key feature of a good product. This makes systems easier to use, builds trust, and helps companies adopt these tools.

Increasingly, businesses, especially HR departments, are using AutoML. These tools promise to make things efficient and accessible. But they also deal with serious ethical and legal issues. AutoML simplifies things like choosing models and fine-tuning them. This saves money and encourages innovation, especially for smaller businesses. However, this ease of use can also lead to bias. Models trained on old, unfair data might continue discrimination based on race, gender, or age. When these "black box" systems are not transparent, they threaten fairness within organizations. They also risk breaking new laws, like the EU AI Act and U.S. EEOC guidelines. So, there's a conflict: AutoML offers efficiency, but in important areas, efficiency without fairness causes harm to organizations, legally, and to their reputation.

The thesis focuses on a specific problem: current AutoML platforms don't do enough to ensure fairness. They are usually built for technical performance. This leaves everyday business users unprepared to find and fix bias. So, the main question is: how can AutoML tools put fairness first in their design, instead of treating it as an afterthought? To explore this, the study asks four questions:

(RQ1) What fairness problems and solutions are there in popular AutoML tools for hiring data?

(RQ2) How do interface design, transparency, and feedback help or hurt fairness?

(RQ3) What are the missing pieces in human involvement (HITL), fairness visuals, and reporting?

(RQ4) What should AutoML providers focus on in product design to make fairness features easy to use and widely adopted by businesses?

The research uses foundational theories including Technology Acceptance Model, Innovation Diffusion Theory, Human Centered AI, Cognitive Load Theory and Affordance Theory. The Technology Acceptance Model (TAM) helps understand fairness features through how useful people perceive them to be or ‘perceived use’ (PU) and how easy they perceive them to be to use or ‘Perceived ease of use’ (PEOU). Things like fairness dashboards and audit reports make tools seem more useful and easier to use. This is because they reduce legal risks and make things simpler for users. Innovation Diffusion Theory (IDT) explains why companies adopt new things. It emphasizes that new tools are adopted if they offer clear advantages, aren't too complicated, and show obvious benefits in terms of fairness. To evaluate AutoML interfaces, the study also uses UX strategy frameworks. These include Human-Centered AI (HCAI), Cognitive Load Theory (CLT), and Affordance Theory. HCAI stresses that tools need to be transparent, controllable, and reliable. CLT shows that information should be revealed gradually with visual cues to prevent users from feeling overwhelmed by fairness data. Affordance Theory helps analyze how fairness tools can be designed so non-experts can easily see, find, and use them.

To answer the research questions, the study uses both qualitative and quantitative methods. The qualitative part includes structured HCI audits, heuristic evaluations, and cognitive walkthroughs. These simulate how non-expert business users try to find bias, apply solutions, and create audit reports for compliance. The audits are also supported by analyzing user contracts and documents to find where fairness disclosures are unclear. The quantitative part tests eight well-known AutoML tools using specially prepared HR datasets with sensitive information. The models created are assessed with legal fairness metrics, such as Demographic Parity Difference and Equalized Odds. This determines how well AutoML outputs match fairness expectations in real hiring situations. This two-part approach, looking at both design features and measurable fairness results, gives a full picture of the AutoML ecosystem.

The findings show consistent problems with fairness transparency, user control, and bias mitigation in the reviewed AutoML tools and libraries. Across different platforms, fairness features often lack clear visuals, guidance, or useful human-in-the-loop options. This leaves non-experts with little ability to meet ethical and legal rules. This research introduces a new framework for evaluating HCI, focused on fairness. It has five parts: user contracts, interface design, information architecture, human support, and accountability. This framework helps critique current tools and also shows how to build fairness directly into product development.

The study's practical suggestion emphasizes making fairness a core part of AutoML tools, using Human Computer Interaction inspired factors. This means offering clear, interactive ways to see bias, improving feedback to support human oversight, and building accountability into both the interface and documentation. These suggestions align with current best practices from companies like IBM and Google. They also broaden the discussion by seeing fairness as a critical business need, not just a compliance burden.

The thesis is structured to go from defining the problem to showing insights and strategic contributions. Chapter 1 provides an overview of the problem space. Chapter 2 places fairness in the context of Human Resources (specifically hiring). Chapter 3 details the theories (foundational) and fairness metrics. Chapter 4 explores the legal environment. Chapter 5 outlines the evaluation framework. Chapters 6 and 7 present qualitative and quantitative findings. Chapter 8 then offers an integrated discussion based on the outcomes from the qualitative and quantitative analysis. Chapter 9 looks at the business importance of fairness in hiring. Chapter 10 provides product design recommendations. Finally, Chapter 11 brings together the contributions, limitations, and future research directions.

In the end, this thesis helps both theory and practice. It represents that fairness in AutoML is a core part of usability, trust, and adoption. It offers a dual evaluation: how AutoML tools are designed to support fairness, and what fairness results they produce.The thesis represents fairness not as an extra feature, but a key factor in whether AI in hiring can be ethical, compliant, and sustainable.

# Executive summary

This executive summary presents key insights from the thesis, which explores fairness and bias in Automated Machine Learning (AutoML) systems, particularly in the human resources (HR) domain. It combines regulatory, business, and user-interface perspectives, grounded in Human-Computer Interaction (HCI) and product design theories. Chapters 1 and Chapter 2 of the thesis set the foundation by discussing the motivation, background, and importance of fairness in AutoML systems.

**Business Value of AutoML**

AutoML simplifies model building for users without deep data science expertise. It helps automate tasks like model selection, hyperparameter tuning, and feature engineering. Chapter 2 discusses how AutoML makes machine learning accessible to non-experts. In HR, this means HR managers and analysts can use AutoML to predict attrition, screen resumes, or make promotion decisions. AutoML's business value lies in saving time, cutting costs, and enabling innovation, especially in small and medium businesses. However, these benefits come with fairness risks. Furthermore, AutoML tools and libraries democratize leveraging Artificial Intelligence and Machine Learning, enabling varied stakeholders including non-data science experts to build, deploy and leverage AI/ ML technology for domain specific needs.

It is pertinent to note that the AutoML market is expanding rapidly, projected to grow over 40% CAGR through 2030; driven by enterprise interest in low-code AI, especially in sensitive areas like HR. However, fairness, bias, and transparency are emerging as top concerns for business leaders, with over 40% of organizations reporting significant worry about AI bias affecting reputation, trust, and compliance. In response, leading platforms like IBM Watson, Google AutoML, DataRobot, and Dataiku are already integrating fairness-by-design features—such as bias detection tools, explainable AI modules, and audit trails—to meet regulatory and ethical demands from HR buyers. These proactive moves signal an industry shift toward responsible AI as both a business necessity and competitive differentiator.

**Challenges in Fairness**

Bias in AI tools can replicate or amplify human discrimination. This includes discrimination based on race, gender, age or other sensitive variables. AutoML tools can inherit these biases through historical training data or lack of human oversight. Models, on some occasions, choose parameters and considerations that also contribute to bias during model development and deployment (Chapter 2).

AutoML tools often act as 'black boxes,' making decisions that are hard to interpret. Further, AutoML tools often lack transparency features or fairness guidance thereby neither taking into account the fairness considerations, nor enabling users to take adequate efforts to address potential biases (Chapter 2 and Chapter 3). Issues like gender bias or indirect discrimination through proxy variables (zip code determining whether a candidate is qualified for hire) were observed in HR datasets across the literature reviews. Tools often lacked built-in support to identify or mitigate these biases.

Further, it is necessary to consider the AutoML from the lens of grounded theories including Technology Acceptance Model, Information Diffusion Theory and UX strategies (including design thinking, affordance and cognitive load theory) in Chapter 3 to consider the perceived ease of use/usefulness and its link to fairness adoption, relevance of innovation features like transparency and auditability influence diffusion and design principles for user-centered fairness controls (nudges, feedback loops, alerts). This approach supports the AutoML providers to enhance their product design thereby reducing the onboarding friction and enhancing adoption of their tools.

**Regulatory Expectations**

Regulatory expectation analysis outlines the requirements from EU AI Act and U.S. laws like EEOC guidance and NYC Local Law 144 from the lens of fairness in human resource context (Chapter 4). These laws require high-risk AI systems, such as those used in hiring, to be fair, explainable, and auditable. The EU AI Act mandates that systems provide human oversight, explainability, bias mitigation, and continuous monitoring (post market deployment monitoring). U.S. laws emphasize auditing and adverse impact assessment. Companies must show that their AI tools do not unfairly harm protected groups. AutoML tools, when used for developing automated employment decisions, become Automated Employment Decision Tools and thereby are required to comply with the obligations.

**Research Methodology**

The research employed a mixed-methods approach to evaluate the fairness capabilities of AutoML tools, integrating both qualitative and quantitative analyses (Chapter 1). The qualitative component was rooted in Human-Computer Interaction (HCI) principles and involved a structured audit of user interfaces to assess how each system supports or hinders human agency in addressing algorithmic bias. Through heuristic interface audits, cognitive walkthroughs, and content analysis, the study examined elements such as fairness affordances, transparency features, user guidance, and the discoverability of ethical tools. These methods captured how well non-expert users could understand, interact with, and influence fairness-related processes within the tools, highlighting both points of friction and best practices in interface design.

In addition, the quantitative component involved running the same AutoML tools on curated HR datasets to assess their fairness outcomes using established metrics like True Positive Rate, False Positive Rate, Demographic Parity Difference, and Predictive Rate Parity. This scenario-driven audit allowed for empirical comparison of group-level fairness, revealing disparities in model behavior across sensitive attributes. By triangulating these two approaches—interface-level audits and outcome-based fairness measurement—the research provided a comprehensive assessment of how AutoML systems support fairness in both their user experience and algorithmic outputs.

**Evaluation Framework**

The learnings from Chapter 2 (bias and fairness in the context of AutoML and hiring) and Chapter 3 (regulatory requirements and theoretical frameworks) are leveraged to build a Human Computer Interaction based evaluation framework for evaluating AutoML tools. The framework is structured around five dimensions, namely, (1) Contracts and User Development: Focuses on clear disclosures, responsible data use, and user training to set expectations and support fairness in AutoML use; (2) User Interface and Experience Design: Emphasizes intuitive interfaces, visual explanations, and usability features that reduce cognitive load and guide users through fairness processes; (3) Information Architecture: Ensures information is logically organized and visualized to help users navigate and understand fairness-related data effectively; (4) Human Augmentation: Supports collaboration between humans and AutoML through features like explainability, feedback loops, and override capabilities; and (5) Care and Responsibility: Addresses ethical and safety considerations through governance tools, guardrails, transparency measures, and mechanisms for continuous system improvement (Chapter 5).

**Selection of Tools, Datasets and Metrics**

The selection of AutoML tools and libraries systematically to make sure they were relevant and suitable for studying fairness in hiring. Firstly, a wide search was done to find the most used tools. For code-based libraries, only those with a strong presence on GitHub—more than 3,500 stars and at least 50 contributors—were included. This led to the choice of AutoGluon, FLAML, PyCaret, and H2O AutoML (Python). For GUI-based tools, the focus was on those that are easy for non-experts to use, widely adopted in business, and available through free or research versions. Tools like Dataiku, DataRobot, H2O AutoML Studio, and Alteryx RapidMiner were selected based on how easy they are to use, how well they handle data, and the quality of their user interface and support materials. The above approach resulted in choosing four GUI-based and four code-based libraries for further analysis (Chapter 3).

The selection approach for choosing appropriate and domain-context aligned dataset related to fairness in hiring decisions also had a systemic approach. Searches were done across various trusted sources like Kaggle, Google Scholar, and UCI Machine Learning Repository using keywords like "fairness" and "employment." The datasets were then filtered based on specific three criteria: they had to be freely available, include demographic and hiring-related information, and be in a tabular format. Tabular data was chosen because most AutoML tools work best with that kind of structured data. Datasets like those focused on text or images were left out to stay focused on the core strengths of the AutoML tools being studied (Chapter 3).

The approach to selection of appropriate metrics for fairness started with eliminating traditional classification metrics like accuracy, precision, recall, and F1 score in assessing algorithmic fairness, as they overlook disparities between demographic groups. Group fairness metrics—such as Demographic Parity, Equalized Odds, and Predictive Rate Parity—were considered as essential for detecting and addressing bias in binary classification, especially with tabular data based on literature reviews. Each of these metrics captures different fairness concerns but comes with inherent trade-offs and theoretical conflicts, making it impossible to satisfy all simultaneously when group base rates differ. The choice of fairness metric should therefore be context-dependent, considering factors like risk distribution, data imbalance, and domain-specific harms. Hence, instead of choosing one or more of these metrics, all metrics were considered to collectively examine fairness considerations in AutoML model outcomes. Together, these metrics offer a more systemic and legally relevant (covers metrics referred to in NYC Local Law 144 on Bias audit) understanding of fairness, enabling a nuanced evaluation of bias that general performance metrics cannot provide (Chapter 3).

**Evaluation Findings**

The evaluation of AutoML tools and libraries using an HCI-based framework and curated HR datasets uncovered several key fairness gaps, especially for non-expert users (Chapter 6, Chapter 7 and Chapter 8). A major issue is the heavy reliance on users to detect and address bias without enough system support. This overdependence is not sustainable, as it expects users to manage fairness concerns on their own, often without the necessary tools, training, or guidance. Many interfaces reinforce automation bias by making users overly trust model outputs. Only a few tools provide visible alerts, explanations, or fairness-related nudges, and most lack clear visualizations to help users understand when and how fairness problems occur during the model-building process.

—--------—------------------------------------------------------------------------------------

Another problem is the limited user control and lack of human-in-the-loop (HITL) features. Most AutoML tools do not allow users to override fairness decisions or adjust fairness settings meaningfully. There are few built-in systems to monitor bias after deployment, and fairness-related actions are often not traceable. Tools also fall short in helping users detect deeper data problems like multicollinearity or proxy variables, which are critical for identifying hidden sources of bias during data preparation. Additionally, fairness explanations are often weak—most tools only show final metrics without explaining the causes of bias or how user feedback affects fairness outcomes.

Lastly, the lack of visual tools for comparing model performance across demographic groups and analyzing fairness-performance trade-offs is a major gap. Very few tools include dashboards showing specific group fairness metrics such as Demographic Parity Difference or Equalized Odds. Code-based tools especially offer little support here, often requiring users to rely on manual coding or external libraries. This lack of built-in visualization and trade-off analysis increases the mental effort required from users and makes it harder to make fair, ethical decisions using AutoML tools.

**Business Implications**

The gaps in fairness features within AutoML tools pose serious business risks for non-expert enterprise users. Biased algorithms in hiring can lead to poor candidate selection, reducing workforce diversity and innovation. They also expose companies to legal issues under regulations like Title VII or the EU AI Act and create ethical concerns due to a lack of transparency. These risks can damage a company's reputation, as seen in high-profile cases involving Amazon, Apple, and HireVue. Additionally, failing to comply with fairness-related laws can result in financial penalties, making it critical for businesses to ensure fairness in AI-driven hiring systems (Appendix 5).

For AutoML tool providers, these shortcomings directly affect product success and adoption. Tools lacking clear fairness, transparency, and auditability features create friction in the buying process, especially with HR, legal, and IT teams. This can delay or halt adoption, increase support costs due to user confusion, and raise the total cost of ownership. Moreover, in a competitive and regulated market, vendors that do not prioritize fairness risk losing customers to those who offer it as a built-in advantage. As more buyers seek trustworthy AI tools, fairness features are becoming essential for both customer retention and long-term market growth (Appendix 5).

**Recommendations**

The research offers a practical and structured set of design and product recommendations for AutoML tool providers, grounded in Human-Computer Interaction (HCI) principles (Chapter 9). It emphasizes that fairness should not only be treated as an ethical goal but also as a practical requirement for enterprise adoption—especially in regulated sectors like hiring. By embedding fairness-by-design into AutoML tools, providers can reduce user complexity, increase perceived usefulness, and align product functionality with non-expert expectations. The thesis highlights the value of incorporating UX Key Performance Indicators (KPIs), such as how often fairness features are used or how quickly users resolve fairness issues, to continuously improve product design and usability. Reducing fairness opacity through clearer interfaces can also lower support costs and build greater trust with enterprise buyers.

The recommendations are organized across five HCI dimensions, including clearer user contracts, more intuitive interface designs with fairness visualisations, better information architecture for accessing fairness settings, and stronger human augmentation features like feedback loops, customisable fairness settings, and documentation. It also advocates for more robust accountability mechanisms and incident response systems to ensure fairness is consistently applied and monitored. Many of these suggestions are already being adopted by industry leaders—IBM, Google, Microsoft, and Apple have incorporated elements like fairness dashboards, explainability tools, and transparency logs—demonstrating that these HCI-driven practices are becoming standard in trustworthy AI product design (Chapter 9).

**Contributions and Future Work**

This research makes key theoretical and practical contributions by reframing fairness in AutoML as a strategic product feature that supports usability, trust, and market adoption, especially for non-expert users. Grounded in established theories like the Technology Acceptance Model and Human-Centered AI, it introduces a fairness-centered HCI design approach that shifts the focus from just model accuracy to user empowerment and responsible AI (Chapter 10). Practically, it provides an actionable evaluation framework, benchmarks eight major AutoML tools, and offers design recommendations that reduce cognitive load, mitigate automation bias, and improve fairness transparency. Its novelty lies in addressing the ethical and social dimensions of AutoML, an area often overlooked in favor of technical performance. The thesis also outlines future research paths, including expanding bias mitigation to other sectors, exploring fairness across the AI lifecycle, and improving transparency in fairness-performance trade-offs through interactive and user-friendly design features.

# Chapter Organization

The thesis is structured into five overarching Parts, each comprising multiple chapters that build on one another to form a coherent research journey.

- **Part I – Foundations and Context** establishes the motivation, problem framing, and conceptual underpinnings of fairness in AutoML within the human resources domain. It reviews existing literature, datasets, and regulatory expectations, thereby laying the groundwork for the study.

- **Part II – Methodology and Framework Development** introduces the Human-Computer Interaction (HCI)-based evaluation framework, outlining its rationale, structure, and alignment with theoretical and regulatory insights.

- **Part III – Empirical Evaluation** applies the framework through systematic audits and fairness metric evaluations, generating empirical findings that highlight strengths and limitations of AutoML tools.

- **Part IV – Business and Design Implications** interprets these findings from both enterprise user and AutoML provider perspectives, emphasizing risks, compliance requirements, and product differentiation. It also offers structured design recommendations to embed fairness-by-design principles.

- **Part V – Contributions and Future Directions** concludes the thesis by summarizing theoretical and practical contributions, reflecting on limitations, and identifying avenues for future research in fairness-centric AutoML.

This approach from context to methods, from evaluation to implications, and finally to contributions enables the thesis to provide both a rigorous academic foundation and actionable insights for practice.

The chapters under each of the parts are organized to reflect a logical progression from problem definition and conceptual grounding to empirical evaluation, business implications, and strategic design recommendations for fairness in AutoML systems used in hiring.

# Part I – Foundations and Context

**Chapter 1: Introduction, motivation and premise of the thesis**

This chapter introduces the thesis, outlines the research motivation, and frames the central problem of fairness in AutoML tools within regulated domains like hiring. It presents refined research questions and provides an overview of the multi-stage research methodology that includes systemic literature review, empirical evaluation, and audit-based benchmarking. It also highlights the theoretical and regulatory framing and concludes with the overall structure of the thesis.

**Chapter 2: Understanding Fairness While Using AutoML in Human Resources Context**

This chapter explores fairness concerns in the context of algorithmic hiring, beginning with a review of existing literature and practices. It identifies key forms of bias, datasets used for benchmarking fairness in hiring, and fairness limitations in AutoML tools. It also explores AutoML adoption trends in the HR domain, market expectations for fairness, and the consequences of failing to address bias.

**Chapter 3: Foundational Considerations for the Thesis**

This chapter consolidates the theoretical, empirical, and dataset-related foundations of the thesis. It introduces and applies key frameworks such as the Technology Acceptance Model (TAM), Innovation Diffusion Theory (IDT), and UX strategy to inform fairness design. It also outlines the selection criteria for tools, libraries, and datasets used in the study, as well as relevant fairness metrics drawn from existing academic and policy literature.

**Chapter 4: Legal Frameworks and Expectations Regarding Bias in Hiring**

This chapter reviews major global legal and regulatory frameworks influencing fairness in algorithmic hiring. It provides an in-depth analysis of the EU AI Act, EEOC guidelines, NYC Local Law 144, and EU directives, synthesizing their expectations for fairness, transparency, oversight, and bias mitigation. These expectations form the compliance lens for the evaluation framework.

## Part II – Methodology and Framework Development

**Chapter 5: Scope and Considerations for Evaluation**

This chapter introduces the Human-Computer Interaction (HCI)-based evaluation framework developed for this study. It discusses the rationale for HCI alignment in regulated tool design, outlines key evaluation categories (e.g., UX design, care and responsibility, human augmentation), and maps them to grounded theories. The chapter concludes with industry-aligned best practices that influenced the evaluation design.

**Chapter 6: HCI Evaluation Framework Applied to AutoML Tools and Libraries**

This chapter presents the empirical application of the HCI-based audit framework across selected AutoML tools. It describes the audit procedures, criteria, and outcomes, including tool-specific findings across UX design, transparency, human-in-the-loop support, content quality, and oversight mechanisms. The audit highlights strengths and gaps across both GUI-based platforms and code-based libraries.

## Part III – Empirical Evaluation

**Chapter 7: Evaluation of AutoML Tools and Libraries on Fairness Metrics Leveraging the Human Resources Datasets**

This chapter evaluates the tools using a set of fairness metrics (e.g., TPR, FPR, Demographic Parity) on hiring-related datasets. It provides comparative performance insights across datasets and tools, identifies consistent disparities in outcomes, and aligns these findings with HCI observations. The chapter supports a multidimensional understanding of fairness risk in tool usage.

**Chapter 8: Discussion on Evaluation Outcomes**

This chapter synthesizes the insights from Chapters 6 and 7, analyzing gaps in fairness-related feedback mechanisms, transparency disclosures, user control, and automation bias. It highlights systemic limitations in current AutoML tool design and introduces critical trade-offs between automation efficiency and fairness. The chapter also reflects on user over-reliance and the need for better interpretability.

—------—--------------------------------------------------------------------------------------

# Part IV – Business and Design Implications

### Chapter 9: Business Implications of Bias in the Recruitment Process

This chapter examines the practical and reputational risks that arise for enterprise users and AutoML tool providers when fairness is not adequately addressed. It draws on case examples, including HireVue, to illustrate the impact of biased automation. It categorizes risks for both downstream users and tool developers and positions fairness as a product differentiator and compliance imperative.

### Chapter 10: Recommended Feature Considerations for AutoML from a Fairness Perspective

This chapter translates the findings into a structured set of design and product recommendations for AutoML tool providers. It proposes improvements across UI, documentation, transparency, oversight, and risk mitigation. The recommendations are categorized by interface layers and mapped to industry practices that support scalable, fair-by-design AutoML development.

# Part V – Contributions and Future Directions

### Chapter 11: Contributions, Limitations, and Future Work

The final chapter presents the theoretical and practical contributions of the research. It reflects on methodological and empirical limitations (e.g., tool scope, generalizability), and outlines future research opportunities. These include empirical studies on fairness feature adoption, design patterns for trade-off transparency, and the development of regulatory-aligned fairness dashboards for AutoML tools.

This dissertation includes peer-reviewed and preprint material that has been modified from first-authored papers by **Sundaraparipurnan Narayanan**. **Sundaraparipurnan Narayanan** produced all the writing that has been reproduced in this dissertation.

The following list serves as a declaration of the Versions of Record for works included in this dissertation:

- Portions of **Chapter 1 and Chapter 9:** *The cost to the pocket: Business implications of discrimination caused by AI-driven recruitment tools.* Published as a preprint. https://doi.org/10.5281/zenodo.17445881
- Portions of **Chapter 5 and Chapter 6:** *Democratise with Care: The need for fairness specific features in user-interface based open source AutoML tools.* Published as a paper and poster at **NeurIPS 2023 conference**. https://arxiv.org/abs/2312.12460
- Portions of **Chapter 5:** *Towards building a comprehensive evaluation framework for Human Computer Interaction with a specific focus on AutoML.* Published as a preprint. https://doi.org/10.5281/zenodo.17445915 and as a Journal article in International Journal of Scientific Research in Engineering & Management. https://doi.org/10.55041/IJSREM53221
- Portions of **Chapter 6:** *Democratise with Care: The need for fairness specific features in user-interface based open source AutoML tools.* Presented as a **poster at AutoML School, September 2024**.

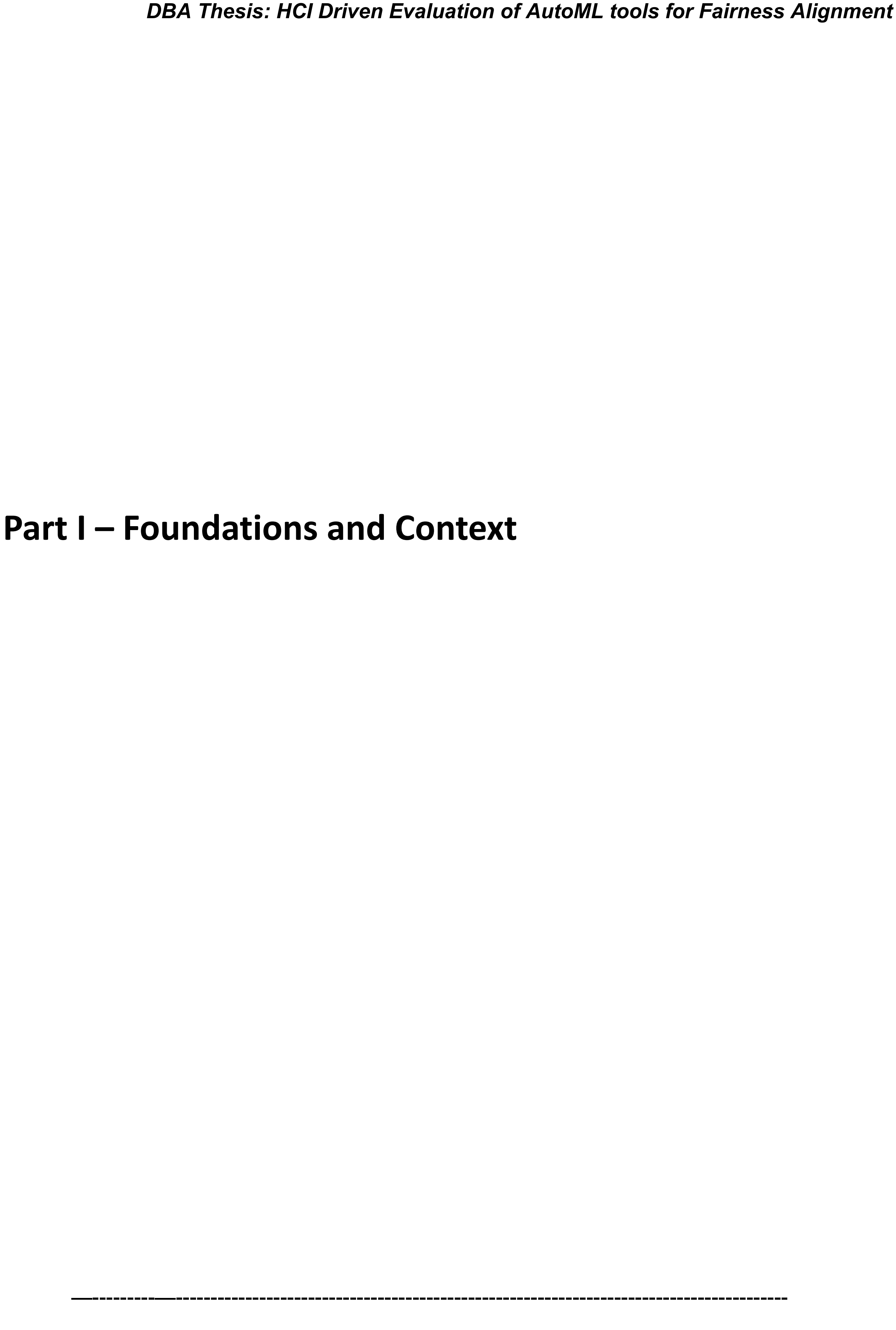

# Part I – Foundations and Context

# Chapter 1: Introduction, Motivation and the Premise of the thesis

**Chapter Overview**

**Purpose:** The chapter provides context towards how AutoML enhances hiring efficiency but introduces bias and opacity risks, potentially conflicting with fairness regulations such as the EU AI Act and U.S. EEOC guidelines. It also introduces that the AutoML tools do not currently consider fairness as a core product feature.

**Context:** The chapter introduces the context of the study as one that examines the fairness considerations within Automated Machine Learning (AutoML) tools when used for human resources tasks

**Problem & Focus Areas:** The chapter introduces that current AutoML platforms prioritize accuracy over fairness, leaving non-expert users unable to detect or mitigate bias. The chapter expresses four focus areas of the study: fairness issues in AutoML, interface design issues in AutoML & fairness, human involvement and visualization gaps in AutoML, and recommendations for fairness in product design priorities.

**Relevant Literature, Regulations & Theoretical Framework Considerations:** The chapter introduces the methodology for the study, involving (a) Developing a framework and (b) evaluating the tools based on the framework. The framework is developed leveraging Human Computer Interaction research studies, regulatory requirements and by leveraging principles from TAM, IDT, and UX frameworks like Human-Centered AI, Cognitive Load Theory, and Affordance Theory to evaluate fairness-oriented design and functionality.

**Evaluation of AutoML tools and libraries:** The chapter explains the mixed-methods approach to the study, combining HCI audits and quantitative testing of AutoML tools using HR datasets and legal fairness metrics to develop a comprehensive understanding of fairness in AutoML ecosystems.

## A. Introduction

Industry 4.0, often called the fourth industrial revolution, is characterized by its cognitive capabilities, where cyber-physical systems interact to enhance manufacturing processes. This revolution integrates various emerging technologies like artificial intelligence, blockchain, and machine learning. Industry 4.0 is considered a disruptive innovation crucial for the development of human resources. HR 4.0 should focus on Smart Human Resource (SHR 4.0) to effectively manage disruptions in recruitment, training, performance management, and other HR responsibilities. (Nirmala & Chitte, 2021). Innovative hiring solutions leverage advanced natural language processing techniques and machine learning algorithms to address these challenges.

They introduce resume parsing and ranking systems that enhance efficiency, reduce bias, and optimize candidate selection outcomes, showcasing high accuracy and significantly improving the efficiency of the process. (Source: "Towards smarter hiring: resume parsing and ranking with YOLOv5 and DistilBERT"). Methods including Automated Machine Learning (AutoML), or automated ML, automates the tedious tasks involved in machine learning model development, enabling non-experts to develop algorithms, thereby supporting development and adoption of such innovative hiring solutions. However, the use of machine learning does not necessarily guarantee bias-free outcomes, as these models are trained on the data derived from biased past human decisions. AI technology carries risks that could negatively affect individuals, groups, organizations, communities, society, and the environment, including perpetuating discrimination and harm, thereby exposing businesses creating or using AutoML to liabilities, reputational impact, and loss of value. This thesis aims at identifying specific implications caused by unfair or discriminatory algorithms or outcomes associated with downstream use (hiring) of widely used AutoML tools.

Like other technology risks, the risk posed by AI can take shape over a long or short term, may have a high or low impact, and be classified as high or low probability events (NIST, 2023[1]). These risks might turn into economic risks eventually due to reputational impact, stock price impact, and legal fines impact. This is true even for automated machine learning tools (AUTOML) users.

Companies may face legal repercussions and liabilities for engaging in discrimination using AI that causes harm to certain classes and groups[2]. Similarly, organizations that engage third parties for AI design, development, and deployment must know the risk mitigation and governance standards each applies. They should also independently test and audit all high-stakes inputs Developers of AutoML will also fall under the broader definition of the third parties for organizations using these tools for downstream applications, thereby exposing themselves to civil liabilities, reputational impact and loss of value.

For enterprises developing or procuring AutoML solutions, fairness features are no longer optional. Increasingly, discriminatory outcomes pose legal, reputational, and operational risks that can disrupt product adoption, delay procurement, or result in compliance failures. Thus, fairness must be considered not only as an ethical concern, but also as a core aspect of product-market fit and user trust in regulated domains.

---

[1] NIST. (2023). *NIST AI Risk Management Framework*. https://nvlpubs.nist.gov/nistpubs/ai/NIST.AI.100-1.pdf

[2] *Identifying and managing your biggest AI risks | McKinsey*. (n.d.). Retrieved February 17, 2023, from https://www.mckinsey.com/capabilities/quantumblack/our-insights/getting-to-know-and-manage-your-biggest-ai-risks

Enterprises shall better equip themselves to identify, assess and manage automation risks in line with their larger mission and business objectives (NIST, 2020[3]). Organizations are becoming aware of the universe of ethical pitfalls (e.g. unfair discrimination) and risk(s) caused by AI and regulators' emerging interest in monitoring failures or adverse incidents pointed out by these risks, in the current environment. The European Union recently proposed AI regulations, if breached, could result in material fines and hold organizations liable for proliferating biases or inequities through AI[4]. Therefore, Enterprises or Corporates are accountable for AI failures.

Market analyses from Gartner (2023) and McKinsey (2022) suggest that the use of AI and AutoML in HR decision-making is rapidly expanding, particularly in areas such as resume screening, employee attrition prediction, and promotion decisions—heightening the urgency of fairness interventions within these tools.

Fairness features are evaluated not just by technical users but also by multiple stakeholders across the buyer ecosystem: HR professionals look for transparency in candidate treatment, compliance teams assess legal exposure, and IT gatekeepers evaluate system trustworthiness and integration overhead. A product that addresses these multi-stakeholder expectations has a greater likelihood of enterprise adoption.

## C. Motivation

This research is motivated by the potential impact algorithmic bias could have on people, when such bias is arising out of models developed using AutoML by non-expert users. This research will help fill that gap and provide practical guidance to organizations developing and/ or using AutoML. For organizations developing and offering AutoML as a product, it's critical to consider how Human-Computer Interaction (HCI), particularly fairness, contributes to product strategy, democratizing reach, and avoiding potential biases for non-expert users. This research aims to bridge that gap by offering practical advice for organizations using AutoML. For companies developing and offering AutoML as a product, it's crucial to consider how Human-Computer Interaction (HCI), especially fairness, impacts product strategy, broadens access, and prevents biases for non-expert users. The impact AutoML can have on varied stakeholders are represented below:

---

[3] NIST. (2020). *Integrating Cybersecurity and Enterprise Risk Management (ERM)*. https://csrc.nist.gov/publications/detail/nistir/8286/final

[4] *Identifying and managing your biggest AI risks | McKinsey*. (n.d.). Retrieved February 17, 2023, from https://www.mckinsey.com/capabilities/quantumblack/our-insights/getting-to-know-and-manage-your-biggest-ai-risks

| Stakeholder | Risks of Unfairness in AutoML | Implications |
|---|---|---|
| Enterprise Users (Non-Expert Practitioners) | Risk of deploying biased models unknowingly due to lack of transparency, explainability, or fairness guidance. | Discriminatory hiring or promotion outcomes, regulatory non-compliance, and reduced trust in AI systems. |
| Compliance, Legal, and Risk Teams | Inability to audit or document fairness features effectively; limited traceability of bias mitigation steps. | Legal exposure under EU AI Act, EEOC guidelines, or NYC Local Law 144; procurement delays; higher cost of compliance. |
| AutoML Tool Providers and Developers | Lack of fairness-by-design features (e.g., dashboards, explainability, oversight). | Reduced enterprise adoption; competitive disadvantage; liability for downstream discriminatory outcomes; loss of customer trust. |
| Regulators and Policymakers | Difficulty ensuring that AI systems in hiring meet fairness and transparency requirements. | Increased burden of oversight and enforcement; potential for fragmented regulations |
| Affected Individuals (Job Applicants & Employees) | Biased decisions in hiring, promotions, or performance evaluation due to unfair AutoML models. | Loss of opportunities, systemic discrimination, reduced workforce diversity, and erosion of trust in employers. |
| Business Leaders & Decision-Makers | Underestimating fairness risks when procuring or deploying AutoML solutions. | Financial penalties, stalled AI adoption, reputational crises (e.g., Amazon, Apple, HireVue cases); and customer confidence. |
| Researchers & Academic Community | Lack of access to fairness-oriented benchmarks and standardized evaluation frameworks. | Gaps between theory and practice; slower progress in responsible AI research; limited influence on industry adoption. |

—------—-----------------------------------------------------------------------------------

Business risks can be examined from two key perspectives:

1. Risks for AutoML Providers: There is a pressing need for specific regulatory guidelines and prior research concerning business risks stemming from bias and discrimination in AutoML tools. Without clear guidance, providers face reputational damage, legal liabilities, and financial penalties if their products perpetuate or amplify existing societal biases. This could also hinder market adoption and trust in their offerings. Implementing robust human-in-the-loop processes, effective feedback mechanisms, and responsible AI practices are crucial for mitigating these risks.

2. Risks for Non-Expert Enterprise Users: For organizations utilizing AutoML, the lack of transparency around fairness-aware features can lead to flawed decision-making, discriminatory outcomes, and potential regulatory non-compliance. Non-expert users, without proper understanding or controls, risk unknowingly implementing biased models that can harm their customers, employees, or operations, leading to significant financial and reputational losses. Clear User Interface and Interaction Design, coupled with well-defined Information Architecture for fairness-related insights, along with guidance on human-in-the-loop and feedback mechanisms, are vital to empower these users in upholding responsible AI practices.

This thesis aims to bridge that gap by providing practical guidance to organizations using and building AutoML. It employs a multi-theoretical lens, rooted in the Technology Acceptance Model (TAM), Innovation Diffusion Theory, and UX strategy principles, to understand how the design of fairness-aware features in AutoML tools influences their adoption by non-expert users, emphasizing the importance of User Interface, Interaction, and Experience Design, Information Architecture, human-in-the-loop processes, feedback mechanisms, and responsible AI practices.

# D. Broad Research Problem

This study aims to determine if automated machine learning tools can produce biased or discriminatory models, and what that means for companies creating or using these tools.

The research problem is explained using an illustration below:

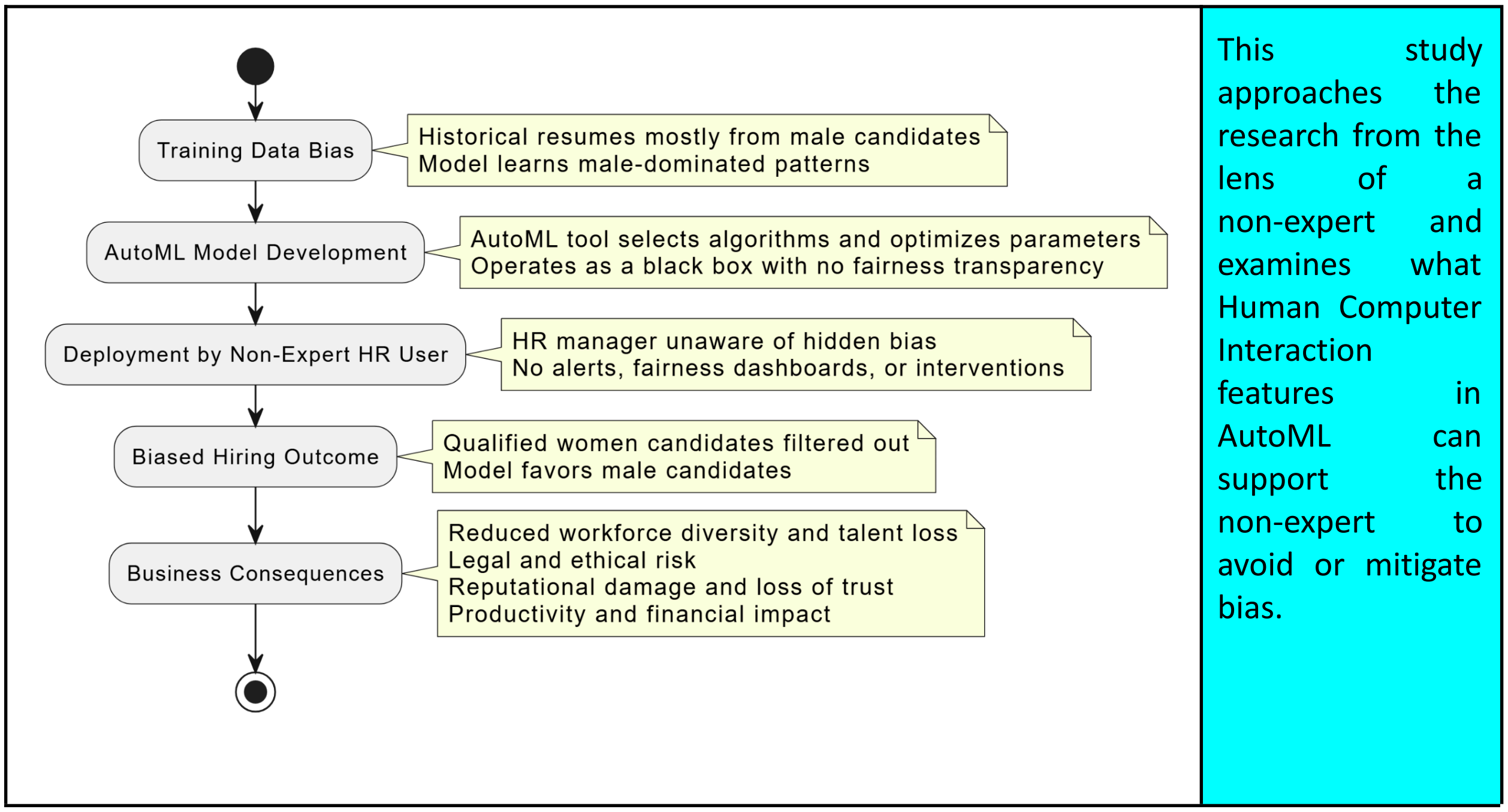

AutoML tools help to speed up and simplify the development of machine-learning systems (Frank Hutter, 2014). AutoML in recent years, has raised questions about, the interplay of data, models and human experts, given its associated risks[5]. Some of the key challenges associated with AutoML are represented below:

[5] *The Risks of AutoML and How to Avoid Them*. (n.d.). Retrieved March 5, 2023, from https://hbr.org/2019/10/the-risks-of-automl-and-how-to-avoid-them

| | | |
|---|---|---|
| Data and Bias Challenges | Datasets often contain historical or proxy biases (e.g., gender, zip code). Benchmark datasets are limited or biased. Models can also be vulnerable to adversarial attacks that reinforce discrimination. | Atabek et al., 2023; S. Biswas & Rajan, 2020; Geden & Andrews, 2021 |
| Fairness Metrics and Trade-offs | Multiple fairness metrics (e.g., demographic parity, equalized odds) conflict with each other and with accuracy. Non-experts struggle to interpret fairness scores, which are highly context-dependent. | Chouldechova, 2017; Canetti et al., 2018; Liao & Naghizadeh, 2022; Han et al., 2023; Souverain et al., 2024; Hegarty et al., 2025; Barocas et al., 2019; Mehrabi et al., 2021 |
| Transparency and Explainability Gaps | AutoML models act as black boxes. Tools provide limited explanations of why bias arises or how interventions change outcomes. Users are expected to self-diagnose fairness without sufficient system support. | Mollas et al., 2023; Sikorski, 2021; Alexander et al., 2023; Miller, 2019; Muthusamy et al., 2018 |
| Human Factors and Interaction Challenges | Users often overtrust automation (automation bias). High cognitive load arises from limited visualizations or trade-off tools. Few mechanisms exist for meaningful human-in-the-loop control over fairness processes. | Raulf et al., 2023; Chromik & Butz, 2021; Pop & Raţiu, 2024; |
| Business, Legal, and Ethical Risks | Tools risk non-compliance with EU AI Act, EEOC, and NYC Bias Audit laws. Failures can lead to reputational damage, financial penalties, and erosion of user trust, as seen in high-profile cases (Amazon, Apple, HireVue). | Yildiz & Beloff, 2020; Baumann et al., 2023; Alves et al., 2024; Grove et al., 2020; Canhoto & Clear, 2020; |

There has been prior research that compares the performance of the AutoML tools from several dimensions, including general performance assessment (Y.-W. Chen et al., 2019), performance in specific context (Halvari et al., 2021), computational efficiency (Ferreira et al., 2020), and optimization methods (Feurer et al., 2015). As seen above, while there is research about the business risks caused by technology and about performance assessment of the AutoML tools and libraries, they do not cover the implications for AutoML companies caused by unfair or discriminatory models or outcomes for downstream uses of AutoML tools.AutoML providers lack standardized fairness-oriented product design guidelines, which limits the usability, trust, and adoption of their tools in regulated domains like hiring. For companies developing AutoML, this results in business risks exposing them to civil liabilities, reputational impact and loss of value. In addition, inadequate transparency, and disclosure in enabling the downstream user to understand how the AutoML considers fairness context, may accentuate the risks.

## E. Research questions

Despite the proliferation of AutoML tools, most fail to offer robust fairness diagnostics, transparency mechanisms, or bias mitigation features; especially in high-stakes applications like hiring. This creates a disconnect between regulatory expectations and current AutoML capabilities. This research addresses this gap by evaluating how fairness features in AutoML affect usability, transparency, and enterprise trust. To that end the research proposes the following research questions from the perspective of fairness-aware product design and usability for non-expert users, particularly in the context of AutoML use in human resources:

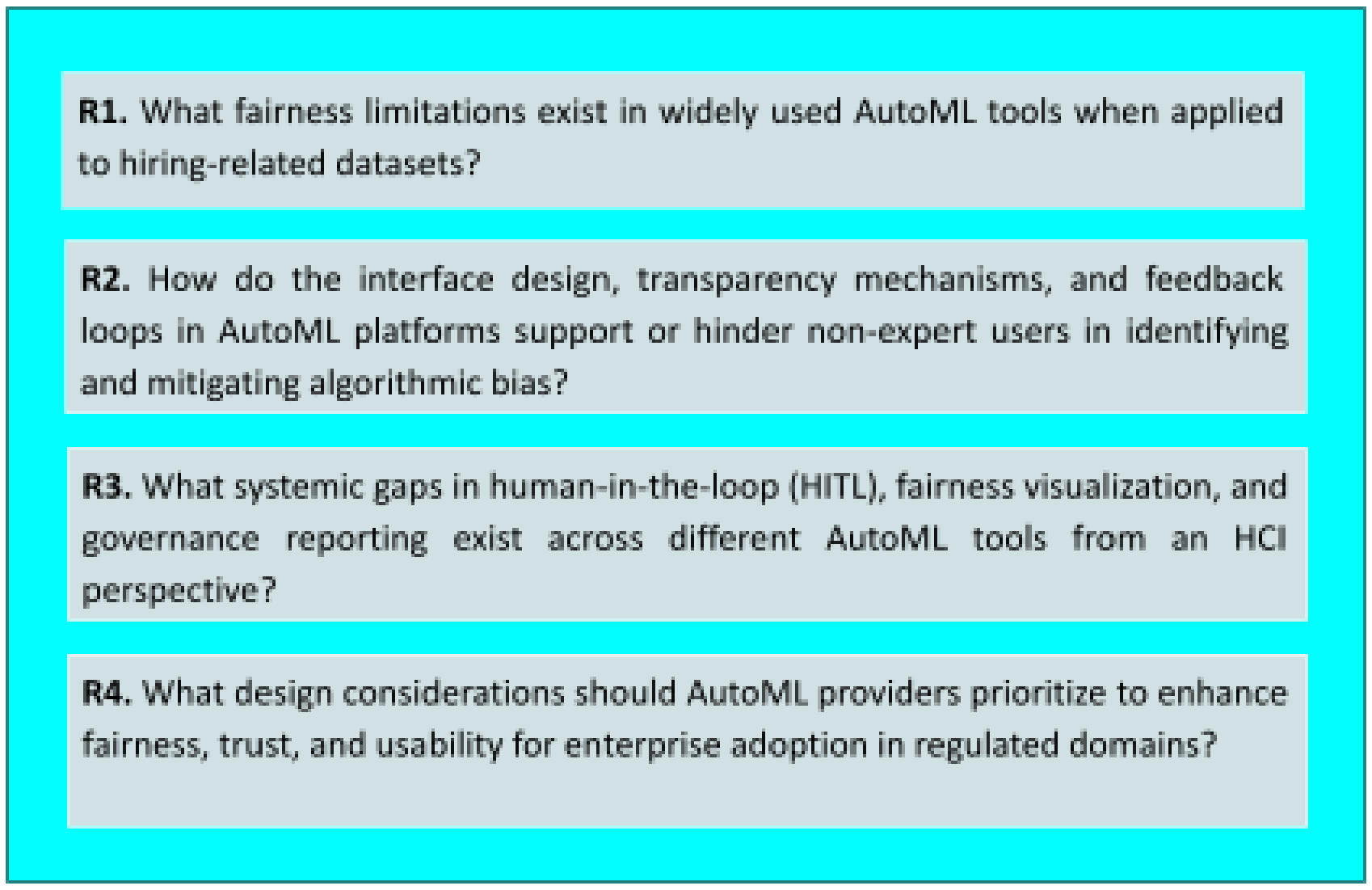

These research questions aim to explore the implications, risks, and mitigation strategies related to unfair or discriminatory algorithms and outcomes while developing downstream hiring algorithms in AutoML, specifically focusing on the product strategy and business implications for companies developing AutoML tools. A mapping of the research questions to the enterprise relevance is provided below:

| Research Question | Enterprise Relevance |
|---|---|
| **RQ1**: What fairness limitations exist in widely used AutoML tools for hiring datasets? | Helps identify product gaps that can hinder adoption in sensitive domains. |
| **RQ2**: How do interface design, transparency, and feedback loops support or hinder fairness? | Guides UX improvements for trust and usability by non-expert users. |
| **RQ3**: What systemic gaps in HITL, fairness visualization, and reporting exist? | Supports compliance readiness and internal audit processes. |
| **RQ4**: What product design considerations should AutoML providers prioritize? | Directs roadmap, differentiation, and enterprise sales readiness. |

**Scope exclusions**

The research is not intended to cover the following:

1. **Limited stakeholder scope:** The thesis primarily addresses AutoML tool providers and non-expert enterprise users, excluding broader stakeholder groups such as organizational change managers, supply chain or third party vendors or governance stakeholders.
2. **Limited implementation depth:** It does not fully explore real-world implementation challenges or enablers such as cost-benefit of fairness-aware HCI

—------—-------------------------------------------------------------------------------------

# F. Research Methodology

The research intends to use algorithm audit approach to case study research method to enable better more comprehensive understanding of the fairness and implications of AutoML tools in real-world contexts.

## F1. Detailed Research methodology

This study adopts a multi-stage, mixed-method research design anchored in a Human-Computer Interaction (HCI) and fairness evaluation perspective. It investigates how fairness affordances and limitations in AutoML tools influence usability, transparency, and adoption in regulated enterprise domains such as hiring.

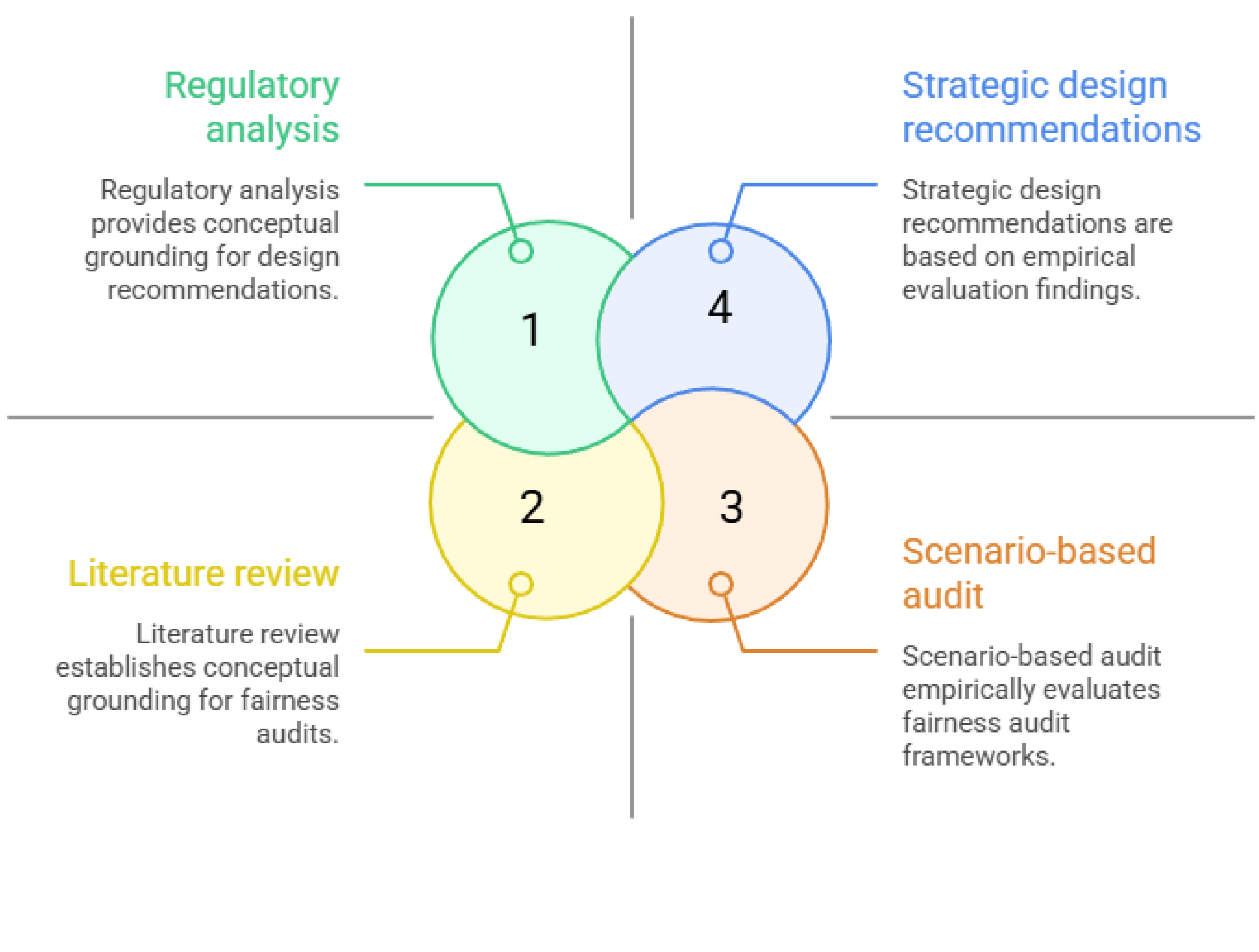

—-------—---------------------------------------------------------------------------------------

## F1.1 Methodology

The research is structured around three cumulative articles that address four research questions (RQ1–RQ4) using a combination of conceptual modeling, empirical audit, and interface evaluation.

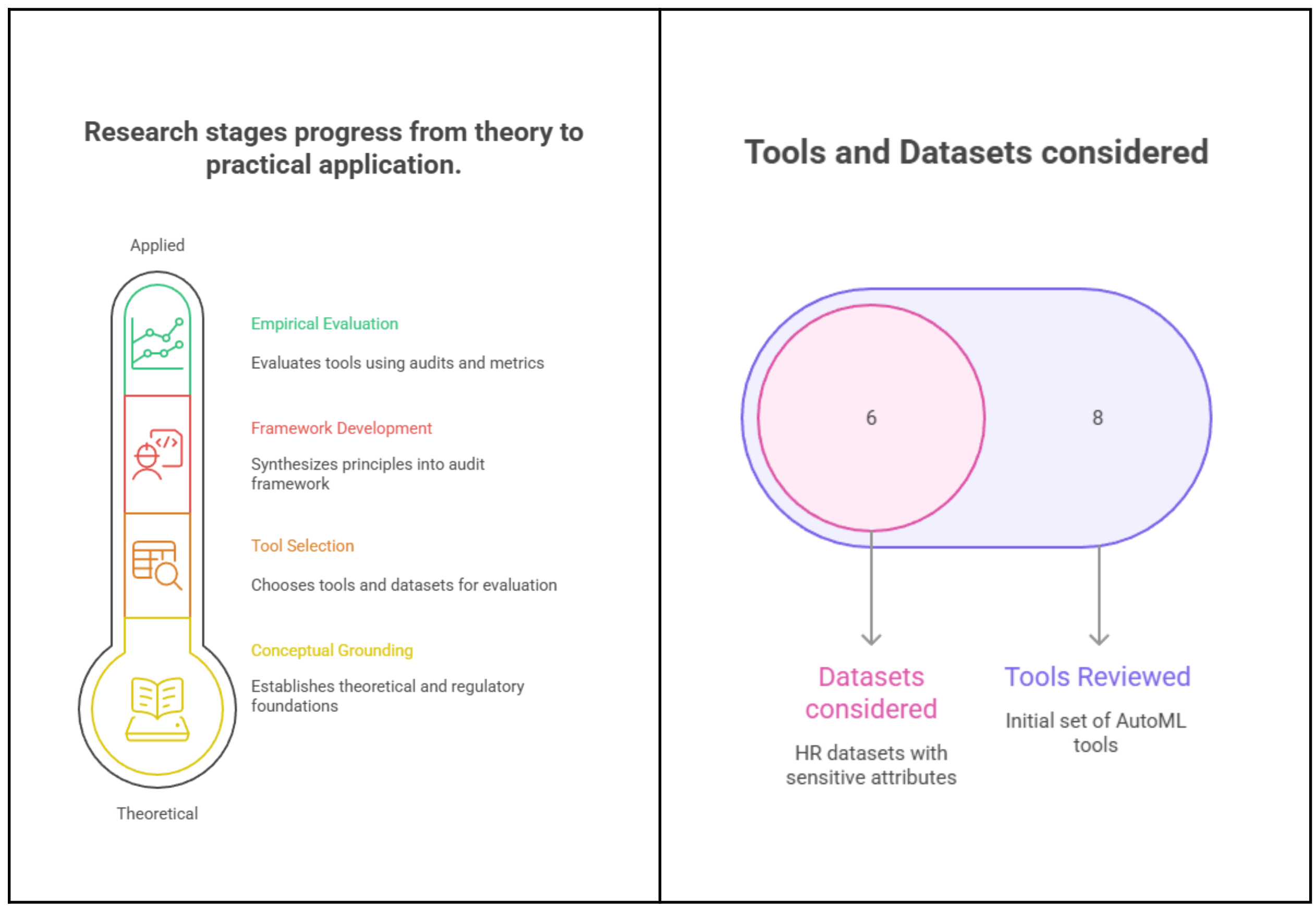

The research will be based on a schematic review of the literature covering business risk management, fairness in Artificial intelligence/Machine learning (AIML), AutoML tools, fairness and non-discrimination expectations laid out by the EU AI Act for high-risk applications, and algorithmic audits. The study will follow secondary resources like magazines, newspapers, and online articles for further reference. The study aims to include literature published between 2018 to May 2024. All papers were searched on the Scopus database. The criteria for selecting relevant papers will be based on relevance and the body of literature. A systemic literature review will also be performed using the SCOPUS database to identify all the relevant literature mappings regarding the context of the paper.

The methodology is organized into four interlinked stages:

| Stage | Purpose | Method |
|---|---|---|
| 1 | Conceptual and regulatory grounding | Literature review (Fairness in hiring, fairness in AutoML, HCI literature, UX theory, TAM/IDT in the context of AutoML)<br><br>Regulatory analysis and compilation of key expectations from EU AI act (for hiring as a high risk AI system), EEOC guidelines and NYC bias audit. |
| 2 | Tool & dataset selection | Popularity and criteria-based tools and library filtering from the broader list collated through secondary research;<br><br>Relevance and domain based identification of HR datasets with sensitive attributes |
| 3 | Developing of audit framework and fairness metrics | Synthesis of HCI principles, fairness laws, and AutoML interface elements in line with the literature review and regulatory expectations from Stage 1 |
| 4 | Empirical evaluation via scenario-based audit | Fairness metric computation; heuristic interface audit; walkthroughs |

### F1.2 Articles plan

The three articles that addresses the four research questions are presented as follows:

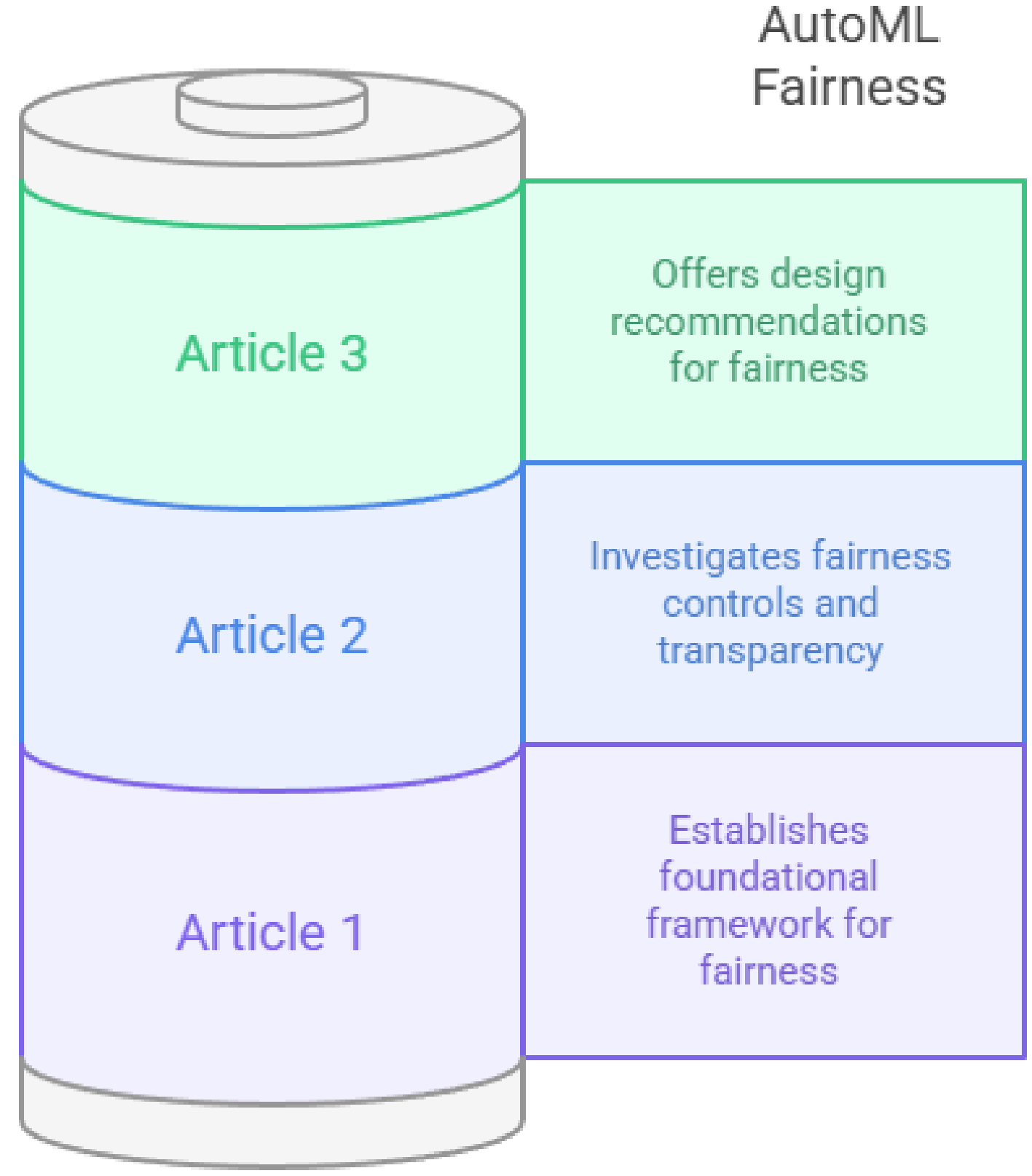

Detailed description of the same is provided in the table below:

| Article | Research questions & Objective | Method |
|---|---|---|
| 1 | RQ1, RQ4<br><br>Establish a foundational framework to evaluate fairness-related capabilities of AutoML tools used in hiring. | Conduct a **systematic literature review** (2018–2024) covering AI fairness, AutoML affordances, algorithmic bias in hiring, and legal frameworks (EU AI Act, EEOC).<br><br>Compile relevant insights from TAM, IDT and UX strategies for AutoML context covering the following:<br><br>● Perceived ease of use/usefulness and its link to fairness adoption.<br>● Relevance of innovation features like transparency and auditability influence diffusion<br>● Design principles for user-centered fairness controls (nudges, feedback loops, alerts). |
| 2 | RQ2, RQ3<br><br>Evaluate how AutoML providers disclose fairness controls, transparency, and usage guidance.<br><br>Assess tool performance and interface support. | ● Review documentation, screen UI, onboarding flows, and usage agreements across eight tools.<br>● Identify customization and override capabilities; disclosures on bias, limitations, and data handling; and presence of alerts, nudges, or interpretability features<br>● Run scenario-based audits on six HR datasets with sensitive attributes.<br>● Measure group fairness metrics (TPR, FPR, DPD, PRP, etc.).<br>● Benchmarking and analysis of the outcomes from the evaluation. |
| 3 | RQ4<br><br>Compile design recommendation | ● Compile the results of the evaluation and identify fairness-centric design considerations for AutoML |

The first article is based on a literature review regarding fairness, AutoML tools, and fairness and non-discrimination expectations laid out by the EU AI Act, EEOC guidelines and NYC Bias audit requirements for high-risk applications. The key learnings from the literature review is compiled to scope the future research direction. This article formulates foundational considerations for evaluating AutoML applications for fairness.

The second article has two parts. The first part of the article is based on a mock audit of information shared with non-expert users by identified AutoML tools. The article focuses on information shared by the AutoML companies with the downstream users in form of contractual terms, terms of use, on screen disclosures, warning, notification, alerts and guidance or instructions of use, to evaluate their adequacy in the context of fairness and thereby the business risk implications. The review covers aspects including ability of users to choose or remove features, change hyperparameters, and assessing comparative impact analysis of these changes on fairness, transparent in disclosing the dataset, preprocessing, and modeling steps, and provides warning labels or indicators to alert users of potential biases or discriminatory practices. It also examines whether AutoML explains how outcomes are arrived at, discloses measures taken to avoid overfitting, notifies users of unintended consequences, and offers guidance on the uses of the model, including limitations. Additionally, the review examines whether AutoML discloses the ethical considerations taken into account when developing a model and seeks confirmation from users regarding the existence of personal data, inferences, proxies, or causal relationships within the data. For the audit, the tools will be installed as a downstream user and appropriate subscription / contractual engagements will be available for commercial tools to gather the information mentioned above.

The second part of the article is based on scenario-driven audit of the AutoML tools. For the said purpose, the AutoML tools will be run for different scenario-driven use cases (curated datasets) and compared for specific fairness metrics. The scenario-driven use cases and associated relevant fairness metrics are identified based on literature review of research focused on fairness evaluations. The article summarizes the learnings into three main themes namely, (a) methods for detecting and addressing bias in training data, (b) developing fairness-aware modelling techniques, and (c) implementing other measures to ensure that AutoML tools produce fair and unbiased models.

The third article focuses on business risk implications for AutoML tool companies based on the findings from the above process including recommendations to address the business risks. The article is structured into 2 parts, one that systematically enumerates the business risk associated with the fairness failures and the other that guides with recommendations. The article expresses the recommendations to address the risks including monitoring, bias mitigation, fairness disclosure, and more from a HCI lens. While all these capabilities can reduce risk somehow, most companies need help to precisely determine the extent of such efforts required (Mckinsey, 2019). The article will further highlight that while safeguarding the organization against an ever-increasing number of AI risks may seem overwhelming, it is possible to mitigate them proactively[6] It would extend to cover the broad plan for potential mitigation. By prioritizing those likely to cause harm, organizations can help prevent or limit such events from ever taking place. This can in turn help in compliance.

### F1.3 Research Question Alignment

| Research Question | Addressed In | Key Outputs |
|---|---|---|
| RQ1 – Fairness limitations | Article 1, 2 | Audit framework, comparative feature map |
| RQ2 – Interface design, transparency, feedback | Article 2 | Walkthrough logs, design gap analysis |
| RQ3 – HITL, visualization, and governance gaps | Article 2 | UX audit summaries, cognitive flow diagrams |
| RQ4 – Product design considerations | Article 3 | Strategic design recommendations for fairness-centric tools |

## F2. The novelty of the work

Research on the implications of bias and discrimination caused by AI automation tools in downstream applications. Although there has been significant research into the development and usage of AutoML, most of it has focused on practical and technical aspects, such as improving performance, efficiency and scaling. AutoML's ethical and social implications have been understudied, especially in terms of regulatory compliance and fairness. This research

[6] *Identifying and managing your biggest AI risks | McKinsey*. (n.d.). Retrieved February 17, 2023, from https://www.mckinsey.com/capabilities/quantumblack/our-insights/getting-to-know-and-manage-your-biggest-ai-risks

——------——-------------------------------------------------------------------------------

topic is timely and relevant in light of increasing usage of AutoML across various industries and the need for ethical and responsible development and use.

## F3. Implications of the Research

The research on implications of bias and discrimination caused by AI automation tools in downstream applications has significant implications for researchers, autoML developers, and regulators. Here are some potential implications for each group:

1. Researchers: Researchers should consider the potential for bias and discrimination when designing AI automation tools and conducting research on AI. They could examine how AI can impact different groups and explore ways to ensure that AI systems are fair and equitable using this research.
2. AutoML developers: They could leverage methods proposed by this research to detect and mitigate bias and discrimination in AutoML systems. Furthermore, they should also ensure that their tools are accessible and easy to use for non-experts. Moreover, they could use the risk register to identify the potential for bias and discrimination and consider ways to prevent it from being built into their systems.

# Chapter 2: Understanding fairness while using AutoML in human resources context

## A. Fairness in the context of human resources

**Section Overview**

**Methodology:** The section illustrates the systematic literature search using keywords like "bias," "fairness," and "hiring," screening 140 papers to retain 90 relevant studies for in-depth synthesis. The extracted insights were organized into themes addressing research objectives.

**AI & ML in HR:** The section covers literature reviewed to identify applications of AI and machine learning in recruitment, including resume screening, interview automation, candidate assessment, personality testing, gamification, and predictive hiring analytics to improve efficiency, reduce bias, and enhance decision-making.

**Ethical considerations:** The section covers ethical issues such as bias amplification, potential violations of human rights, algorithmic accountability gaps, and the need for fairness, privacy, and explainability in AI-driven recruitment.

**Bias in algorithmic hiring:** The section Identifies major bias categories (gender, race, disability), risks from historical data, and the role of human oversight. It also emphasized the importance of equitable frameworks, governance, and ongoing training for fair AI deployment.

**Approaches to mitigate bias:** The section summarizes strategies (based on literature review) including data preprocessing, fairness-aware modeling, counterfactual analysis, adversarial debiasing, human-in-the-loop systems, explainability, transparency, and continuous monitoring for ethical and fair decision-making.

**Business implications:** The study introduces consequences of bias in hiring AI systems, including inefficiency, legal and ethical risks, reputational damage, privacy and security concerns, and financial penalties, highlighting the importance of fairness for organizational success.

This section reviews literature on AI and machine learning in HR, focusing on fairness, bias, and ethical considerations. It examines applications of AI in recruitment, candidate assessment, and talent management, highlighting associated risks and challenges. The section also explores approaches to mitigate bias, promote transparency, and ensure ethical and equitable outcomes in AI-driven hiring processes.

### A1. Methodology

To investigate the topic of bias and fairness in hiring, a comprehensive literature search was conducted using the Scopus database. The search was performed using the keywords "bias," "fairness," and "hiring." A total of 140 papers were initially identified through this search. To ensure the relevance of the papers to the research topic, each paper was carefully reviewed. Out of the 140 papers, 50 were found to be unrelated to hiring and were excluded from further analysis. The remaining 90 papers were considered for in-depth analysis and synthesis.

The selected papers covered a range of topics related to bias and fairness in hiring. These topics included a brief overview of machine learning in human resources, the use of new techniques for machine learning in HR, ethical implications of machine learning in hiring, issues relating to bias in hiring processes, failed definitions of fairness, discrimination in job advertisements, metrics for dealing with bias, approaches to addressing bias in hiring, toolkits for bias management, datasets for algorithmic fairness, and the role of explainability as a solution to bias and fairness concerns.

**Literature Review Process for Bias in Hiring**

Initial Paper Identification

Relevance Screening
140 papers were identified

In-Depth Analysis
90 papers were analyzed

Theme Synthesis

Research Insights

The included papers were reviewed, and relevant information and findings were extracted and synthesized to form the basis of the research analysis. The extracted data and insights from the papers were then analyzed and organized into meaningful themes to address the research objectives.

# A2. Literature Review

## A2.1 Artificial intelligence and machine learning in HR

HR professionals are concentrating on maximizing the integration of human and automated tasks to establish a smooth and user-friendly work setting. This approach provides additional time for creativity, intelligence, and empathy to improve the experience for candidates and employees. There is a strong connection between talent and value creation, as a considerable part of market value is associated with intangible human capital (Achchab & Temsamani, 2022). Utilizing AI in HRM boosts efficiency in recruiting and selection, diminish biases like nepotism and favoritism, and promote staff development, retention, and usage. Utilizing AI techniques can also have a beneficial effect on HRM practices in developing countries (Kshetri, 2020). Machine learning may revolutionize HRM through data-driven decision-making, personalizing employee experience, and optimizing procedures. Real-world case studies showcase benefits like objectivity, customization, cost savings, and data-driven decision-making (Saxena et al., 2023). Research shows that algorithm-based data screening helps decrease cognitive bias and enhance candidate-organization compatibility (Peisl & Edlmann, 2020)

Research publications recommend using machine learning and natural language processing to improve the recruitment process for companies. Key themes of using artificial intelligence in recruitment include:

- Using natural language processing (NLP) in resume screening and candidate profiling offers a data-driven approach that saves time, eliminates human biases, and improves recruiting decisions. NLP approaches are used to assess and rank candidate resumes, streamlining the process for recruiters to identify the best prospects for a job (Alamelu et al., 2021).
- Interview bots are used to conduct multiple interviews simultaneously, capturing facial expressions and speech patterns of the candidates. These systems use unique scoring algorithms to evaluate candidate responses, resulting in increased efficiency and cost savings when recruiting top-tier employees (Caldera et al., 2023).
- Furthermore, additional research has explored the utilization of artificial intelligence in conducting interviews, evaluating candidate responses, and analyzing emotional cues. Systems utilize natural language processing, deep learning, facial expression identification, and speech recognition to automate interviews and assess candidates' abilities, experience, and body language. These systems aim to reduce bias and enhance the accuracy and efficiency of candidate selection (Ghadekar et al., 2023).
- Automated methods are recommended to streamline application selection by evaluating eligibility criteria and personality traits. The systems utilize online personality quizzes, evaluate resumes using machine learning algorithms, and generate predictions based on Twitter data to determine candidates' personality attributes. The models exhibit high accuracy in efficiently and quickly evaluating candidate personality attributes (Sudha et al., 2021) and (Parameswara et al., 2023). Personality traits can be successfully assessed from interview responses using natural language processing and machine learning approaches. Regression models analyzing interview material demonstrate a significant correlation between trait descriptors produced by the model and human ratings. Implementing such measures can enhance the decision-making process for recruiters and hiring managers (Jayaratne & Jayatilleke, 2020).

- Researchers used gamification and image-based assessments in talent acquisition evaluations to increase engagement, reduce testing time, and address issues linked to applicant fatigue and poor data quality. Image-based assessments of personality traits have great validity and psychometric characteristics, demonstrating high agreement with traditional questionnaire-based evaluations (Hilliard et al., 2022).
- Researchers have explored employing automated techniques to assess a candidate's social media presence to evaluate their employability score and emotional intelligence. Social media profiles, APIs, web crawlers, machine learning, and natural language processing tools can speed up the recruitment process, improving accuracy, efficiency, and minimizing bias. This offers an in-depth evaluation of potential candidates that surpasses traditional resume screening methods (Pendyala et al., 2022).

A broader categories of use of AI and ML in hiring are illustrated below:

| Category | Description | References |
|---|---|---|
| Targeted job advertisement or candidate search | AI can be used to analyze job requirements and candidate profiles to create targeted job advertisements or conduct candidate searches. By leveraging AI algorithms, businesses can match job postings with the most relevant candidates based on skills, experience, location, or other criteria, increasing the chances of finding qualified candidates. | (Greif & Grosz, 2023) |
| Resume scraping and scoring | AI can automate the process of extracting information from resumes or CVs through techniques like natural language processing. AI algorithms can then analyze and score the extracted data based on predetermined criteria, such as education, work experience, or skills. This automated process saves time and enables recruiters to efficiently screen a large volume of resumes. | (Alamelu et al., 2021) |
| Predictive hiring analysis | AI can utilize historical data and machine learning algorithms to predict the likelihood of a candidate's success in a particular role. By analyzing factors like past job performance, education, skills, and other relevant data points, predictive hiring analysis can provide insights into a candidate's potential performance, retention rates, or cultural fit, aiding decision-making during the hiring process. | (Kinger et al., 2024) |
| Applicant screening chatbot | An AI-powered applicant screening chatbot can engage and interact with job applicants through chat interfaces. These chatbots can ask pre-determined questions, assess candidates' qualifications, and gather relevant information. The chatbot's AI capabilities enable it to provide personalized responses, assess candidate fit, and streamline the initial screening process, reducing manual effort for recruiters. | |
| Video interview | AI can enhance video interviews by providing automated features such as facial expression analysis, tone detection, and sentiment analysis. These AI-driven tools can help evaluate non-verbal cues, communication skills, and other behavioral aspects of candidates during the video interview process. This technology enables recruiters to make more informed assessments and comparisons between candidates. | (Booth et al., 2021)<br><br>(Escalante et al., 2017) |
| Pre-employment job function test | AI can be employed to create and administer pre-employment job function tests. These tests can be tailored to simulate real-world scenarios or assess specific skills required for a particular role. AI algorithms can automatically score and evaluate the test results, providing objective and standardized assessments of candidates' abilities in performing job-related tasks. | |
| Personality tests | AI can assist in administering and analyzing personality tests to evaluate a candidate's personality traits, work preferences, and behavioral tendencies. By leveraging AI algorithms, these tests can provide insights into a candidate's compatibility with a role or team dynamics. AI can help identify patterns and correlations between personality traits and job success, aiding in the selection process. | (Sudha et al., 2021) |

## A2.2 Ethical issues in Algorithmic hiring

Using algorithms in hiring may raise ethical considerations, specifically around bias amplification. Biases found in candidate ranking software and chatbot interactions emphasize the importance of addressing fairness in AI-driven recruitment for equitable treatment of candidates (Mujtaba & Mahapatra, 2024). Hiring algorithms may inadvertently violate individual human rights such as the rights to employment, equality, non-discrimination, privacy, freedom of expression, and freedom of association. This could create an algorithmic accountability gap, leading to discrimination and unequal access to employment prospects. (Yam & Skorburg, 2021). Algorithms, depending on their design and development, have the potential to compromise privacy, introduce biases, lead to unjust discrimination in job decisions, and result in harm owing to inaccurate data produced by Artificial intelligence and Machine learning (Lukaszewski & Stone, 2024). Some algorithms are designed to predict competency scores that candidates would receive from human reviewers, potentially inheriting biases from recruiters (Fabris et al., 2023). Algorithmic bias has the capacity to magnify and sustain societal bias, leading to significant ethical concerns for society. Gender bias in employment advertising and recruitment tools is mostly caused using language processing and recommendation algorithms (Leavy et al., 2020)

## A2.3 Bias in Algorithmic Hiring

Gender bias in AI recruitment tools (Nadeem et al., 2021) and (Alexander, 2022), and biases against marginalized groups including people from a certain race or people with disability (Tilmes, 2022) are considered as two major categories of bias contributed by recruitment tools.

Business process management including algorithmic decision aids can automate procedures but may also magnify algorithmic biases. Cooperation between humans and decision aids is successful, although complacency and authority bias might reduce choice quality if the recommender is incorrect. It is essential to assess and analyze the performance and prejudices in augmented business decision systems (Baudel et al., 2021). Machine learning models can exhibit bias because they may reflect human biases present in the training data, leading to possible discrimination. Identifying and assessing bias in models is essential for developing more transparent and equitable models (Alelyani, 2021). Biases, especially those that promote perceived masculine norms, must be addressed to achieve fairness (Cheng et al., 2023).

“Recruitment and HR recommendation systems are built on data that is generally historical data of successful applicants. Of course, historically speaking you’re talking about CVs of mostly men, especially for the more technologically-oriented profiles…… A historical data set of successful applicants will essentially be a male-dominated data set…..To give women equal opportunity in these areas, the research needs to look at how to de-bias the systems".

Dr Marielza Oliveira, a director in communications and information at UNESCO (News)

Moreover, employing supervised machine learning algorithms in the recruitment process prompts legal and ethical concerns because of the possible discriminatory outcomes. It is essential to differentiate purposeful from non-intentional discrimination when examining the legal consequences of prejudice in algorithmic decision-making technologies (Páez, 2021). Organizations require equitable and ethical internal frameworks, business strategy, and governance processes, as well as ongoing training on fairness and ethics. Workplace diversity, algorithmic transparency, and accountability are crucial for combating gender prejudice in recruitment (Nadeem et al., 2021). Hence, the implementation of AI in human resources requires the creation of algorithms that are precise, unbiased, and just. It is crucial to include ethics, such as fairness and explainability, in AI-driven recruitment to reduce biases and guarantee fairness in decision-making (Mujtaba & Mahapatra, 2019)

### A2.4 Approaches to handle bias

Fairness, trust, and justice are key in developing and using AI for recruitment. Ensuring system fairness promotes diversity, prevents discrimination, and improves organizational outcomes (Mujtaba & Mahapatra, 2024). Research suggests using different methods to detect and assess bias in recruitment algorithms, improve models with synthetic datasets, incremental learning, and causal models, ensure fair decision-making and bias control, investigate different treatment of job seekers using counterfactuals, apply adversarial de-biasing techniques, utilize fairness-aware models, and stress the significance of user engagement, transparency, and explainability to reduce biases.

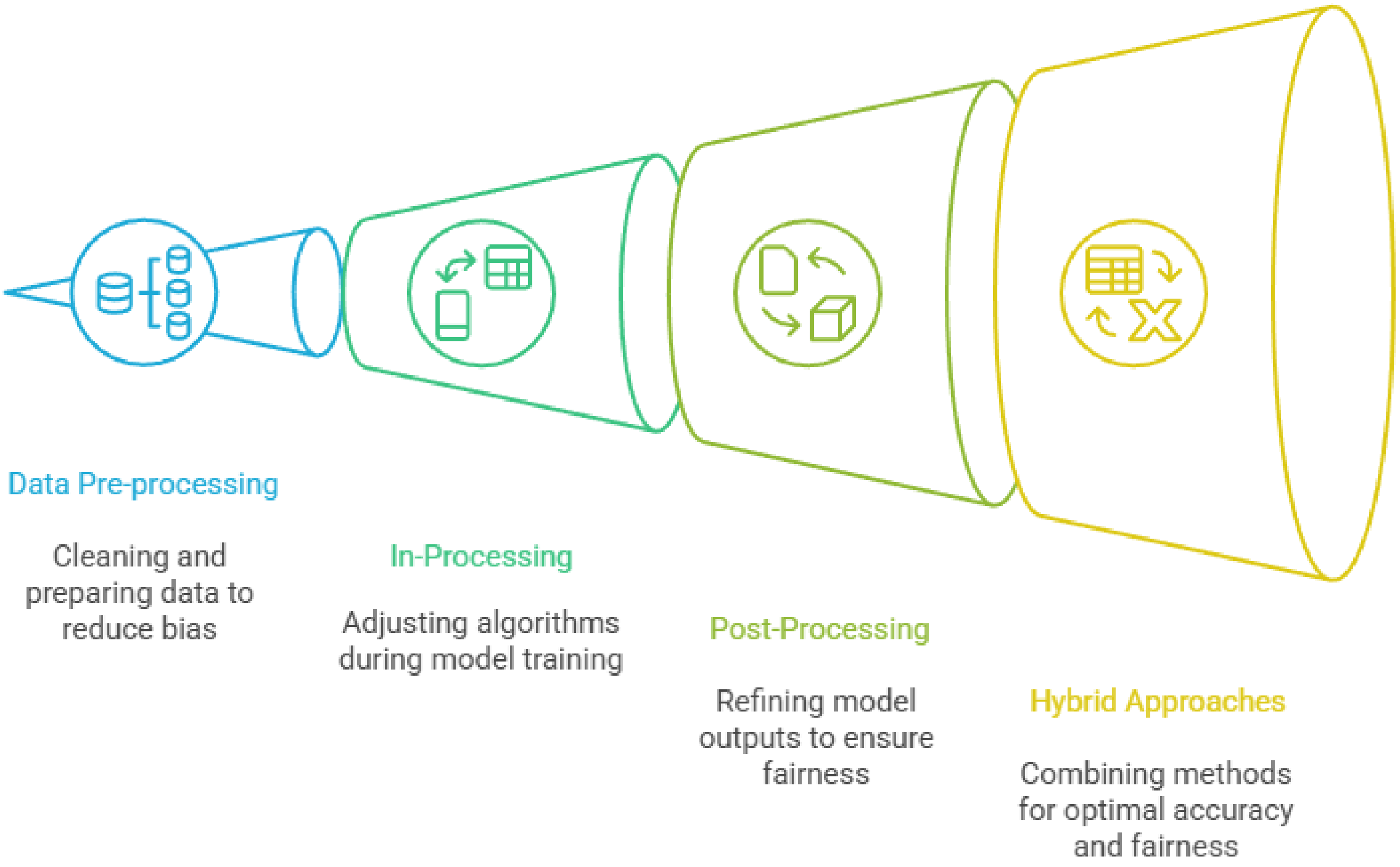

Some of such approaches are highlighted below:

**A2.4.1 Data management and modelling**

Methods such as pre-processing, in-processing, post-processing, and hybrid approaches have been developed to mitigate bias and address the fairness-accuracy trade-off in machine learning models. Empirical results indicate that pre-processing and post-processing methods efficiently mitigate bias, while the hybrid approach achieves the best accuracy (G. Alves et al., 2023).
In specific, methods using synthetic datasets, incremental learning, evolutionary algorithms, and causal models are suggested to adjust variable associations, mitigate bias, and attain fair and impartial decision-making (Barbierato et al., 2023). Another study examined the influence of BERT's architecture and tokenization protocol on the representation of minority-related data, while certain other research explored AI methods to mitigate algorithmic prejudice in job recruitment. Several methods, such adjusting vector space and utilizing data augmentation, are suggested to reduce prejudice and bias and enhance justice and inclusivity in the recruiting process (Albaroudi et al., 2024; Ramezanzadehmoghadam et al., 2021). Another research explored the need for human-readable explanations of machine learning systems and their application in fair recruitment.

Techniques such as learning from interpretation transition (LFIT) within inductive logic programming (ILP) show promise in automatically learning declarative theories to provide accurate explanations for black-box systems. This approach aims to address biases and ensure fairness in the recruitment process (Ortega et al. 2021). One of the research explored fairness by removing correlations between data and protected variables and adjusting the strength of debiasing; popular models such as linear regression, random forest, and multilayer perceptrons can achieve both improved fairness and accurate predictions. The research further proposed a geometric method that focuses on debiasing data has been shown to be a simple and effective approach to improving fairness in predictions (He et al., 2020). Further research needs to evolve to understand how causal-based fairness notions can effectively reconcile different fairness requirements (G. Alves et al., 2023). Another research expressed the need for provably fair solutions in machine learning, particularly in real-world applications, specifically algorithms that can provide fair outcomes with reliable guarantees (Lohaus et al., 2020).

**A2.4.2 Differential treatment or counterfactual approaches**

Examining differential treatment of job seekers based on gender, conceptions of justice, and adversarial de-biasing approaches are essential (Bied et al., 2023; Goretzko & Israel, 2022). Another research suggested counterfactual representations are used to improve diversity in ranking while ensuring compliance with data protection laws. The pre-processing technique trains a model on fairer representations without accessing sensitive attributes, allowing for fair ranking in recruitment (Rus et al., 2023).

**A2.4.3 User engagement and transparency**

It is crucial to have fairness and transparency, which includes providing the candidate with an explanation of the reasoning, in machine learning-based candidate selection (Bied et al., 2023; Goretzko & Israel, 2022). Emphasizing knowledge exchange, collaboration between HR managers and AI, and implementing fairness control mechanisms are crucial to overcome biases and improve fairness (Soleimani et al., 2022). Integrating computer-human interaction (CHI) into black-box models helps eliminate bias and increase diversity in hiring by adding a human element to the evaluation and functioning of AI systems (Gilbert, 2021).

**A2.4.4 Explainability**

Model explainability is crucial for establishing confidence and acceptance of AI systems in critical sectors including healthcare, automated transportation, and industrial applications. Researchers and practitioners are emphasizing explainable AI to enhance comprehension and confidence in models on a large scale (Gade et al., 2020) and (Gade et al., 2019). Trustworthy explanations are not just essential for ensuring fairness, they are also core for reliability of machine learning technologies (Chakraborty et al., 2020).

—-------—-----------------------------------------------------------------------------------

### A2.4.5 Effective Human-in-the-loop

To combat age bias in AI hiring, a combination of Human-in-the-Loop systems and AI Fairness Toolkits is emphasized as crucial for creating fairer job search processes. Age bias can inadvertently be perpetuated by AI systems, making it important to employ HITL systems and AI fairness toolkits to mitigate such bias. HITL systems involve human reviewers in the decision-making process, enhancing fairness through human oversight and expertise. AI fairness toolkits proactively identify and rectify bias within AI systems, promoting fairness and equity. By combining HITL systems and AI fairness toolkits, a robust strategy can be implemented to reduce biases in AI hiring. Continuous monitoring and evaluation of AI systems are essential to proactively detect and address biases (Harris, 2024).

### A2.4.6 Automated detection and evaluation of bias

Methods like wrapper techniques and publicly accessible datasets are employed to detect and evaluate bias in algorithms (Alelyani, 2021). Another research suggested an utility called 'BiasTrap' which is a runtime verification tool that combines data augmentation and bias detection components to detect and prevent discrimination in machine learning systems. It enables real-time monitoring and ensures fairness in ML models trained on different datasets using various algorithms (Mamman et al., 2024). Furthermore, research also suggested search-based fairness testing (SBFT) approaches to evaluate the fairness of regression-based machine learning systems. The approach significantly outperforms existing methods, providing more effective and efficient fairness testing for regression-based ML systems, particularly in the context of emergency department wait-time prediction (Perera et al., 2022).

### A2.4.7 Fairness aware models

Researchers have suggested fair algorithms that stress fairness, diversity, and transparency in the recruitment process (Delecraz et al., 2022). Emerging techniques like distance correlation minimization, counterfactual and observational measures, and multi-objective optimization are being used to develop fair machine learning algorithms, fairness-aware clustering, fairness in data streams, and fairness in federated learning. These techniques aim to achieve fair representations, fair labeled clustering, and fair aggregation (Esmaeili et al., 2022) , (Ezzeldin et al., 2023) and (Dutta et al., 2021). Another research suggested, a kernel density estimation (KDE) methodology is developed to ensure fairness in machine learning classifiers by quantifying fairness measures based on KDE as differentiable functions. This approach allows for a high accuracy-fairness trade-off and optimization using gradient descent (Cho et al., 2020). In addition, Selection Parity is another fairness notion for selection tasks with a pre-defined budget.

The Fair Selection with the Differentiable Distribution Difference (FS-DD) framework integrates a differentiable constraint into the training process to ensure fair decisions in selection problems. Such frameworks guarantee fairness and outperforms existing methods in terms of fairness-accuracy trade-off leveraging KDE methodology (Jiang et al., 2023). One other research suggested, "An Assure AI" (AAAI) Bot which utilizes deep Neural Networks (NN) and the Detection Transformer (DETR) framework to monitor and detect internal biases within deep learning models. AI bots are trained to detect biases and errors in AI software applications, particularly in systems that sort and classify people based on various attributes (Tellez et al., 2022)

### A2.5 Challenges associated with detecting and addressing bias

Detecting bias in a machine learning dataset is a complex task that involves understanding and quantifying different types of biases, such as sampling bias and label bias (Hinnefeld et al., 2018). In some instances inappropriate use of proxy variables to sensitive variables could lead to more bias in the model (Datta et al., 2017). The prevailing challenges in the algorithmic hiring process, such as imprecise data extraction and bias, necessitate better human-centered recruitment strategies. However, implementing an AI system that supports human-centred recruitment strategies, while eliminating bias and improving efficiency has its challenges. Primary among them, is the ability to detect and eliminate biases in models. Detecting and eliminating bias in models employed by firms is challenging, particularly biases related to attributes such as gender and address (D’Souza et al., 2023).

Challenges in Algorithmic Fairness

Fairness Control

Managing optimization strategies to prevent bias

Metrics and Measures

Evaluating fairness using various metrics and their conflicts

Tools and Exposures

Using tools to identify bias and addressing vulnerabilities

Intersectional Fairness

Ensuring fairness across overlapping groups

Real-World Concepts

Integrating real-world fairness experiences into ML

Fairness Notions

Different concepts of fairness and their trade-offs

Some of the key contributors to the challenge are provided below:

### A2.5.1 Metrics and measures

Furthermore, fairness and mitigation approaches are assessed using real-world machine learning models. Various fairness measurements and mitigation approaches are utilized to evaluate fairness and their influence on performance. Considering a wide range of fairness metrics and comprehending their agreements and conflicts is crucial for establishing reliable and trustworthy fairness assessments (Atabek et al., 2023). Many studies view the use of group fairness as a suitable metric where outcomes are representative of each group individually. Nevertheless, social norm bias, a type of algorithmic discrimination, can still exist in algorithms created for group fairness.

Ensuring fairness and objectivity in algorithmic decision-making is essential due to the ethical considerations involved (Cheng et al., 2023). Various concepts of fairness, such as statistical parity and equal opportunity, have been suggested to counteract bias. However, conflicts arise between these concepts and other important factors like privacy and classification accuracy (G. Alves et al., 2023). Challenges also persist in selecting appropriate fairness indicators and integrating fairness into machine learning and artificial intelligence models for specific applications (Gilbert, 2021). Another study highlighted the importance of selecting appropriate fairness metrics to effectively detect bias in models trained on potentially biased datasets. It also demands that practitioners are aware of the limitations of different fairness metrics and consider the context in which they are applied to ensure accurate assessments of bias (Hinnefeld et al., 2018).

#### A2.5.2 Intersectional and overlapping fairness

The concept of intersectional fairness, wherein fairness is examined across multiple overlapping groups, plays a crucial role in ensuring fairness in algorithmic decision-making. Recognizing that intersectional fairness implies overlapping group fairness under specific and evolving conditions highlights the complexity of fairness considerations. However, the idea of generalizing fairness that enforces static overlapping fairness does not always guarantee intersectional fairness (Yang et al., 2020).

#### A2.5.3 Fairness notion

Different notions of fairness, such as statistical parity and equal opportunity, have been defined to address fairness in machine learning models. However, these notions can conflict with other important properties like privacy and classification accuracy, leading to a fairness-accuracy trade-off that requires careful management (G. Alves et al., 2023). Additional notions were proposed in recent years including proportional equality and bias parity to handle bias in AI systems. Proportional Equality (PE) is introduced as a formalized notion of fairness for evaluating prediction models' outcomes for discriminatory traits. PE is suggested as a more suitable criterion for assessing fairness compared to the commonly used Disparate Impact (DI) notion (A. Biswas & Mukherjee, 2019). Bias Parity (BP) score is introduced as a fairness measure, and a feature-rich representation derived from temporal aspects is used to select models with lower bias. The approach is demonstrated to improve accuracy in predicting recidivism while achieving higher fairness scores (Jain et al., 2020). However, these increased definitions and complexity impacts approaches to unify fairness considerations. Further, there is a disparity in the comprehension of definitions between researchers and the general public. Efforts are being made to address this gap by assessing comprehension and exploring relevant aspects (Saha et al., 2020). However, the issue remains unresolved at present.

#### A2.5.4 Fairness concepts in real world and machine learning

Fairness concepts in hiring have a long history and have been evaluated for many years. Real-world experiences in dealing with fairness and its evaluation are sometimes disregarded in the realm of machine learning, highlighting the significance of comparing past and present concepts to steer future study (Hutchinson & Mitchell, 2019). Moreover, conversations around algorithmic bias frequently exclude considerations of disability and accessibility, underscoring the deficiencies of existing de-biasing methods (Tilmes, 2022). While fair machine learning methods can help address disparities, they may fall short in ensuring accessibility and inclusivity for disabled individuals (Tilmes, 2022). Given that marginalized groups are not constant across domains or dataset, there is a need to have a fairness approach that is tailored to the specific domain and use case environment considering the specific need for fairness.

#### A2.5.5 Tools & emerging adversarial exposures

Toolkits like Fairness360 can assist in identifying bias in both the training data and model results (Bellamy et al., 2019). These methods currently lack the ability to incorporate the context of the use case to effectively evaluate bias. The limitations of present differentially private machine learning methods and the susceptibility of transformer-based NLP models to backdoor attacks introducing bias are also concerns that need to be resolved (Atabek et al., 2023).

#### A2.5.6 Fairness control

It is crucial to recognize that optimization strategies might result in bias in models. In addition, lack of adequate description of fairness control mechanisms could also contribute to bias (S. Biswas & Rajan, 2020) .  In addition, using Evolutionary Many-Objective Optimization (EMOO) models, offer a promising approach to developing fair, interpretable, and legally compliant hiring algorithms that can outperform human experts and industry baselines. However, addressing challenges like limited demographic representation in datasets and covariate shift is essential through ongoing research to ensure the fairness and effectiveness of algorithmic hiring practices (Geden & Andrews, 2021).

### A2.6 Business implications of bias and fairness

The business implications highlight the importance of addressing bias in hiring processes to ensure fairness, diversity, and the overall success of organizations. Review of relevant literature revealed several aspects of business implications including performance issues, trust factors, user privacy and security and penalties. The details are provided below:

### A2.6.1 Root causes of Bias and their effect on business

The lack of interpret-ability (Mollas et al., 2023), non-generalisability, and the black box nature of AI systems (Sikorski, 2021), which can lead to challenges in understanding and explaining how these technologies work. The failure of AI projects (Schlegel et al., 2023), the impact of the AI system's limitations, unreliable results (Alexander et al., 2023) (Miller, 2019) and potential negative effects on deployment performance (Muthusamy et al., 2018) can all contribute to and/ or result in such ineffectiveness (Sikorski, 2021). Further, the presence of bias or unfairness in AI algorithms (Baumann et al., 2023) raises legal and ethical concerns (I. Alves et al., 2024). Additionally, privacy exposure (Alexander et al., 2023) (Kikuchi et al., 2018) and security impacts (Siddiqui et al., 2021) (Barta & Görcsi, 2021) can have ethical issues and make the business vulnerable to liabilities. With regulations emerging governing artificial intelligence, it also emphasizes the effects of a lack of accountability (Yildiz & Beloff, 2020), transparency (Muthusamy et al., 2018) (Yildiz & Beloff, 2020) and legal compliance (Barta & Görcsi, 2021) In addition, the negative consequences associated with the black box nature of AI and the potential adverse effects (Muthusamy et al., 2018) of AI deployment can harm a business's reputation (Grove et al., 2020). Stakeholders may also be negatively impacted (Güngör, 2020), as their trust in the business is eroded. Lastly, failure to comply with legal and compliance requirements can result in financial penalties for the business. The value destruction (Canhoto & Clear, 2020) for the business caused by ineffective AI systems and the negative impact on customer due to privacy (Kikuchi et al., 2018) (Luo et al., 2019) or other legal exposures can further lead to financial losses.

### A2.6.2 Key Incidents of Bias in AI Hiring Tools

Several studies and reported incidents highlight significant concerns about the effectiveness and scientific basis of AI-driven recruitment tools in reducing bias and promoting diversity in the hiring process. These incidents demonstrate how algorithmic biases can perpetuate and even amplify existing discrimination, leading to various business, legal, and ethical implications.

| Incident | Details | References |
| --- | --- | --- |
| Amazon AI hiring tool | Amazon has shut down an artificial intelligence (AI) tool it was developing for recruiting after it was found to be discriminating against women. The tool, which was intended to help with recruitment by searching for candidates online, downgraded résumés containing the word "women's" and filtered out potential hires who had attended women-only colleges. | *Amazon Shuts Down AI Hiring Tool for Being Sexist* |

—------—------------------------------------------------------------------------------------

| | | |
|---|---|---|
| | The tool was built using past résumés submitted to Amazon over a 10-year period, which were predominantly submitted by male applicants, perpetuating a bias against female hires. The case study highlights the challenges of developing fair and unbiased AI systems. | |
| Apple card algorithm | This article discusses the gender bias allegations against Goldman Sachs regarding its credit card practices for the Apple Card. The allegations arose when software developer David Heinemeier Hansson highlighted the differences in credit lines between male and female customers in a viral Twitter thread. Hansson revealed that his wife, Jamie Hansson, was denied a credit line increase despite having a higher credit score than him. The article highlights the regulatory investigation sparked by these allegations and emphasizes the need for transparency and fairness in credit card practices. | *Apple Card Algorithm Sparks Gender Bias Inquiry - The Washington Post* |
| Fired by Bot/ algorithm | This article discusses how algorithms used by Amazon to manage its contract drivers, including those in the Flex program, are making decisions about hiring and firing with little human oversight. The algorithms track drivers' performance and determine their eligibility for routes, often leading to terminations without clear explanations. Flex drivers are monitored for factors such as punctuality and delivery quality, but the algorithms do not always account for real-world challenges faced by drivers, such as traffic or access issues. Drivers who are wrongly terminated have limited recourse, and the appeal process is often ineffective. Amazon's use of algorithms in its human-resources operations raises concerns about transparency and fairness, as drivers face the consequences of automated decisions. | *Payout for Estée Lauder Women 'Sacked by Algorithm,'*<br><br>*Fired by Bot: Amazon Turns to Machine Managers And Workers Are Losing Out - Bloomberg* |
| | Estée Lauder has reached an out-of-court settlement with three make-up artists who lost their jobs after being assessed by an algorithm in a video interview. The women, who worked for Estée Lauder subsidiary MAC, were initially required to reapply for their positions before being informed that they were being made redundant based partly on the algorithm's judgment. The software, created by recruiting platform HireVue, analyzed the content of their answers, expressions, and other data about their job performance. The women began legal proceedings against Estée Lauder, claiming they were not informed about the nature of the assessment. The case highlights the potential issues and biases associated with automated hiring software. In an interview, one of the women expressed the importance of speaking out about the issue and stopping such practices. The women received an out-of-court settlement. | |
| Hirevue's discrimination | This article discusses the use of an artificial intelligence (AI) hiring system developed by HireVue, which analyzes facial movements, word choice, and speaking voice of job candidates to generate an "employability" score. The system has gained widespread use among prominent employers, but some experts argue that it lacks scientific reasoning and could result in unfair discrimination. Critics argue that the system may penalize nonnative speakers, nervous interviewees, or those who do not fit the model for look and speech. The article highlights concerns about the system's objectivity and the lack of transparency in its decision-making process. An official complaint has been filed against HireVue, urging the Federal Trade Commission to investigate its practices. In an update, HireVue has stated that it no longer uses visual analysis in its software since 2020, as advances in natural language processing have made it more effective in assessments. | *HireVue's AI Face-Scanning Algorithm Increasingly Decides Whether You Deserve the Job* |
| Algorithms to evaluate Phd's | The University of Texas at Austin has announced that it will no longer use a machine-learning system called GRADE to evaluate applicants for its Ph.D. program in computer science. The system, which predicted the likelihood of an applicant's approval and assigned a numerical score, has faced criticism for potentially exacerbating existing inequalities in the field. Critics argue that the system encoded biases into its algorithms, which could have a negative impact on underrepresented groups. The creators of GRADE maintain that the system was designed to replicate the admissions committee's decision-making process and was not programmed to use race or gender in its predictions. However, detractors argue that biases can still be present in other factors used by the system. The university cited difficulties in maintaining the system as the reason for its discontinuation. | *U of Texas Will Stop Using Controversial Algorithm to Evaluate Ph.D. Applicants* |
| Flawed AI interview tools | The article discusses the testing of AI interview tools by MIT Technology Review. The tools, MyInterview and Curious Thing, use AI algorithms to evaluate candidates for job positions. The article highlights concerns about the accuracy and reliability of these algorithms. One specific issue raised is that MyInterview gave a candidate a high score for English proficiency even though she spoke only in German. The article also mentions challenges in assessing personality traits through AI interviews and the potential for bias in the hiring process. The companies behind these tools are often reluctant to share details about their algorithms, making it difficult to assess their accuracy. | We Tested AI Interview Tools. Here's What we Found. \| MIT Technology Review |
| False representation that tools are 'Bias Free' | This academic paper highlights a specific complaint against Aon, a major hiring technology vendor, regarding their online hiring tests. The complaint, filed by the ACLU, alleges that Aon deceptively markets these tests as "bias-free" while discriminating against job seekers based on their race or disability. | The Long History of Discrimination in Job Hiring Assessments \| ACLU of Florida |

---

| | | |
|---|---|---|
| | The complaint also includes charges filed with the Equal Employment Opportunity Commission against both Aon and an employer using Aon's assessments. The paper discusses how Aon's personality assessment test and automated video interviewing tool, which integrate AI, assess general personality traits that are not directly related to job performance and can unfairly screen out people with disabilities. The paper also raises concerns about cognitive ability assessments, which can disadvantage Black job candidates and individuals with disabilities. The paper emphasizes the need for employers to thoroughly vet assessments for compliance with anti-discrimination laws and hold vendors accountable for designing inclusive and non-discriminatory products. It calls for a future where skills and potential, rather than bias, determine job opportunities. | \| we Defend the Civil Rights and Civil Liberties of All People in Florida, by Working through the Legislature, the Courts and in the Streets |

### A2.6.3 Analysis of Algorithmic Bias in Hiring

The recurring theme across these incidents is the inherent risk of algorithmic bias, often stemming from historical data, lack of transparency, and insufficient human oversight. Claims by recruitment AI companies that their tools can objectively assess candidates by removing gender and race are misleading, as they often fail to understand broader systemic power imbalances (Drage & Mackereth, 2022). This thesis compiles the implications of algorithmic discrimination in hiring, encompassing the following dimensions: inefficacy in hiring practices, legal and ethical considerations and ramifications, reputational harm, and financial penalties.

Ineffective Hiring: Algorithmic recruiting can lead to biased candidate selection, excluding qualified individuals and limiting workforce diversity. This hinders innovation and creativity, impacting a firm's ability to adapt and prosper. For example, automated hiring systems, including CV scanners, are estimated to prevent 27 million people from finding full-time work by focusing on perceived candidate deficiencies rather than potential value, disproportionately affecting caregivers, veterans, immigrants, and people with disabilities (Fuller et al., n.d.).

Legal and Ethical Concerns: Algorithmic biases can result in significant legal repercussions and ethical issues related to privacy, freedom of speech, and fair employment opportunities. The American Civil Liberties Union (ACLU) has filed complaints against vendors like Aon Consulting, Inc., alleging unfair and deceptive practices that discriminate based on disability and race.

Regulatory bodies are increasingly mandating employer notification when AI is used for assessment and regular algorithm audits. The proliferation of AI-generated content, including fake profiles on LinkedIn, also raises concerns about online deception and the reliability of job applications. Many AI hiring tools lack independent testing, and developers are often hesitant to disclose algorithm details, making it difficult to assess their accuracy and impact on recruiting decisions.

Reputational Impact: Publicly reported incidents of discrimination can severely damage an organization's reputation. Analysis of publicly traded companies like Amazon and Estée Lauder suggests a correlation between negative news regarding algorithmic bias and a decline in their stock prices. For instance, Amazon's share price fell 8% following news of its sexist recruitment algorithm being scrapped, and Estée Lauder's dropped 11% after a documentary on algorithmic bias was announced. While Apple's stock was not negatively affected by gender bias allegations against its credit card algorithm, this was attributed to a concurrent dividend notification. Refer to Appendix 5 for details.

Fines and Penalties: Discrimination in hiring is prohibited by various employment laws, leading to legal action, fines, and penalties for non-compliant organizations. Refer to Chapter 4 for more details. The Bologna Labour Court, in the case of Filcam VGIL Bologna and others v Deliveroo Italia SRL, found Deliveroo's algorithm indirectly discriminatory for failing to consider reasons for cancellation or non-participation in shifts, emphasizing the need for human oversight and transparency. The George Washington University's list of AI-related litigations includes cases like Equal Employment Opportunity Commission v. iTutorGroup, Inc., where iTutorGroup was fined $365,000 for using software that discriminated against older applicants based on age. Other cases, such as In the Matter of HireVue, Inc. and Mobley v. Workday, Inc., highlights ongoing investigations and lawsuits regarding programmer bias and discriminatory screening algorithms. Refer to Appendix 5 for details.

### A2.6.4 Business Implications for AutoML Providers

For AutoML providers, fairness opacity not only raises ethical concerns but also creates friction for enterprise buyers, who often view tools lacking interpretability as compliance risks. The absence of intuitive fairness workflows increases support burdens and inflates the total cost of ownership. Integrating fairness-by-design features is therefore a strategic imperative, as it reduces procurement friction, lowers the need for custom compliance documentation, and builds buyer trust. This proactive approach can accelerate market entry in highly regulated domains like HR and reduce post-sale support costs.

—-------—------------------------------------------------------------------------------------

# B. Fairness in the context of AutoML

**Section Overview**

**Methodology:** The section illustrates a systematic Scopus search of AutoML-related papers, screening 1,156 papers to arrive at 167 studies focused on governance, ethics, fairness, interpretability, and performance evaluation for in-depth analysis. Extracted insights were organized into themes addressing research objectives.

**Introduction to AutoML:** The section covers literature on AutoML tools, libraries, and frameworks, including automated model selection, hyperparameter tuning, feature engineering, and deployment strategies. It highlights the growing adoption of AutoML across industries, its efficiency benefits, and its potential to democratize AI for users with varying expertise.

**Bias and discrimination in AutoML:** The section discusses risks of biased outcomes due to imbalanced datasets, limitations in bias detection, lack of transparency, and black-box operations. It emphasizes the need for contextual human oversight and data preprocessing to ensure fairness in model outputs.

**Gaps and emerging needs:** The section identifies limitations in current AutoML tools, including weak handling of imbalanced datasets, insufficient data preprocessing, underdeveloped feature engineering, lack of transparency, inconsistent performance across datasets, generalization challenges, limited meta-learning support, and the importance of human-in-the-loop systems to maintain user agency and trust.

**Approaches to handle bias:** The section summarizes fairness-aware strategies in AutoML, including fairness-aware search and optimization, constrained Bayesian optimization, fairness metrics, human oversight, blockchain-based trust mechanisms, and interactive evaluation tools. Guidelines emphasize documentation, contextual benchmarks, and iterative user support to enhance fairness in AutoML systems.

This section reviews literature on AutoML, focusing on governance, fairness, transparency, and performance evaluation. It examines the benefits, limitations, and adoption trends of AutoML tools, emphasizing their role in automating machine learning processes and supporting non-expert users. The section also identifies gaps in current tools and explores approaches for mitigating bias and ensuring fairness in AutoML systems.

## B1. Methodology

A scopus search of "AutoML" papers resulted in 1156 papers. The core theme of the papers included the following topics:

- Automated model selection and tuning techniques.
- Various AutoML frameworks and tools, considering their capabilities and limitations.
- Application of AutoML in specific domains, exploring how it has been utilized across various industries.
- Automated feature engineering and selection methods.
- Governance considerations, including ethics, security, safety, fairness, accountability, and audits.

———------------------------------------------------------------------------------------------

- Interpretability and explainability techniques to enhance transparency.
- Model deployment and integration strategies and scalability & efficiency considerations for large-scale deployments.
- Performance evaluation and benchmarking approaches and
- Transfer learning and meta-learning techniques and their impact on AutoML.

**AutoML Research Themes and Governance**

AutoML Research Themes

AutoML Frameworks and Tools

Governance Considerations
Ethics
Security
Safety
Fairness
Accountability
Audits

Application in Specific Domains

Interpretability and Explainability
Transparency
Explainability

Model Deployment and Integration
Scalability
Efficiency

Performance Evaluation and Benchmarking
Evaluation Metrics
Benchmarking Standards

Automated Model Selection and Tuning

Transfer Learning and Meta-Learning
Knowledge Transfer
Learning from Experience

Automated Feature Engineering and Selection

Of the above topics, 167 papers pertaining to Governance considerations, including ethics, security, safety, fairness, accountability, and audits, interpretability and explainability techniques to enhance transparency, and Performance evaluation and benchmarking approaches were considered for further review and analysis.

# B2. Literature Review

## B2.1 Introduction to AutoML

AutoML tools help to speed up and simplify the development of machine-learning systems (Frank Hutter, 2014). Categories of AutoML include automated machine learning tools, automated machine learning libraries, code-assist tools, low-code or no-code tools, automated hyper-parameter optimization tools[7], and automated data labeling, feature engineering, or model monitoring tools.

According to a Forrester survey in 2020, 61% of data-and analytics decision makers in companies using AI have adopted automated machine learning (AutoML). Furthermore, another 25% said they will do so within the year next (2021). AutoML market size is estimated to be USD 15 bn in 2030.

One of the main benefits of AutoML tools is that they can significantly reduce the time and resources required to build and optimize machine learning models. AutoML tools can enable developers and data scientists to quickly and efficiently develop machine learning models that perform well on a given task by automating the process of model selection, hyperparameter tuning, and feature engineering.

Several AutoML tools and libraries are available, including Google AutoML, H2O.ai, DataRobot, TPOT, and Auto-Keras (Ferreira et al., n.d.). These tools and libraries support a variety of machine learning tasks, including classification, regression, and clustering, and can be used with a wide range of data types and formats. AutoML tools and libraries have been widely adopted in the industry and have demonstrated strong performance on various tasks. AutoML frameworks are a valuable strategy for machine learning experiments and projects.

The demand for machine learning applications is increasing rapidly, but the number of knowledgeable data scientists is not scaling accordingly (Eldeeb et al., 2024). Data scientists and application developers can use AutoML to automate the process of designing, training, and deploying machine learning models, reducing the need for specialized knowledge and expertise in machine learning. They often outperform human data scientists or deliver similar results, making them a beneficial approach for both beginners and veterans in the field (Hanussek et al., 2020).

[7] *Auto-tuning data science: New research streamlines machine learning | MIT News | Massachusetts Institute of Technology*. (n.d.). Retrieved March 5, 2023, from https://news.mit.edu/2017/auto-tuning-data-science-new-research-streamlines-machine-learning-1219

Such efforts can make it easier for people with various backgrounds and skill levels to use machine learning to solve problems and create new applications. AutoML has also made it possible for individuals without a background in data science to use machine learning effectively. This is especially beneficial for small and medium-sized enterprises (SMEs) who can now independently harness the power of machine learning (Bender et al., 2022).

AutoML can also help to democratize AI by providing tools and resources for building and deploying machine learning models at scale. 91% of leaders surveyed by HBR-Google in 2020, expressed that democratizing access to data and analytics across the organization is important to business success. For example, many AutoML platforms offer cloud-based services that allow users to train and deploy machine learning models on large amounts of data without expensive hardware or infrastructure.

AutoML can also help to democratize AI by providing tools and resources for building and deploying machine learning models at scale. For example, many AutoML platforms offer cloud-based services that allow users to train and deploy machine learning models on large amounts of data without the need for expensive hardware or infrastructure.

### B2.2 Bias and discrimination in AutoML

Machine Learning can exhibit discrimination due to unfairness bugs in the models, resulting in biased decisions based on protected attributes. The machine learning community has developed testing and verification strategies to detect unfairness, and the machine learning literature offers research on defining fairness criteria and bias mitigation methods. However, current bias mitigation techniques often come at the cost of decreased accuracy, and there is a need to improve fairness in AutoML systems given their increasing adoption.

AutoML in recent years, has raised questions about the interplay of data, models and human experts, given its associated risks. For instance, Pre-modelling is a method used by professional data scientists to detect biases in existing datasets and then devise approaches for "augmenting" the information to balance it out. This creates an optimal learning environment that achieves the desired outcome without exacerbating preexisting imbalances within the dataset. While some AutoML tools have features to detect data imbalance, it may not always be effective (Xin et al., 2021) or most appropriate approach in cases where there is a need to have positive bias (having more women candidates for hire), thereby requiring a specific construct before using AutoML. This clarifies the need for contextual consideration and need for human augmentation while using AutoML.

—-------—----------------------------------------------------------------------------------------

To ensure that AutoML tools do not perpetuate bias or discrimination, it is crucial to consider the data used for training the models. Biases present in the training data can be propagated to the ML models, leading to biased predictions and discriminatory outcomes. Hence, it is essential to carefully curate and preprocess the training data to mitigate bias and ensure fairness in the models (Drozdal et al., 2020). In AutoML context, this represents the need to examine if the tools have features to support this process.

One of the other critical factors to earning trust in AutoML is explainability. While AutoML providers have tried to address bias and explainability issues, these are not optimal. Transparency mechanisms like visualization can increase understanding and trust within AutoML but aren't enough for humans to build that trust and identify problems more quickly (Drozdal et al., 2020)

AutoML often operates in a "black box," where it's unclear which steps are taken to reach the final result[8]. Therefore, it can be challenging to establish if cleaning processes eliminated important data or created biases. If an algorithm produces discriminatory results, claiming ignorance of factors involved won't be adequate; transparency and understanding the methods taken by AutoML are necessary for guaranteeing fairness in outcomes. Hence, examining the risks of unfair AutoML systems and whether there are sufficient fairness mitigation techniques to increase overall fairness is essential (Wu & Wang, n.d.)

In addition, it is also pertinent to note that, some of the commercial tools like Google AutoML or Microsoft Azure does not provide the code used for training the model, thereby providing low visibility and less customization or cross-validation options in each context (Doris Xin, 2022).

Based on the above its relevant to examine the implications of unfairness contributed by AutoML due to four key reasons:

These tools make ethical choices on behalf of the developers

They have preset considerations that cannot be changed or modified or in some cases not disclosed explicitly

They do not explain how they arrive at a choice or a decision

They do not always understand use cases that are domain-specific or requires detailed manual feature engineering.

To understand the bias implications caused by AutoML the relevant and pertinent limitations associated with AutoML are discussed in the section below.

[8] *Cracking open the black box of automated machine learning | MIT News | Massachusetts Institute of Technology*. (n.d.). Retrieved March 5, 2023, from https://news.mit.edu/2019/atmseer-machine-learning-black-box-0531

## B2.3 Gaps and emerging needs in current AutoML tools / libraries

At the outset, key gaps in AutoML tools/ libraries in the context of fairness are represented below:

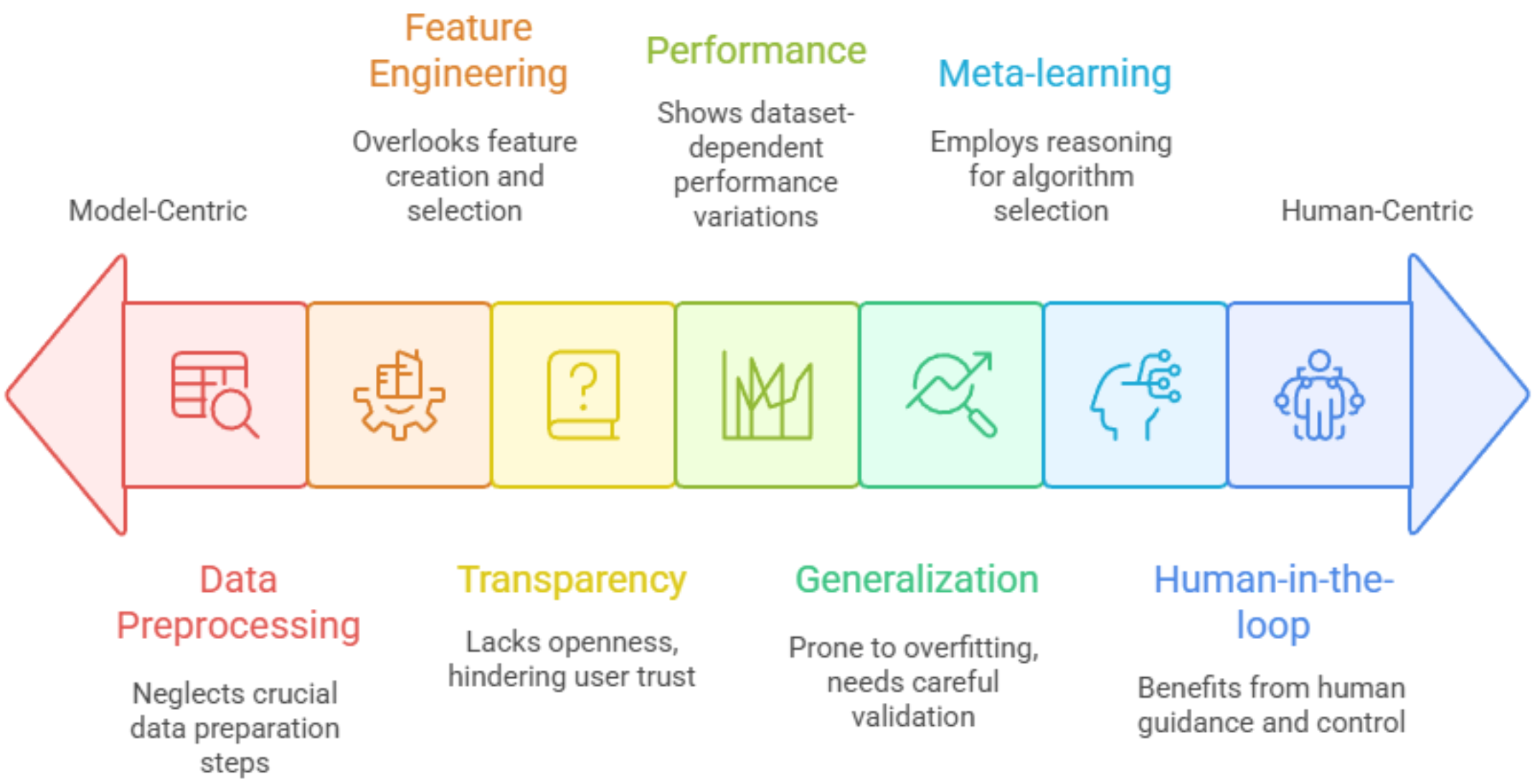

### B2.3.1 Dataset imbalance

Research reveals that most AutoML tools show weak performance on imbalanced datasets. It is pertinent to note that an imbalance dataset is also a core issue that impacts the bias in the outcomes (Singh & Vanschoren, n.d.) (Gijsbers et al., 2019).

### B2.3.2 Data preprocessing gaps in current AutoML

Another research highlighted that most functions used by ML practitioners are model-oriented, with a significant emphasis on hyperparameter optimization (40.9%) and model training (11.5%) stages of the ML pipeline (Majidi et al., 2022). It also suggested that current AutoML tools prioritize effective model training and the resulting models, while paying less attention to the data preprocessing process (Majidi et al., 2022). It is pertinent to note that data preprocessing has been identified as the least focused area in AutoML tools, aligning with previous research (*State of Data Science 2020)* (Truong et al., 2019) (Yao et al., 2018). However, it is important to recognize that the quality of an ML model is highly dependent on the quality of the data used, emphasizing the significance of data preprocessing in the overall ML pipeline (Breck et al., 2019).

**B2.3.3 Feature engineering**

It is observed that most of the current work on AutoML primarily focuses on automating preprocessing, algorithm selection, and hyperparameter tuning, while overlooking the crucial aspect of feature engineering. However, in practice, feature engineering plays a substantial role in ML pipeline development and consumes a significant amount of time for engineers. The quality of feature engineering significantly impacts the overall performance of machine learning models (Eldeeb et al., 2024). In addition, accurate feature type inference is critical for achieving high downstream model accuracy in AutoML (Shah et al., 2021) .

**B2.3.4 Transparency**

The utilization of AutoML tools in the machine learning domain may present obstacles to transparency and interoperability, as well as limited adaptability to intricate scenarios, potentially impeding their widespread adoption. Moreover, the effectiveness of AutoML in facilitating the adoption of machine learning practices is contingent upon the specific tool employed and the context in which it is applied (Azevedo et al., 2024). Further, research notes that data scientists perceive AutoML as a supplementary tool rather than a substitute for human expertise (Wang et al., 2019), and lack of transparency in AutoML-generated models impacts user trust (Drozdal et al., 2020). Improving transparency through metrics, visualization (Narkar et al., 2021), and opening the black box (Weidele et al., 2020) enhances user trust and comprehension of AutoML.

**B2.3.5 Performance differences across different datasets**

Another research revealed that the performance disparities among AutoML tools were less pronounced for churn datasets but more significant for other datasets like diabetes and credit. This suggests that the effectiveness of AutoML tools can vary based on the unique characteristics and complexities of the datasets under analysis (Ferreira et al., 2021). Similarly research also revealed that performance variations (Truong et al., 2019) exist in different computational environments (including AWS SageMaker and Azure ML) (Choi et al., 2023)(Zöller & Huber, 2021). This underscores the importance of carefully selecting both the algorithm and the computational environment to achieve optimal predictive accuracy (Choi et al., 2023).

**B2.3.6 Generalization of AutoML performance**

Research also indicates that AutoML tools provide better performance than the traditional models (Paladino et al., 2023). While the goal of automated machine learning is to eliminate the burden of pipeline creation for developers, the issue of overfitting can persist and can become more critical issue when attempting to iteratively optimize the performance of an internal cross-validation, typically done through k-fold validation (Evans et al., 2020).

#### B2.3.7 Meta-learning and reasoning

Automated machine learning (AutoML) allows the user to choose from a range of learning algorithms, each associated with a hypothesis space. AutoML is currently being studied to examine if it can support finding the optimal algorithm and set of hyperparameters that minimize the error on the data, taking into consideration resource constraints and the splitting of data into training and validation sets. Various AutoML tools use metareasoning techniques, such as warm starting and reasoning about the evaluation process, to improve efficiency. Hyperparameter optimization also plays a role, with the allocation of computational resources and cost considerations. The perspective of AutoML as a deliberative agent can help improve existing methods and inspire new approaches based on rational metareasoning and bounded optimality (Hullermeier et al., 2021). Researchers also theorise that AutoMl can be primed for incremental learning by using automated curriculum learning to design task sequences, neural architecture search for task-specific architectures, and optimization techniques to tune hyperparameters dynamically (Kilickaya & Vanschoren, 2023).

#### B2.3.8 Human in the loop and human agency

Human involvement and guidance in AutoML tools improve performance, user trust, and address deficiencies like system failures and lack of transparency. Balancing human agency and automation is essential, with a focus on human-centered approaches that enhance productivity while maintaining user control (Sun et al., 2023). Supporting user agency in ML involves providing interface mechanisms, user feedback (Holzinger et al., 2016), customizability (Winter & Jackson, 2020), and the ability to modify AutoML's search space, allowing users to adapt and work around limitations for effective ML implementation (Cai et al., 2019). That said, involving humans does not guarantee there won't be bias; humans need agency and understanding to trust AutoML fully (Doris Xin, 2022).

### B2.4 Approaches to handle bias in AutoML

AutoML simplifies model building but requires structured approaches for fairness evaluation and mitigation of algorithmic bias, particularly in educational data mining (Griep et al., 2023). The development of AutoML systems can introduce unfairness, requiring fairness-aware approaches. Fairness is a complex subject that cannot be fully automated, but efforts are being made to optimize both fairness and predictive performance (weerts et al., 2023). It is also pertinent to note that NAS-based outcomes are vulnerable to attacks due to their architectural properties, resulting in adversarial attempts to make the model discriminate (Pang et al., 2021).

Fair-AutoML is an AutoML approach that mitigates bias, incorporating a fairness-aware search space and optimization function. It successfully repairs buggy cases, outperforming existing techniques (Nguyen et al., 2023). In addition, researchers have proposed that the FairAutoML framework incorporates fairness definitions, unfairness mitigation techniques, and hyperparameter search methods to prioritize fairness in AutoML. It allows for the incorporation of existing fairness definitions into the model search and evaluation process (Wu & Wang, 2021). In addition, constrained Bayesian optimization (BO) methods are proposed which optimizes the performance of ML models while enforcing fairness constraints, showing competitiveness with specialized fairness techniques (Perrone et al., 2021). Researchers have also proposed 'Trustless AutoML' , a framework that combines AutoML techniques with blockchain technology to decentralize the design and training process, establishing transparency and trust (Bathen & Jadav, 2022).

In the recent publication (weerts et al., 2023) provide guidance on a set of considerations in building fairness-aware AutoML much beyond integrating fairness constraints in the process. The considerations include: (1) Clearly state assumptions and limitations of the system, emphasizing that fairness metrics cannot guarantee fairness. (2) Support users in identifying potential fairness-related harm through interventions beyond modeling, such as improved data collection and problem formulation. (3) Incorporate principles of seamless design to counteract automation bias and encourage users to critically evaluate solutions. (4) Support users in statistically sound fairness evaluation, including warning users if fairness evaluation results lack sufficient data support. (5) Account for inherent limitations of fairness metrics and encourage comprehensive evaluations beyond simple metrics. (6) Ensure well-substantiated system design by documenting design decisions and incorporating user requirements. (7) Evaluate the system against contextualized benchmarks that reflect real-world requirements. (8) Support users in performing quick iterations through the development of fast and interactive fairness-aware AutoML systems. These guidelines aim to improve the development and evaluation of fairness-aware AutoML systems.

# C. AutoML Market assessment and business context of fairness

**Section Overview**

**Market Overview:** The section illustrates the global AutoML market as valued at approximately **$2.5–2.7 billion in 2023**, with projected growth exceeding **40% CAGR through 2030**, emphasizing its role in democratizing AI and enabling predictive modeling with minimal coding.

**HR Adoption:** The section illustrates that the AI in HR market ranges from **$3.2–6.0 billion in 2023**, with forecasts of **$14–26 billion by 2029–2033**, highlighting applications in talent acquisition, employee retention, performance management, and workforce planning, and identifying key tools including IBM Watsonx, Google AutoML/Gemini, DataRobot, Dataiku, Alteryx, and open-source libraries (FLAML, AutoGluon, H2O, PyCaret).

**Risks and Concerns:** The study highlights that **42% of organizations** express high concern about AI bias, while **75% of HR leaders** cite bias as a primary concern in adoption, illustrating the importance of addressing transparency, fairness, and trust to mitigate reputational, operational, and legal risks.

**Enterprise Expectations:** The section illustrates enterprise demands for transparency logs, audit trails, fairness benchmarks, explainable predictions, and governance features, emphasizing that legal and compliance teams can delay or block adoption if these features are insufficient.

**Implications of Inaction:** The study illustrates that failure to address risks can lead to lost customers and staff, legal penalties, reputational damage, and stalled AI adoption, particularly in HR where employment and anti-discrimination laws are stringent.

**Tool Features:** The section illustrates that leading platforms embed HR-specific responsible AI features, including bias mitigation, explainability, auditability, and governance controls, supporting ethical AI deployment and alignment with regulatory requirements.

This section examines the current state of Automated Machine Learning (AutoML) tools in the HR sector, focusing on market growth, adoption trends, and enterprise priorities. It highlights both the opportunities these tools provide for efficiency and democratization of AI, as well as the risks associated with bias, transparency, and trust. The section also outlines how leading AutoML platforms address these challenges through HR-specific fairness, explainability, and governance features.

## C1. Market Assessment: AutoML Tools in the HR Sector

The global market for Automated Machine Learning (AutoML) platforms is experiencing rapid growth, driven by increasing enterprise demand for scalable, accessible, and efficient AI solutions. According to Gartner's 2025 Magic Quadrant for Data Science and Machine Learning Platforms, the AutoML market was valued at approximately $2.5–2.7 billion in 2023 and is projected to grow at a compound annual growth rate (CAGR) exceeding 40% through 2030. AutoML's appeal lies in its ability to democratize machine learning, enabling business analysts and non-experts to build and deploy predictive models with minimal coding.

Within the broader AI landscape, the HR sector is emerging as a key adopter of AI-driven automation and analytics. The AI in the HR market was estimated at $3.2–6.0 billion in 2023, with forecasts suggesting this could reach $14–26 billion by 2029–2033 (IBM, 2023). AutoML is a significant subset of this market, particularly as HR departments seek to automate tasks such as talent acquisition, employee retention analysis, performance management, and workforce planning.

IBM Watsonx and Google Gemini/AutoML have positioned themselves as leaders in HR-specific AI integration. IBM Watsonx, for example, offers dedicated HR agents that automate workflows across payroll, benefits, onboarding, and talent management, with prebuilt connectors for over 80 HR applications (IBM watsonx, 2024). Google's Gemini and AutoML tools are widely used for HR document automation, recruiting, onboarding, and policy management, leveraging the ubiquity of Google Workspace in enterprise environments (Google Cloud, 2024). DataRobot and Dataiku are prominent in the enterprise AutoML landscape, actively marketing to HR users with solutions for attrition prediction, workforce analytics, and structured hiring in addition to serving finance, healthcare, manufacturing, and retail (DataRobot, 2022; Dataiku, 2024).

Alteryx is recognized for its strong presence in HR analytics, providing tools for engagement tracking, retention modeling, and dashboard automation (Alteryx, 2024). Open-source and low-code tools such as FLAML, AutoGluon, H2O AutoML, and PyCaret are popular among technical users and researchers, hence, the proportion of human resources use of such tools cannot be estimated (H2O.ai, 2024; PyCaret, 2024).

## C2. Business Leaders' Perception of Risks: Fairness, Trust, and Transparency

As AI and AutoML adoption accelerates, business and HR leaders are increasingly concerned about the risks of bias, lack of transparency, and erosion of trust. According to DataRobot's State of AI Bias Report, 42% of organizations are "very" to "extremely" concerned about AI bias, with the most significant perceived risks being compromised brand reputation and loss of customer trust. This concern is echoed in IBM's 2023 CEO Study, where nearly half of CEOs surveyed cited accuracy and bias as top issues in AI adoption. The Warden AI State of AI Bias in Talent Acquisition 2025 report found that 75% of HR and talent acquisition leaders cite bias as a primary concern when adopting AI, with many demanding greater clarity, transparency, and responsible use of AI in hiring and promotion. Deloitte's 2024 State of AI in the Enterprise survey similarly reports that a majority of executives see ethical risks—including bias, fairness, and lack of transparency—as major barriers to scaling AI initiatives.

## C3. Enterprise Expectations from Tool Providers

Given these risks, enterprise buyers now expect AI and AutoML vendors to deliver robust features for transparency, explainability, and fairness. Gartner's AI Trust, Risk and Security Management (AI TRiSM) research highlights that procurement teams increasingly require:

- Transparency logs and audit trails for all AI-driven decisions, enabling traceability and accountability.
- Fairness benchmarks and bias detection tools, especially for regulated domains like HR.
- Explainable predictions that business users, auditors, and regulators can understand and interrogate.
- Governance and compliance features to support legal and ethical requirements, such as documentation, model versioning, and access controls.

Legal and compliance teams are empowered to delay or block adoption if these features are lacking. Gartner further notes that high-friction tools—those that are difficult to use or lack clear onboarding—hurt SaaS conversion and slow enterprise rollout, especially in environments where "citizen data scientists" (non-experts) are expected to use these platforms (Gartner SaaS Conversion, 2024).

## C4. Implications of Not Addressing the Risks

The implications of failing to address fairness, trust, and transparency in AI and AutoML tools are profound:

- Loss of customer trust and revenue: Over a third of technology leaders report negative impacts from AI bias, including lost customers and staff attrition. In HR, this can translate to poor employee experiences, reduced engagement, and higher turnover (DataRobot, 2022).
- Brand and regulatory risk: Organizations deploying biased or non-transparent AI face reputational damage, lawsuits, and regulatory penalties. In HR, where employment and anti-discrimination laws are stringent, the consequences can be severe (Gartner, 2024).
- Operational and ethical risk: Unchecked bias and lack of transparency undermine fairness, equity, and diversity efforts. This can erode employee morale and trust, and damage the organization's standing with stakeholders (Deloitte, 2024).
- Stalled AI adoption: Gartner reports that legal and compliance concerns are a leading cause of delayed or abandoned AI projects, particularly in regulated sectors like HR (Gartner, 2024).

## C5. Features for HR context by prominent tools

In response to these enterprise demands, leading AutoML and AI platforms are investing heavily in features that address transparency, trust, and fairness—particularly for HR users. Below is an overview of how prominent tools are meeting these needs:

| Tool | Fairness and Transparency Features in HR Context |
|---|---|
| **IBM watsonx** | IBM Watsonx has made responsible AI a core principle, embedding trust, transparency, and fairness throughout its platform (IBM watsonx, 2024):<br>● **End-to-End Trust & Transparency**: All AI-driven HR decisions are logged, with transparency logs and audit trails available for compliance and governance.<br>● **Explainable Recommendations**: HR managers receive clear, contextual explanations for AI-driven recommendations (e.g., pay decisions, promotions), supporting fact-based, transparent choices.<br>● **Bias & Fairness Controls**: IBM's AutoAI toolkit includes fairness and bias mitigation, allowing HR teams to validate models for adverse impact and ensure equitable outcomes.<br>● **Governance & Auditability**: Comprehensive documentation and audit trails support regulatory compliance and internal reviews. |
| **Google AutoML / Gemini** | Google's AI tools are built on responsible AI principles, with a strong focus on fairness, privacy, and transparency (Google Cloud, 2024):<br>1. **Fairness & Responsible AI**: Google provides best practices for bias detection, sensitive feature exclusion, and fairness benchmarking in HR contexts.<br>2. **Explainable AI**: Tools such as Explainable AI in Google Cloud help HR users understand model predictions and ensure decisions are justifiable and auditable.<br>3. **Bias Evaluation**: Google recommends using "golden datasets" and intentional bias correction techniques for sensitive HR applications. |
| DataRobot | DataRobot positions trust and explainability as core differentiators, with features tailored for HR users (DataRobot, 2022):<br>● **Automatic Bias & Fairness Testing**: Users can flag protected features (e.g., gender, ethnicity), select relevant fairness metrics, and receive guided workflows to assess and mitigate bias.<br>● **Explainable Insights**: The platform provides clear explanations for each prediction (e.g., what factors contributed to an employee attrition risk score), supporting HR transparency and auditability.<br>● **Model Monitoring**: Ongoing monitoring for fairness and drift ensures models remain compliant and trustworthy after deployment. |
| Dataiku | Dataiku integrates responsible AI and DEI (diversity, equity, inclusion) principles into its platform (Dataiku, 2024):<br>● **Model Transparency Features**: Customizable explainability tools at both the global and individual prediction level help HR teams understand and communicate how models make decisions. - **Bias Mitigation in HR Processes**: Dataiku supports bias-reducing HR workflows (e.g., structured interviewing, competency-based evaluation) and ongoing DEI improvements. - **Auditability**: Visual ML pipelines and data lineage features enable traceability and compliance. |
| Alteryx | Alteryx is widely used for HR analytics, with a focus on transparency and auditability (Alteryx, 2024):<br>● **Visual, No-Code Workflows**: HR analysts can build transparent data pipelines and dashboards without coding, making data transformations and model logic clear.<br>● **Audit Trails**: All data and model steps are logged, supporting compliance and review in HR processes. |

These features are specifically aimed at HR's regulatory and ethical needs, empowering HR teams to deploy AI responsibly and with confidence.

# Chapter 3: Foundational considerations for the research

## A. Theoretical frameworks regarding product design

**Section Overview**

**Multi-Theoretical Framework**: The section illustrates the use of multiple theories, including the Technology Acceptance Model (TAM), Innovation Diffusion Theory, and UX strategy principles, to examine how fairness-aware features influence adoption by non-expert users with AutoML considerations in focus.

**Technology Acceptance Model (TAM)**: The study explains TAM's focus on perceived usefulness and ease of use, highlighting the role of fairness dashboards, transparency features, and human-in-the-loop interfaces in shaping user trust and adoption.

**Innovation Diffusion Theory (IDT)**: The section covers Innovation Diffusion Theory, emphasizing relative advantage, complexity, and observability as determinants of adoption and the importance of workflow alignment for usability.

**UX Strategy Principles**: The study discusses UX strategy, including Human-Centered AI, design thinking, affordance theory, and cognitive load theory, as guiding principles for developing accessible, actionable, and trustworthy fairness features.

**Application to HCI Evaluation**: The section summarizes how these theories inform interface design, fairness feature evaluation, cognitive load management, and HCI-based audits to support responsible adoption in enterprise contexts.

**Integration for Responsible Adoption**: The study concludes that integrating these theoretical lenses provides a foundation for designing, assessing, and improving AutoML fairness features for non-expert users while ensuring practical alignment with industry needs.

This chapter endeavors to identify suitable principles from established theories applicable to Human-Computer Interaction considerations for bias management in AutoML. These theories serve as a reference point for extracting the pertinent principles. This research does not aim to validate or invalidate the theories but rather to apply selected principles within the research context.

### A1. Theoretical Underpinnings: Adoption and Usability of Fairness Features in AutoML

This study adopts a multi-theoretical lens rooted in the Technology Acceptance Model (TAM), Innovation Diffusion Theory, and UX strategy principles to understand how the design of fairness-aware features in AutoML tools influences their adoption by non-expert users.

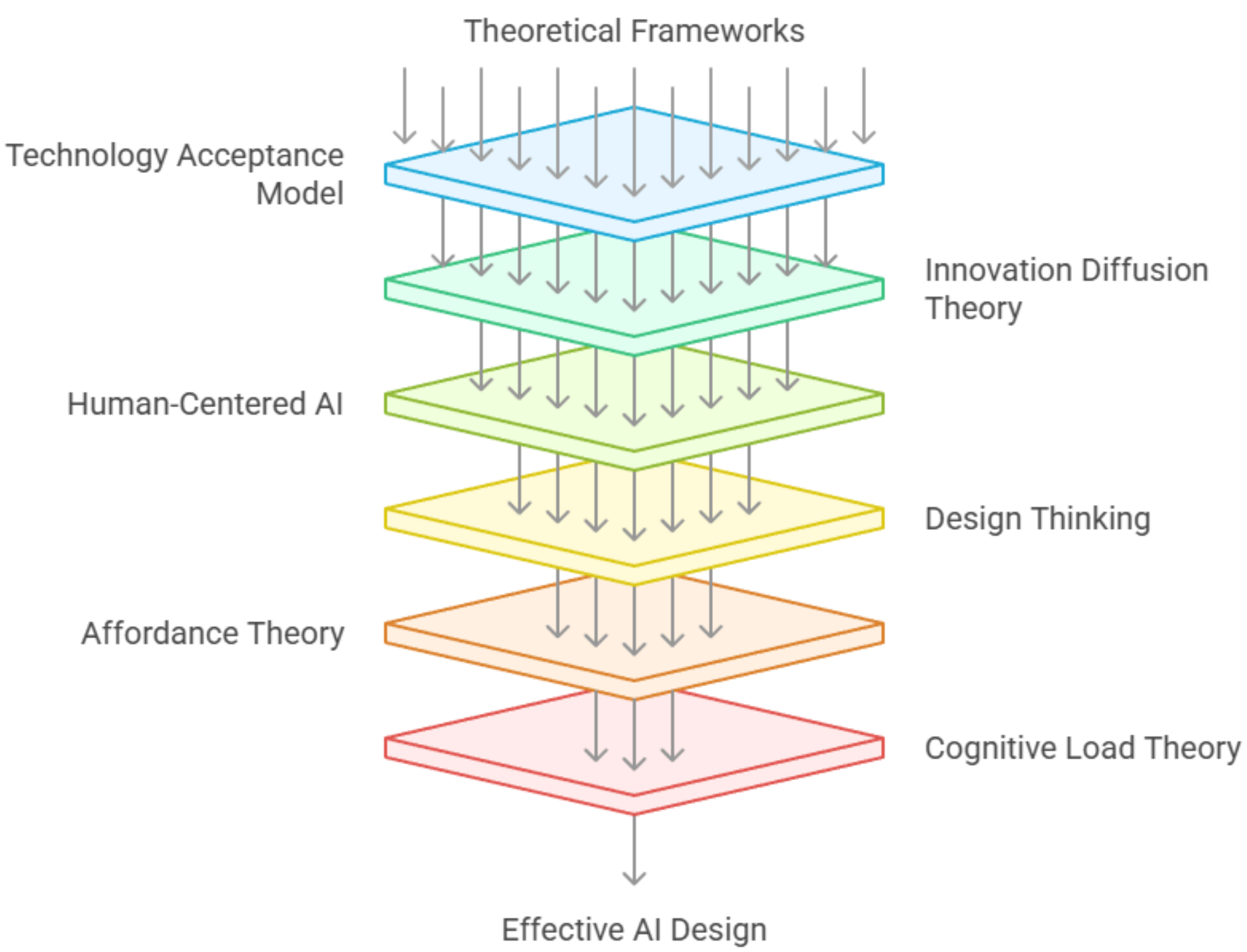

These theories provide a strategic foundation for assessing the usability, perceived trust, and utility of fairness-related functionalities, which are increasingly vital to product differentiation in high-risk domains such as hiring.

# A2. Detailed understanding of the theoretical lens

## A2.1 Technology Acceptance Model

The Technology Acceptance Model (TAM) posits that perceived usefulness and perceived ease of use are the primary drivers of technology adoption among users (Wicaksono & Maharani, 2020). In the context of AutoML (Automated Machine Learning), the integration of fairness dashboards, transparency features, and human-in-the-loop interfaces can significantly influence how non-expert users perceive both the usability and trustworthiness of these systems.

- Fairness Dashboards: These tools help users understand and monitor bias and fairness in model outcomes. For non-experts, clear visualization and actionable insights about fairness can enhance trust and perceived usefulness, as users feel more confident that the system is not propagating harmful biases (Heidrich et al., 2023) (Narayanan, 2023).
- Transparency Features: Features that explain model decisions (such as explainability modules or algorithmic transparency) directly address concerns about the "black box" nature of AI. When users can see how decisions are made, their perception of the system's ease of use and reliability increases, which in turn supports adoption (Pagano et al., 2023).
- Human-in-the-Loop Interfaces: Allowing users to intervene, provide feedback, or adjust model parameters gives them a sense of control and agency. This interactive approach can make the technology feel more approachable and easier to use, especially for those without deep technical expertise (Zender et al., 2024) (Narayanan, 2023) .

Empirical studies of open-source AutoML platforms indicate that while these features are increasingly recognized as essential, many tools still lack robust implementations that fully support fairness-aware model development and user understanding. Addressing these gaps is crucial, as the presence of such features can measurably improve user experience, trust, and ultimately, the adoption of AutoML systems by non-experts (Zender et al., 2024).

Hence, fairness dashboards, transparency, and human-in-the-loop features in AutoML are not just technical add-ons—they are central to shaping non-expert users' perceptions of usefulness and ease of use, which are the core determinants of technology acceptance according to TAM (Wicaksono & Maharani, 2020).

## A2.2 Innovation Diffusion

Rogers' Innovation Diffusion Theory identifies relative advantage, complexity, and observability as key determinants of innovation adoption. According to Rogers' Innovation Diffusion Theory, three principal factors—relative advantage, complexity, and observability—are instrumental in facilitating the adoption of innovations (Chen, 2024) (Miller, 2015). When applied to fairness-aware features in AutoML tools, these factors can be interpreted as follows:

- Relative Advantage: Fairness-aware features confer a competitive advantage by enabling users to develop models that are not only accurate but also ethically sound and compliant with regulatory standards. This perceived benefit enhances the likelihood of adoption, particularly as concerns regarding algorithmic bias become increasingly prominent (Weerts et al., 2024) (Narayanan, 2023).

- Complexity: The interpretability and user-friendliness of fairness features significantly influence their adoption. If these features are excessively complex or necessitate advanced expertise, non-expert users may be discouraged. Conversely, user-friendly interfaces and clear guidance (such as dashboards or conversational interfaces) can reduce perceived complexity, thereby enhancing accessibility (Narayanan, 2023) (Guo et al., 2024).

- Observability: The visibility of fairness outcomes—such as dashboards that display bias metrics or transparency features that elucidate model decisions—enables users to observe the tangible benefits of fairness-aware tools in practice. High observability fosters trust and encourages broader adoption, as users can witness the positive impact of these features on their workflows (Narayanan, 2023) (Guo et al., 2024).

Furthermore, alignment with user workflows is essential. Features that integrate seamlessly into existing processes and provide real-time, actionable feedback are more likely to be adopted, as they minimize disruption and learning curves (Guo et al., 2024). Hence, the adoption of fairness-aware features in AutoML is strongly influenced by their visibility (observability), interpretability (complexity), and integration with user needs (relative advantage and workflow alignment), in line with Rogers' framework

## A2.3 UX Strategy

UX strategy bridges business goals and user needs. By aligning fairness affordances with non-expert cognitive models and workflows, AutoML providers can improve user engagement and reduce friction, ultimately influencing product success. UX strategy can take four forms in the context of AutoML, namely, Human-Centered AI (HCAI) principles, design thinking based principles, affordance theory based considerations, and cognitive load theory based considerations.

**A2.3.1 HCAI Principles**

Aligning Human-Centered AI (HCAI) principles to the context of AutoML and fairness means moving beyond technical optimization to explicitly prioritize the needs, values, and capabilities of non-expert users throughout the design and operation of AutoML systems. HCAI is not a rigid framework but a set of guiding principles that extend traditional Human-Centered Design (HCD) to address the unique challenges of AI, including fairness, transparency, and user empowerment. This includes:

- Transparency & Explainability: AutoML tools should make fairness-related decisions and processes understandable to non-expert users. This can be achieved through intuitive dashboards, clear explanations of bias metrics, and visualizations that demystify how fairness interventions affect outcomes (Ghai, 2023) (Weerts et al., 2024). Such transparency is essential for building trust and enabling users to meaningfully engage with fairness features.
- Controllability: Users must have practical means to influence fairness settings and outcomes. This includes adjustable parameters for fairness constraints, the ability to select or prioritize different fairness definitions, and interfaces that allow users to intervene or provide feedback during the model selection and evaluation process (Rahman et al., 2024) (Ghai, 2023).
- Reliability & Robustness: Fairness features in AutoML should be designed to perform consistently and safely, even as data or user requirements evolve. This ensures that fairness interventions do not inadvertently introduce instability or degrade model performance in unpredictable ways (Cohen et al., 2025) (Zhang et al., 2020).
- Bias Mitigation & Fairness: HCAI-informed AutoML systems should proactively identify, prevent, and mitigate biases. This involves surfacing fairness metrics, highlighting potential trade-offs, and making corrective actions clear and actionable for users—especially those without deep technical expertise (Wu & Wang, 2021) (Weerts et al., 2024).
- User Feedback Loops: Effective AutoML tools incorporate mechanisms for users to provide feedback, both on fairness outcomes and the usability of fairness features. This feedback should be used to iteratively improve the system, ensuring it remains aligned with user expectations and ethical standards (Ghai, 2023) (Weerts et al., 2024).

Hence, applying HCAI to fairness in AutoML means designing for transparency, controllability, reliability, bias mitigation, and feedback, ensuring that fairness features are understandable, actionable, and aligned with the real-world needs and values of non-expert users.

---

**A2.3.2 Design Thinking**

Design thinking presents a robust, empathy-driven, and iterative framework for developing fairness features in AutoML that are technically robust and human-centered. This methodology is particularly effective in addressing user challenges related to explainability, trust, and oversight, all of which are critical for ensuring fairness in AI. The relevance from the context of AutoML includes:

- Empathy and User Understanding: The process commences with a profound empathy for non-expert users, aiming to uncover their genuine concerns regarding fairness, transparency, and control within AutoML systems (Stone et al., 2022). This involves engaging with users to comprehend their mental models, pain points (such as the lack of clarity on how fairness is measured), and the factors that would enhance their trust and adoption of fairness features.
- Problem Definition: Design thinking emphasizes the clear articulation of user-centric problems, such as "How might we assist non-expert users in understanding and controlling fairness outcomes in AutoML?". This step ensures that technical development remains anchored in real user needs, rather than abstract fairness metrics.
- Ideation: Teams collaboratively generate a variety of solutions, such as interactive fairness dashboards, context-sensitive explanations, or workflow-integrated feedback loops (Stone et al., 2022). The emphasis is on fostering creativity and inclusivity by drawing on diverse perspectives to conceptualize features that genuinely resonate with users.
- Prototyping: Rapid, low-fidelity prototypes of fairness features (e.g., mockups of bias alerts, transparency toggles, or user-adjustable fairness constraints) are developed and iteratively refined (Stone et al., 2022). This approach facilitates early testing of ideas without significant investment, ensuring that solutions evolve in response to user feedback.
- Testing and Iteration: Prototypes are evaluated with real users to gather feedback on usability, clarity, and trustworthiness (Stone et al., 2022). Insights from testing drive continuous refinement, ensuring that fairness features are not only technically robust but also intuitive and empowering for non-experts.

Hence, design thinking's empathy-driven, iterative development process provides a framework for building fairness features that could support in responding to real user challenges in explainability, trust, and oversight.

—---------—-------------------------------------------------------------------------------------

### A2.3.3 Affordance Theory

Affordance Theory covers the design of fairness features in AutoML systems, ensuring they are both accessible and actionable for users without specialized expertise. Within this framework, an affordance pertains to the cues and properties present in the AutoML interface that assist users in understanding how and when to engage with fairness-related components, the significance of these components, and the potential consequences of their interactions. In the context of AutoML it can include:

- Perceptible Fairness: Affordances Designers are required to develop explicit visual, textual, or interactive indicators that denote the relevance of fairness considerations. This may involve emphasizing datasets or model outputs where fairness metrics are either available or necessary (Feng & Mcdonald, 2023) (Narayanan, 2023). For instance, a fairness icon or a color-coded alert could serve to indicate the potential presence of model bias, thereby directing user attention at the appropriate juncture.
- Mapping Complex Concepts to Familiar Patterns: Complex fairness concepts, such as disparate impact and equal opportunity, should be translated into familiar user interface elements. Examples include sliders for adjusting fairness thresholds, tooltips for metric explanations, or comparison charts illustrating the trade-offs between accuracy and fairness (Feng & Mcdonald, 2023) (Narayanan, 2023). This strategy capitalizes on users' pre-existing mental models, thereby reducing cognitive load and rendering fairness features more accessible to non-experts.
- Actionable Interactivity: Users should be afforded the capability to inspect, adjust, or simulate the impact of fairness constraints through intuitive controls, such as toggles, dropdowns, or interactive dashboards (Narayanan, 2023). For example, enabling users to select which fairness metric to optimize or to preview the effects of changes on model outcomes empowers them to make informed decisions without requiring extensive technical expertise.
- Explicit Communication of Trade-offs: Affordances should explicitly communicate the trade-offs between accuracy and fairness, utilizing visualizations or summary statements to elucidate the impact of user decisions (Narayanan, 2023). This transparency enhances user agency and trust, facilitating non-experts' comprehension of the consequences associated with prioritizing different objectives.
- Consistency and Feedback: Consistent layout, iconography, and feedback mechanisms ensure that users can reliably interpret and engage with fairness affordances throughout the AutoML workflow. Immediate feedback, such as confirmation messages and updated metrics, reinforces correct actions and bolsters user confidence.

—-------—-----------------------------------------------------------------------------------

In the context of AutoML, research underscores the inadequacy of current fairness affordances and the necessity for more intuitive, user-aligned features to support fairness-aware model development for non-experts (Feng & Mcdonald, 2023) (Narayanan, 2023). Hence, applying Affordance Theory in AutoML fairness design means making fairness features perceptible, understandable, and actionable for non-expert users. This involves mapping technical concepts to familiar UI patterns, providing clear cues and feedback, and ensuring users can easily inspect, influence, and understand fairness settings and their implications

#### A2.3.4 Cognitive Load Theory

Cognitive Load Theory (CLT) is of significant importance in the design of AutoML interfaces that present fairness features to non-expert users. This theory posits that humans possess a limited capacity for information processing, necessitating that interfaces minimize extraneous cognitive load and effectively manage the intrinsic complexity of fairness concepts (Fox & Rey, 2024) (Kompaniets & Chemerys, 2019).

To achieve this, information should be simplified. Fairness concepts and metrics ought to be presented in digestible, progressive layers, employing progressive disclosure to reveal additional details only when necessary, thereby preventing users from becoming overwhelmed by technical jargon or data (Fox & Rey, 2024). For instance, an initial summary view might display a straightforward fairness Analysis Remarks, such as "No major bias detected," with options available to expand for more detailed metrics or definitions. Furthermore, visualization should be prioritized to avoid overwhelming users. Clear, intuitive visualizations—such as color-coded indicators, charts, or sliders—should be utilized to represent fairness metrics instead of raw numbers or complex tables (Fox & Rey, 2024) (Kompaniets & Chemerys, 2019). Visual cues facilitate users' quick comprehension of fairness status and trends without requiring deep statistical knowledge.

Implement guided workflows and prompts to assist users in navigating fairness-related tasks, such as selecting appropriate fairness metrics or addressing identified biases (Roundtree, 2025). The use of tooltips, wizards, and contextual help can mitigate uncertainty and support users at critical decision points. To alleviate cognitive load, design interfaces that enable users to recognize options and definitions rather than recall them. This can be achieved by incorporating dropdown menus with explanations or maintaining persistent glossaries for fairness-related terminology (Faudzi et al., 2022)(Yamani et al., 2024). This approach aligns with Nielsen's heuristic of "Recognition rather than recall," facilitating user interaction with fairness features without necessitating the memorization of complex procedures or terminology (Faudzi et al., 2022) (Yamani et al., 2024).

Furthermore, such suggestions employ language and user interface patterns that correspond to the mental models of non-expert users, reflecting their expectations and real-world understanding rather than relying on technical or abstract terms (Roundtree, 2025). For instance, instead of using the term "equalized odds," opt for "Ensuring similar accuracy for all groups," accompanied by a plain-language explanation.Evidence from Research

Studies confirm that poorly designed interfaces increase cognitive load and disengage users, especially when presenting complex content like AI fairness (Faudzi et al., 2022). Applying UX heuristics—especially those focused on clarity, feedback, and reducing memory load—improves usability, trust, and fairness perceptions in AI tools (Roundtree, 2025)

Hence, applying Cognitive Load Theory and related UX heuristics to AutoML fairness features could result in simplifying, visualizing, and guiding user interactions to minimize cognitive overload. This ensures fairness affordances are not just technically available, but actually usable and trustworthy for non-expert users (Fox & Rey, 2024) (Yamani et al., 2024).

## A3. Relevance of the theories for the research

The theories presented above including TAM, Design Thinking, Innovation Diffusion, Affordance Theory, and Human-Centered AI jointly inform the thesis' HCI-based evaluation framework. The summary of key insights from the theories are represented below:

| Theory | Relevance to AutoML Fairness Design | Influence on HCI Evaluation Framework |
| --- | --- | --- |
| Technology Acceptance Model (TAM) | Explains user acceptance via perceived usefulness and ease of use | Guides interface design audits and fairness feature adoption likelihood |
| Innovation Diffusion Theory (IDT) | Helps analyze adoption curves and enterprise buyer friction | Informs fairness nudges, onboarding needs, and feature discoverability |
| Design Thinking | Promotes iterative, human-centered fairness feature design | Informs UI/UX and interaction heuristics, HITL (Human-in-the-Loop) design |
| Affordance Theory | Focuses on how interface elements suggest their function | Shapes feedback mechanism evaluation and user control over fairness settings |
| Human-Centered AI / UX Strategy | Emphasizes ethical, transparent, inclusive systems | Underpins entire HCI framework—transparency, explainability, oversight, accountability |

—-------—----------------------------------------------------------------------------------

Specifically, these theories shape the criteria used in heuristic audits (e.g., ease of use, affordance of fairness controls), the assessment of user trust and cognitive load, and the analysis of fairness feature adoption by enterprise buyers.

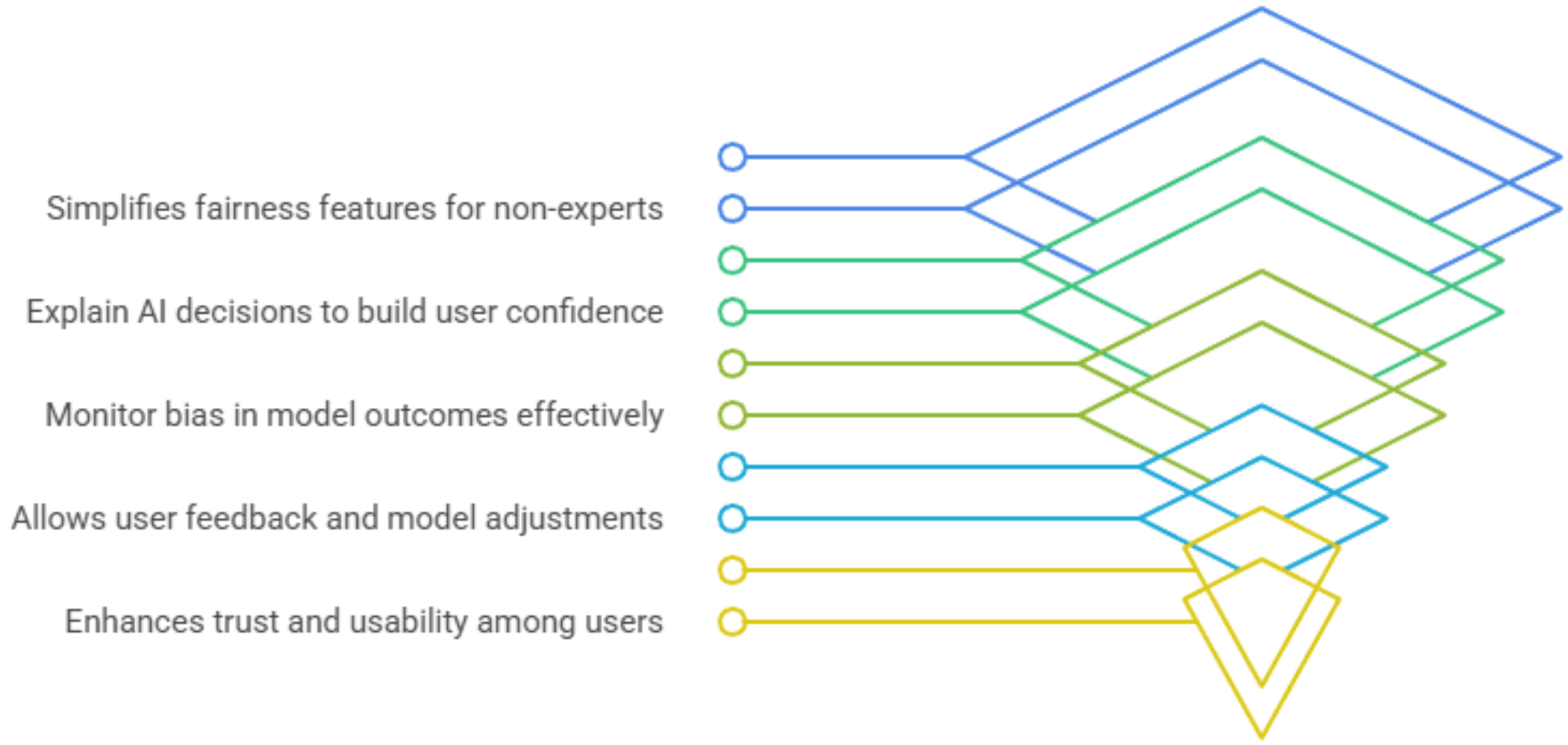

Their integration ensures that the proposed framework is both academically grounded and practically aligned with the needs of AutoML tool providers seeking responsible, inclusive adoption.

# B. AutoML tools and libraries in the context of the thesis

### Section Overview

**Tool Overview:** The section presents a range of AutoML tools and libraries (e.g., Google AutoML, H2O.ai, DataRobot, AutoGluon, TPOT, Auto-Keras) used for classification, regression, and clustering, noting strong industry adoption and performance.

**Selection Sources:** The section identifies popular code libraries and UI-based tools through GitHub and web searches, focusing on widely cited, actively maintained platforms with significant community support.

**Feature Coverage:** The section examines feature coverage for end-to-end ML workflows, including data management, feature engineering, pipeline management, model training, hyperparameter optimization, visualization, logging, and user interfaces.

**Feature Analysis Limitations:** The section highlights that feature-based analysis alone does not sufficiently filter tools suitable for evaluation.

**Criteria-Based Selection:** The section applies criteria-based selection considering usability, documentation quality, data handling, platform familiarity, interface accessibility, and support for tabular data and hyperparameter optimization.

**Evaluation Candidates:** The section identifies libraries for evaluation (AutoGluon, FLAML, H2O AutoML, PyCaret) and UI-based tools for benchmarking (RapidMiner Studio, H2O AutoML, DataRobot, Dataiku).

This section provides an overview of prominent AutoML tools and libraries, highlighting their capabilities, adoption, and relevance for enterprise and research use. It examines both code-based libraries and UI-driven platforms, focusing on features that support end-to-end machine learning workflows. The section also outlines the methodology for selecting tools suitable for evaluation and benchmarking, emphasizing usability, accessibility, and comprehensive functionality.

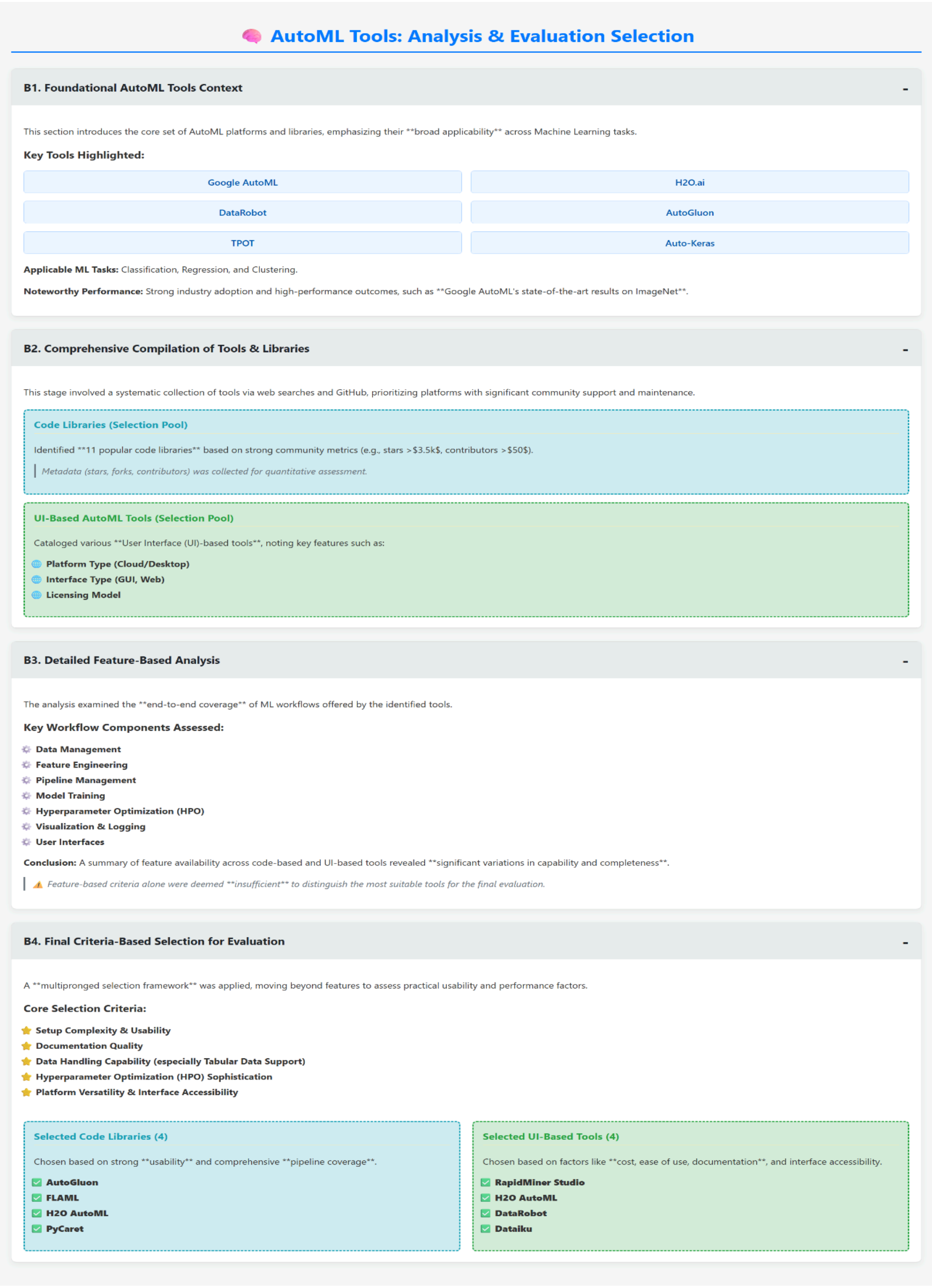
AutoML Tools: Analysis & Evaluation Selection
B1. Foundational AutoML Tools Context
This section introduces the core set of AutoML platforms and libraries, emphasizing their **broad applicability** across Machine Learning tasks.
Key Tools Highlighted:
Google AutoML
H2O.ai
DataRobot
AutoGluon
TPOT
Auto-Keras
Applicable ML Tasks: Classification, Regression, and Clustering.
Noteworthy Performance: Strong industry adoption and high-performance outcomes, such as **Google AutoML's state-of-the-art results on ImageNet**.
B2. Comprehensive Compilation of Tools & Libraries
This stage involved a systematic collection of tools via web searches and GitHub, prioritizing platforms with significant community support and maintenance.
Code Libraries (Selection Pool)
Identified **11 popular code libraries** based on strong community metrics (e.g., stars >$3.5k$, contributors >$50$).
Metadata (stars, forks, contributors) was collected for quantitative assessment.
UI-Based AutoML Tools (Selection Pool)
Cataloged various **User Interface (UI)-based tools**, noting key features such as:
Platform Type (Cloud/Desktop)
Interface Type (GUI, Web)
Licensing Model
B3. Detailed Feature-Based Analysis
The analysis examined the **end-to-end coverage** of ML workflows offered by the identified tools.
Key Workflow Components Assessed:
Data Management
Feature Engineering
Pipeline Management
Model Training
Hyperparameter Optimization (HPO)
Visualization & Logging
User Interfaces
Conclusion: A summary of feature availability across code-based and UI-based tools revealed **significant variations in capability and completeness**.
Feature-based criteria alone were deemed **insufficient** to distinguish the most suitable tools for the final evaluation.
B4. Final Criteria-Based Selection for Evaluation
A **multipronged selection framework** was applied, moving beyond features to assess practical usability and performance factors.
Core Selection Criteria:
Setup Complexity & Usability
Documentation Quality
Data Handling Capability (especially Tabular Data Support)
Hyperparameter Optimization (HPO) Sophistication
Platform Versatility & Interface Accessibility
Selected Code Libraries (4)
Chosen based on strong **usability** and comprehensive **pipeline coverage**.
AutoGluon
FLAML
H2O AutoML
PyCaret
Selected UI-Based Tools (4)
Chosen based on factors like **cost, ease of use, documentation**, and interface accessibility.
RapidMiner Studio
H2O AutoML
DataRobot
Dataiku

---

## B1. AutoML tools context

Several AutoML tools and libraries are available, including Google AutoML, H2O.ai, DataRobot, AutoGluon, TPOT, and Auto-Keras. These tools and libraries support a variety of machine-learning tasks, including classification, regression, and clustering, and can be used with a wide range of data types and formats. AutoML tools and libraries have been widely adopted in the industry and have demonstrated strong performance in a variety of tasks. For example, in a study published in Nature, researchers used Google AutoML to develop a machine learning model that achieved state-of-the-art performance on the ImageNet image classification task.

## B2. Tools and libraries compiled

### B2.1 Code libraries

The AutoML coding libraries examined in this study were identified through web-based search queries (e.g., "top AutoML tools" and "top AutoML libraries") using sources such as Google. This approach surfaced widely referenced libraries commonly featured in academic discussions, technical blogs, and open-source repositories.

To streamline our list, this research focused on identifying AutoML libraries with an associated GitHub repository. During this process, the research discovered a valuable resource called the AutoML GitHub repository, which cataloged an impressive collection of 56 AutoML code libraries.

To optimize the analysis, a criterion based on popularity was established. Specifically, the research prioritized the most widely recognized 11 AutoML code libraries listed on the AutoML GitHub repository; specifically, those with more than 3.5k stars and over 50 contributors were considered for the analysis. These libraries were considered for further evaluation, as described in the subsequent section.

| Name | Description | No of stars | Forks | Contributors |
|---|---|---|---|---|
| AutoGluon | AutoML Toolkit for Deep Learning | 7.3k | 864 | 119 |
| AutoSklearn | an automated machine learning toolkit and a drop-in replacement for a scikit-learn estimator | 7.5K | 1.3K | 80 |
| FLAML | Fast and lightweight AutoML (paper) | 3.7k | 492 | 86 |
| H2O AutoML | Automatic Machine Learning by H2O.ai | 6.8K | 2k | 176 |
| Hyperopt | Distributed Asynchronous Hyperparameter Optimization in Python | 7.1k | 1k | 103 |
| Ludwig | A toolbox built on top of TensorFlow that allows to train and test deep learning models without the need to write code | 10.9k | 1.2k | 155 |
| Microsoft Neural Network Intelligence (NNI) | An open source AutoML toolkit for neural architecture search and hyper-parameter tuning | 13.8k | 1.8k | 184 |
| Optuna | Define-by-run hyperparameter optimization framework | 10k | 971 | 237 |
| Ray Tune | Scalable Hyperparameter Tuning¶ | 31.7K | 5.4K | 978 |
| TPOT | One of the very first AutoML methods and open-source software packages | 9.6k | 1.6k | 78 |
| Pycaret | An open-source, low-code machine learning library in Python | 8.6k | 1.7k | 124 |

### B2.2 AutoML tools

The AutoML tools included in this study were identified through an exploratory search using publicly available online sources, including Google search results for “top AutoML tools.” This approach prioritizes widely referenced and actively maintained platforms that are frequently cited in academic literature, practitioner blogs, and technology review repositories. The resulting list reflects tools that are both prominent in public discourse and accessible for research purposes. These tools were considered for further evaluation, as discussed in the subsequent section.

| Name of the tool | Name of the automl feature | UI based or code based | Desktop or cloud | Type of license |
|---|---|---|---|---|
| Rapidminer Studio | Auto Model | UI Based | Desktop | Educational |
| H2O Studio | AutoML | UI Based | Desktop | Open source |
| Google | AutoML cloud | UI based | Cloud | Paid (with some free credits) |
| Data Robot | Automated Machine Learning | UI Based | Cloud | 30 day free trial |
| IBM watsonX | AutoAI | UI Based | Cloud | Free trail |
| Dataiku | AutoML | UI Based | Desktop | Free trail |
| KNIME | Automated Machine Learning | UI Based | Desktop | Open source |
| Alteryx | AutoML | UI Based | Desktop | Free trail |

There are no deterministic ways to identify the extent of use or downloads of these tools.

## B3. Feature based selection of tools and libraries

The research further analyzed their features to understand if the repository/tool supports the workflow covering all stages of the machine learning pipeline. These include the following, as highlighted by (Majidi et al., 2022):

| Features | Descriptions |
|---|---|
| Data management | Using AutoML for tasks related to handling and processing data used for building ML models, including transformation, analysis, and labeling. |
| Feature engineering | Using AutoML to automate feature extraction from data using domain knowledge. |
| Pipeline management | Using AutoML to automate the end-to-end management of ML models or a set of models. |
| Utility functions | Functions provided by AutoML tools to perform common tasks and simplify the ML pipeline. |
| Model management | Using AutoML for tasks related to training ML algorithms and managing models, such as evaluation, testing, and monitoring. |
| Hyperparameter optimization (HPO) | Using AutoML to find the best combination of hyperparameters for model performance. |
| Visualization | Graphically representing data and information for interpretability and analysis. |
| Logging | Tracking and storing records related to the execution of ML models, data inputs, processes, and outputs. |
| Storage management | AutoML tools having options for data storage and management. |
| User interface | The interactive element of AutoML tools that allows ML practitioners to easily interact with features and code. |

## B3.1 Code libraries

A compilation of features in these libraries is tabulated below:

| Code based libraries | Data Management | Feature Engineering | Pipeline Management | Utility Functions | Model Management | HPO | Visualization | Logging | Storage Management | User Interface (GUI) |
|---|---|---|---|---|---|---|---|---|---|---|
| AutoGluon | Y | Y | L | Y | Y | Y | Y | N | N | N |
| AutoSklearn | Y | Y | Y | Y | Y | Y | L | N | N | N |
| FLAML | N | L | N | Y | Y | Y | N | N | N | N |
| H2O AutoML | Y | Y | Y | Y | Y | Y | L | N | N | N |
| Hyperopt | N | N | N | L | L | Y | N | N | N | N |
| Ludwig | Y | L | L | Y | Y | L | L | N | N | N |
| Microsoft NNI | L | L | Y | Y | Y | Y | N | Y | N | N |
| Optuna | N | N | L | L | L | Y | N | N | N | N |
| Ray Tune | N | N | N | Y | Y | Y | Y | Y | N | N |
| TPOT | Y | Y | L | Y | Y | Y | L | N | N | N |
| PyCaret | Y | Y | Y | Y | Y | Y | Y | Y | N | N |
| Features available in | 6/11 | 5/11 | 4/11 | 9/11 | 9/11 | 10/11 | 3/11 | 3/11 | 0/11 | 0/11 |

(Y – Feature exists in the application; N – Feature does not exist in the application; L – Limited features in the application)

Specific details of the features are compiled in the table below

| Tool | Feature compilation | Sources |
|---|---|---|
| AutoGluon | Data loading (files, pandas); automated/customizable feature engineering; limited pipeline customization; model explainability and evaluation; leaderboard, saving/loading; automated hyperparameter optimization; leaderboard and explainability visualizations; no explicit logging; no storage management; Python/CLI interface. | 1, 2, 3 |

———————————————————————————————

| | | |
|---|---|---|
| AutoSklearn | scikit-learn data loading; automated preprocessing and feature selection; automated pipeline construction; model evaluation and scoring; saving/loading support; automated hyperparameter tuning; basic CLI output visualizations; no logging; no storage management; Python/CLI interface. | 4 |
| FLAML | No internal data handling; basic preprocessing; no pipeline builder; model evaluation and explainability; model saving/loading; automated hyperparameter tuning; no visualization; no logging; no storage management; Python/CLI interface. | 5 |
| H2O AutoML | Data loading (files, pandas); automated preprocessing and feature selection; pipeline construction; model explainability and evaluation; leaderboard and model export/import; automated hyperparameter tuning; limited visualizations; no logging; no storage management; Web/CLI/Python interface. | 6 |
| Hyperopt | No data handling; no feature engineering or pipeline management; basic parameter space optimization; limited model support; hyperparameter tuning only; no visualization, logging or storage; CLI interface. | 7 |
| Ludwig | Data loading (files, pandas); basic preprocessing; limited pipeline flexibility; model explainability and evaluation; model saving/loading; basic hyperparameter search; basic visualizations; no logging or storage management; GUI available. | 8 |
| Microsoft NNI | Limited data support; basic feature engineering; pipeline experiments; experiment management and evaluation; model saving and tracking; automated hyperparameter optimization; some tracking, no rich visualizations; experiment logging; no storage management; CLI/GUI interface. | 9, 10, 11 |
| Optuna | No data handling or feature engineering; limited pipeline tracking; experiment tracking only; automated hyperparameter optimization; basic plots; no true logging; no storage; CLI/Web Dashboard interface. | 12, 13 |
| Ray Tune | No data or feature engineering; no pipeline building; experiment management and evaluation; checkpointing and tracking; automated hyperparameter optimization; TensorBoard integration; experiment logging; no storage management; CLI/Python/Web interface. | 14 |

——-------——------------------------------------------------------------------------------

| | | |
|---|---|---|
| TPOT | Data loading (files, pandas); automated preprocessing and feature selection; limited pipeline customization; model evaluation and export; model saving/loading; automated hyperparameter optimization; basic visual output; no logging or storage management; CLI/Python interface. | 15 |
| Pycaret | Data handling and cleaning; automated feature engineering; pipeline building and reuse; model training, tuning, and comparison; explainability tools; automated hyperparameter search; visual outputs; logging of experiments; model storage; Python/CLI interface with low-code syntax. | 16, 17 |

The above approach of feature-based criteria did not help in filtering specific libraries for evaluation.

## B3.2 AutoML tools

A compilation of features in these tools is tabulated below:

| UI based tool | Data Management | Feature Engineering | Pipeline Management | Utility Functions | Model Management | HPO | Visualization | Logging | Storage Management | User Interface (GUI) |
|---|---|---|---|---|---|---|---|---|---|---|
| RapidMiner Studio | Y | Y | Y | Y | Y | Y | Y | N | Y | Y |
| Google AutoML | Y | Y | L | Y | Y | Y | L | N | N | Y |
| DataRobot | Y | Y | Y | Y | Y | Y | Y | L | N | Y |
| H2O AutoML | Y | Y | Y | Y | Y | Y | Y | N | N | Y |
| IBM Watson AutoML | Y | Y | Y | Y | Y | Y | Y | N | N | Y |
| Dataiku | Y | L | L | Y | L | L | L | N | N | Y |
| KNIME | Y | Y | Y | Y | Y | Y | Y | L | Y | Y |
| Alteryx | Y | Y | L | Y | L | L | Y | L | Y | Y |

(Y – Feature exists in the application; N – Feature does not exist in the application; L – Limited features in the application)

The specific features of these tools are compiled below:

| Tool | Feature compilation | Sources |
|---|---|---|
| RapidMiner Studio | Data import, cleaning, enrichment; transformation and feature creation; visual workflow interface; data prep, scripting; model building, evaluation, deployment; hyperparameter tuning; robust data/model visualization; integrations; drag-and-drop GUI; no logging. | 1, 2 |
| H2O AutoML (Open Source) | Data upload, cleaning, preparation; automated preprocessing (imputation, encoding, standardization), feature selection/extraction; automatic pipeline generation; model explanation, export, evaluation; leaderboard-based model management; grid search and cross-validation for hyperparameter tuning; explainability visualizations; no built-in logging; file handling; no-code GUI. | 3, 4 |

—-------—-------------------------------------------------------------------------------------

| Google AutoML | Data import, preprocessing; automated/user-guided feature engineering; end-to-end pipeline automation; cleaning, formatting, deployment; model training/deployment; automated hyperparameter search; dashboard visualizations; no logging; file handling; no-code GUI. | 5, 6 |
|---|---|---|
| DataRobot | Data preparation, management; automated/user-guided feature engineering; pipeline creation and management; interpretability tools, model deployment; training/evaluation; automated hyperparameter tuning; performance visualizations; model monitoring and alerts; file handling; web-based GUI. | 7 |
| IBM Watson AutoML (AutoAI) | Data upload, preparation, preprocessing; automatic feature detection, transformation, encoding; pipeline generation and ranking; model evaluation, deployment, monitoring; leaderboard and promotion for model management; automated hyperparameter search; performance visualizations; monitoring for deployed models; file handling; drag-and-drop GUI. | 8, 9 |
| Dataiku | Data integration and management; plugin/script-based feature engineering; visual workflow interface; data prep and integration tools; AutoML-driven model management; AutoML-based hyperparameter search; dashboards and visual workflows; no logging; file handling; drag-and-drop GUI. | 10 |
| KNIME | Data import, cleaning, blending; node-based feature engineering; workflow design for pipelines; wide utility node coverage; model building, training, deployment; hyperparameter optimization via nodes; comprehensive visualizations; execution logs; storage integration; drag-and-drop GUI. | 11 |
| Alteryx | Data blending, enrichment, management; built-in tools for feature engineering; workflow automation; data prep, blending, analysis; predictive analytics; search tools for hyperparameter tuning; dashboarding and reporting; workflow execution monitoring; storage integrations; drag-and-drop GUI. | 12 |

The above approach to feature-based criteria does not help in filtering specific tools for evaluation.

# B4. Criteria based selection of AutoML tools and libraries

## B4.1 Code libraries

To identify suitable AutoML libraries for analysis, seven key criteria were considered, and the tools were reviewed accordingly:

- Setup and platform familiarity: The library setup and familiarity requirements must be low to facilitate non-expert use.
- Documentation Quality: The library should provide comprehensive guidance materials, such as tutorials, manuals, and community forums, to support onboarding and usage.
- Data Handling Capability: The library must support convenient mechanisms for data loading, cleaning, and preparation.
- Support for tabular data: The library shall support tabular data for classification or regression tasks; and
- Hyper-parameter optimization: The library has a hyper-parameter optimization framework integrated in it.
- Supports a complete autoML pipeline: The library should be a complete automated machine learning library.

| Tool | Set-up & Platform Familiarity | Documentation Quality | Data Handling Capability | Support for Tabular Data | Hyper-parameter Optimization | Complete AutoML Tool |
|---|---|---|---|---|---|---|
| AutoGluon | Y | Y | Y | Y | Y | Y |
| AutoSklearn | L | Y | Y | Y | Y | Y |
| FLAML | Y | Y | L | Y | Y | Y |
| H2O AutoML | Y | Y | Y | Y | Y | Y |
| PyCaret | Y | Y | Y | Y | Y | Y |
| Ludwig | Y | Y | Y | Y | L | Y |
| TPOT | Y | Y | Y | Y | Y | Y |
| Microsoft NNI | L | Y | L | L | Y | N |
| Hyperopt | Y | L | N | L | Y | N |
| Optuna | Y | Y | N | L | Y | N |
| Ray Tune | Y | Y | N | L | Y | N |

(Y –exists in the library ; N – Does not exist in the library ; L – Limited in the library)

—-------—---------------------------------------------------------------------------------

Libraries that were not considered owing to operational constraints: Among the tools that support tabular data for classification or regression tasks, TPOT, Ludwig, and AutoSKlearn could not be tested owing to configuration issues, need for detailed technical knowledge, and lack of recent updates or issue resolutions to the libraries.

Libraries considered for evaluation: AutoGluon, FLAML, H2O AutoML, and Pycaret were considered for evaluation.

### B4.2 AutoML tools

To identify suitable AutoML tools for analysis, a multipronged selection framework was applied. The tools were shortlisted based on the following methodological criteria:

- Cost and Access: The tool should be freely available or offer a research trial to support non-commercial academic use.
- Set-up and platform familiarity: The tool's set-up and platform familiarity requirements (e.g., enterprise setup or platform-specific dependencies) must be low to facilitate non-expert use.
- Documentation Quality: The tool should provide comprehensive guidance materials, such as tutorials, manuals, and community forums, to support onboarding and usage.
- Ease of Use: The interface and workflow should be sufficiently intuitive for users with minimal machine learning expertise.
- Data Handling Capability: The tool must support convenient mechanisms for data loading, cleaning, and preparation.
- Platform Versatility: This tool should be usable in multiple environments (including cloud-based and local/desktop systems).
- Interface Accessibility: The tool should provide a user-friendly interface suitable for non-expert users, preferably with visual or low-code design elements.

| Tool | Cost & access | Set-up and platform familiarity | Documentation quality | Ease of use | Data handling capability | Platform versatility | Interface accessibility |
|---|---|---|---|---|---|---|---|
| RapidMiner Studio | Y | N | Y | Y | Y | Y | Y |
| H2O AutoML | Y | N | Y | Y | Y | Y | Y |
| Google AutoML | L | Y | Y | Y | Y | L | Y |
| DataRobot | L | N | Y | Y | Y | Y | Y |
| IBM Watson AutoML (AutoAI) | L | Y | Y | Y | Y | L | Y |
| Dataiku | L | N | Y | Y | Y | Y | Y |
| KNIME | Y | N | L | Y | Y | Y | Y |
| Alteryx | L | N | Y | Y | Y | Y | Y |

(Y –exists in the application / free or freemium; N – Does not exist in the application; L – Limited in the application/ Limited trial/ cloud only)

Tools that were not considered owing to operational constraints: KNIME was not considered for analysis owing to insufficient guidance documentation and bug resolution. Google and IBM were not considered because of the need for enterprise setup and platform awareness. Alteryx was not considered due to a lack of response from the organization for research trials.

Tools considered for feature analysis and benchmarking: RapidMiner Studio, H2O AutoML, DataRobot, and Dataiku were considered for analysis and benchmarking. While H2O Studio was considered for feature analysis, the tool could not be used for testing due to the bugs and failures on the UI.

# C. Fairness datasets in hiring for further research

**Section Overview**

**Fairness Context:** The section emphasizes the need to assess algorithmic hiring systems across all recruitment stages, accounting for sensitive attributes and using fairness metrics such as Disparate Impact.

**Data-Centric Focus:** The section stresses the importance of dataset selection, curation, and transparency for fairness evaluation, highlighting challenges from incomplete documentation and ethical data use.

**Literature Insights:** The section reviews datasets across multiple domains, including hiring, finance, and criminal justice, noting limitations of common fairness datasets like Adult, COMPAS, and German Credit, and the need for domain-specific evaluation.

**Dataset Selection Methodology:** The section outlines the search and filtering process to identify open-source, tabular datasets with demographic and employment-related variables relevant for AutoML benchmarking.

**Compiled Datasets:** The section presents six shortlisted datasets, including Utrecht Fairness Recruitment, Employees Evaluation for Promotion, Employee Satisfaction Index, and IBM HR Analytics, chosen for accessibility, relevance, and tabular structure.

**Preliminary Data Analysis:** The section details dataset quality, missing values, outliers, target distributions, feature correlations, and fairness disparities across demographic groups to support benchmarking and evaluation.

This section examines fairness in machine learning within the context of algorithmic hiring, emphasizing the need to assess bias across all stages of recruitment. It highlights the importance of high-quality, transparent datasets for evaluating fairness, focusing on tabular data with demographic and employment-related variables. Six publicly available datasets are compiled and preliminarily analyzed to support benchmarking of AutoML tools and libraries in fair and ethical decision-making.

## C1. Fairness in machine learning context

Algorithmic hiring requires a broader perspective to consider all stages of recruitment and the biases embedded in these technologies. The broader perspective includes understanding fairness in the context of data and processes. It should account for sensitive attributes and evaluate systems using measures such as Disparate Impact to address potential biases and discrimination (Fabris et al., 2022).

## C2. Datasets: The baseline for any evaluation or benchmarking

Algorithmic fairness requires a data-centric perspective, focusing on dataset selection, dataset curation, and measuring fairness in practical scenarios where access to testing sensitive attribute information may be unavailable (Fabris et al., 2022).

However, insufficient documentation of datasets used in algorithmic fairness research presents a significant challenge, resulting in collective data documentation debt within the algorithmic fairness community. It is essential to address this issue to ensure the transparency, accountability, and ethical utilization of data-driven algorithms in decision-making processes that have a profound impact on people's research (Fabris et al., 2025).

## C3. Literature review

Datasets used in algorithmic fairness research encompass a wide range of domains, including criminal justice, education, search engines, online marketplaces, emergency response, social media, medicine, hiring, and finance. This diversity highlights the complexity of fairness challenges across sectors and emphasizes the need to consider domain-specific nuances when developing fair machine learning algorithms. A detailed analysis of over 200 datasets exhibited the limitations of popular fairness datasets such as Adult, COMPAS, and German Credit, indicating that they may not be appropriate as general-purpose fairness benchmarks. This underscores the importance of critically evaluating the suitability and biases present in commonly used datasets to ensure fair and unbiased algorithmic decision-making.

(Paullada et al., 2021) highlighted the limitations of current dataset collection practices. They express concerns about the potential biases and flaws in datasets resulting from the unrestricted collection of large amounts of data from the web and reliance on non-expert crowdworkers. Additionally, the study emphasized the ethical implications of biased dataset annotations, pointing out the harm that can arise from disproportionate associations between certain identities and negative labels. To address these challenges, the research advocated educational interventions that promote data literacy and self-advocacy, empowering individuals, especially students, to collect and analyze their own data using machine-learning pipelines. This also fostered technical literacy while encouraging the critical evaluation of ethical considerations surrounding data usage and model development (Paullada et al., 2021).

## C4. Methodology

Conducted a search on Google Scholar, Kaggle, HuggingFace, UCI Machine learning repository, Papers with code, and University of Padua with "fairness datasets" and filtered for content containing "hiring" or "employment" related content to identify the relevant datasets. In addition, datasets referred to in recent papers relating to algorithmic hiring, discrimination, and bias mitigation were also compiled for the purposes of the research. HuggingFace, the UCI Machine Learning Repository, and Papers with Code did not have tabular data and/or data relating to hiring. The University of Padua contained multiple datasets, including the Pymetrics bias group dataset; however, these datasets were not publicly accessible and hence were not considered for further analysis or research. After identifying the related sources, the datasets were analyzed for their relevance to the research using the following conditions: (1) the dataset is open source, (2) the dataset contains demographic and other variables relating to hiring/employment, and (3) the dataset is in tabular format. Hence, Datasets including WinoBias (which is a text dataset) (Zhao et al., 2018) and datasets containing images of faces of Caucasian individuals were not considered for the analysis, as they were not tabular datasets.

Most AutoML tools and libraries are equipped to support tabular data, and Automated Machine Learning (AutoML) tools have shown success in tabular data tasks (Blohm et al., 2021). Hence, to facilitate the benchmarking and evaluation of fairness considerations of these tools and libraries, only Tabular datasets were considered.

## C5. Datasets compiled

Six publicly available datasets were shortlisted after reviewing relevant literature on hiring and employment practices. Each dataset was assessed for inclusion based on three criteria: open-source accessibility, presence of demographic and employment-related variables, and tabular data structure. This ensured alignment with the study's focus on transparent, structured, and demographically diverse hiring data. List of the datasets compiled along with their contents are provided below:

| Dataset | Description & source | Contents | Reference |
| --- | --- | --- | --- |
| Utrecht Fairness Recruitment dataset | This dataset contains the recruitment decisions of four companies with over 500 candidates. For each candidate, the research had a few general descriptions (gender, age, sport) and indicators. The actual decision was also included. The dataset can be used to obtain basic data science experience, but also to gain a deeper understanding of fairness.<br><br>Source: Kaggle | Contains entries with values covering demographic information including gender, age, nationality, and other information including university grade, programming experience, and international exposure. | 1<br><br>2 |

—------—--------------------------------------------------------------------------------------

| | | | |
|---|---|---|---|
| Employees Evaluation for Promotion | The HR team stored data of the promotion cycle last year, which consists of details of all the employees in the company working last year and also whether they got promoted or not, but every time this process gets delayed due to so many details available for each employee, it becomes difficult to compare and decide.<br><br>Source: Kaggle | The dataset contains entries with key fields, including employee_id, department, region, education, gender, recruitment_channel, no_of_trainings, age, previous_year_rating, length_of_service, awards_won, avg_training_score, and is_promoted. | 3 |
| Employee Satisfaction Index Dataset | A fictional dataset was created to help the data analysts play around with the trends and insights on the employee jab satisfaction index. | The dataset consists of entries with fields, including emp_id, age, Dept, location, education, recruitment_type, job_level, rating, onsite, awards, certifications, salary, and satisfied. Note that this fictional dataset does not represent any real organization. | 4 |
| Detailed Analysis on campus recruitment & Campus Placement Prediction: Binary Classification | This dataset consists of placement data of students in an XYZ campus. It includes secondary and higher secondary school percentages and specialization. It also includes degree specialisation, type and Work experience and salary offers to the placed students | The dataset contains entries with variables including gender, school scores, graduation topic and work experience | 5, 6 |
| Employee Dataset(All in One) | The Synthetic Employee Records Dataset is a simulated dataset created to explore various data analysis and machine learning techniques in the context of human resources and employee management. This synthetic dataset mirrors the structure and characteristics of real employee data, while all the information contained within is entirely fictional and generated for illustrative purposes | The dataset comprises of entries with key fields including Employee ID, First Name, Last Name, Start Date, Exit Date, Title, Supervisor, Email, Business Unit, Employee Status, Employee Type, Pay Zone, Employee Classification Type, Termination Type, Termination Description, Department Type, Division Description, Date of Birth (DOB), State, Job Function, Gender, and Location. | 7 |
| IBM HR Analytics Employee Attrition & Performance | Uncover the factors that lead to employee attrition and explore important questions such as ‘show me a breakdown of distance from home by job role and attrition’ or ‘compare average monthly income by education and attrition.’ This is a fictional dataset created by IBM data scientists. | The dataset consists of entries with key fields including Education (ranging from 'Below College' to 'Doctor'), Environment Satisfaction, Job Involvement, Job Satisfaction, Performance Rating, Relationship Satisfaction, and Work-Life Balance. Each field is represented by coded values that indicate different levels of satisfaction or achievement. | 8 |

# D. Fairness metrics in hiring for further research

**Section Overview**

**Introduction to Fairness Metrics:** The section emphasizes the importance of evaluating algorithmic fairness in binary classification, particularly for tabular data in critical domains, and explains why standard performance metrics are insufficient for capturing bias.

**Basic Classification Metrics:** The section explains definitions and limitations of accuracy, precision, recall, and F1 score are presented, highlighting their inability to capture group-level disparities and potential to mask discrimination against protected groups.

**Group Fairness Metrics:** The section explains foundational fairness metrics such as Demographic Parity, Equalized Odds, Equality of Opportunity, and Predictive Rate Parity are defined, including their formulas, fairness notions, and key limitations.

**Trade-offs and Metric Incompatibilities:** The section discusses theoretical and empirical challenges, including impossibility theorems and trade-offs between fairness goals, emphasizing the need for careful metric selection.

**Systemic Perspective on Fairness:** The section also supports examining multiple metrics collectively provides a broader understanding of bias, connecting fairness metrics to regulatory and legal contexts for detecting disparate impact and treatment.

**Implementation and Contextual Considerations:** The section also provides guidance on choosing fairness metrics based on application context, data characteristics, and potential harms, stressing that standard classification metrics alone are insufficient for fairness evaluation.

## D1. Introduction

Algorithmic fairness in binary classification, particularly with tabular data, is a critical concern in domains such as healthcare, finance, and criminal justice. While traditional performance metrics like accuracy, precision, recall, and F1 score are foundational for model evaluation, they are insufficient for assessing fairness. This section provides a structured, systemic analysis of both basic and group fairness metrics, highlighting their definitions, limitations, and interrelationships, and explains why fairness-specific metrics are essential in research and practice.

## D2. Basic Classification Metrics: Definitions and Limitations

| Metric | Formula | Description | Key Limitation in Fairness Context |
|---|---|---|---|
| **Accuracy** | (TP + TN) / (TP + FP + TN + FN) | Overall correctness | Masks group disparities, sensitive to class imbalance |
| **Precision** | TP / (TP + FP) | Correctness of positive preds | Ignores group-specific errors |
| **Recall** | TP / (TP + FN) | Coverage of actual positives | Does not guarantee group-level equity |
| **F1 Score** | 2 × (Precision × Recall) / (Precision + Recall) | Harmonic mean of prec/recall | Aggregates over all groups, hides bias |

TP: True Positive; TN: True Negative; FP: False Positive; FN: False Negative

These metrics aggregate performance across all samples, ignoring how errors are distributed among different demographic groups (e.g., race, gender, age). A model can achieve high accuracy or F1 score while systematically creating disadvantage for a protected group (Barocas et al., 2019). Similarly, high accuracy can be achieved by predicting only the majority class, which often corresponds to privileged groups, thus masking poor minority group performance (Hinduja et al., 2024; PMC10550141, 2023). Also, it is pertinent to note that precision, recall, and F1 score do not measure whether these rates are equitable across groups. Group-specific disparities can persist even when these metrics are high overall. These metrics do not detect or quantify discrimination, especially when positive prediction rates differ between groups. While these are considered "basic" metrics in the literature due to their origin in the confusion matrix and their ability to summarize overall performance (Suresh & Guttag, 2021), they are not suitable for fairness analysis, which requires group-level comparison.

## D3. Group Fairness Metrics: Definitions, Properties, and Systemic Roles

Researchers have extensively used True Positive Rate (TPR)(Sensitivity, Recall) False Positive Rate (FPR), Selection Rate, Demographic Parity Difference (DPD)(Statistical Parity), Equalized Odds Difference (ΔEO), Equality of Opportunity (EOpp), and Predictive Rate Parity (PRP)(Precision Parity) across binary classification problems. These group fairness metrics (Demographic Parity,

Equalized Odds, and Predictive Rate Parity) are considered foundational in fairness research because they explicitly measure and address disparate impact and treatment.

| Metric | Fairness Notion Captured | Key Limitation / Challenge |
|---|---|---|
| **True Positive Rate (TPR)** | Equality of Opportunity (EOpp) | Sensitive to threshold, misleading if risk distributions differ (Hegarty et al., 2025) |
| **False Positive Rate (FPR)** | Used in Equalized Odds (ΔEO) | Threshold-sensitive, can be manipulated |
| **Selection Rate** | Demographic Parity (DPD) | Ignores ground truth, masks error disparities |
| **Demographic Parity Difference (DPD)** | Statistical Parity | May incentivize ignoring relevant features; does not ensure equal error rates (Han et al., 2023) |
| **Equalized Odds Difference (ΔEO)** | Error Rate Parity | Hard to achieve with calibrated models; may conflict with other metrics (Souverain et al., 2024) |
| **Equality of Opportunity (EOpp)** | Opportunity Parity | Can still mislead if risk distributions differ |
| **Predictive Rate Parity (PRP)** | Precision Parity | Affected by base rates; conflicts with calibration and error rate parity (Canetti et al., 2018) |

An illustrative representation of the metrics in the context of AutoML use in hiring is provided below:

## Fairness Metrics in AI-Powered Hiring

**Illustrative Case Example: AI-Powered Hiring Bias**

An organization uses an AI-driven resume screening tool to predict whether an applicant should be invited to an interview ("Positive") or not ("Negative"). Fairness is analyzed across two demographic groups:

- Group A: Male (Protected Group)
- Group B: Female (Reference Group)

The model was tested on 1,000 applications (500 from each group).

**Contextual Definitions (Prediction Outcomes):**

- True Positive (TP): A qualified candidate who was correctly shortlisted for an interview.
- True Negative (TN): An unqualified candidate who was correctly rejected.
- False Positive (FP): An unqualified candidate who was incorrectly shortlisted (an undeserved interview).
- False Negative (FN): A qualified candidate who was incorrectly rejected (a missed opportunity for the organization and candidate).

### Model Outcome Data

| Group | True Positives (TP) | False Positives (FP) | True Negatives (TN) | False Negatives (FN) | Total |
|---|---|---|---|---|---|
| Male (A) | 180 | 70 | 200 | 50 | 500 |
| Female (B) | 120 | 40 | 250 | 90 | 500 |

### Fairness Metric Calculations

#### 1. True Positive Rate (TPR)

*Formula:*

$$\text{TPR} = \frac{TP}{TP + FN}$$

Group A (Male): $180/230 \approx 0.78$

Group B (Female): $120/210 \approx 0.57$

**TPR Difference:** $0.78 - 0.57 = \mathbf{0.21}$

The model identifies qualified female candidates (Group B) 21% less often than male candidates (Group A). This is the key component of Equality of Opportunity.

#### 2. False Positive Rate (FPR)

*Formula:*

$$\text{FPR} = \frac{FP}{FP + TN}$$

Group A (Male): $70/270 \approx 0.26$

Group B (Female): $40/290 \approx 0.14$

Male applicants (Group A) are incorrectly shortlisted (False Positive) more often ($0.26$ vs. $0.14$), suggesting a higher risk of being 'over-promoted' by the model.

#### 3. Selection Rate (Demographic Parity)

*Formula:*

$$\text{Selection Rate} = \frac{TP + FP}{\text{Total}}$$

Group A (Male): $250/500 = 0.50$

Group B (Female): $160/500 = 0.32$

**DPD (Difference):** $0.50 - 0.32 = \mathbf{0.18}$

Men are shortlisted 18 percentage points more frequently than women, showing a clear Demographic Parity Difference (DPD) in overall outcomes.

#### 4. Equalized Odds Difference (ΔEO)

*Formula:*

$$\Delta EO = |TPR_A - TPR_B| + |FPR_A - FPR_B|$$

Calculation: $|0.78 - 0.57| + |0.26 - 0.14|$

**ΔEO:** $0.21 + 0.12 = \mathbf{0.33}$

A ΔEO of $0.33$ suggests a substantial disparity in both error rates (True Positive and False Positive rates), indicating systematic unfairness across both metrics.

#### 5. Equality of Opportunity (EOpp)

**EOpp is achieved when $\text{TPR}_A \approx \text{TPR}_B$.**

**EOpp Difference:** $\mathbf{0.21}$

This metric emphasizes that qualified female candidates are significantly less likely to be correctly shortlisted than equally qualified males, reflecting a core failure in providing equal opportunity for the positive outcome.

**Summary Insight: Systemic Bias Detected**

The collective metric analysis reveals a systemic gender bias in the AI hiring model:

- **Selection Bias (DPD = 0.18):** Men are far more likely to be shortlisted overall.
- **Accuracy Bias (EOpp = 0.21):** Qualified female candidates are significantly less likely to be correctly shortlisted than equally qualified males.
- **Error Inflation:** False positives are higher for males, while false negatives are higher for females.

This model requires immediate intervention, such as rebalancing training data or applying fairness constraints during retraining, to close the significant fairness gap (ΔEO = 0.33).

Theoretical and empirical studies highlight key incompatibilities and trade-offs in fairness metrics that challenge their simultaneous optimization. Impossibility theorems demonstrate that achieving both predictive parity (equal positive predictive value) and error rate balance (equal false positive and true positive rates) is mathematically impossible when base rates differ across groups (Chouldechova, 2017); (Kleinberg et al., 2016). Similarly, calibration and equal error rates can only be aligned under strict conditions, such as group-blind thresholds or specific score transformations (Reich & Vijaykumar, 2020). From an empirical standpoint, optimizing for demographic parity often leads to a degradation in equalized odds performance, underscoring the inherent tension between these goals (Souverain et al., 2024); (Han et al., 2023). Additionally, while true positive rate (TPR) parity is commonly used, it may misrepresent fairness if risk distributions vary between groups, prompting researchers to propose adjusted metrics like aTPR to account for these discrepancies (Hegarty et al., 2025). Finally, some fairness metrics—such as equalized odds difference (ΔEO) and false positive rate (FPR) parity—are shown to be more robust to label bias than others, including demographic parity and TPR parity (Liao & Naghizadeh, 2022). These findings collectively emphasize the need for context-aware fairness metric selection and cautious interpretation of fairness trade-offs in practice.

While individually some of the above metrics have limitations, when examined in context collectively these metrics provide a systemic view of bias. These metrics explicitly compare model performance across sensitive groups, making them foundational for fairness analysis (Mehrabi et al., 2021; Barocas et al., 2019). **Further,** metrics like Equalized Odds and Demographic Parity are closely tied to legal definitions of discrimination and are used in regulatory contexts (Hardt et al., 2016). Also, they are specifically designed to detect and quantify disparate impact and treatment, which general performance metrics cannot do.

## D4. Implementation and Contextual Considerations

Choice of fairness metric must be guided by application context, data characteristics (e.g., imbalance, sensitive attribute distribution), and the nature of potential harms (Suresh & Guttag, 2021). For example, in clinical risk prediction, adjusting for risk distribution is crucial to avoid misleading fairness assessments, while in credit or criminal justice, post-processing and deferral strategies are often necessary (Hegarty et al., 2025; Canetti et al., 2018).

While accuracy, precision, recall, and F1 score are essential for understanding overall classifier performance, they are insufficient for fairness analysis because they do not account for group-level disparities or guarantee non-discrimination.

# Chapter 4: Legal frameworks and expectations regarding bias in hiring

## Chapter Overview

**Regulatory Context:** This chapter explores the evolving legal and institutional frameworks governing algorithmic decision-making in employment, emphasizing the balance between innovation, transparency, accountability, and the protection of fundamental rights.

**Methodological Basis:** The analysis draws on a targeted review of four key regulatory instruments — the EU AI Act, EU Non-Discrimination Directives, U.S. Equal Employment Opportunity Commission (EEOC) Guidance, and New York City Bias Audit Law — identified through systematic searches of Google, Scopus, and official EU publications.

**Pre-EU AI Act Landscape:** The chapter explains the state of affairs prior to the EU AI Act, specifically covering how European jurisdictions approached algorithmic discrimination. It explains that the regulators applied primarily data protection and transparency principles, with limited case law on gender or algorithmic bias. However, this scenario seems to be changing with several emerging regulations across both EU and United States.

**European Union AI Act:** The analysis provides contextual reference to the **EU AI Act** comprehensive risk-based regulation for artificial intelligence, which classifies employment-related AI systems as *high-risk*. It mandates stringent requirements for **data governance, human oversight, transparency, risk management, and documentation**, ensuring accountability across the AI value chain.

**U.S. Regulatory Developments:** The analysis also cover the **EEOC Guidance. It** interprets Title VII of the Civil Rights Act to include algorithmic discrimination, requiring employers to assess **adverse impact** through metrics such as *selection rate* and the *four-fifths rule*. It clarifies that **New York City Bias Audit Law (Local Law 144/2021)** further institutionalizes bias audits and public transparency for automated employment decision tools (AEDTs).

**EU Non-Discrimination Directives:** The analysis also covers the **EU Directives 2000/78/EC** and **2006/54/EC** which establish foundational protections against discrimination based on religion, disability, age, sexual orientation, and gender, mandating **equal treatment, accessibility, and enforcement mechanisms** across employment contexts. These directives underpin the broader ethical and legal framework within which AI fairness regulations operate.

This chapter provides an overview of algorithmic fairness expectations in regulation, specifically the ones that address bias in AI-driven employment decisions. It examines how international, regional, and national laws particularly in the EU and the United States are adapting to the challenges of automated decision-making systems. This analysis covers regulations such as the EU AI Act, EEOC Guidance, and New York City Bias Audit Law, and highlights the convergence of ethics and compliance expectations in shaping responsible AI deployment in recruitment and employment contexts.

---

## A. Regulatory frameworks

Employers have a wide range of algorithmic decision-making tools available to assist them in making employment decisions. Employers utilize these tools including artificial intelligence driven tools to save time, increase objectivity, optimize employee performance, or decrease bias. Artificial intelligence poses significant risks to privacy, national security, and social and economic stability. To ensure regulatory compliance and foster trust, transparency, safety, and enhancements in the quality of life provided by AI-based solutions, it is necessary to make alterations to the business, legal, and institutional frameworks (Sikorski, 2021).

## B. Methodology

To gather the key expectations in the context of algorithmic discrimination in employment decision making, the following two steps were followed:

1. Conducted a google search on employment decision making and algorithmic discrimination laws. The search revealed prominently 4 key regulations (2 in the U.S. and 2 in the European Union). Additionally, a report titled “Challenges for EU States in relation to Algorithmic Discrimination” by the European commission that specifically dealt with the subject of the research was also considered to gain additional context of regional regulations.
2. Searched Scopus database for papers with keywords of the regulations, “algorithmic discriminations” and “hiring” or “employment decisions” for key research insights with reference to these regulations.

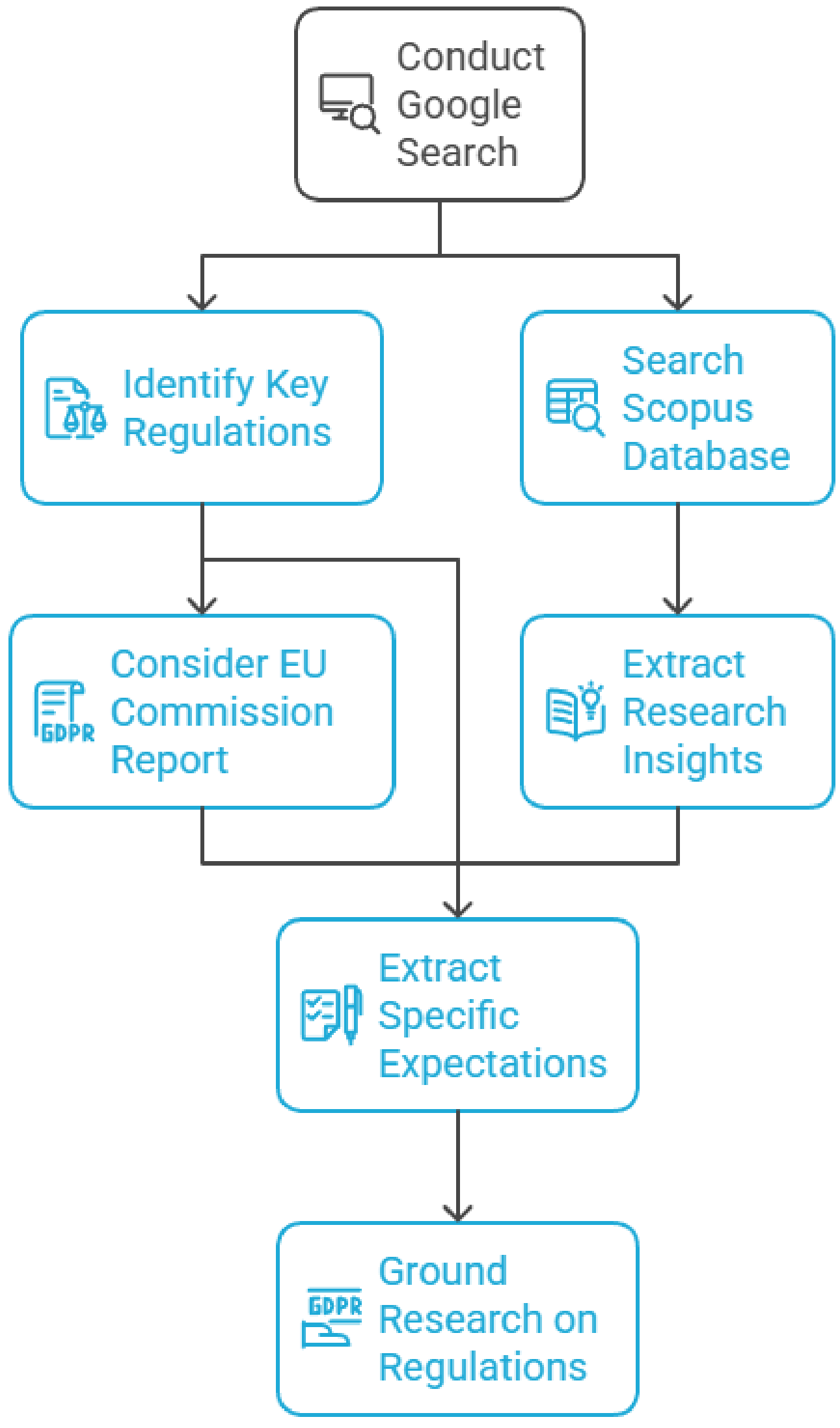

While the research papers were considered for relevant insights, the bare acts and associated guidance were used for extracting specific expectations regarding algorithmic discrimination in employment decisions. Therefore, the research is grounded on the expectations from two regulations in the United States and two from the European Union as follows: (1) the EU AI Act, (2) Non-discrimination law in EU, (3) U.S. Equal Employment Opportunity Commission guidelines, and (4) New York City Bias Audit.

## C. Prior to EU AI Act

In a report titled Challenges for EU States in relation to Algorithmic Discrimination published by the European Commission, the enforcement mechanisms adopted by different jurisdictions within the European Union have been explained as semi-judicial applications and enforcements. It highlights that, currently, there are limited case laws specifically addressing gender inequality or discrimination caused by algorithms. However, cases brought before national courts often focus on aspects such as transparency and data protection.

In the Netherlands, the highest administrative court has established a general obligation for public authorities to ensure explainability, transparency, and accessibility of algorithms. This position has been supported by the Supreme Court. In a recent Dutch case, the legislation allowing the use of a predictive profiling algorithm in detecting social security fraud was found to be incompatible with the right to privacy. Although discrimination complaints were made, they did not play a significant role in the judgment. Similarly, in Poland and France, cases have been brought regarding profiling systems and algorithms, but discrimination issues were not central to the cases. The judgments primarily focused on accuracy, transparency, and monitoring content on social media platforms. Other cases in various European countries have addressed automated decisions in areas such as targeted advertising, credit ratings, facial recognition, higher education admissions, and employment-related decisions.

The judgments and decisions in both equality and non-equality related cases indicate concerns about decisions solely based on algorithms without human involvement. Inaccurate outcomes and the disclosure of algorithm functioning are also contested issues. Different courts in different countries hold varying views on the need for transparency and information disclosure regarding algorithmic decision making. Some emphasize the importance of explainability and information provision, while others prioritize intellectual property rights and business secrets. There was limited litigation specifically addressing algorithmic discrimination. However, judgments and decisions related to other aspects of algorithmic decision making demonstrate divergent perspectives among courts in different countries (*Algorithmic Discrimination in Europe*). However, this scenario evolved with the introduction of the European Union Artificial Intelligence Act.

Prior research also suggested the need for legislative and policy measures are necessary to address gender-based biases and algorithmic discrimination in the context of artificial intelligence in Europe (Lütz, 2022). Research also highlighted that discrimination by AI systems is caused by lack of diversity in training data, bias in training data, or errors in the underlying modeling algorithm. Additionally, the legal framework, including data protection laws like the EU General Data Protection Regulation (GDPR) and the EU Unfair Competition Law, may not be sufficient to prevent discrimination. However, there are emerging regulations like the Digital Service Act (DSA) and the AI Act (AIA) that are taking steps to address this issue through measures such as regular monitoring and audit obligations and the development of an information model to enable autonomous decision-making by users (Gössl, 2023). In addition, integrating fairness considerations throughout the entire machine learning lifecycle is a complex yet crucial measure to ensure the responsible development and deployment of AI systems in alignment with EU AI Act and European General Data Protection Regulation. This integration is essential for building AI systems that are genuinely trustworthy, compliant with regulations, and ultimately beneficial to society (Elisa et al., 2023).

It is also pertinent to note that the European Parliament's Resolution on a framework of ethical principles and legal obligations for artificial intelligence emphasizes the importance of a human-centric approach, ensuring full human oversight in the development, deployment, and use of AI technologies. This human-centric approach aims to prioritize human control and decision-making in AI systems to safeguard fundamental rights and safety rules (Nikolinakos, 2023).

## D. Synthesised expectations from specific regulations

The overall synthesis of the regulatory expectations from (1) the EU AI Act, (2) Non-discrimination law in EU, (3) U.S. Equal Employment Opportunity Commission guidelines, and (4) New York City Bias Audit regarding algorithmic bias can be structured as follows:

**Regulatory Expectations for Algorithmic Bias**

Regulatory Expectations for Algorithmic Bias

Transparency and Explainability
- EU AI Act Article 13
- New York City Local Law 144

Human Oversight and Intervention
- EU AI Act Article 14
- Competence and Training

Foundational Non-Discrimination Principles
- EU Charter of Fundamental Rights
- Directive 2000/78/EC
- Directive 2006/54/EC

Measuring and Auditing Bias
- EEOC Anti-Discrimination Guidance
- Local Law 144 Audits
- EU AI Act Articles 15 and 16

Risk Management and Bias Identification
- EU AI Act Article 9
- Fundamental Rights Impact Assessment

Data Quality and Processing
- EU AI Act Article 10
- Special Categories of Personal Data

Refer Appendix 2 for detailed information regarding each of the regulations referred above.

## H1. Foundational Non-Discrimination Principles

The EU Charter of Fundamental Rights prohibits discrimination on grounds including sex, race, ethnic origin, disability, age and sexual orientation. Directive 2000/78/EC seeks to eliminate discrimination in employment and occupation based on religion or belief, disability, age or sexual orientation (mandating reasonable accommodation for workers with disabilities). Directive 2006/54/EC requires equal treatment of men and women in employment (including protection for gender reassignment) and obliges employers to combat all forms of discrimination. Both directives establish effective enforcement mechanisms—such as reversing the burden of proof in prima facie discrimination cases—which in turn compel AI system designers and deployers to anticipate, identify and mitigate bias proactively.

—----------—-------------------------------------------------------------------------------------

## H2. Risk Management and Bias Identification

Regulatory frameworks mandate a continuous, iterative risk-management cycle throughout an AI system's life cycle to detect, analyse and evaluate bias. Under Article 9 of the EU AI Act, providers of high-risk AI systems must assess potential harms to vulnerable groups and conduct empirical testing under real-world conditions to calibrate appropriate mitigation measures. Deployers must perform a Fundamental Rights Impact Assessment (Article 27) prior to deployment, detailing anticipated harms (including those arising from bias), the populations at risk and the human-oversight and remedial controls planned.

## H3. Data Quality and Processing for Bias Mitigation

High-risk AI systems must be trained on datasets that are relevant, representative, error-free and complete, reflecting the geographical, contextual and functional diversity of the intended application (EU AI Act, Article 10). The Act also permits processing of special categories of personal data (for example racial or ethnic origin, sexual orientation) exclusively for bias detection and correction, subject to strict technical, privacy and documentation safeguards and prompt deletion once remediation is complete. Moreover, Article 26 obliges deployers to ensure input data remain relevant and representative for the system's specified purpose, directly contributing to ongoing bias mitigation.

## H4. Transparency and Explainability

Providers must furnish deployers with comprehensive documentation of system characteristics, capabilities, limitations and explainability features that clarify decision rationales for individuals or groups (EU AI Act, Article 13). New York City's Local Law 144 of 2021 complements these requirements by mandating public disclosure of audit outcomes, including summaries of data sources, performance metrics and disparate impact ratios (posted on an employer's website), thereby enabling external scrutiny and reinforcing deployment-context accountability.

## H5. Human Oversight and Intervention

Effective human oversight constitutes a critical safeguard against algorithmic bias. Article 14 of the EU AI Act stipulates that high-risk systems be designed to facilitate meaningful human intervention, including user training to counter automation bias and the authority to interpret and override system outputs. Article 26 further requires deployers to assign oversight responsibilities to personnel possessing documented competence and training, ensuring continuous monitoring and prompt reporting of bias-related incidents.

---

## H6. Measuring and Auditing Bias

The EEOC's Anti-Discrimination Guidance (under Title VII of the Civil Rights Act of 1964) prescribes the selection-rate metric and the four-fifths rule (flagging potential bias when a protected group's selection rate falls below 80 percent of the majority group's rate), while emphasising the need for supplementary statistical analyses. Employers remain ultimately liable for adverse impacts and should scrutinise vendor bias-assessment methodologies.

Local Law 144 of 2021 requires annual, intersectional bias audits of Automated Employment Decision Tools (AEDTs), assessing disparate impact across race, ethnicity and sex/gender for both standalone and combined demographic groups (e.g., Black women). Key metrics include impact ratio (group selection rate versus highest-scoring category), selection rate (proportion advanced or classified) and scoring rate (proportion assigned a positive outcome in regression models), with results made publicly available.

Under Articles 15 and 16 of the EU AI Act, providers of high-risk AI systems—particularly those with continuous learning capabilities—must minimize bias feedback loops and ensure robustness and accuracy over time. They must also establish a quality-management system that enforces compliance with data governance, risk management, accuracy and human-oversight requirements, reflecting a systemic approach to bias mitigation.

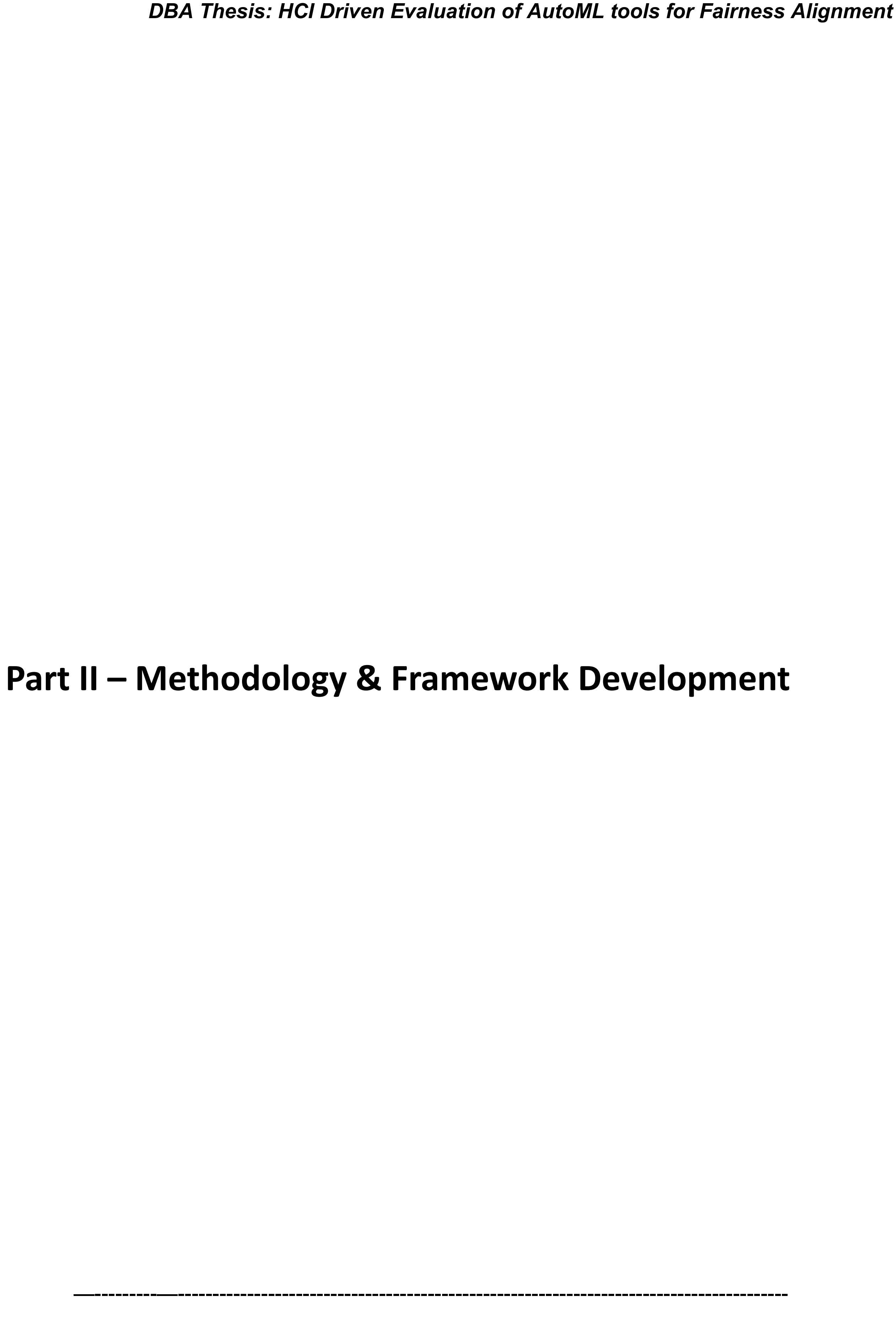

# Part II – Methodology & Framework Development

# Chapter 5: Scope and considerations for evaluation

**Chapter Overview**

**Human–Computer Interaction Context:** This chapter explores how Human–Computer Interaction (HCI) principles enhance fairness, transparency, and trust in Automated Machine Learning (AutoML) systems, emphasizing usability and ethical human–AI collaboration.

**Rationale for HCI-Based Evaluation:** The chapter argues that despite AutoML's automation goals, human oversight remains essential for explainability, fairness, and responsible deployment, as users evolve from operators to supervisors and auditors.

**Framework for Fairness-Aware Evaluation: The chapter also introduces** a five-dimensional framework—Contracts & User Development, Interface Design, Information Architecture, Human Augmentation, and Care & Responsibility—provides a structured lens to assess fairness and accountability in AutoML tools.

**Theoretical Integration: The chapter provides an analysis of how the framework integrates with the principles from grounded theories mentioned earlier in Chapter 1 and Chapter 2,** such as TAM, IDT, and Human-Centered AI, linking usability, explainability, and fairness within coherent user experience and governance structures.

**Human-Centered AutoML Evolution:** The chapter extends the discussion highlighting the transition toward Human-Centered AI and AutonoML in research, where systems act as partners and fairness is ensured through transparent, interactive, and explainable design.

**Industry Alignment:** The chapter also provides brief introduction to leading industry frameworks and tools IBM AI Fairness 360, Google Responsible AI Toolkit, and Microsoft Fairlearn that illustrate practical adoption of HCI-based fairness principles through transparency, user education, and ethical accountability.

## A. Human Computer Interaction based evaluation

The rapid advancement of Automated Machine Learning (AutoML) has revolutionized the field of Artificial Intelligence (AI), enabling the automation of complex machine learning workflows. However, the increasing autonomy of AutoML systems has highlighted the critical importance of human–computer interaction (HCI) principles in ensuring their usability, transparency, and trustworthiness. Most comparisons of AutoML systems focus on quantitative criteria such as predictive performance and execution time, but there is a need to evaluate and compare these services from a user's perspective, essentially Human Computer Interaction perspective, going beyond just predictive performance. The research further mentions that the future of data analysis may involve the use of AutoML services or libraries and the success of it may be determined by interpretation of results and ease-of-use rather than just predictive performance (Xanthopoulos et al., 2020). In addition, (Narayanan, 2023) mentions that essential features promoting fairness in AutoML tools are necessary to be able to support appropriate bias management in downstream uses.

This chapter proposes a comprehensive evaluation framework for HCI in AutoML (with an emphasis on fairness), focusing on five key dimensions: Contracts and User Development, User Interface, Interaction and Experience Design, Information Architecture, Human Augmentation Factors Design, and Care and Responsibility.

## B. Reason for Human Computer Interaction based evaluation

Human–computer Interaction (HCI) is a multidisciplinary field that combines knowledge from computer science, psychology, cognitive science, and social sciences to create interactive computing systems that are both useful and usable (*Human-Computer Interaction*, 2003) (Olson & Olson, 2002). It focuses on designing, evaluating, and implementing interfaces that facilitate effective communication between users and computers, with the goal of improving user experience and system efficiency (Fallman, 2007) (Kheder, 2023). HCI has given rise to numerous sub-disciplines, including Human Computation (HCOMP), human-AI collaboration (HAI), human-robot interaction (HRI), and HITL (Human-in-the-loop) (Lakkshmanan et al., 2024).HCI research has evolved from its initial focus on functionality to encompass user-friendliness, learnability, efficiency, enjoyment, and emotional aspects of interaction (Kheder, 2023). This shift has led to the development of various methodologies and approaches, such as user-centered design (UCD), usability testing, and prototyping techniques (Kheder, 2023). Additionally, emerging technologies such as augmented reality, virtual reality, and gesture-based interfaces have expanded the scope of HCI research (Kheder, 2023) (Kosch et al., 2023).

In the context of AutoML, HCI focuses on understanding the 'how' and 'why' of human interaction within these frameworks, which is crucial for optimal system design and identifying both opportunities and risks presented by increasing machine autonomy (Khuat et al., 2022).

Automated Machine Learning (AutoML) automates aspects of the ML application workflow. Initially, HCI was not considered fundamental to AutoML, which primarily aims to operate with minimal human interaction. Currently, however, there is a shift back towards incorporating HCI as a necessity to support technical users who configure and control semi-AutoML packages, including interactions involving basic operations such as selection, value entry, exploration, and reconfiguration, with visualization naturally enhancing understanding and engagement. Key challenges in the present include building trust, ensuring explainability, mitigating bias for fairness, and increasing transparency, supported by emerging tools such as Data Cards for dataset documentation. The future trajectory points towards greater AutoML autonomy (AutonoML), shifting human roles from direct operation to supervision, auditing, and ultimately collaboration with the system as a partner. Designing for this future collaboration, guided by frameworks such as Human-Centered AI (HCAI), requires optimizing interactions for shared understanding, trust, and leveraging complementary human and machine strengths (Khuat et al., 2022).

AutoML aims to make machine learning accessible to non-experts and improve efficiency (Karmaker ("Santu") et al., 2021). Despite the aim of full automation, AutoML systems still require human intervention to be practically applicable (Crisan & Fiore-Gartland, 2021). Humans are involved in various stages of the ML workflow, providing inputs, feedback, or oversight (Mathewson, 2019) (Khuat et al., 2022). Human involvement in critical steps of AutoML includes understanding domain-specific data, defining prediction problems, and creating suitable training datasets (Karmaker ("Santu") et al., 2021). This human-machine interaction is crucial, yet current AutoML systems often lack transparency, making it difficult for users to understand and trust the decision-making process. Many current AutoML tools have become black-box systems, obscuring their internal working. This development highlights the need for further research into human–computer interaction (HCI) to address this weakness. Hence, the fields of AI and HCI share common roots, particularly in early work on conversational agents (Li et al., 2020). Recent advancements in deep learning have revolutionized AI, creating new opportunities for machines and humans to interact.

The field is increasingly recognizing the importance of a human-centered approach to AutoML by recognizing the need to address user interaction, considering the diverse roles, expectations, and expertise of humans involved (Lindauer et al., 2024). Human–computer interaction in AutoML is evolving towards factors including domain knowledge and context awareness, evaluation and interpretation, handling uncertainties and novelty, collaboration and oversight, interface design, interaction modalities, and visualization (Khuat et al., 2022). A human-centered paradigm promotes the collaborative design of ML systems that integrate the complementary strengths of human expertise and AutoML methodologies, partly triggered by an increasing awareness of the social and ethical (including trust, explainability, transparency, fairness, accountability, and causality) implications of ML technologies (Lindauer et al., 2024)(Khuat et al., 2022).

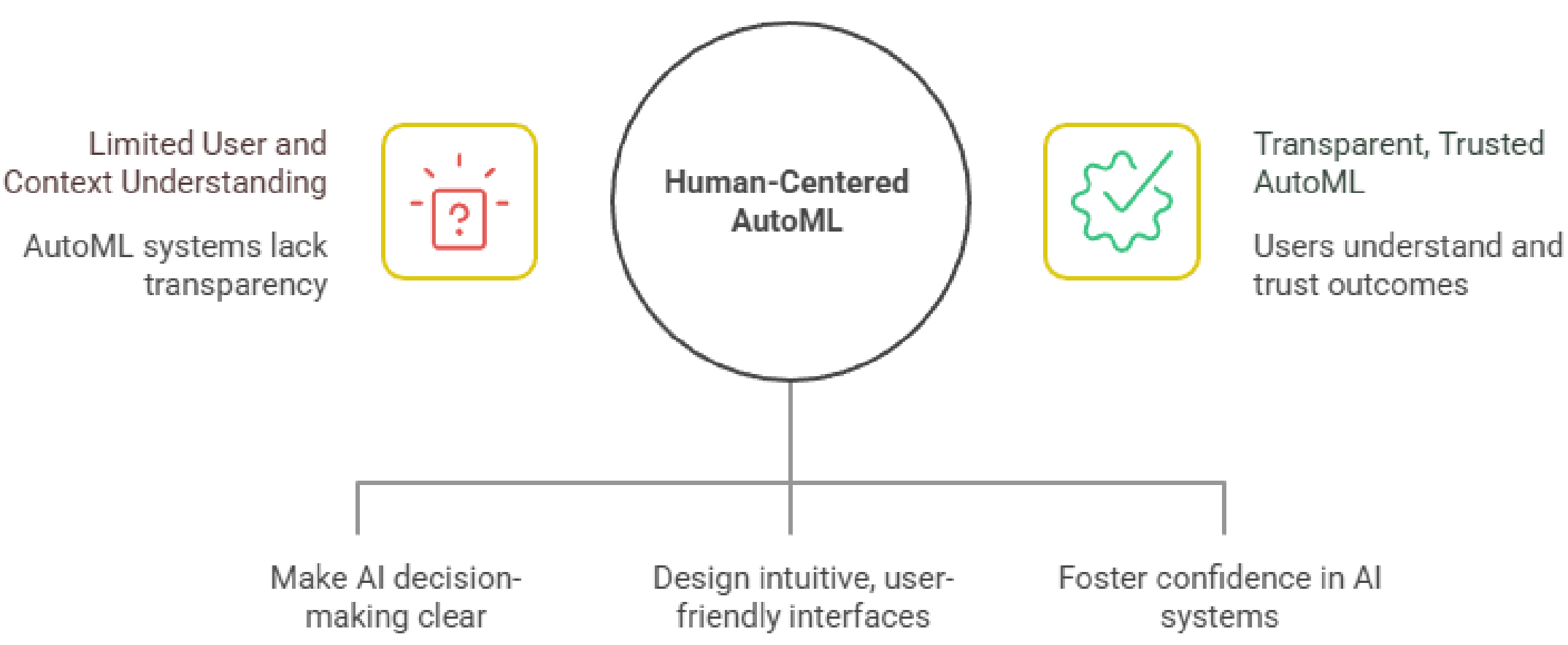

With the above, it is clear that trust and transparency have emerged as critical factors in HCI, particularly in the context of AI-enabled systems. Studies have shown that user trust is influenced by socio-ethical considerations, technical features, and user characteristics, highlighting the need for tailored approaches to system design (Bach et al., 2022).User-centric design remains a cornerstone of HCI, with frameworks such as the User-Centered Design Process (UCDP) prioritizing users' goals and characteristics throughout the design process (Kheder, 2023). Such an approach extends to explainable AI (XAI), where human-centered XAI focuses on addressing the distinct needs of non-expert end users, emphasizing usability, trust, and safety (Veitch & Alsos, 2021). Explainability and interpretability have gained prominence, particularly in the context of AI-powered systems. Social Transparency has been proposed as a sociotechnically informed perspective that incorporates socio-organizational context into explaining AI-mediated decision-making, potentially improving trust calibration and decision-making processes (Ehsan et al., 2021).Usability and human factor engineering continue to play crucial roles in HCI. Researchers have developed innovative frameworks that combine expert cognitive walkthroughs with user surveys to evaluate website UI/UX, thereby providing actionable insights for design improvements (Whaiduzzaman et al., 2023). Additionally, the integration of HCI principles into healthcare systems has shown promise in enhancing patient safety and optimizing processes (Mishra et al., 2023).The expanding scope of HCI encompasses governance, accountability, and risk management. As intelligent systems (including AutoML) have become more prevalent in high-stakes domains, there is a growing need to address moral and ethical concerns and develop transparency frameworks to enhance trust and acceptance (Vorm & Combs, 2022). This broader view of HCI emphasizes the importance of considering not only technical aspects but also the social, ethical, and organizational implications of human-system interactions.

## C. Framework for evaluation of Human Computer Interaction

Given the above context, this paper proposes a consistent framework approach towards establishing a consistent framework for human–computer interaction evaluation. The effective integration of human–computer interaction (HCI) principles is essential for the successful development and deployment of AI systems in general and AutoML systems in particular, particularly in enabling non-expert users to engage with complex AI. The framework contains five dimensions: (1) contracts and user development that aims at setting expectations, clarifying responsibilities and limitations, and enabling the user; (2) user interface, interaction, and experience design that enables intuitive, usable, and engaging interfaces; (3) information architecture that supports organizing, simplifying, and visualizing complexity for the user; (4) human augmentation factors design that empower users and enables control; and Care and Responsibility that enables ethical, safe, and accountable AI.

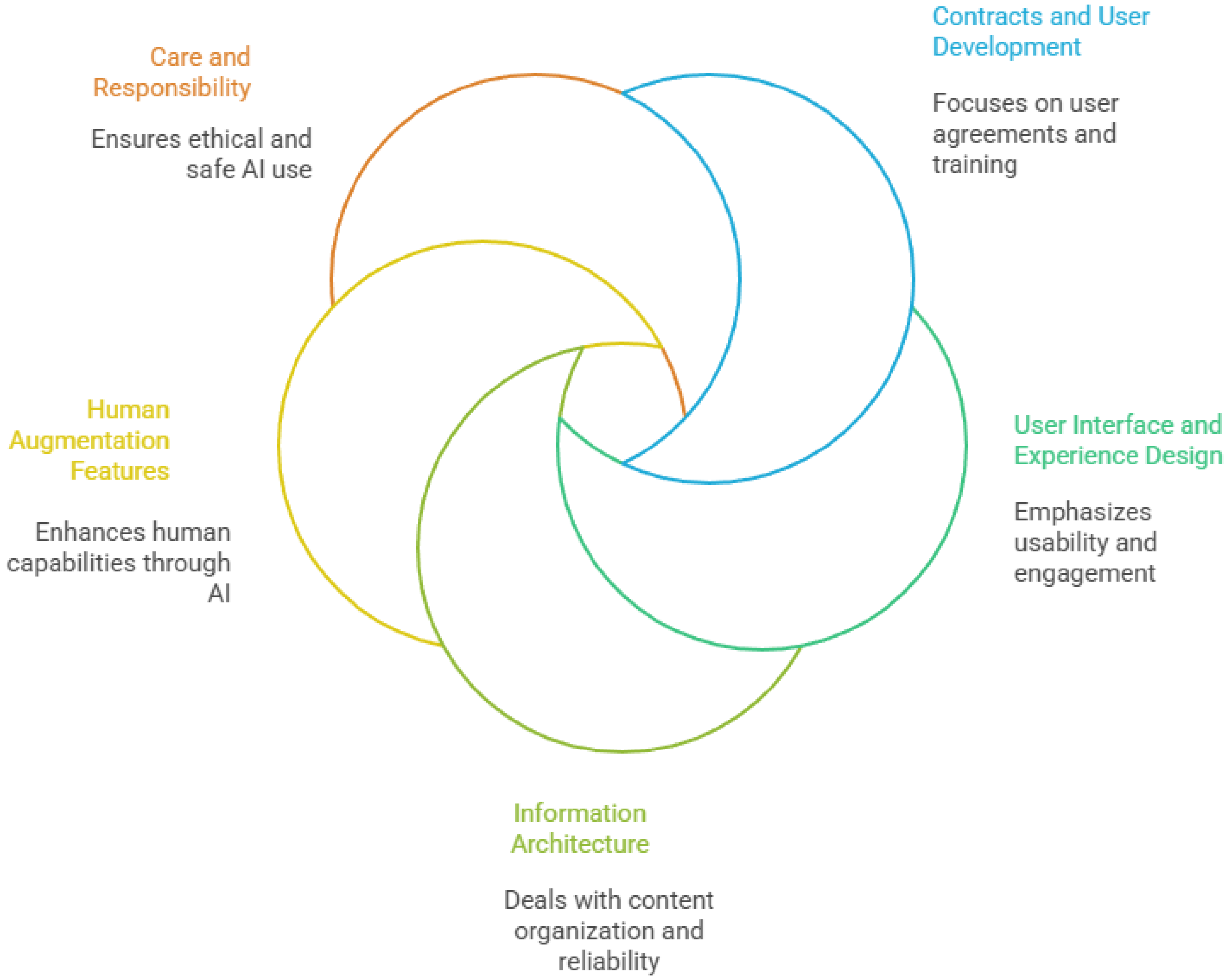
HCI Evaluation Framework for AutoML
Care and Responsibility
Ensures ethical and safe AI use
Contracts and User Development
Focuses on user agreements and training
Human Augmentation Features
Enhances human capabilities through AI
User Interface and Experience Design
Emphasizes usability and engagement
Information Architecture
Deals with content organization and reliability

The detailed list of aspects covered against each dimension are provided below:

| Dimension | Integrated Categories and Key Aspects |
|---|---|
| **Contracts & User Development** | • **Acceptable uses policy:** Acceptable uses and limitations of the tool.<br>• **User data related disclosure:** Clear information regarding the use of user data for model training and improvement<br>• **User training:** User training, guidance, and detailed instructions for appropriate use.<br>• **User responsibilities:** Defined user responsibilities and expected behavior when deploying or interacting with the system. |
| **User Interface, Interaction, and Experience Design** | • **UI/UX Design:** Consistency, intuitiveness, and aesthetic coherence of the interface.<br>• **Usability:** Ease of navigation, task efficiency, and error recovery support.<br>• **Transparency:** Visibility of model decisions, system processes, and feature influence.<br>• **Human Factors Engineering:** Ergonomic alignment with user needs, minimizing cognitive load.<br>• **User Engagement:** Sustained interaction through feedback loops and interactive design.<br>• **Nudges:** Subtle interface cues to guide responsible or optimal user behavior. |
| **Information Architecture** | • **Navigation System:** Logical, consistent, and predictable structuring of menus and pathways.<br>• **Content Hierarchy:** Clear prioritization and labeling of content elements for easy comprehension.<br>• **Information Visualization:** Effective data representation using charts, dashboards, and summaries for clarity and insight. |
| | • **Content Reliability:** Assurance of data accuracy, version control, and source credibility. |
| **Human Augmentation Features** | • **Iterative Engagement with Tool, Data, and Model/Automation:** Support for repeated experimentation, adjustment, and feedback loops.<br>• **Iterative Interpretability & Explainability:** Continuous model transparency through progressive insight into data and decisions.<br>• **Feedback Exchange on Functionality or Performance:** Mechanisms for users to provide or receive performance-related feedback.<br>• **Download Reports, Models, and Benchmarks:** Access to exportable assets for audit, evaluation, and comparative analysis.<br>• **Human Oversight and Override:** Explicit control features enabling user intervention, supervision, and override of automated outputs. |
| **Care and Responsibility** | • **Decision Governance:** Defined governance structure for responsible decision-making and model deployment.<br>• **Accountability:** Traceable responsibility mechanisms for users and developers.<br>• **Guardrails:** Built-in safety features to prevent misuse, bias propagation, or regulatory violations.<br>• **Disclosure:** Transparent communication of model purpose, limitations, and ethical considerations.<br>• **Patches and Updates:** Ongoing maintenance, bug fixes, and iterative updates to ensure reliability.<br>• **Adverse Incident Reporting System:** Structured process for reporting, reviewing, and responding to failures or unintended consequences. |

## D. Mapping the framework to the grounded theories

The above framework also ties in with the grounded theories discussed in chapter 2. The table below maps key theoretical frameworks to the Human-Computer Interaction (HCI) evaluation dimensions relevant to fairness-aware AutoML systems. Each theory(refer Chapter 2 for details) ranging from Technology Acceptance Model to UX strategies like Human-Centered AI and Cognitive Load Theory, offers unique insights into designing systems that are usable, transparent, and ethically aligned. This mapping highlights how different theoretical lenses inform critical aspects of system design, such as user interaction, explainability, human oversight, and responsible deployment.

| HCI Evaluation Framework Element | Technology Acceptance Model (TAM) | Innovation Diffusion Theory (IDT) | UX Strategy: HCAI | UX Strategy: Design Thinking | UX Strategy: Affordance Theory | UX Strategy: Cognitive Load Theory |
|---|---|---|---|---|---|---|
| Contracts and User Development | Perceived usefulness & trust through transparency and feedback | Relative advantage via fairness-aligned contracts & value communication | Controllability, feedback loop design | User understanding & empathy in defining fairness needs | Explicit affordances for user responsibilities | Progressive disclosure, recognition vs. recall for fair use instructions |
| User Interface, Interaction and Experience Design | Ease of use through fairness dashboards | Observability via visual feedback & interactive UI | Explainability & user control via visual & accessible tools | Iterative prototyping & testing of fairness features | Actionable interactions, visual cues | Minimal cognitive load, guided interaction |
| Information Architecture | Information clarity improves usability | Workflow alignment reduces complexity | Transparency & reliability via well-structured fairness info | Problem framing & ideation to improve structure | Mapping fairness to familiar patterns | Chunking, layering, simplified terminology |

———-------——-------------------------------------------------------------------------------

| Human Augmentation Features | Human-in-the-loop interfaces enhance usefulness | User-driven fairness control encourages adoption | Bias mitigation & feedback incorporation | Interactive feedback systems for fairness tuning | Direct manipulability of fairness settings | Guided fairness workflows, support prompts |
|---|---|---|---|---|---|---|
| Care and Responsibility | Trust building via perceived system fairness | Compliance as relative advantage supports adoption | Bias mitigation, reliability, transparency | Clear problem articulation around fairness oversight | Fairness warnings & trade-off visibility | Plain language summaries, error reporting clarity |

## E. HCI converging towards AutoML

As AutoML continues to advance towards greater autonomy, a human-centered approach to HCI will remain paramount, transforming human roles from direct operation to strategic supervision and collaborative partnership with intelligent systems. The evolution of human–computer interaction (HCI) from a focus on basic functionality to encompassing user experience, ethical considerations, and emerging technologies has profoundly impacted the development of Artificial Intelligence (AI) systems, particularly Automated Machine Learning (AutoML). Although AutoML initially aimed to minimize human intervention, the growing recognition of the need for human oversight, collaboration, and trust has underscored the critical role of HCI (Holzinger et al., 2025).

As AutoML systems advance towards greater autonomy (AutonoML), the nature of human interaction is expected to evolve. The relationship may transform from direct instruction to collaboration, where the system is seen more as a partner or teammate. This collaboration leverages the complementary strengths of humans and machines. New human roles, such as "explainers" and "sustainers," may emerge to bridge the human-system gap, interpret system behaviors, ensure ethical compliance, and validate outcomes. Optimizing collaborative interactions involves strategically distributing tasks based on the strengths of humans and the autonomous system. Modern viewpoints advocate for a Human-Centered AI (HCAI) framework that treats automation and human control as orthogonal axes, ensuring that humans retain the option to intervene or oversee (Khuat et al., 2022).

Convergence of HCAI views presents an opportunity to address the black-box nature of AutoML systems by incorporating HCI principles to enhance user understanding and control. Therefore, emergent approaches are attempting to address the greatest weakness of modern AutoML offerings – their black-box nature – which serves as a significant motivating factor for further research into HCI (Mueller et al., 2023). To improve the interaction between humans and AutoML systems, researchers can create more transparent, user-friendly, and trustworthy automated machine learning tools that balance automation with user understanding and control (Karmaker ("Santu") et al., 2021) (Li et al., 2020). The framework proposed in the chapter will be used as a consistent way to evaluate the fairness considerations in AutoM, across five crucial dimensions: Contracts and User Development, User Interface, Interaction and Experience Design, Information Architecture, Human Augmentation Factors Design, and Care and Responsibility. The continued integration of HCI principles is paramount for realizing the full potential of AutoML, ensuring that these powerful tools are developed and deployed in a manner that benefits all users and society (Nakao et al., 2022) (Yu, 2023) .

# F. Best Practices and Evidence from Industry Reports

The framework also aligns well with the emerging efforts towards Human Computer Interaction by the industry. Illustrative efforts by enterprise in line with the above framework include:

## F1. Contracts and User Development

IBM AI Fairness 360 includes fairness metrics, explanations, and extensive documentation to guide users. Google's Responsible AI Toolkit urges defining model behavior and sharing artifacts like model cards to communicate clearly. Microsoft's Responsible AI Standard emphasizes users must understand intended uses and output interpretation, using templates like Impact Assessments. HBR 2022[9] highlights the importance of clear user onboarding and contracts to prevent misuse. Forbes (2023)[10] survey found 73% of executives cite unclear expectations and lack of user education as top AI adoption barriers.

## F2. Interface and Interaction Design

Google's What-If Tool, Microsoft Fairlearn, and IBM AIF360 enable fairness metric visualizations and bias mitigation strategies. Azure ML's interactive visualizations support fairness-aware model comparison and regulatory compliance. Meta's Inclusive AI and Apple's Human Interface Guidelines highlight inclusive, accessible, and transparent design with user-first principles. Stanford HAI (2023)[11] finds interactive visual fairness metrics significantly increase user trust and satisfaction. Apple (2024)[12] emphasizes user control, transparency, and error prevention in trustworthy AI design.

## F3. Information Architecture

Google's Model Cards and Microsoft's Responsible AI Scorecard communicate fairness and safety in layered, digestible formats. Siemens promotes the use of dashboards and bias detection to support industrial AI applications. Wall Street Journal (2023)[13] reports layered AI information increases understanding and boosts adoption rates. Harvard SEAS (2022)[14] finds interactive ethical visualizations improve comprehension and ethical awareness.

[9] https://hbr.org/2022/07/how-to-improve-human-ai-collaboration

[10] https://www.forbes.com/sites/forbestechcouncil/2023/01/11/five-ways-to-improve-human-ai-collaboration/

[11] https://hai.stanford.edu/news/ai-and-human-computer-interaction

[12] https://developer.apple.com/design/human-interface-guidelines/foundations/machine-learning/

[13] https://www.wsj.com/articles/how-to-make-ai-more-understandable-11661411060

[14] https://www.seas.harvard.edu/news/2022/10/visualizing-ai-ethics

## F4. Human Augmentation

Stanford's research introduces context- and difference-aware evaluation starting points. IBM, Google, and Microsoft provide robust HITL frameworks for iterative feedback and fairness evaluation. FairCompass showcases a mixed visual analytical system combining subgroup discovery and HITL audit guidance. Stanford HAI (2024)[15] confirms HITL systems improve fairness and reliability, especially in high-stakes domains. IBM Research (2023) notes feedback-enabled systems reduce error rates and improve user trust.

## F5. Care and Responsibility

IBM's Factsheets call for standardization and transparency to build trust in AI. Apple promotes transparency, privacy, and inclusive design through its civil rights disclosures. Meta's datasets and proposed incident reporting frameworks support fairness auditing and harm tracking. HBR (2023)[16] shows that 80% of transparent organizations report greater trust and reduced regulatory risk. Forbes (2023)[17] links structured incident reporting to faster issue resolution and better public perception.

| HCI Framework element | Best Practice | Real-World Example |
|---|---|---|
| Contracts & user development | Transparent disclosures & training | IBM AIF360, Google Responsible AI |
| UI Design | Visual fairness dashboards, alerts | Google What-If Tool, Azure ML |
| Information Architecture | Layered info, warnings | Model Cards, Siemens MindSphere |
| Human Augmentation Features | HITL, feedback loops | IBM AutoAI, Vertex AI |
| Care and Responsibility | Incident logs, fairness factsheets | Meta AIR, Microsoft RA Dashboard |

[15] https://hai.stanford.edu/news/human-loop-ai
[16] https://hbr.org/2023/03/the-transparency-imperative-in-ai
[17] https://www.forbes.com/sites/forbestechcouncil/2023/01/11/five-ways-to-improve-human-ai-collaboration/

# Chapter 6: HCI evaluation framework applied to AutoML tools and libraries

**Chapter Overview**

**Introduction and Context:** This chapter examines how Automated Machine Learning (AutoML) platforms enable fairness through human–computer interaction (HCI), emphasizing transparency, interpretability, and user agency. It highlights the ethical tension between automation and accountability, showing that fairness depends not only on algorithms but on how interfaces guide human engagement and oversight.

**Methodological Design:** Based on the framework from Chapter 5, the study adopts heuristic audits, cognitive walkthroughs, and content analysis to evaluate fairness-supporting features. Each method traces how interfaces reveal, explain, or obscure fairness cues, documenting patterns that affect user understanding, control, and trust.

**Evaluation Outcomes:** The chapter covers the results of the analysis. Overall the results indicate that GUI-based tools demonstrate higher maturity in fairness visibility through guided workflows, fairness dashboards, and feedback loops. Code-based tools, though flexible, lack consistent transparency, fairness visualization, and user guidance. Common weaknesses include limited bias mitigation support, incomplete documentation of fairness boundaries, and insufficient mechanisms for human-in-the-loop control. At a feature level fairness in AutoML remains uneven and fragmented than a coherent design philosophy, underscoring the need for standardized, human-centered frameworks in future system development.

## A. Introduction

As machine learning systems become increasingly automated through AutoML platforms, questions of *fairness*, *transparency*, and *human agency* have moved to the forefront of ethical AI research. While automation promises to streamline model development, it also risks abstracting away crucial decisions that affect model bias, representational harm, and downstream accountability. In this evolving context, ensuring that AutoML tools meaningfully support fairness—not merely through backend algorithms but through the user interfaces and experiences they facilitate—has become a pressing imperative.

This chapter presents a qualitative inquiry into how AutoML systems afford fairness-aware decision-making for users. Specifically, it evaluates eight prominent AutoML tools through a human-centered lens, analyzing how their interfaces surface, support, and sometimes obscure fairness considerations. Leveraging a bespoke evaluation framework grounded in Human–Computer Interaction (HCI) and action research principles, the study explores the ways in which users can identify, intervene in, and potentially mitigate algorithmic bias within these platforms.

Building on the theoretical grounding introduced in Chapter 4, which articulated fairness as an emergent property of socio-technical interaction, this chapter transitions from conceptual framing to methodological application. The evaluation aims not to score or rank platforms, but to illuminate recurring design patterns, critical omissions, and interface features that shape fairness outcomes. Through heuristic audits, cognitive walkthroughs, and content analysis, the study sheds light on the material ways AutoML interfaces mediate ethical reasoning, inviting reflection on the responsibilities and affordances embedded in automated decision-making systems

# B. Methodology

## B1. Research Design

This study adopts a *qualitative, audit-driven methodology* rooted in action research principles to examine the human-centered fairness affordances of AutoML platforms. The inquiry is designed to uncover how these systems facilitate—or inhibit—human agency in identifying, mitigating, and addressing algorithmic bias through their interfaces, guidance mechanisms, and transparency features.

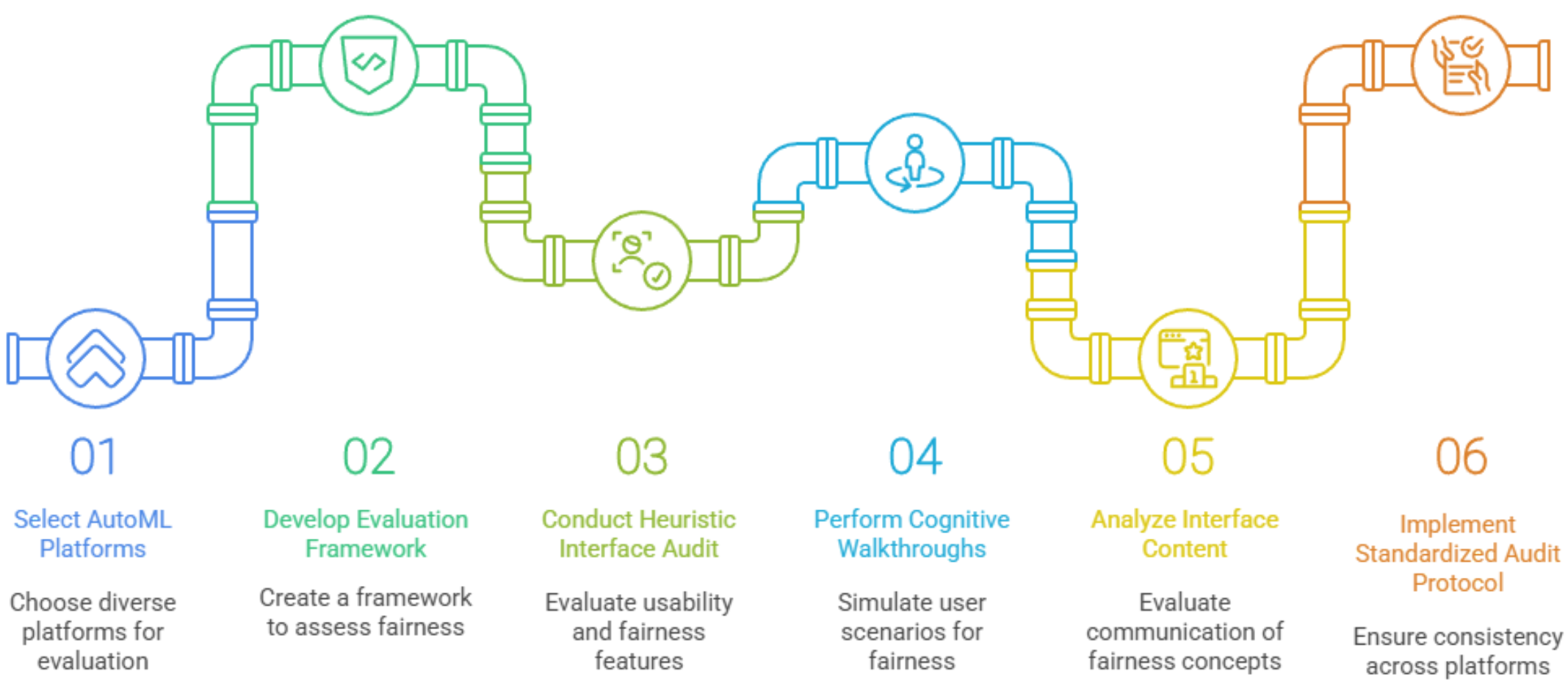

**R1.** What fairness limitations exist in widely used AutoML tools when applied to hiring-related datasets?

**R2.** How do the interface design, transparency mechanisms, and feedback loops in AutoML platforms support or hinder non-expert users in identifying and mitigating algorithmic bias?

**R3.** What systemic gaps in human-in-the-loop (HITL), fairness visualization, and governance reporting exist across different AutoML tools from an HCI perspective?

**R4.** What design considerations should AutoML providers prioritize to enhance fairness, trust, and usability for enterprise adoption in regulated domains?

This design directly supports the research questions outlined in Chapter 1. Specifically:

- RQ1 is addressed through comparative fairness audits of AutoML tool outputs;
- RQ2 and RQ3 are addressed via heuristic interface audits, content analysis, and cognitive walkthroughs of UI-based tools;
- RQ4 is addressed through synthesizing product design patterns and implications for enterprise AutoML deployment.

Drawing from the Human–Computer Interaction (HCI) framework referred in Chapter 4, the study investigates *design patterns, usability constraints,* and *explanatory affordances* that shape fairness-aware user experiences. Emphasis is placed on how interface elements mediate user understanding, control, and intervention with respect to fairness concerns.

## B2. Tool Selection

Eight AutoML platforms were purposely selected to reflect a diversity of technical affordances and user audiences. Selection criteria include widespread adoption in applied machine learning workflows, accessibility for users without extensive data science expertise, and availability without requiring enterprise infrastructure among others. The tools considered for evaluation include GUI-Based Platforms: Dataiku, DataRobot, H2O AutoML Studio, Altair RapidMiner and Code-Based Libraries: FLAML, AutoGluon, H2O AutoML (Python), PyCaret. Refer Chapter 2 for details regarding the tool selection

## B3. Evaluation Framework

A bespoke, HCI-grounded framework from chapter 5 was leveraged to assess each platform's treatment of fairness taking into account considerations including the following:

1. Bias Identification: Mechanisms through which the system surfaces potential bias—visually, narratively, or diagnostically
2. Bias Mitigation Support: Availability of guidance, heuristics, or automated interventions to address fairness concerns
3. Fairness Warnings and Alerts: Presence and clarity of risk signals when fairness thresholds may be compromised
4. Fairness Reporting: Transparency, accessibility, and communicative design of fairness metrics or evaluation results
5. Human Feedback and Override: Extent to which users can modify fairness configurations, introduce corrective measures, or override system decisions

Evaluation was qualitative and interpretive, centering on the HCI evaluation framework specifically addressing *design rationales*, and *user experience insights* rather than on numerical metrics.

## B4. Evaluation Procedures

Three complementary qualitative methods were employed:

- Heuristic Interface Audit: Evaluation was conducted using *open-ended heuristic walkthroughs*, documenting usability features, fairness affordances, and points of opacity. Emphasis was placed on discoverability, procedural documentation and transparency, and cognitive support for ethical decision-making.
- Cognitive Walkthroughs: Simulated user scenarios (e.g., building a classification model under fairness constraints) were used to trace how each tool guides users in fulfilling fairness-aware objectives. Analysts noted breakdowns, ambiguities, and moments of effective alignment between system guidance and ethical intent.
- Interface Content Analysis: Interface components—tooltips, tutorials, documentation, warnings, and visual outputs—were coded to evaluate how fairness concepts were communicated, prioritized, or downplayed. Special attention was paid to tone, placement, and linguistic accessibility.

## B5. Standardized Audit Protocol

To ensure consistency across platforms, the following procedural steps were implemented: (1) Created independent user accounts for a clean perspective; (2) Reviewed official tutorials, documentation, and example workflows; (3) Simulate and analyse fairness-critical modeling tasks on curated datasets; (4) Apply the evaluation framework during heuristic and cognitive walkthroughs; (5) Capture interface content, visual design cues, and narrative feedback and (6) Analyze and compare qualitative observations

## B6. Ethical Considerations

All evaluations were conducted using publicly accessible or open-source versions of the selected platforms, strictly adhering to their terms of use. No human subjects were involved, and only publicly available or synthetic datasets were employed to avoid exposing sensitive data.

## C. Outcomes from the evaluation

The evaluation of AutoML tools revealed that GUI-based platforms such as Dataiku and DataRobot provide structured support for fairness across multiple dimensions, including user guidance, visualization of fairness metrics, iterative feedback, and human oversight. Tools with code-based interfaces generally lack accessibility to fairness information, interactive analysis, and structured governance mechanisms. Key gaps across most tools include limited mechanisms for preventing misuse, partial transparency on fairness limitations, and restricted support for human-in-the-loop interventions. Visualization and reporting of fairness metrics are often present in UI-based tools but remain inconsistent and not fully integrated across all modeling steps. Overall fairness readiness is still emerging; the data reflects an ecosystem where commercial platforms lead in implementation, however, across tools fairness features are fragmented. The detailed analysis is provided below:

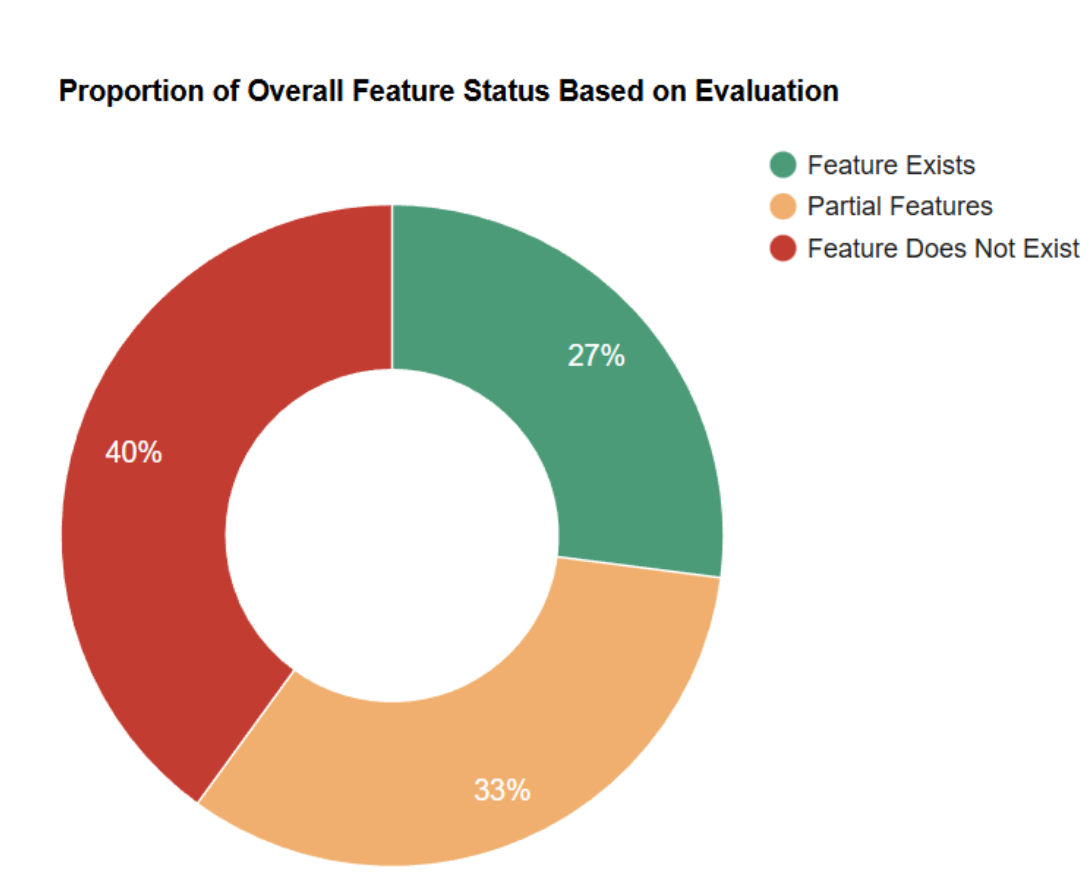

At an overall level, the fairness and responsibility landscape across all evaluated tools shows a wide spread across features existence and partial or missing features. Out of the total evaluation, 173 features (29%) exist, 211 (35%) are partial, and 256 (36%) are missing entirely—signifying that fewer than one-third of fairness-relevant features are fully realized. This distribution underscores that ethical governance and equitable functionality remain underdeveloped. The dominance of partial and missing features illustrates a systemic challenge: most frameworks recognize fairness as a design objective but lack consistent operational frameworks for it.

Across framework categories, fairness integration remains inconsistent and fragmentary, with the highest overall feature maturity appearing in Information Architecture (33% existing, 35% partial) and UI/UX Design (32% existing, 31% partial). These domains contribute most to equitable system design by enabling transparency, structured accessibility, and inclusive human–AI interaction. Human Augmentation (29% existing, 43% partial) also exhibits moderate progress, suggesting growing recognition of human interpretability and control as fairness enablers. In contrast, Contracts & User Development (19% existing, 56% partial) and Care & Responsibility (19% existing, 42% partial) reveal substantial fairness implementation gaps specifically in governance, ethical oversight, and accountability. While technical transparency and usability receive increasing attention, fairness maturity remains uneven across categories reflecting a field still transitioning from ethical intent to embedded, enforceable practice across all AutoML frameworks.

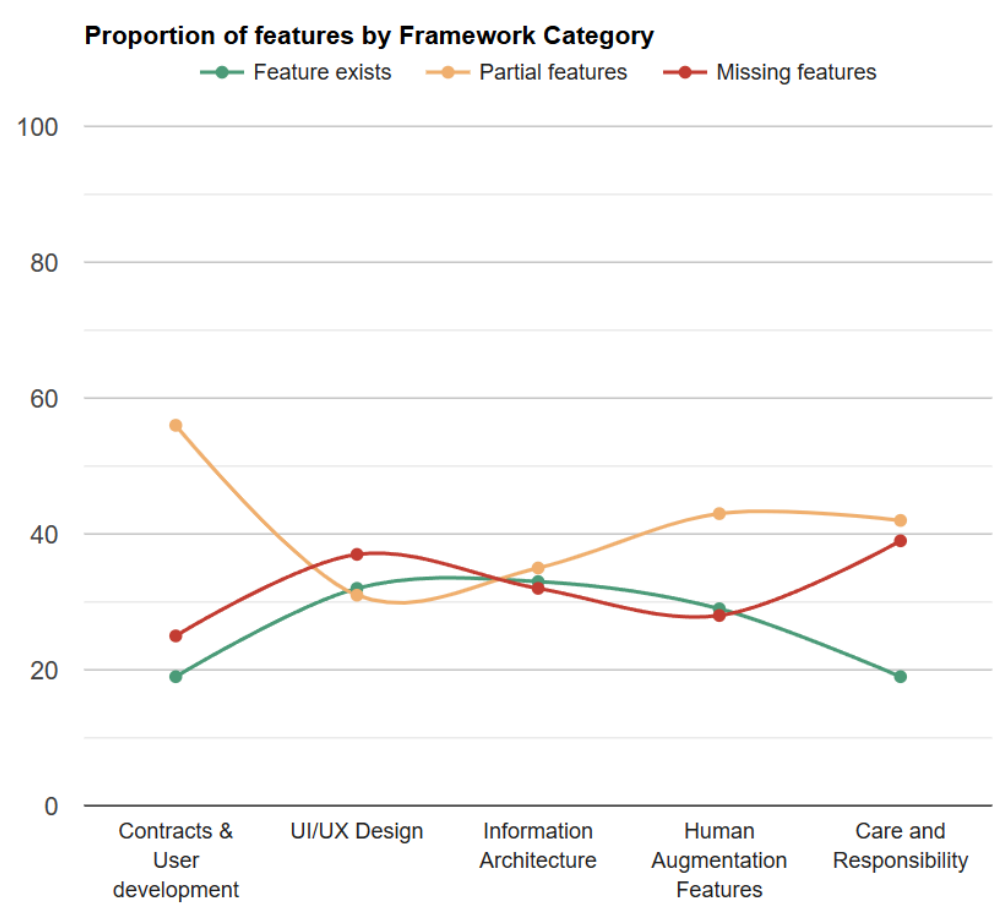

Category wise analysis:

The evaluation of AutoML systems for fairness can be organized into several interrelated categories, based on the relevant framework components first and then thematically divided into specific criteria along with its associated fairness consideration. For instance, Contracts and user development framework component considers whether the system guides acceptable use, prevents misuse, enables user accountability, and provides mechanisms to contest potentially biased outcomes. It further extends to consider user support and education examines whether the system informs users about fairness risks, offers actionable guidance, and facilitates human intervention to address unfairness. The interface design and usability covers visualizations, interaction patterns, and navigation support clear understanding, exploration, and control of fairness-related information without introducing cognitive biases. Information architecture and human augmentation features considers the system's capabilities to handle sensitive data responsibly, apply bias-aware preprocessing, enable iterative model refinement, and provide transparent explanations of fairness-related decisions. The care and responsibility component covers decision governance, accountability and auditability, and transparent disclosure including the tool limitations, and supports traceability over time.

——-------——-----------------------------------------------------------------------------------

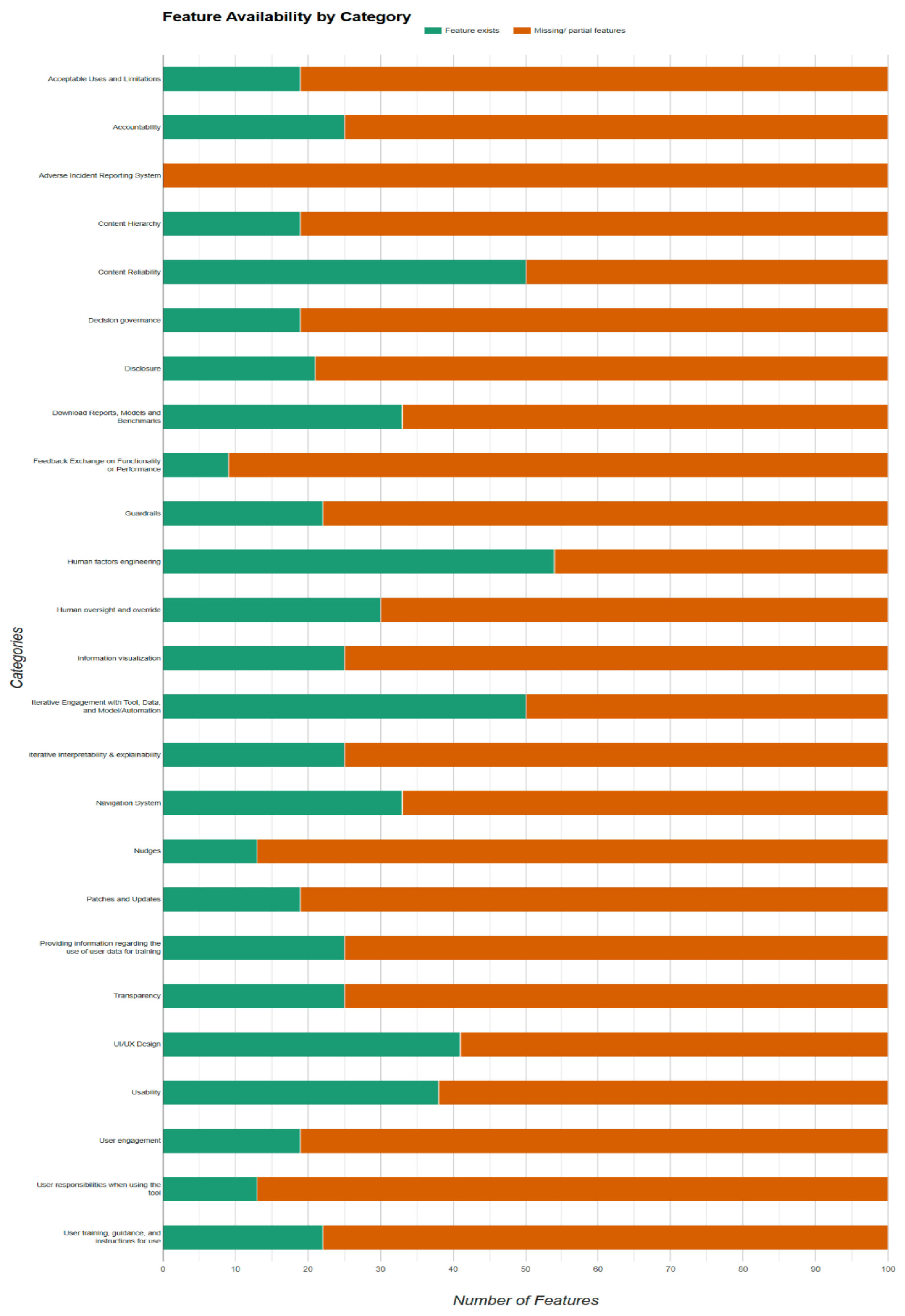
Feature Availability by Category
Feature exists
Missing/ partial features
Acceptable Uses and Limitations
Accountability
Adverse Incident Reporting System
Content Hierarchy
Content Reliability
Decision governance
Disclosure
Download Reports, Models and Benchmarks
Feedback Exchange on Functionality or Performance
Guardrails
Human factors engineering
Human oversight and override
Information visualization
Iterative Engagement with Tool, Data, and Model/Automation
Iterative interpretability & explainability
Navigation System
Nudges
Patches and Updates
Providing information regarding the use of user data for training
Transparency
UI/UX Design
Usability
User engagement
User responsibilities when using the tool
User training, guidance, and instructions for use
Categories
0
10
20
30
40
50
60
70
80
90
100
Number of Features

**Acceptable Uses and Limitations:** Most tools provide general guidance; few explicitly warn against fairness-critical misuse or detailed regulatory alignment. Only DataRobot fully enforces misuse prevention; governance is mostly manual or advisory.

**Transparency and Disclosure:** Dataiku and DataRobot offer transparency regarding the use of training data and system limitations, including fairness-related disclosures. Other tools provide minimal or no explicit visibility into fairness considerations, with limited support for TEVV results or ambiguity reporting.

**User Training, Guidance, and Instructions:** DataRobot and Dataiku provide bias education, detailed instructions, and tailored explanations for fairness. Other tools offer minimal or no structured guidance for users.

**User Responsibilities and Oversight:** Dataiku and DataRobot provide tools and guidance for monitoring fairness, contesting decisions, and supporting human oversight. H2O Studio and Rapidminer provide partial support, while code-based tools generally lack structured mechanisms for oversight or contestation.

**UI/UX Design and Usability:** GUI-based tools support interactive exploration of models, some fairness visualizations, and interfaces for non-experts. Libraries such as FLAML, AutoGluon, and PyCaret lack GUIs, making fairness metrics and guidance less accessible.

**Transparency of Processes:** Dataiku, DataRobot, H2O Studio, and Rapidminer reveal aspects of model search and evaluation processes, but comprehensive and intuitive fairness visualizations are rare. Open-source libraries provide limited transparency and reporting.

**Human Factors and Cognitive Support:** Dataiku and DataRobot provide dashboards, subgroup metrics, and visualizations that reduce cognitive load, support fairness trade-off analysis, and minimize bias in interpretation. Other UI-based tools offer partial support, while code-first libraries provide minimal guidance.

**User Engagement and Feedback:** DataRobot supports structured feedback mechanisms, fairness-aware domain knowledge input, and nudges to guide ethical decision-making. Dataiku provides some feedback functionality, but most other tools lack built-in mechanisms for iterative fairness improvement.

**Navigation and Information Architecture:** GUI-based tools provide structured navigation, layered views, and process mapping to access fairness metrics, with DataRobot leading in surfacing fairness-relevant controls. Code-based tools generally lack navigational support for fairness exploration.

**Content Presentation and Prioritization:** Dataiku and DataRobot highlight fairness alerts, prioritize key information, and support layered presentation with drill-downs. Other tools provide limited or no structured prioritization.

**Information Visualization:** Dataiku and DataRobot enable interactive fairness analysis, subgroup comparisons, and simplified displays for non-experts, mitigating cognitive bias. Other tools provide basic or no fairness visualization.

**Data Reliability and Quality:** Dataiku, DataRobot, and H2O Studio emphasize data quality, consistent workflows, and highlight fairness-related data issues. Code-based tools provide minimal or manual checks.

**Iterative Engagement and Human-in-the-Loop Integration:** Only Dataiku and DataRobot integrate fairness-aware preprocessing, iterative refinement, and HITL feedback loops. Other tools support automation but offer limited fairness-specific guidance.

**Interpretability and Explainability:** Dataiku and DataRobot provide clear explanations of fairness-related decisions, support multiple explanation types, and non-linear workflows. Other tools offer limited interpretability, often requiring technical expertise.

**Performance Feedback and Reporting:** DataRobot enables users to see how feedback impacts fairness outcomes, while Dataiku provides partial visibility. Other tools lack structured feedback loops linking user input to fairness improvements.

**Download and Oversight Capabilities:** GUI-based tools allow exporting fairness reports, comparing models, and reviewing metrics. Libraries generally lack built-in fairness-specific downloads or comparison features.

**Decision Governance and Accountability:** Dataiku and DataRobot support ethical alignment, logging of fairness interventions, and clarify responsibility for fairness outcomes. H2O Studio and Rapidminer provide limited governance, while code-based tools lack accountability.

**Guardrails and Control Mechanisms:** Dataiku and DataRobot offer fairness-related guardrails, verification before deployment, and adjustable controls. Other tools provide no fairness guardrails.

**Disclosure and Reporting:** Dataiku, DataRobot, H2O Studio, and Rapidminer disclose system capabilities, but only DataRobot and Dataiku communicate limitations, fairness failures, and uncertainty. Code-based tools provide minimal transparency.

**Patches and Updates:** Dataiku and DataRobot support fairness-driven updates and ethical oversight, though user review is needed to ensure fairness improvements. Other tools provide no structured update support.

**Adverse Incident Reporting:** Only Dataiku offers an incident reporting feature, though not fairness-specific, while DataRobot helps users understand the limits of fairness.

---

**Framework element wise analysis of feature evaluation**

Across all five dimensions, fairness-oriented capability is concentrated heavily in commercial AutoML tools, with DataRobot and Dataiku consistently demonstrating structured governance, transparency, and user inclusivity. Mid-tier platforms like Altair Rapidminer and H2O AutoML Studio show partial progress, offering limited interpretability and data architecture support that could evolve into fairer frameworks with refinement. Open-source tools such as FLAML, AutoGluon, PyCaret, and H2O AutoML Library remain largely do not address fairness considerations in specific and lack embedded safeguards, interpretability, and oversight.

Examined through a fairness lens, the availability of user governance and contractual frameworks directly impacts equitable access and accountability. DataRobot (89%) and Dataiku (67%) lead with features supporting transparent role management, data lineage, and controlled user interactions—foundational for fair AI deployment. In contrast, H2O AutoML Studio, Altair Rapidminer, FLAML, AutoGluon, PyCaret, and H2O AutoML Library (all 0%) lack structural means to ensure consistent user accountability or fair usage practices.

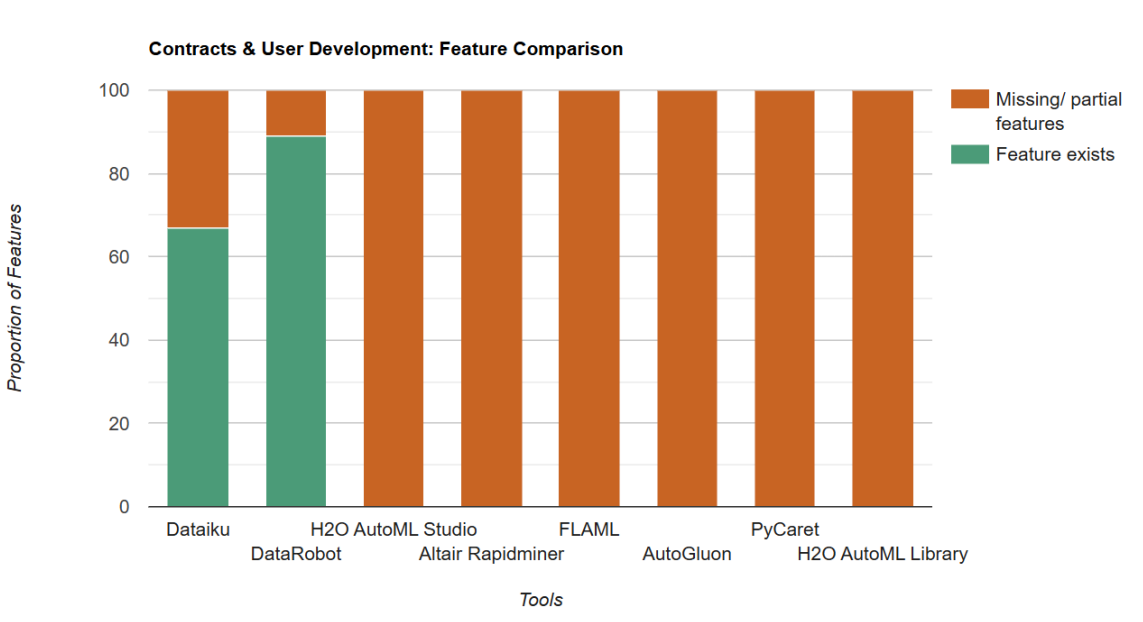

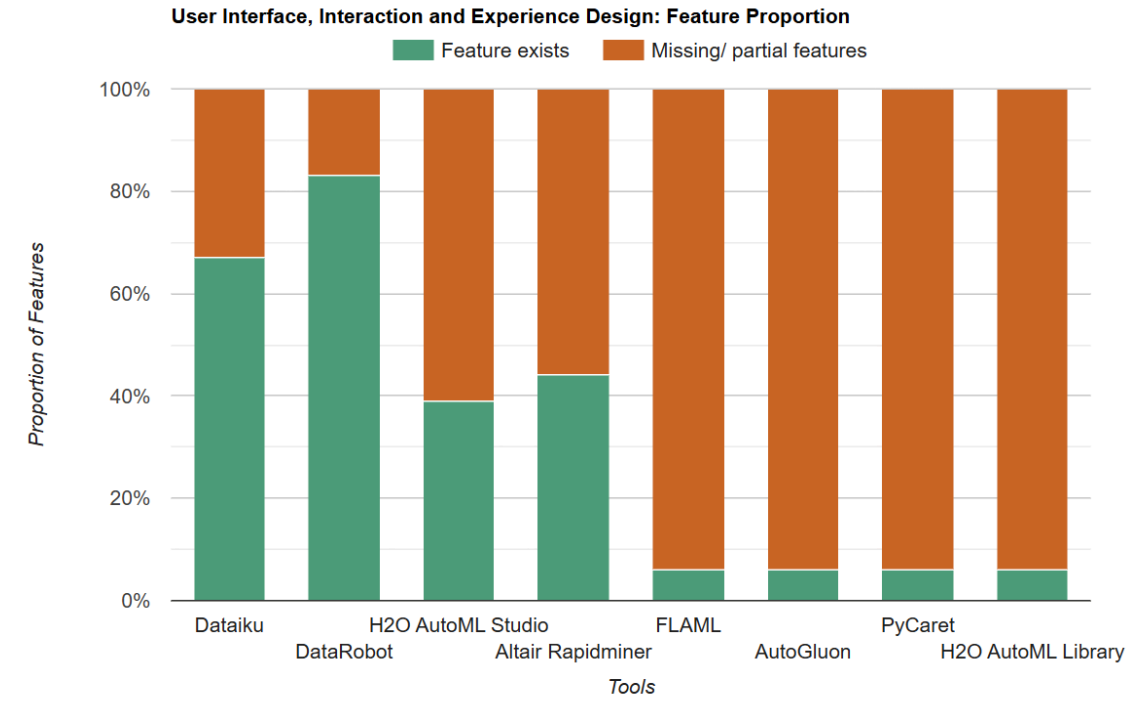

Fairness in AI is also shaped by user accessibility, inclusion, and the ability to engage meaningfully with systems. DataRobot (83%) and Dataiku (67%) demonstrate superior fairness alignment through intuitive, interactive design that empowers users of varying technical abilities to engage equally. Altair Rapidminer (44%) and H2O AutoML Studio (39%) provide moderate inclusivity by supporting both visual and technical interactions. In contrast, FLAML, AutoGluon, PyCaret, and H2O AutoML Library (6%) rely on code-driven access, limiting participation to technically skilled users and creating barriers to fair usage.

A fair AI system depends on equitable data architecture, ensuring transparency, traceability, and balanced access to information. Here, DataRobot (93%) and Dataiku (80%) exemplify fairness-supportive infrastructure, offering robust integration and metadata governance that enable fair data provenance and consistent oversight. Altair Rapidminer (33%) and H2O AutoML Studio (20%) provide partial support, while FLAML, AutoGluon, PyCaret (7%), and H2O AutoML Library (13%) lack sufficient metadata handling capabilities from a fairness perspective.

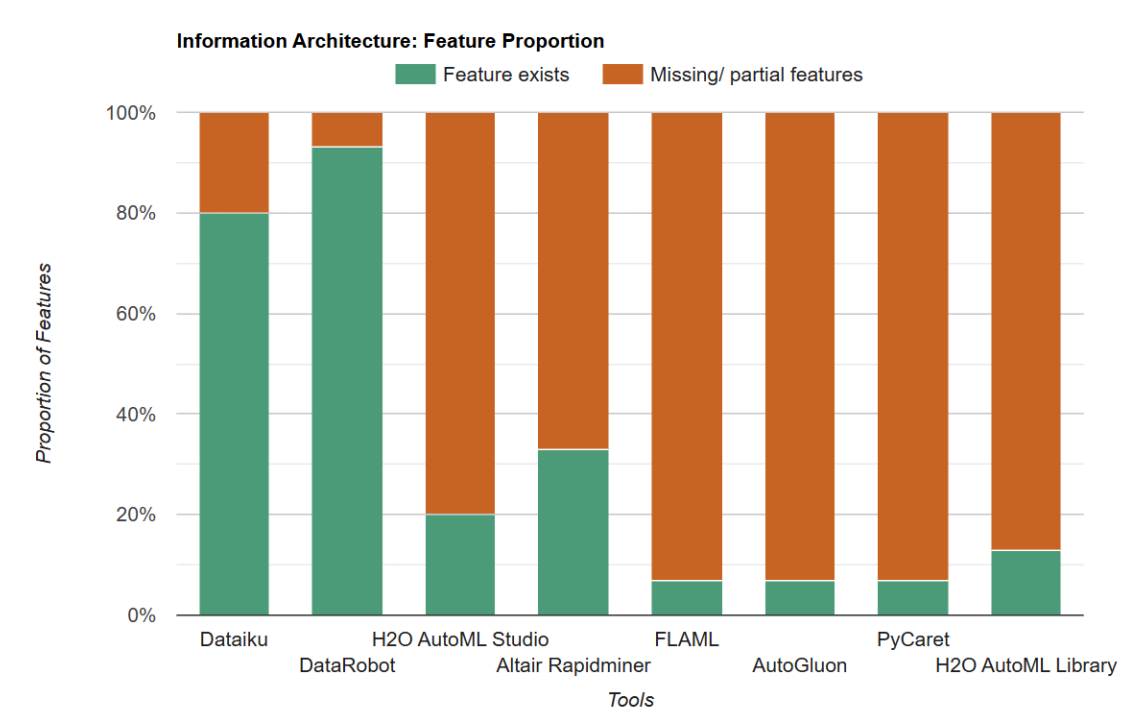

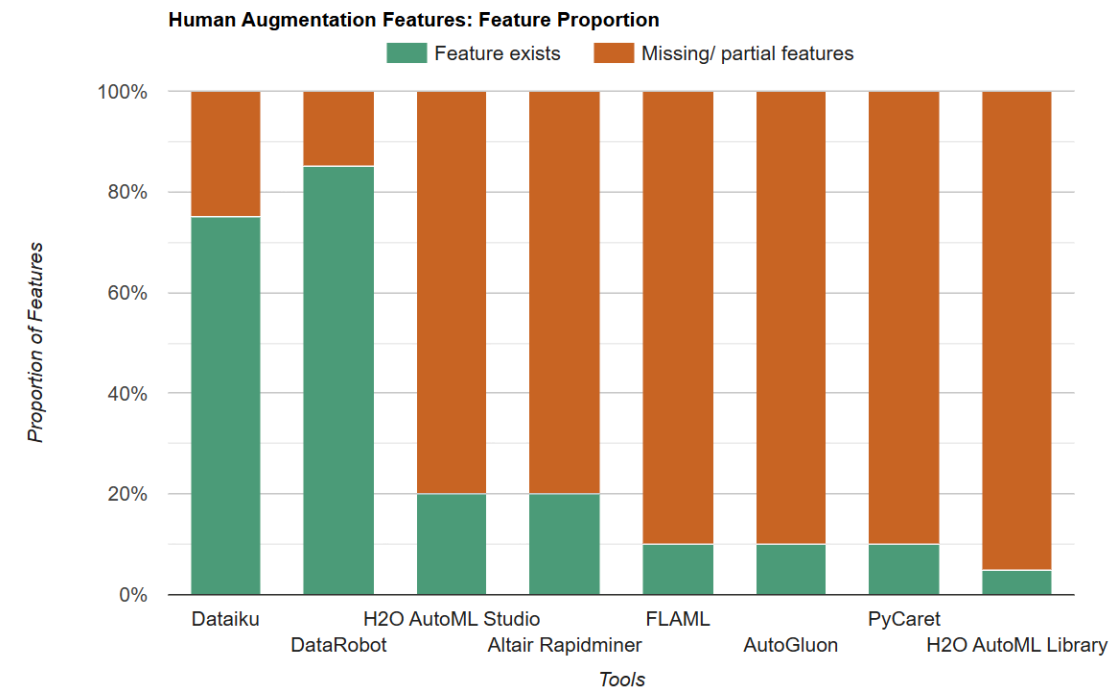

Human augmentation capabilities are central to fairness, as they allow users to interpret, question, and adjust automated decisions. DataRobot (85%) and Dataiku (75%) perform strongly, facilitating interpretability and decision transparency that help identify bias and encourage equitable reasoning. H2O AutoML Studio and Altair Rapidminer (20%) offer emerging, though limited, support for fairness-aware interpretability. In contrast, FLAML, AutoGluon, and PyCaret (10%), along with H2O AutoML Library (5%), provide minimal assistance for user oversight

From a fairness perspective, the uneven integration of care and responsibility features across tools underscores major disparities in ethical AI readiness. DataRobot (78%) and Dataiku (61%) show tangible commitment to fairness principles including embedding transparency, accountability, and bias detection functions that support equitable outcomes. H2O AutoML Studio and Altair Rapidminer (6%) exhibit limited ethical scaffolding, and open-source tools lack mechanisms or have limited mechanisms to identify or mitigate model bias.

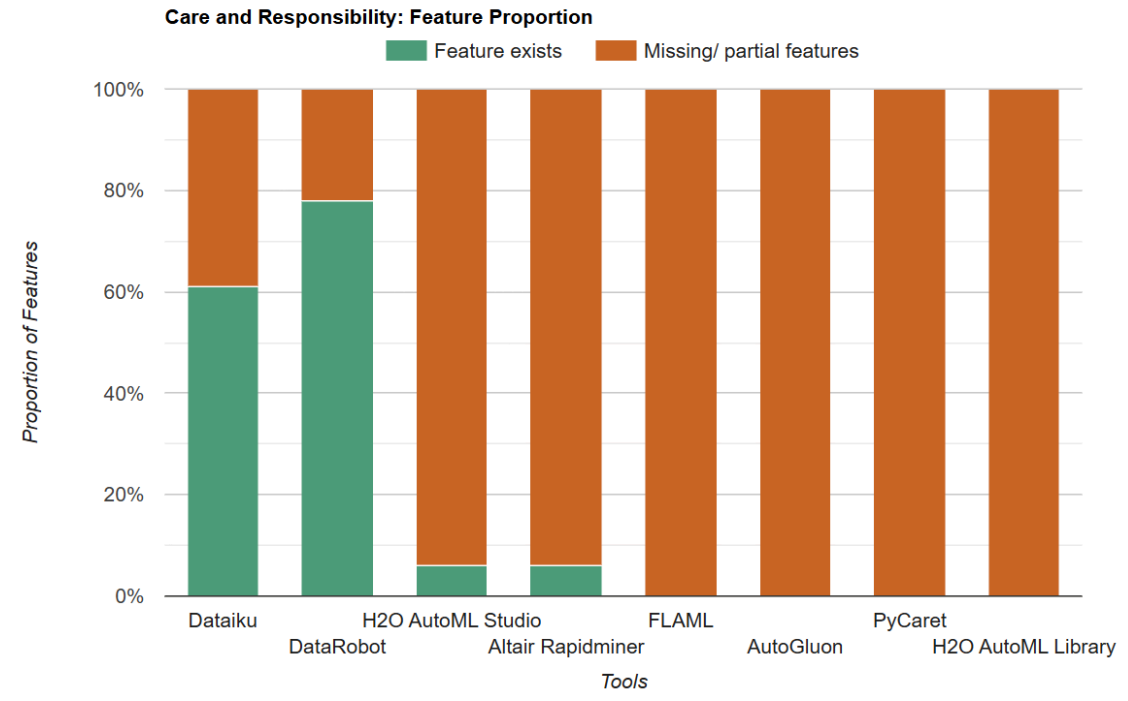

### UI based tools

UI-driven platforms demonstrate strong overall feature coverage, averaging 49%–85% feature existence, which reflects maturity in design, fairness integration, and enterprise readiness.

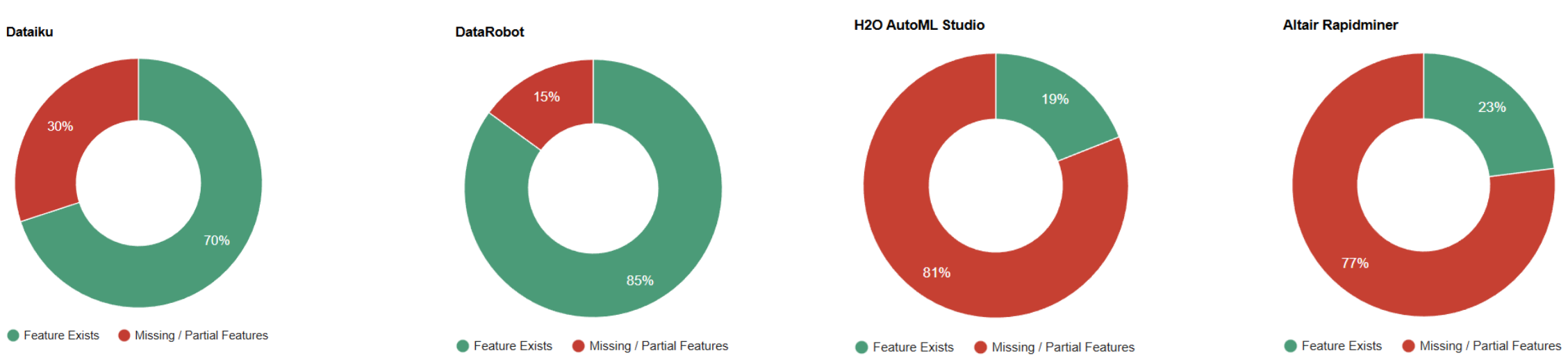

DataRobot (85%) and Dataiku (70%) consistently lead with comprehensive coverage across fairness, responsibility, information architecture, and human augmentation, emphasizing their investment in accessible, ethical AI ecosystems. Altair Rapidminer (23%) and H2O AutoML Studio (19%), showing partial implementations but retaining some essential features for structured analysis and interpretability. These tools generally exhibit well-developed governance frameworks, consistent user interfaces, and strong human-in-the-loop augmentation, all of which reinforce fairness by enabling transparency and oversight. However, their enterprise-centric architectures can limit adaptability and accessibility for smaller, open communities given the cost constraints. Further, the high feature maturity among UI-based tools illustrates a fairness advantage derived from design inclusivity, interpretability, and governance but it also reveals a dependence on proprietary ecosystems and also exposes the lack of all relevant features in any tool.

### Library based tools

Library-driven or code-first AutoML frameworks exhibit minimal feature integration, averaging 5% feature existence, reflecting their focus on computational flexibility rather than governance or fairness infrastructure.

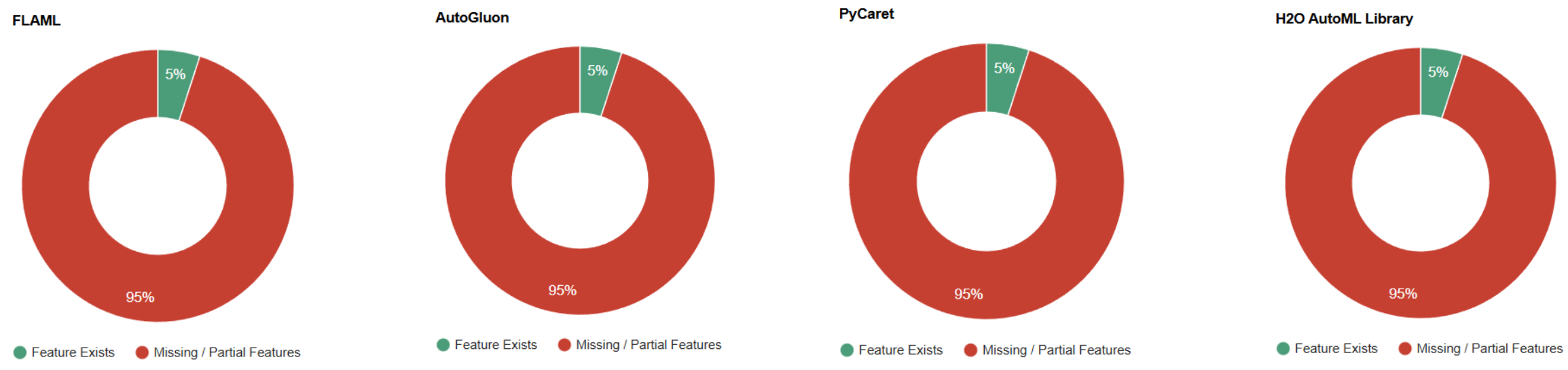

Tools such as FLAML, AutoGluon, PyCaret, and H2O AutoML Library prioritize performance optimization and developer control, but lack embedded ethical safeguards, user management, or fairness-aware augmentation. Their code-based accessibility supports technical inclusivity for skilled users yet creates a fairness gap for non-experts, as interpretability, accessibility, and accountability mechanisms are not natively embedded. While code-based tools democratize experimentation and innovation, absence of built-in fairness governance and interpretability frameworks highlights a systemic gap, wherein fairness depends on user responsibility rather than tool design. This further exasperates the complexity where the user is a non-expert.

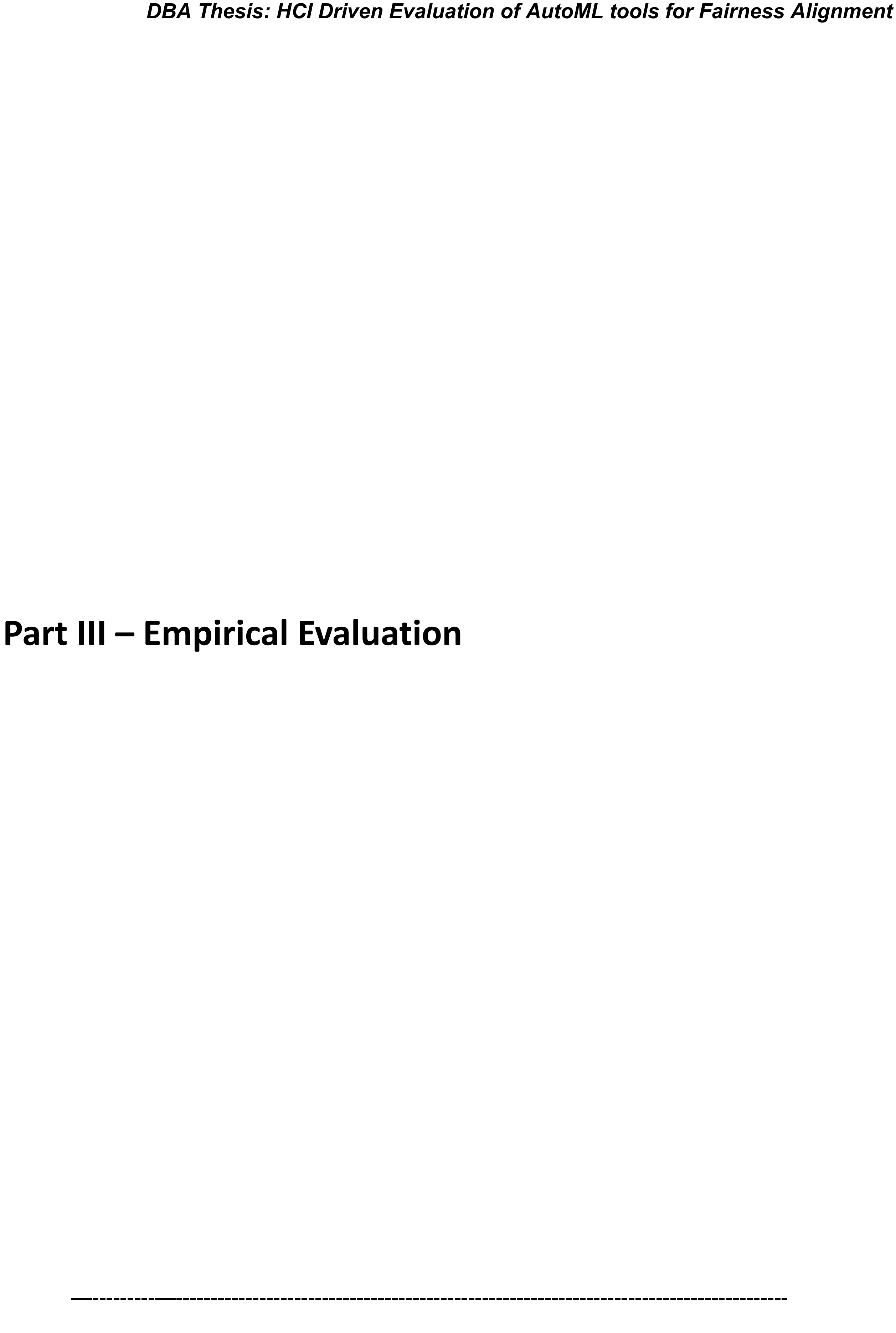

# Part III – Empirical Evaluation

# Chapter 7: Evaluation of AutoML tools and libraries on fairness metrics leveraging the human resources datasets

## Chapter Overview

**Quantitative Fairness Evaluation Context:** This chapter presents a scenario-driven quantitative audit of Automated Machine Learning (AutoML) tools and libraries using curated Human Resources (HR) datasets to examine fairness in model outcomes beyond accuracy. It operationalizes fairness evaluation through empirical testing, emphasizing the ethical implications of bias in hiring, promotion, satisfaction, and attrition predictions.

**Analytical Focus and Framework:** The chapter applies evaluation using the HCI-based fairness framework established in the previous chapter, integrating group fairness metrics such as True Positive Rate (TPR), False Positive Rate (FPR), Demographic Parity Difference (DPD), and Predictive Rate Parity (PRP). The objective is not ranking models, but understanding disparities in fairness behavior across different AutoML platforms.

**Evaluation Design and Process:** The chapter adopts separate pipelines for GUI-based and code-based tools evaluation, each performing model training, prediction, and fairness metric computation using the *fairlearn* library. The process included heuristic auditing of prediction outcomes, compiling comparative fairness results, and assessing metric stability across multiple HR datasets.

**Model Performance and Fairness:** The chapter expresses that the accuracy levels varied widely across tools and datasets, underscoring that high predictive accuracy does not guarantee fairness. GUI-based tools like DataRobot and RapidMiner provided structured fairness visibility, while code-based tools such as AutoGluon and FLAML demonstrated strong quantitative fairness with limited interpretive support.

**Fairness Outcomes Across Contexts:** The chapter analysis also exhibits that no tool achieved full fairness consistency across datasets. AutoGluon frequently emerged as the most balanced performer, while DataRobot and Dataiku occasionally exhibited overfitting or gender-based bias. Tools such as PyCaret and H2O displayed selective fairness patterns—fair in promotion but biased in satisfaction or recruitment contexts.

**Overall Observations and Implications:** The chapter concludes that fairness readiness in AutoML remains uneven and context-dependent. Results reveal fragmented fairness integration, with technical parity achieved in some tools but limited transparency and accountability across the ecosystem. The findings emphasize the need for embedded fairness diagnostics, interpretability mechanisms, and standardized human-centered evaluation practices within AutoML pipelines.

The primary objective of this chapter is to execute a quantitative, scenario-driven audit by running models developed through the identified Automated Machine Learning (AutoML) tools and libraries on curated Human Resources (HR) datasets. This analysis leverages established group fairness metrics—specifically True Positive Rate (TPR), False Positive Rate (FPR), Demographic Parity Difference (DPD), and Predictive Rate Parity (PRP)—to empirically compare model outcomes across platforms and identify consistent disparities in model behavior.

# A. Introduction

Performance of models developed through the AutoML tools and libraries on human resources curated datasets (Refer Chapter 2.3 for details), were further analysed from a fairness perspective focusing on seven specific metrics (Refer Chapter 2.5 for details). Out of the above referred tools and libraries, H2O Studio was considered for feature analysis, the tool could not be used for testing due to the bugs and failure to execute the model creation process.

To simulate and analyse fairness-critical modeling tasks on curated datasets (Refer Chapter 2.3 for details), the following process was adopted:

| For code libraries | For UI based tools |
| --- | --- |
| 1. Created code blocks on Google collab (notebooks allow to combine executable code along with text in the same place) adopting the illustrations provided on the code libraries.<br>2. Uploaded the datasets to Google colab using the user interface functionality.<br>3. Ran the code blocks to generate predictions.<br>4. Compiled the prediction results and fairness metrics using fairlearn python library<br>5. Combined the results and analyzed them for performance of the models from a fairness perspective. | 1. Created user accounts on each of the UI based tool.<br>2. Uploaded the datasets to tool using the user interface functionality.<br>3. Ran the AutoML process to generate predictions.<br>4. Compiled the prediction results and fairness metrics using fairlearn python library<br>5. Combined the results and analyzed them for performance of the models from fairness perspective. |

The outcomes of the analysis are provided below:

# B. Overall accuracy

Experimental results suggested that the accuracy levels of the models modulated based on the tool used for each dataset. In a hiring dataset, accuracy is a representation of how well the model predicted the person to be hired as against the actual hiring.

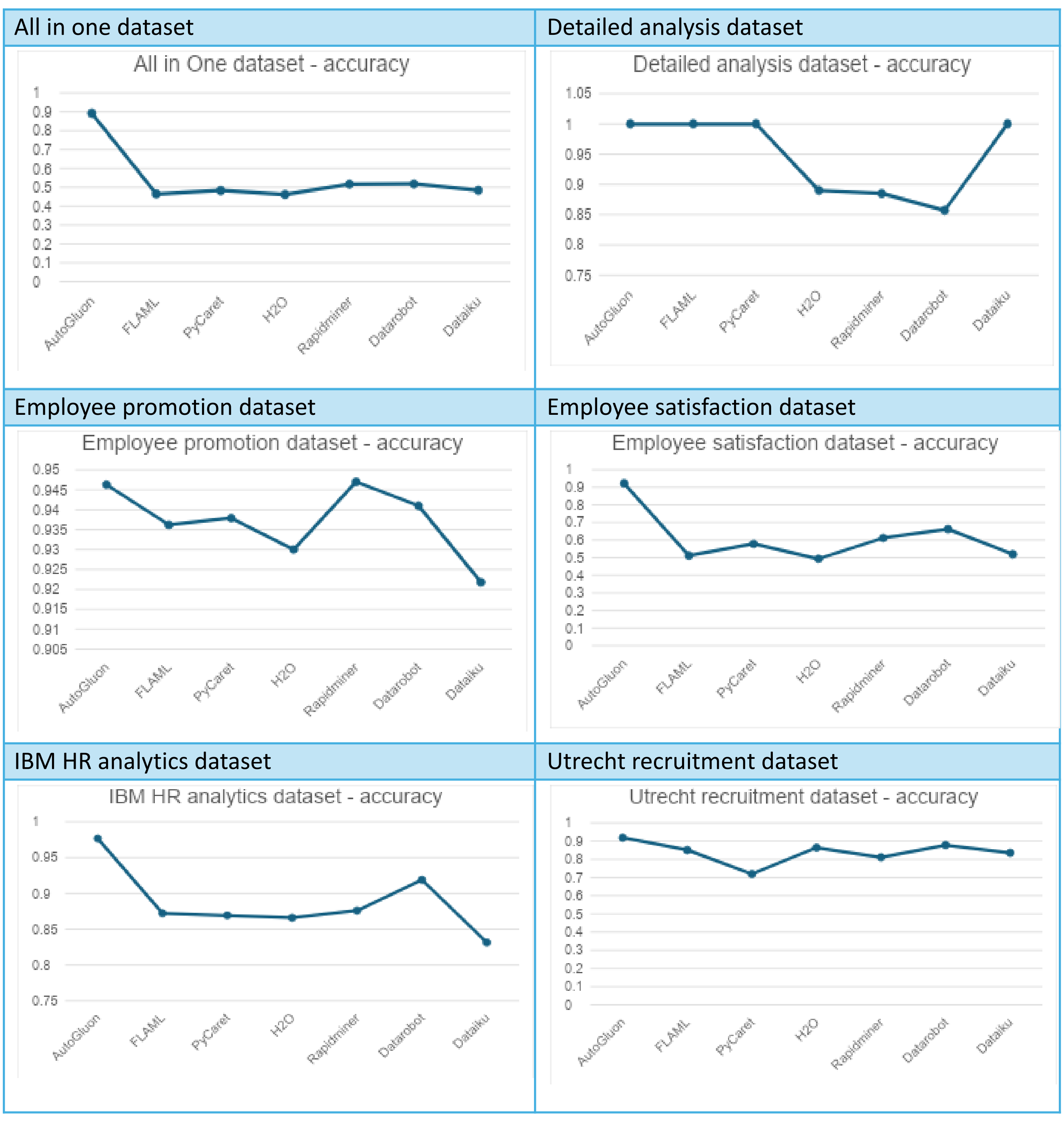

However, accuracy cannot be the only determining factor in making decisions regarding sensitive areas, e.g. hiring. To understand the fairness implications of predictions done by models leveraging each tool, specific fairness metrics were examined. Fairness metrics considered for the analysis are based on the literature review referred in Chapter 2.2 and regulatory expectations referred in Chapter 3.

—-------—------------------------------------------------------------------------------

# C. Fairness metrics in human resources context

The following measures, including the Demographic Parity Difference (DPD) and Equality of Opportunity, are critical group fairness metrics derived from the comprehensive literature review (Chapter 2) and synthesis of regulatory expectations (Chapter 3). These metrics are foundational for algorithmic auditing in high-stakes domains like human resources because they explicitly measure and quantify disparate impact and treatment across sensitive groups, fulfilling key requirements such as those detailed in the NYC Bias Audit and EEOC guidance

| Metric | Definition | Example in Hiring | Focus | Assessment threshold |
|---|---|---|---|---|
| True Positive Rate (TPR)(Sensitivity, Recall) | Proportion of actual qualified candidates correctly hired by the model. | Of all truly qualified candidates, what fraction does the algorithm hire? | Focuses only on identifying the qualified; ignores unqualified applicants. | (F) ≤ 0.10, (PB)≤ 0.25, (HB) > 0.25 |
| False Positive Rate (FPR) | Proportion of unqualified candidates wrongly hired. | Of all truly unqualified candidates, what fraction does the algorithm mistakenly hire? | Captures model errors involving hiring unqualified candidates. | (F) ≤ 0.10, (PB)≤ 0.20, (HB) > 0.20 |
| Selection Rate | Overall proportion of applicants who are hired. | Out of all applicants (qualified and unqualified), how many does the model hire? | Measures hiring frequency, not correctness. | Used with DPD; large differences across groups imply bias |
| Demographic Parity Difference (DPD)(Statistical Parity) | Difference in hiring rates between demographic groups. | If 73.7% of females and 69.6% of males were hired, then DPD = 0.041. | Focuses on equal treatment regardless of qualification. | (F) ≤ 0.10, (PB)≤ 0.20, (HB) > 0.20 |

—---------—-----------------------------------------------------------------------------------------

| Equalized Odds Difference (ΔEO) | Difference in both TPR and FPR between groups. | If TPR difference is 0.098 and FPR difference is 0.097, ΔEO = 0.195. | Enforces fairness in both opportunity and error. | (F) ≤ 0.10, (PB)≤ 0.25, (HB) > 0.25 |
|---|---|---|---|---|
| Equality of Opportunity (EOpp) | Difference in TPR between groups. | If TPR for females is 0.789 and for others is 0.691, ΔEOpp = 0.098. | Focuses only on fairness in hiring qualified candidates. | (F) ≤ 0.10, (PB)≤ 0.20, (HB) > 0.20 |
| Predictive Rate Parity (PRP)(Precision Parity) | Difference in precision (accuracy of hires) between groups. | If precision is 0.533 for females and 0.490 for others, PRP = 0.043. | PRP evaluates whether hired candidates are equally likely to be qualified across groups, focusing on precision rather than selection rates. | (F) ≤ 0.05, (PB)≤ 0.15, (HB) > 0.15 |

{TP: True Positives (qualified and hired); FP: False Positives (unqualified but hired); FN: False Negatives (qualified but not hired); TN: True Negatives (unqualified and not hired)}

## D. Overall summary of fairness metrics comparison based on results from tools and libraries across datasets

On analyzing the fairness metrics for each dataset across all the models from the identified tools or libraries, the following key insights were noted:

- None of the tools had models that were fair across all datasets. On the contrary only in one dataset all the models had consistent fair results.
- Each model from the identified tool or library had partial biased results or heavily biased results based on the metrics.
- The models were not independently reliable on fairness unless examined for specific fairness metrics

**Fairness Summary Across AutoML Tools**

| Dataset | AutoGluon | FLAML | PyCaret | H2O | RapidMiner | DataRobot | Dataiku |
|---|---|---|---|---|---|---|---|
| All-in-One (Hiring) | F | PB | HB | PB | F | HB | NA |
| Detailed Analysis (Hiring) | F | F | F | F | F | F | F |
| Employee Promotion | F | F | F | F | HB | F | PB |
| Employee Satisfaction | F | HB | HB | PB | HB | PB | HB |
| IBM HR Analytics | F | F | F | PB | PB | F | HB |
| Utrecht Recruitment | PB | PB | HB | HB | PB | HB | F |

**Legend:** Fair (F) Partially Biased (PB) Highly Biased (HB) Not Applicable (NA)

Detailed analysis of each of the dataset with reference to the fairness metrics collated from respective tools are provided below:

### D1. All in One dataset

The All in One dataset includes 1,204 rows with a balanced target variable ('Offered' vs. 'Rejected'), good data quality, and no missing values. However, it showed notable outliers in 'Desired Salary' and inconsistent date formats. While no significant multicollinearity was detected (VIF < 7), placement rates varied by education level with Master's degree holders being more likely to be offered a position (Refer Chapter 2.3 for details). Fairness analysis across gender revealed that AutoGluon performed the most equitably, with nearly equal TPRs and low differences in PRP and error rates.

---

In contrast, FLAML and H2O were technically fair due to uniformly low selection rates but failed to identify even highly qualified candidates, suggesting utility issues. PyCaret and DataRobot introduced group-level biases, either over-hiring females or under-hiring males, while Dataiku achieved perfect predictions for all groups, indicating possible overfitting or unrealistic behavior. Overall, fairness varied widely across tools, with some needing stronger fairness-aware calibration and others requiring validation for predictive robustness. The details are provided below:

### D1.1 Fairness Summary Table – Gender in Hiring

Across all tools, fairness analysis using gender as a sensitive variable in hiring shows that AutoGluon and RapidMiner are the fairest with balanced outcomes, FLAML and H2O exhibit mild to moderate bias, PyCaret and DataRobot show strong female-favoring bias, and Dataiku's results are invalid due to incorrect predictions.

| Tool | DPD | ΔEO | ΔEOpp | PRP | Analysis Remarks |
|---|---|---|---|---|---|
| AutoGluon | 0.026 | 0.039 | 0.021 | 0.039 | (F) Very fair |
| FLAML | 0.021 | 0.078 | 0.063 | 0.159 | (PB) Weak at identifying talent |
| PyCaret | 0.224 | 0.246 | 0.201 | 0.043 | (HB) Unfair, over-hiring females |
| H2O | 0.190 | 0.161 | 0.136 | 0.161 | (PB) Biased hiring patterns |
| RapidMiner | 0.041 | 0.195 | 0.098 | 0.043 | (F) Generally balanced |
| DataRobot | 0.140 | 0.442 | 0.280 | 0.198 | (HB) Biased against male candidates |
| Dataiku | 0 | 0 | 0 | 0 | (NA) Incorrect prediction results |

*Note: DPD* measures demographic parity difference, *ΔEO* measures equalized odds gap, *ΔEOpp* measures equality of opportunity gap, and *PRP* measures predictive rate parity — lower values indicate fairer outcomes.

## D1.2 Tool-wise Remarks in Hiring Context

Overview of the tool wise analysis provided below:

| Tool | Remarks |
|---|---|
| AutoGluon | AutoGluon is like a fair hiring manager. It evaluates male, female, and non-binary candidates nearly equally. For example, if 100 qualified candidates from each group applied, it would hire about 90 from each — showing consistent fairness and accuracy. |
| FLAML | FLAML behaves like a very cautious recruiter who is afraid to make hiring mistakes — so it hardly hires anyone. Even highly qualified candidates are rejected. For instance, it might only hire 5 out of 100 strong applicants from each group. It's technically fair because it treats all poorly. |
| PyCaret | PyCaret shows a clear preference — it tends to over-hire women. For example, it might hire 7 out of 10 qualified women but only 4 or 5 out of 10 equally qualified men or non-binary individuals. This favoritism creates unfairness and inflates hiring of unqualified candidates from the favored group. |
| H2O | H2O is generous in hiring but not equally so. It hires a large number of female candidates — even if some aren't fully qualified — while being more conservative with other groups. Think of a recruiter who favors one demographic unintentionally, leading to bias and inconsistent outcomes. |
| RapidMiner | RapidMiner is a fairly balanced recruiter. It gives most candidates from each gender a reasonable and nearly equal chance of being hired. There are some small gaps, but overall, if you're a strong candidate, your chances don't depend heavily on gender. |
| DataRobot | DataRobot seems biased in the opposite way — it hires a significantly lower proportion of qualified male candidates. For instance, it might hire 7 out of 10 qualified women but only 4 out of 10 qualified men. This could appear discriminatory and introduces fairness and legal risks. |
| Dataiku | Dataiku doesn't hire anyone. It rejects every application, no matter how qualified the candidate is. It's like an HR system that's completely broken or turned off. There's no hiring, so technically there's no bias — but also no value. |

## D2. Detailed Analysis dataset

The Detailed analysis dataset had 215 rows with ~31% missing salary values, which may introduce selection bias if not handled carefully. Placement rates varied sharply by education and work experience (Refer Chapter 2.3 for details). Despite the small size, tools like AutoGluon, FLAML, PyCaret, H2O, and Dataiku all returned near perfect or perfect predictions with equal TPRs across genders, indicating either excellent generalization or potential overfitting. RapidMiner and DataRobot showed slightly uneven TPRs and FPRs, with minor tilts toward female candidates. Overall, most tools appeared fair, though the "too perfect" results warrant further robustness checks. The details are provided below:

### D2.1 Fairness Summary Table – Gender in Hiring (Detailed Analysis Dataset)

Across all tools, fairness analysis using gender as a sensitive variable in hiring shows that FLAML, PyCaret, H2O, and Dataiku achieve near-perfect fairness with zero bias, AutoGluon and DataRobot remain highly fair with minimal variation, while RapidMiner, though generally fair, shows a slight female-favoring bias.

| Tool | DPD | ΔEO | ΔEOpp | PRP | Analysis Remarks |
|---|---|---|---|---|---|
| AutoGluon | 0.088 | 0.045 | 0.000 | 0.012 | (F) Very fair and effective |
| FLAML | 0.159 | 0.000 | 0.000 | 0.000 | (F) Extremely accurate and fair |
| PyCaret | 0.150 | 0.000 | 0.000 | 0.000 | (F) Very accurate and balanced |
| H2O | 0.062 | 0.000 | 0.000 | 0.000 | (F) Excellent parity |
| RapidMiner | 0.019 | 0.247 | 0.047 | 0.129 | (F) Fair but female-biased FPR |
| DataRobot | 0.095 | 0.150 | 0.043 | 0.023 | (F) Generally balanced |
| Dataiku | 0.053 | 0.000 | 0.000 | 0.000 | (F) Fair, but unrealistic results |

*Note: DPD* measures demographic parity difference, *ΔEO* measures equalized odds gap, *ΔEOpp* measures equality of opportunity gap, and *PRP* measures predictive rate parity — lower values indicate fairer outcomes.

—-------—-----------------------------------------------------------------------------------

## D2.2 Tool-wise Remarks in Hiring Context with Examples

Overview of the tool wise analysis provided below:

| Tool | Remarks |
|---|---|
| AutoGluon | AutoGluon is like a smart recruiter who hires all qualified candidates equally across genders. For example, if 100 qualified men and 100 qualified women apply, it hires all 200. It does make slightly more false positive hires among men — so out of 20 unqualified male applicants, 3 might get hired vs. 2 among women. |
| FLAML | FLAML behaves like an expert who never makes a hiring mistake — it hires every qualified applicant and never hires an unqualified one. If 150 people apply and 120 are qualified (60 men, 60 women), it hires exactly those 120, no more, no less. Flawless across genders. But such perfect performance should be checked for overfitting. |
| PyCaret | PyCaret also plays the perfect hiring game — hiring every qualified man and woman and ignoring unqualified ones. Imagine a shortlist of 50 strong and 10 weak candidates — PyCaret selects only the 50, treating everyone equally. |
| H2O | H2O hires every qualified applicant just like FLAML and PyCaret. For instance, all 80 deserving candidates (40 women, 40 men) get hired, and none of the unqualified ones do. Its fairness metrics are clean, but again, results should be checked to ensure it's not just perfect on paper. |
| RapidMiner | RapidMiner is mostly fair but tends to favor women in borderline hiring calls. Suppose there are 50 marginal candidates (not obviously qualified), it might hire 17 women and only 7 men from this group. |
| DataRobot | DataRobot gives slightly more opportunity to women. For instance, if 100 equally qualified men and women applied, it might hire 90 women and 85 men. However, it’s also more prone to hiring unqualified women (e.g., 5 out of 20) than unqualified men (3 out of 20). |
| Dataiku | Dataiku is mathematically perfect: it hires all qualified applicants and none of the unqualified ones, treating every gender identically. Imagine 100 great candidates and 20 weak ones — it hires the 100 and rejects the 20, without exception or bias. |

## D3. Employee Promotion

With over 54,000 rows, Employee promotion dataset was large but showed significant imbalance in promotions (only 8.5% were promoted). Missing values in education and performance rating, along with outliers in several numerical fields, demanded attention (Refer Chapter 2.3 for details). None of the tools specifically explain about how they handled the missing values or outliers, except DataRobot which provides a detailed downloadable report for how it handled missing values in each of the variables. Gender-based fairness analysis showed that most tools (AutoGluon, FLAML, PyCaret, H2O, DataRobot) performed well, identifying qualified employees across genders with minimal bias. The details are provided below:

### D3.1 Fairness Summary Table – Gender in Employee Promotion

Across all tools, fairness analysis using gender as a sensitive variable shows that AutoGluon, FLAML, PyCaret, H2O, and DataRobot perform very fairly with minimal bias, while Dataiku shows a mild preference toward female promotion, and RapidMiner demonstrates high bias by under-promoting males.

| Tool | DPD | ΔEO | ΔEOpp | PRP | Analysis Remarks |
|---|---|---|---|---|---|
| AutoGluon | 0.0019 | 0.006 | 0.007 | 0.0065 | (F) Very fair and consistent |
| FLAML | 0.0039 | 0.005 | 0.004 | 0.0041 | (F) Accurate, minor variations |
| PyCaret | 0.0019 | 0.009 | 0.009 | 0.0084 | (F) Reliable, slight skew |
| H2O | 0.0055 | 0.011 | 0.012 | 0.0118 | (F) Acceptably fair, moderate FPR |
| RapidMiner | 0.0050 | 0.159 | 0.157 | 0.2000 | (HB) Unfair: under-promotes males |
| DataRobot | 0.0020 | 0.027 | 0.027 | 0.0020 | (F) Balanced, low bias |
| Dataiku | 0.0270 | 0.088 | 0.073 | 0.0330 | (PB) Slightly favors female promotion |

*Note: DPD* measures demographic parity difference, *ΔEO* measures equalized odds gap, *ΔEOpp* measures equality of opportunity gap, and *PRP* measures predictive rate parity — lower values indicate fairer outcomes.

——————————————————————————————

## D3.2 Tool-wise Remarks in Promotion Context with Examples

Overview of the tool wise analysis provided below:

| Tool | Remarks |
|---|---|
| AutoGluon | AutoGluon is like a promotion committee that treats men and women almost identically. For example, if 100 deserving male and 100 deserving female employees apply, it recommends around 37 from each. It also rarely promotes the unqualified, keeping mistakes minimal. A great example of gender-neutral promotion. |
| FLAML | FLAML also promotes fairly — if 50 eligible men and 50 eligible women seek promotion, it promotes around 29–30 from each. While slightly conservative, it applies the same standards to both genders and avoids promoting unqualified employees. It's an example of a consistent and cautious promoter. |
| PyCaret | PyCaret leans a little toward women but still keeps it mostly fair. If 200 men and women apply, and 100 from each are strong candidates, it might promote 36 women and 35 men. It's reliable, but female promotion rates are just a bit higher — still acceptable. |
| H2O | H2O promotes many deserving employees, especially women. For instance, from 100 strong candidates, 39 women might be promoted compared to 38 men. However, it's also slightly more likely to promote borderline female candidates. Still, overall balanced performance. |
| RapidMiner | RapidMiner clearly under-promotes men. If 100 qualified male and female employees are up for promotion, it promotes only 13 men vs. 29 women. This big gap in opportunity could create workplace trust issues or bias claims. |
| DataRobot | DataRobot performs well across the board. If you had 100 qualified applicants of each gender, it would promote around 32–35 from both groups, making it a dependable and equitable tool for promotion decisions. |
| Dataiku | Dataiku tends to promote more women than men, even among unqualified applicants. For example, if 20 borderline candidates include 10 men and 10 women, it might promote 4 women but only 2 men. This might be unintentional bias creeping in. It's relatively fair overall, but needs closer tuning. |

## D4. Employee Satisfaction

Employee satisfaction dataset had no missing values but presented high multicollinearity between job level and salary, which may distort fairness unless properly handled. Satisfaction rates were slightly in favor of suburban employees (Refer Chapter 2.3 for details). Fairness metrics showed that PyCaret and FLAML were highly biased, consistently underestimating satisfaction among suburban workers. RapidMiner and DataRobot favored suburban locations more noticeably. Dataiku showed unrealistic perfection for city-based employees, indicating probable overfitting. AutoGluon and H2O were the most balanced, though even they showed slight favor toward city dwellers. The details are provided below:

### D4.1 Fairness Summary Table – Location as a sensitive variable in Employee Satisfaction

Across all tools, fairness analysis using location as a sensitive variable shows that AutoGluon and H2O exhibit the fairest outcomes with only slight or mild bias, while FLAML, PyCaret, RapidMiner, and Dataiku show high bias favoring either City or Suburban employees, and DataRobot displays moderate but acceptable fairness with a mild suburban advantage.

| Tool | DPD | ΔEO | ΔEOpp | PRP | Analysis Remarks |
|---|---|---|---|---|---|
| AutoGluon | 0.0586 | 0.067 | 0.018 | 0.0500 | (F) Fair with slight bias |
| FLAML | 0.0715 | 0.291 | 0.199 | 0.1993 | (HB) Uneven satisfaction judgment |
| PyCaret | 0.2727 | 0.534 | 0.291 | 0.2915 | (HB) Biased toward City employees |
| H2O | 0.0582 | 0.118 | 0.098 | 0.0972 | (PB) Mild favor to City |
| RapidMiner | 0.2170 | 0.433 | 0.215 | 0.0190 | (HB) Over-satisfied suburban bias |
| DataRobot | 0.0290 | 0.160 | 0.107 | 0.0930 | (PB) Better satisfaction for suburb |
| Dataiku | 0.1150 | 0.238 | 0.071 | 0.1030 | (HB) Unrealistically favors City |

*Note: DPD* measures demographic parity difference, *ΔEO* measures equalized odds gap, *ΔEOpp* measures equality of opportunity gap, and *PRP* measures predictive rate parity — lower values indicate fairer outcomes.

—--------—----------------------------------------------------------------------------------

## D4.2 Tool-wise Remarks in Employee Satisfaction Context

Overview of the tool wise analysis provided below:

| Tool | Remarks |
|---|---|
| AutoGluon | AutoGluon is like a manager who views satisfaction levels of employees in city and suburb offices almost equally. If 100 satisfied employees from each location were assessed, it would correctly identify ~93 from the city and ~91 from suburbs. A small tilt toward city workers, but not enough to be concerning. |
| FLAML | FLAML treats city employees as more likely to be satisfied. For example, out of 100 city employees who report high satisfaction, 55 are flagged correctly. In contrast, only 35 out of 100 from suburbs are identified, even if equally satisfied. Suburb workers are undervalued, raising concern about geographic bias. |
| PyCaret | PyCaret heavily favors city-based employees. If you had 10 truly satisfied staff from each location, 6 from the city would be recognized, while only 3–4 from the suburbs would be acknowledged. |
| H2O | H2O does a moderately fair job. It detects satisfaction in ~55% of happy city employees and ~45% of their suburban counterparts. It also falsely marks some dissatisfied city workers as satisfied. |
| RapidMiner | RapidMiner over-assigns satisfaction to suburban staff. For instance, nearly 96 out of 100 satisfied suburb employees are correctly flagged, but only 74 out of 100 city employees are. |
| DataRobot | DataRobot is slightly better at understanding suburban satisfaction. For example, it might recognize 79 out of 100 satisfied suburb staff and only 68 of the city-based ones. |
| Dataiku | Dataiku is too perfect with city staff — it marks everyone in the city as fully satisfied, including those who aren't. If 10 city workers are surveyed, all get flagged as happy, even if 2 aren't. In contrast, suburb staff are seen as less satisfied. This is a red flag for overfitting or bias toward urban locations. |

## D5. IBM HR Analytics

IBM HR analytics 1,470-row dataset had no missing values but many features with extreme outliers and low overall attrition (~16%), making it hard to train balanced models (Refer Chapter 2.3 for details). Tools like AutoGluon, FLAML, and PyCaret delivered strong, fair performance, identifying attrition across both genders nearly equally. H2O and DataRobot showed slight biases, with H2O marking more men falsely at risk. Dataiku leaned toward identifying more female attrition while also raising more false alarms for men. RapidMiner had the weakest performance, missing most attrition cases overall. Still, fairness metrics were generally moderate, with no extreme violations except for Dataiku's imbalance. The details are provided below:

### D5.1 Fairness Summary Table – Gender in Employee Attrition

Overall, the employee attrition fairness analysis indicates that most tools perform fairly with minimal gender bias, with AutoGluon, FLAML, and PyCaret showing the most balanced outcomes, H2O and RapidMiner displaying moderate disparities, and Dataiku exhibiting the highest bias, tending to over-predict attrition for females.

| Tool | DPD | ΔEO | ΔEOpp | PRP | Analysis Remarks |
|---|---|---|---|---|---|
| AutoGluon | 0.0119 | 0.045 | 0.000 | 0.0451 | (F) Very accurate & fair |
| FLAML | 0.0068 | 0.038 | 0.007 | 0.0099 | (F) Fair, with minor variance |
| PyCaret | 0.0039 | 0.067 | 0.013 | 0.0132 | (F) Mostly fair and reliable |
| H2O | 0.0404 | 0.134 | 0.003 | 0.1346 | (PB) Acceptable, but male FPR high |
| RapidMiner | 0.0010 | 0.063 | 0.055 | 0.1100 | (PB) Lower TPR for both, slight edge for females |
| DataRobot | 0.0190 | 0.046 | 0.029 | 0.1330 | (F) Decent balance, slight female favor |
| Dataiku | 0.0510 | 0.175 | 0.094 | 0.2320 | (HB) More attrition detected in females |

*Note: DPD* measures demographic parity difference, *ΔEO* measures equalized odds gap, *ΔEOpp* measures equality of opportunity gap, and *PRP* measures predictive rate parity — lower values indicate fairer outcomes.

—------—----------------------------------------------------------------------------------------

## D5.2 Tool-wise Remarks in Attrition Prediction Context

Overview of the tool wise analysis provided below:

| Tool | Remarks |
|---|---|
| AutoGluon | AutoGluon performs like an HR analytics engine that flags attrition perfectly — it identifies every employee who is going to leave, across both genders, while keeping false alarms lower for women. If 100 women and 100 men plan to leave, it correctly flags all 200. A strong and equitable model. |
| FLAML | FLAML is highly precise, catching almost every person intending to leave across both genders. For example, if 50 women and 50 men are about to leave, it identifies 49 of each. It does trigger slightly more false positives for women, so may occasionally overestimate female attrition. |
| PyCaret | PyCaret acts with high recall, identifying almost every at-risk employee, but is a bit more generous with men when marking attrition. Out of 100 women and 100 men planning to leave, it finds all women and 98–99 men, with more false flags for men who weren't planning to leave. |
| H2O | H2O performs well, but is more likely to misclassify men who won't leave as at-risk. Think of a scenario where 100 employees are flagged for leaving: some men may be wrongly marked, potentially causing confusion or misallocated retention efforts. |
| RapidMiner | RapidMiner is more conservative — it catches only 33–39% of those about to leave. It's slightly better with women, but both genders are under-detected. Good for avoiding false positives, but may miss many real attrition risks. |
| DataRobot | DataRobot moderately identifies leavers — about 54% of men and 57% of women are correctly flagged. It doesn't falsely flag many, especially women. It's like an HR analyst who is cautious but generally even-handed. |
| Dataiku | Dataiku flags more women than men as at-risk and also raises more false alarms for men. For instance, out of 10 women planning to leave, it catches 6, but for 10 men, it finds only 5. However, it wrongly flags more men than women who weren't planning to leave, indicating possible gender imbalance. |

## D6. Utrecht Recruitment

The Utrecht recruitment dataset, with 4,000 applicants, showed clean structure but notable imbalance in acceptance rates (~68% rejected) and strong multicollinearity between age and university grades (Refer Chapter 2.3 for details). Certain features like international experience and entrepreneurship were disproportionately linked to success, raising fairness flags. Gender-based fairness metrics revealed significant disparities: tools like DataRobot completely excluded “Other” gender candidates, and PyCaret heavily favored males with over twice the TPR for them compared to females. Only Dataiku showed relatively balanced performance, while most other tools exhibited moderate to high bias favoring male applicants.

### D6.1 Fairness Summary Table – Gender in Recruitment (Utrecht Dataset)

Overall, the Utrecht recruitment fairness analysis shows that most tools exhibit moderate to high gender bias, with DataRobot and H2O showing the greatest disparities, while Dataiku performs most fairly and AutoGluon, FLAML, and RapidMiner show only slight male preference.

| Tool | DPD | ΔEO | ΔEOpp | PRP | Analysis Remarks |
|---|---|---|---|---|---|
| AutoGluon | 0.1091 | 0.148 | 0.120 | 0.1187 | (PB) Slight male preference |
| FLAML | 0.0753 | 0.221 | 0.096 | 0.0950 | (PB) Slight skew, male favored |
| PyCaret | 0.1155 | 0.319 | 0.226 | 0.2256 | (HB) Heavily biased, under-selects women |
| H2O | 0.2021 | 0.479 | 0.331 | 0.3310 | (HB) High disparity in selection |
| RapidMiner | 0.1010 | 0.132 | 0.071 | 0.0740 | (PB) Minor imbalance, favors males |
| DataRobot | 0.2090 | 0.948 | 0.823 | 0.8230 | (HB) Extremely biased against “Other” |
| Dataiku | 0.0860 | 0.107 | 0.082 | 0.1040 | (F) Acceptable performance |

*Note: DPD* measures demographic parity difference, *ΔEO* measures equalized odds gap, *ΔEOpp* measures equality of opportunity gap, and *PRP* measures predictive rate parity — lower values indicate fairer outcomes.

### D6.2 Tool-wise Remarks in Recruitment Context

Overview of the tool wise analysis provided below:

---

| Tool | Remarks |
|---|---|
| AutoGluon | AutoGluon gives a mild edge to male applicants. For instance, if 100 highly qualified male and female candidates apply, about 88 males and 78 females may be selected. It also marks slightly more unqualified male candidates as suitable compared to others. While relatively fair, it may benefit from bias calibration. |
| FLAML | FLAML leans slightly toward male candidates. If 50 male and female candidates each are qualified, FLAML selects about 39 men vs. 35 women, which is not dramatic. However, "Other" gender group shows varied results, and false positive rates are noticeably higher. |
| PyCaret | PyCaret exhibits major gender bias. Female candidates are less than half as likely to be selected than male candidates of equal merit. For example, only 27 out of 100 qualified women might get selected vs. 49 men. Such outcomes could reinforce systemic hiring discrimination. |
| H2O | H2O shows a significant skew: male candidates have a 33% higher chance of selection than non-binary or female applicants. For example, if 10 candidates from each group are equally qualified, H2O might select 9 men, 7 women, and only 6 from “Other”. It risks discriminatory recruitment behavior. |
| RapidMiner | RapidMiner performs reasonably but gives a slight edge to male and “Other” candidates. For instance, if 10 candidates from each gender are equally qualified, it might pick 5–6 men, 4–5 women, and 5–6 from Other. Not egregious, but not ideal. |
| DataRobot | DataRobot shows a very large fairness gap, completely excluding “Other” gender candidates and substantially favoring males. For example, in a group of strong applicants, males are highly likely to be picked, females moderately so, but non-binary candidates are left out entirely. |
| Dataiku | Dataiku is among the fairest here. It selects candidates in a relatively balanced way: if 100 qualified people from each gender apply, it might select 94 men, 86 “Other,” and 87 women. Minor variances exist, but overall promotes equitable recruitment. |

# Chapter 8: Discussion on evaluation outcomes

## Chapter Overview

**Chapter Context:** This chapter synthesizes insights from the qualitative HCI evaluation and quantitative fairness benchmarking to examine how Automated Machine Learning (AutoML) tools enable or constrain fairness-aware, human-centered interaction. It highlights recurring design patterns and systemic limitations shaping users' ability to achieve equitable outcomes in sensitive domains such as hiring and promotion.

**Fairness-Aware Data Handling:** The clapter clarifies that while most tools provide basic data preprocessing support, only platforms like DataRobot and Dataiku integrate fairness-aware data handling through bias detection and guided preprocessing interfaces. However, the lack of support for identifying multicollinearity and proxy variables increases cognitive load and leaves fairness interventions largely user-dependent.

**User Guidance and Feedback Mechanisms:** The chapter explains variations in approaches by different tools. For example, DataRobot demonstrates strong, structured fairness feedback loops, allowing users to iteratively adjust fairness settings and observe results. In contrast, other tools—especially code-based and semi-automated platforms—lack interactive feedback or guidance, leaving non-experts unable to trace or refine fairness decisions effectively.

**Transparency, Explainability, and Guardrails:** The chapter highlights that DataRobot and Dataiku stand out in providing clear fairness explanations, visual feedback, and human-in-the-loop (HITL) safeguards. Most other tools fail to connect fairness inputs to model behavior, offering limited interpretability and weak oversight. The absence of visual guardrails and fairness alerts contributes to automation bias and reduced user trust.

**Visualization and Fairness Trade-offs:** The chapter also highlights that Dataiku and DataRobot excel in presenting fairness metrics, bias diagnostics, and fairness–performance trade-offs through interactive dashboards. Other tools offer limited or no fairness-specific visualization, requiring technical expertise to interpret bias metrics and hindering accessibility for non-experts.

**Concluding Insights:** The chapter concludes that current AutoML systems place excessive responsibility on users to identify and correct bias, often without adequate transparency, feedback, or control mechanisms. True democratization of AI requires reimagining HCI design toward fairness-by-default—embedding intuitive fairness guidance, human oversight, and transparent feedback loops that empower non-experts to make equitable and informed modeling decisions.

This chapter synthesizes the qualitative and quantitative insights derived from the Human-Computer Interaction (HCI) evaluation (Chapter 6) and the scenario-based fairness metric benchmarking (Chapter 7) to illuminate systemic limitations and recurring design patterns in contemporary Automated Machine Learning (AutoML) tools. This discussion evaluates how existing interfaces and workflows either enable or inhibit non-expert users from achieving equitable outcomes in high-stakes domains like hiring.

# A. Discussion

The proliferation of AutoML tools promises to democratize machine learning, making sophisticated model development accessible to a broader user base, including non-experts. However, the successful and ethical deployment of these automated systems, particularly in sensitive domains, critically hinges on how non-experts are guided and supported in ways that reduce complexity, enhance perceived usefulness, and align with their cognitive expectations and workflows. This discussion evaluates various AutoML tools and libraries from a Human-Computer Interaction (HCI) perspective based on the aforementioned evaluation, focusing on how their interfaces and workflows enable non-experts to address fairness in tabular classification and prediction problems.

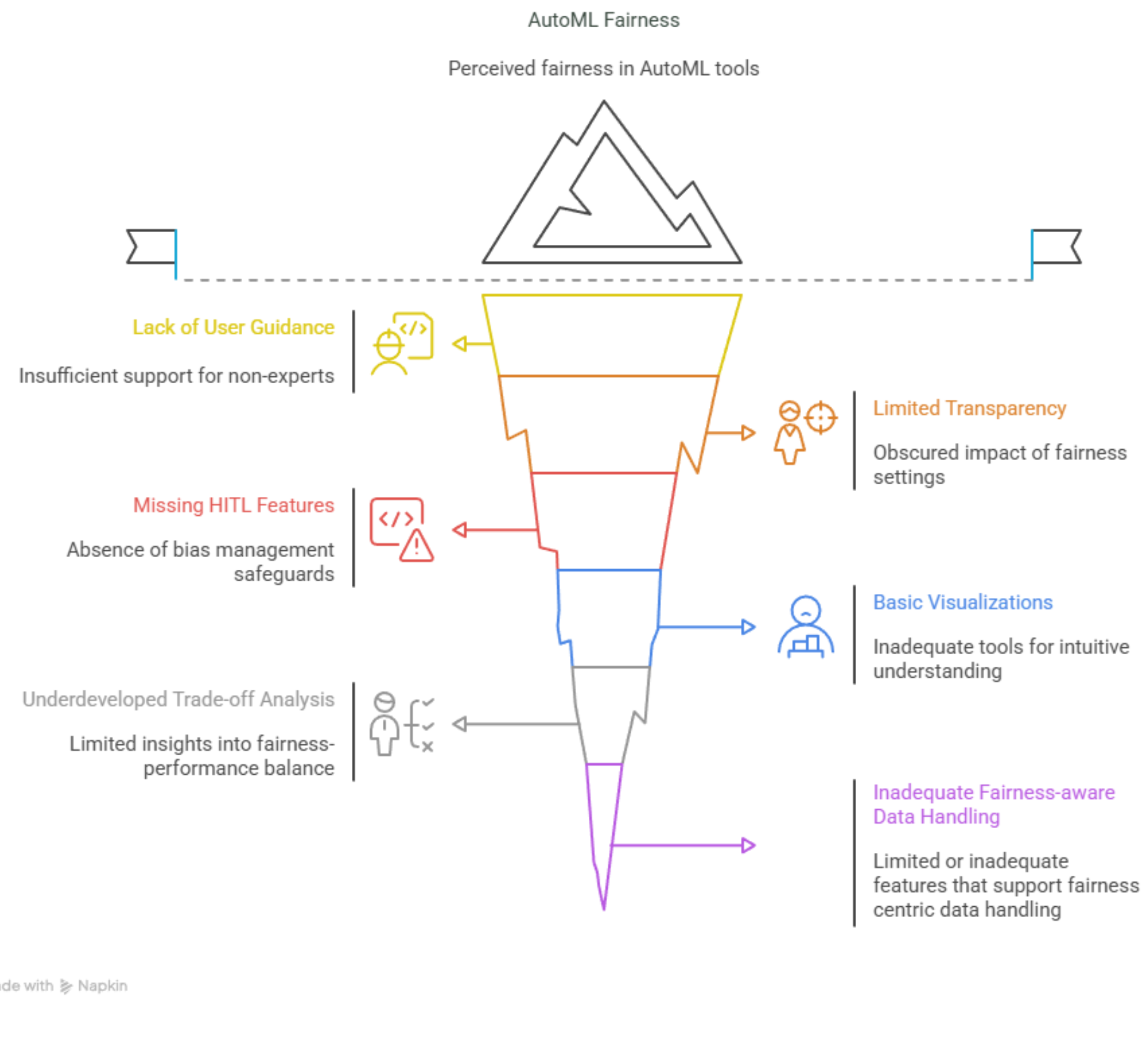

## A1. Data Preprocessing and Fairness-Aware Data Handling

Effective bias mitigation often begins at the data preprocessing stage. All AutoML tools, including Dataiku, DataRobot, H2O Studio, RapidMiner, and code-based libraries, inherently offer capabilities for fundamental data preprocessing tasks, such as handling missing values, outlier removal, encoding, and scaling. This foundational support is universal across landscapes.

However, a key difference emerges in the integration of fairness-aware data handling. Platforms such as Dataiku and DataRobot distinguish themselves by enhancing preprocessing with user-friendly graphical interfaces, automated data quality checks, and, crucially, specific bias detection mechanisms integrated directly into the preprocessing stages. For instance, DataRobot's Bias Mitigation features can include bias detection and mitigation within the pipeline with robust GUI support. Dataiku's custom preprocessing within a visual ML tool and advanced data quality rules actively promote responsible AI by identifying potential biases in the data source. This interface-level support exemplifies effective affordance design, where bias detection is made perceptible and actionable for users (Affordance Theory), enhancing ease of use and usefulness (TAM), and reducing complexity. These platforms explicitly include fairness-aware features, such as bias detection, fairness metric selection, and automated or manual bias mitigation techniques at this early stage. This proactive approach, guided by user interfaces, significantly assists non-experts in identifying and addressing potential data-driven biases before model training begins. However, none of the tools or libraries are able to guide the user to identify multicollinearity and proxy variables explicitly. This absence places a higher cognitive burden on users and reflects a missed opportunity for progressive disclosure or guided simplification (Cognitive Load Theory).

In contrast, while H2O Studio, RapidMiner, and other tools provide basic bias detection or allow for flagging sensitive features, these functionalities are often more limited and typically require manual intervention or additional configurations. While Rapidminer identifies certain variables like Gender and Age as sensitive variables, it too does not identify multicollinearity or proxy explicitly. Code-based tools, while automating many preprocessing steps, generally lack built-in bias safeguards or fairness-aware preprocessing checks without substantial manual coding, placing a higher burden on non-expert users to implement these crucial steps. In absence of structured feedback systems, users must mentally map input changes to outputs, violating the cognitive economy principle and risking overload (Cognitive Load Theory). Further none of the evaluated tools surfaced preprocessing warnings or addressed these structural data issues.

## A2. User Guidance and Feedback Mechanisms for Fairness

A fundamental component of effective human–computer interaction (HCI) in the realm of responsible Artificial Intelligence (AI) is the delivery of explicit guidance and comprehensive feedback. Within the fairness framework, this entails aiding non-experts in comprehending potential biases, establishing fairness criteria, and evaluating the consequences of their interventions.

Platforms such as DataRobot place significant emphasis on structured user input and feedback concerning fairness. Users are guided in configuring fairness metrics, selecting protected features, and defining thresholds, thereby enabling traceable input on both the general model and fairness-specific aspects. This structured approach helps to demystify complex fairness concepts for non-experts by providing clear interaction points. This structured feedback aligns with Design Thinking principles of iterative user involvement and supports user agency through explainable, controllable interfaces (HCAI). Furthermore, DataRobot's capacity to integrate human feedback into fairness-oriented model refinement constitutes a crucial strength of the human-computer interaction (HCI). Their workflows permit users to set and adjust fairness constraints, apply mitigation techniques, and immediately observe the impact of these changes on model outcomes. This iterative feedback loop, supported by visual documentation such as bias mitigation badges and before/after bias metrics, offers a tangible understanding of how fair inputs translate into changes in model decisions. Such visual feedback loops enhance observability and facilitate perceived usefulness, two critical factors for technology adoption (TAM, IDT). This direct and visual linkage assists non-experts in comprehending the consequences of their fair choices, which are essential for effective bias mitigation. DataRobot explicitly facilitates structured feedback on both the model and fairness features, enabling users to configure and provide inputs on protected features, metrics, and thresholds. This formalized process ensures the traceability and integration of user feedback into fairness-driven refinement. DataRobot's support for feedback-driven refinement of fairness explanations allows users to iteratively adjust settings and immediately observe the impact through updated visualizations and explanations.

In contrast, Dataiku, while accommodating user feedback and permitting adjustments based on input (e.g., via the Model Fairness Report plugin), demonstrates a less formalized and systematic approach to fairness feedback. Although users can assess and report fairness and modify pipelines, the platform currently lacks formal systematized channels for structured fairness feedback that are directly integrated into fairness-driven model-refinement loops. Consequently, while non-experts can respond to fairness reports, the system does not inherently guide them through a continuous cycle of feedback and refinement, specifically for fairness.

—-------—------------------------------------------------------------------------------------

Such actions / expectations potentially impose a greater cognitive burden on the user to interpret and act upon findings. Although iterative model refinement is a fundamental feature of Dataiku, its fairness-specific feedback loops, while present, are less deeply integrated than those of DataRobot, relying more on the user interpretation of updated reports rather than a direct, automated feedback-to-explanation cycle. This reflects a high intrinsic cognitive load and limited support for recognition-based interaction, contrary to best practices in cognitive load reduction.

Various AutoML tools and libraries, such as H2O Studio and RapidMiner, and code-based tools, such as FLAML, AutoGluon, and PyCaret, typically offer limited or no integrated, user-friendly feedback mechanisms for fairness. Although some tools permit manual editing or sensitivity checks, they frequently lack comprehensive systems for structured feedback or fair governance within their user interfaces. For individuals without expertise, this requires a more profound comprehension of the underlying code and statistical concepts, thereby impeding their capacity to actively engage in fairness considerations. The absence of bidirectional feedback features in these tools implies that users cannot readily observe how their fairness inputs directly influence the model behavior or outcomes, rendering the process of bias mitigation less intuitive and more challenging. For instance, models with serious fairness violations (e.g., PyCaret under-selecting qualified women; DataRobot excluding “Other” gender entirely) did not issue any user-facing warnings or triggers. Further there existed no interactive feedback loop prompts users to reevaluate biased results or retrain with rebalancing fairness priorities. These tools generally lack integrated fairness feedback loops and depend on manual checks or external analyses.

## A3. Transparency, Explainability, and Limitations Disclosure

Ensuring fairness in human-computer interaction necessitates transparency concerning model behavior, the effects of fairness interventions, and the explicit elucidation of decisions related to fairness. It is imperative that non-experts receive clear communication regarding the presence and reasons for any ambiguity or uncertainty surrounding fairness, as well as how their actions may influence these factors.

DataRobot demonstrates ease of use by visually documenting the effects of fair intervention. Features such as per-class bias charts and before-and-after metrics explicitly illustrate the impact of modifications to fairness settings or mitigation techniques on model predictions and fairness outcomes. This transparent feedback mechanism enables non-experts to comprehend the implications of their decisions and fosters trust in the system. The platform's dedication to clearly documenting the influence of user-defined fairness settings and protected features on model outputs and fairness metrics further enhances transparency and traceability.

This improves user trust and supports the core tenets of Human-Centered AI—transparency, reliability, and user empowerment. DataRobot offers comprehensive explanations for fairness-related decisions through guided workflows for metric selection, root-cause analysis, and visual insights into testing and mitigation, clearly indicating which protected features and metrics are employed and how bias is addressed. It provides a variety of explanations, including permutation-based feature impact, SHAP explanations, prediction explanations, and specific visualizations for fairness metrics and bias-mitigation actions.

Dataiku prioritizes transparency and explainability through its Model Fairness Report plugin and interactive subpopulation analysis. These tools enable users to measure and visualize fairness metrics and assess uncertainty, thereby enhancing their understanding of the model behavior. Dataiku offers comprehensive explanations for fairness-related decisions, including detailed reports on selected metrics, harm types, and subpopulation analysis, thereby elucidating the rationale behind fairness metrics and implications of model choices. It supports various explanation types such as feature importance, partial dependence plots, subpopulation analysis, and error analysis, all of which are contextualized for fairness. This platform provides transparency regarding the selection of fairness metrics, detection of bias, and mitigation steps undertaken, allowing users to comprehend the logic and data underpinning fairness interventions.

In other AutoML tools, the articulation of feedback's impact on fairness is frequently limited or nonexistent. Although some tools may offer basic explainability features or mechanisms for handling sensitive features, they typically do not provide clear documentation or visual indicators that explicitly connect user input on fairness to alterations in model behavior or outcomes. For instance, while performance metrics were available, none of the tools explained the causes behind bias (e.g., why male candidates had higher TPR or FPR). Some of them were potentially overfitting the results at the time of prediction as referred above. While H2O Studio and RapidMiner provide certain model interpretability features, such as feature importance and global/local explanations, they lack the same depth or clarity regarding fairness-specific decisions, as observed in Dataiku and DataRobot. Furthermore, they do not offer the same breadth and integration of fairness-specific explanations. Also, while some tools like DataRobot and Dataiku provided feature relevance and importance charts for prediction, they did not allow the users to analyze feature importance from the lens of protected or sensitive variable (e.g. Gender). These platforms, along with code-driven libraries, generally lack built-in, user-friendly fairness explanation features and comprehensive support for diverse explanation formats specifically related to fairness. Consequently, they provide minimal or no iterative explanation functionality or meaningful insights into automated fairness logic, and do not effectively communicate how user feedback influences fairness decisions.

## A4. Guardrails and Human-in-the-Loop (HITL) Fairness Integration

The implementation of fairness-related safeguards and explicit override options is essential to enable non-experts to effectively manage and mitigate bias. These mechanisms serve as a protective framework, directing users towards more equitable outcomes, and are intrinsically linked to the concept of human-in-the-loop (HITL) fairness integration.

Both Dataiku and DataRobot incorporate explicit safeguards within their AutoML workflows and offer comprehensive human-in-the-loop (HITL) fairness integration. Dataiku provides features, such as built-in assertions, automated model documentation, and alerts for unexpected model behavior, all of which are designed to assist users in identifying potential biases. Its fairness reports and subpopulation analysis facilitated the identification and mitigation of bias. However, the enforcement of fairness constraints in Dataiku frequently relies on user intervention and configuration, rather than strict automated pre-deployment enforcement. Similarly, DataRobot offers robust safeguards, including configurable fairness monitoring, notifications of fairness breaches, and root-cause analysis tools. Users can establish thresholds for protected features and receive alerts when models are at risk or fail to meet predefined fairness criteria. These safeguards are intended to prevent unintended biases in production models, although enforcement still depends on user configuration and oversight. For instance, even tools with extreme fairness violations (e.g., DataRobot's $\Delta$EOpp = 0.823) permitted automated output without constraint. Both platforms prioritize HITL processes for fairness, allowing users to intervene, override, and guide fairness constraints directly within the AutoML pipeline, as exemplified by DataRobot's bias-mitigation workflows and Dataiku's governance and model override features. This design aligns with HCAI principles, emphasizing user controllability, feedback, and ethical agency.

A significant distinguishing factor is the capacity to permit explicit overrides in both general and fairness-specific AutoML decisions. Dataiku enables users to establish explicit override rules, including those specific to fairness (e.g., enforcing outcomes for protected groups), directly within its visual interface. This feature introduces a "human layer" over automated predictions, which is invaluable for critical or regulated domains. Similarly, DataRobot offers advanced configuration options for bias and fairness, allowing users to specify protected features, select metrics, and select mitigation techniques with detailed control. Users can override default behaviors, set custom thresholds, and explicitly apply fairness interventions to ensure alignment with fairness objectives. Both Dataiku and DataRobot support nonlinear, flexible workflows with branching, manual overrides, and iterative governance steps, which are well suited for complex fairness governance and collaboration. These features reduce perceived complexity and support adoption by enhancing workflow integration and observability (IDT).

By contrast, H2O Studio and RapidMiner, while allowing some manual edits (e.g., adjusting preprocessing steps or handling sensitive features), are generally less "fairness-aware" in their override interfaces and lack comprehensive human-in-the-loop (HITL) fairness integration. Their capabilities for explicitly overriding AutoML decisions, particularly concerning fairness, are less developed and integrated into user-friendly systems, often requiring more technical expertise to implement fairness-specific interventions. These platforms tend to follow more linear, pipeline-style workflows with less flexibility and greater reliance on code scripting, in contrast to the nonlinear, flexible workflows with branching, manual overrides, and iterative governance steps supported by Dataiku and DataRobot for complex fairness governance and collaboration. Furthermore, while Dataiku and DataRobot provide extensive tools (visualizations, reports, root-cause analysis, alerts) to augment human fairness judgment, other tools offer minimal augmentation, often limited to basic bias metrics or feature importance, without integrated guidance or alerts. The absence of visual affordances or explanatory cues limits perceptible fairness and impairs usability for non-experts (Affordance Theory, TAM).

## A5. Visualization and Interactive Analysis for Fairness

The efficacy of an AutoML tool for non-experts to address fairness is determined by its visual interface and the level of interactivity it offers for fairness analysis. Simple and intuitive visualizations are essential for effectively communicating complex fairness concepts and facilitating informed decision-making.

Dataiku and DataRobot offer intuitive and interactive visualizations to assess the model performance, fairness, and bias diagnostics. These platforms provide comprehensive explanation types that extend beyond basic feature importance, incorporating partial dependence plots, SHAP explanations, prediction explanations, and visual representations of fairness metrics and bias mitigation actions, all contextualized within a fairness framework. They include fairness diagnostics by subgroup, enabling users to compare model performance and fairness metrics across subpopulations defined by sensitive attributes, such as Dataiku's subpopulation analysis and DataRobot's per-class bias charts. This functionality is essential for identifying where bias may manifest within specific groups. Additionally, they support interactive fairness analysis, allowing users to explore fairness metrics, adjust sensitive groups, and dynamically analyze model performance across various subpopulations.

Moreover, these platforms are proficient in illustrating the implications of bias and effectively elucidating the consequences of bias for users through comprehensive bias and fairness visualizations, including both per-class and aggregate bias charts. Notably, both Dataiku and DataRobot are engineered to demystify fair data for non-experts by employing clear visualizations, explanations, and guided workflows.

DataRobot further assists in mitigating cognitive bias in visualization by offering guided workflows and distinct visual cues, such as color-coded bias indicators, to facilitate the accurate interpretation of fairness and bias results. Similarly, Dataiku provides support through its responsible AI documentation and fairness report explanations, aiding users in comprehending the limitations and context of fairness metrics. These platforms also feature visual dashboards that effectively present fair information and subgroup metrics. However, none of the tools provide an option for users to visually compare performance across demographic groups or understand fairness trade-offs across models. Also, none of the tools have built in dashboard (except for Dataiku) for fairness specific metrics, namely Demographic Parity Difference (DPD)(Statistical Parity) , Equalized Odds Difference (ΔEO) , Equality of Opportunity (EOpp) or Predictive Rate Parity (PRP)(Precision Parity) .

H2O Studio and RapidMiner offer fundamental visualizations for model results and some degree of interpretability; however, these features are generally less sophisticated, particularly in terms of fairness. While they provide a certain level of model interpretability, they often lack the depth or clarity necessary for fairness-specific decisions, as observed in Dataiku and DataRobot. Their subgroup fairness diagnostics are limited, and their interactive fairness analysis is more constrained, particularly concerning fairness. These tools provide basic visualizations but do not emphasize visualizing bias implications as comprehensively as Dataiku and DataRobot do. Although they offer some explanations, they often require more technical expertise to interpret fairness or bias results. Generally, they do not explicitly address cognitive bias in visualization design and offer less intuitive fairness guidance than Dataiku and DataRobot. While they support performance visualization, their dashboards provide less detail and interactivity for fairness and subgroup analysis.

Code-based tools (e.g., PyCaret, AutoGluon, and H2O library) typically lack built-in visual GUIs for fairness. Users are required to code their own visualizations for model results, fairness, and bias diagnostics, which inherently limits their suitability for non-experts. These tools do not typically offer out-of-the-box subgroup-level fairness diagnostics or interactive fairness analysis GUIs. Visualizing the implications of bias also necessitates manual coding. Owing to the absence of GUI and explanation features, these tools are often unsuitable for non-experts, and they do not explicitly address cognitive bias or provide safeguards in visualization design. The H2O AutoML library offers a minimal fairness explanation, primarily through logs or code outputs, and mainly logs fair information with limited visual or dashboard-based presentation.

## A6. Automation, Efficiency, and Fairness Trade-off Analysis

Although all AutoML tools prioritize automation and efficiency in model construction and optimization, their capabilities in supporting fairness trade-off analysis vary significantly, which is crucial for practical and ethical deployment. Although all platforms aim to automate model development and tuning to enhance efficiency, Dataiku and DataRobot are notable for their ability to enable users to explicitly explore and visualize the trade-offs between various fairness metrics and model performance. This feature is essential for informed decision-making, allowing non-experts to understand the necessary compromises between achieving optimal model performance and ensuring equitable outcomes across different subgroups. By providing tools to visualize these trade-offs, these platforms empower users to make deliberate choices that align with their ethical and business objectives. Such features enhance the relative advantage of the tool and improve user acceptance by making trade-offs transparent (IDT, TAM). In contrast, H2O Studio and RapidMiner offer basic automation and performance visualization; however, their functionality for fairness-specific trade-off analysis is notably limited. While they may provide some basic bias detection, they lack integrated tools or visualizations that allow users to easily assess the impact of fairness interventions on performance, or vice versa. The absence of clear trade-off visualization can pose challenges for non-experts in navigating the complexities of balancing competing objectives.

FLAML, AutoGluon, PyCaret, and the H2O AutoML library primarily emphasize automation and efficiency in model development. For instance, libraries like AutoGluon, despite showing mild bias (PRP = 0.119), do not let users examine trade-offs between fairness and performance and do not provide ability to tune thresholds dynamically with fairness feedback. Although these tools excel in automating the core machine-learning pipeline, they generally offer minimal to no support for a fairness trade-off analysis. For users of these code-first tools, conducting an analysis of fairness-performance trade-offs necessitates substantial manual coding, the use of external libraries, and expert knowledge. This effectively renders this critical aspect inaccessible to non-expert users within the tool's native environment, thereby highlighting a significant gap in supporting responsible AI development for a broader user base. Moreover, the absence of trade-off visualization reflects a lack of empathy-driven UX design (Design Thinking) and increases cognitive effort required to interpret ethical consequences (CLT).

## B. Conclusion and Future Directions for HCI in AutoML Fairness

The thematic analysis of contemporary AutoML tools highlights a critical need for a paradigmatic shift within the domain of human–computer interaction (HCI), particularly concerning the integration of fairness. Despite notable advancements in automating the model development process, a recurring pattern is evident: the prevailing design of these systems relies disproportionately on human users to identify, mitigate, and manage algorithmic bias. This reliance gives rise to a constellation of challenges, namely, human overreliance, automation bias, lack of transparency, limited user control, and inadequate oversight, which collectively impede the ethical and responsible deployment of AI technologies.

### B1. Human Overreliance: Reassigning the Burden of Bias Mitigation

A recurring and significant theme is the overdependence on human users to ensure fairness within AutoML workflows. This is inconsistent with HCAI and UX strategy principles, which advocate for shared control and feedback loops to augment rather than burden users. Across multiple evaluative dimensions, such as “Misuse Prevention” and “Error Prevention and Recovery”many tools demonstrate minimal or entirely absent automated mechanisms for preventing discriminatory or biased outcomes. Consequently, end users are often thrust into the role of de facto fairness auditors, frequently without adequate training, resources, or technical support. For example, the lack of explicit warnings concerning fairness-critical misuse and the absence of structured protocols for identifying and correcting fairness-related errors exemplifies this shortcoming. This approach is inherently unsustainable and insufficient for ensuring equitable AI systems at scale.

### B2. Automation Bias and Transparency Gaps: Encouraging Informed User Engagement

These findings indicate that the design of many AutoML interfaces unintentionally reinforces automation bias, wherein users may overly trust and uncritically accept automated outputs. The sparse implementation of nudges, on-screen disclosures, and fairness-related alerts, except in a limited subset of tools, points to a significant design deficiency. In the absence of explicit prompts that communicate trade-offs, caveats, or uncertainties related to fairness, users are less likely to scrutinize the ethical implications of system-generated results. This exemplifies automation bias, which could be mitigated through strategic affordance cues, nudges, and visualized uncertainty disclosures (Affordance Theory, HCAI). This issue is further exacerbated by persistent transparency gaps, such as the widespread lack of accessible and comprehensive visualizations that communicate fairness metrics and bias mitigation strategies throughout the model development lifecycle. The opacity surrounding fairness-related decisions and the mechanisms behind them hinders users' ability to engage critically and meaningfully with the systems they operate.

## B3. Control Deficiencies and Fragmented Oversight: Restoring Human Agency

The analysis further reveals substantial deficiencies in user control and inconsistent oversight mechanisms stemming from limited support for human-in-the-loop (HITL) paradigms. Many AutoML systems lack features that allow users to override or adjust fairness-specific decisions as well as interfaces that permit meaningful human input into fairness assessments. Such features align with TAM and UX strategy, ensuring that users perceive fairness tools as useful, accessible, and controllable. Consequently, users are often unable to guide or revise fairness objectives, particularly when automated processes yield ethically questionable outcomes. Furthermore, the general absence of structured tools for post-deployment bias monitoring, along with the limited auditability and traceability of fairness-related interventions, undermines sustained governance. These limitations pose significant barriers to ensuring accountability and iterative refinement of fairness strategies over time.

The evaluation underscores that although AutoML tools are progressively integrating fairness features, their efficacy from a human–computer interaction (HCI) perspective, particularly for non-experts, exhibits significant variability. Platforms that emphasize structured user feedback, clear bidirectional transparency, robust governance frameworks, intuitive override mechanisms, comprehensive human-in-the-loop (HITL) integration, and powerful yet straightforward visualizations are more adept at enabling non-experts to address bias in tabular classification and prediction tasks. Moreover, the capacity to clearly visualize and analyze fairness-performance trade-offs is crucial for the practical and ethical deployment of these tools.

In essence, the true promise of AutoML in democratizing AI hinges on its ability to not only automate complex model building, but also empower all users, including non-experts, to develop and deploy AI systems that are both effective and fair. The HCI design of these tools plays a pivotal role in achieving this critical objective.

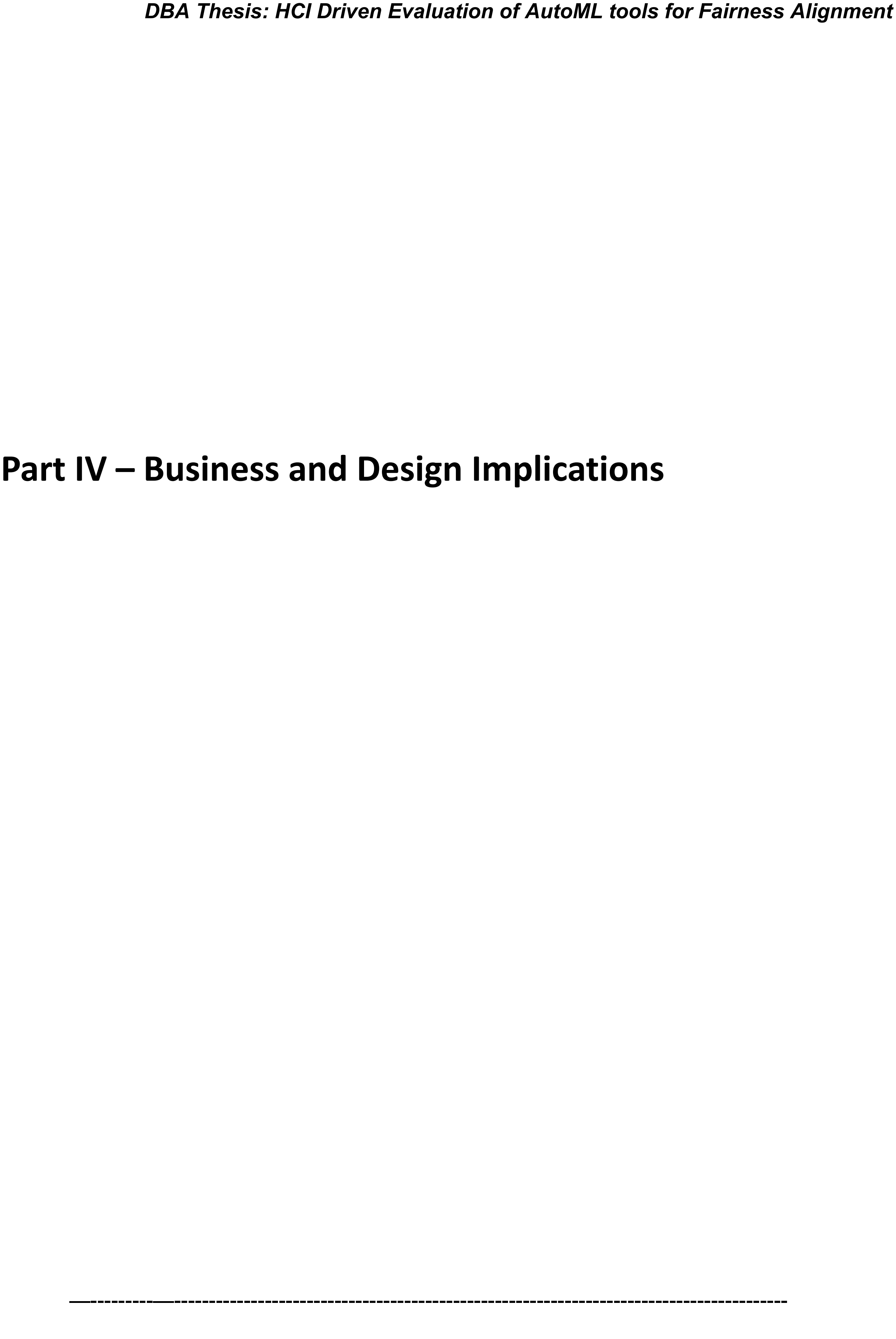

# Part IV – Business and Design Implications

# Chapter 9: Recommended feature considerations for AutoML from fairness perspective

### Chapter Overview

**Business Context and Value:** The chapter establishes fairness not only as an ethical imperative but as a business differentiator for AutoML providers. Fairness-aware design directly influences enterprise trust, regulatory compliance, and adoption in high-stakes domains like hiring and HR analytics.

**Integrating Fairness into Product Strategy:** The chapter clarifies that fairness features enhance buyer confidence, reduce compliance risk, and strengthen brand credibility. Embedding UX-driven fairness metrics—such as feature activation and resolution time for fairness issues—supports continuous improvement and measurable business impact.

**Human–Computer Interaction–Driven Recommendations:** Using the HCI framework from Chapter 4, the chapter proposes structured recommendations across five dimensions—Contracts & User Development, User Interface & Experience, Information Architecture, Human Augmentation, and Care & Responsibility—to guide fairness-centered AutoML design.

**Actionable Design Practices:** The recommendations in the chapter emphasize fairness-by-design principles, including bias detection dashboards, interactive visual fairness tutorials, human-in-the-loop safeguards, feedback-driven refinement, and transparency mechanisms for explainability and accountability.

**Governance, Oversight, and Reporting:** The chapter stresses standardized fairness reporting, continuous auditing, and traceable decision governance to embed accountability in AutoML workflows. Developers are urged to integrate fairness targets, bias monitoring, and clear communication of limitations into product lifecycles.

**Industry Alignment and Future Outlook:** The chapter also provides examples of leading organizations IBM, Google, Microsoft, and Apple who have already adopted or adopting such recommendations as part of their tools, demonstrating convergence toward these practices. The chapter also expresses that industry adoption validates that fairness, transparency, and human oversight are now competitive and regulatory necessities. The chapter concludes that fairness-aware HCI design is both a trust enabler and a market differentiator for next-generation AutoML systems.

## A. Business value for the AutoML tool providers

Fairness is not merely an ethical concern but a practical necessity for achieving enterprise-grade adoption of AutoML tools. In regulated domains like hiring, fairness features influence the confidence of not just users but also enterprise decision-makers — including HR leadership, compliance teams, and IT stakeholders. Designing fairness-aware systems that align with these roles' expectations is critical to realizing AutoML's democratizing potential at scale.

Improving Fairness in AutoML through HCI

Automated Fairness Checks
Implement automated bias detection

Transparency and Visualization
Communicate fairness-related trade-offs clearly

Enhanced User Control
Support human-in-the-loop interventions

AutoML Fairness Issues
Redesigning HCI for Fairness
Equitable AI Systems

Further, to ensure sustainable adoption of fairness-aware AutoML systems, developers must integrate UX KPIs into design and evaluation cycles. Metrics like feature activation rates, time-to-resolution for fairness issues, and user interaction depth can guide design improvements. Moreover, proactively addressing fairness opacity through interface design reduces downstream support burden and strengthens enterprise buyer confidence — particularly in regulated sectors where explainability is a procurement requirement.

## B. Structured recommendations

This Chapter outlines key recommendations for the development and implementation of AutoML systems that support fairness, categorized leveraging the Human Computer Interaction evaluation framework (Refer Chapter 4) elements. These recommendations aim to promote fairness, transparency, and user trust in AI-driven decision-making processes.

## B1. Contracts and User Development

The recommendations focus on establishing clear guidelines, transparent communication, targeted bias-mitigation training, and strong monitoring and redress mechanisms to ensure fairness and accountability in AutoML systems.

| Fairness Strategy Area | Actionable Recommendations |
|---|---|
| Establish Foundational Guidelines and Contracts | 1. Develop clear AutoML guidelines that specify acceptable use cases, promoting fairness and transparency in implementation.<br>2. Draft contracts that explicitly prohibit discrimination and enforce compliance rigorously. |
| Foster Trust Through Clear Communication | 1. Communicate the implications of data use clearly to build user trust and promote fairness. |
| Provide Comprehensive Training and Guidance | 1. Include specific guidelines in training materials for identifying and mitigating algorithmic biases.<br>2. Include specific examples in the guidance materials that demonstrate how to identify and mitigate bias in AutoML systems. |
| Ensure Equitable Treatment and Clear Explanations | 1. Create clear guidelines for equitable treatment in fairness scenarios for all users.<br>2. Make explanations of fairness issues clear, concise, and accessible for all users. |
| Implement Robust Monitoring and Redressal Mechanisms | 1. Establish baselines, conduct regular audits, use fairness metrics, visualize outputs, implement feedback mechanisms, and document monitoring guidelines for AutoML bias.<br>2. Create clear guidelines that allow users to contest AutoML decisions and report instances of discrimination. Provide a straightforward process for users to submit their concerns, including specific steps to follow and contact information for support. Regularly review and update these guidelines based on user feedback and emerging best practices. |

## B2. User Interface, Interaction, and Experience Design

The recommendation emphasizes accessible fairness metrics, interactive visualizations, integrated bias detection tools, transparent user feedback mechanisms, and deployment guardrails to ensure informed, equitable, and accountable AutoML decision-making.

| Fairness Strategy Area | Actionable Recommendations |
|---|---|
| Prominent Display and Accessibility of Fairness Metrics | 1. Design AutoML interfaces that prominently display fairness metrics with visual indicators or dashboards for quick assessment.<br>2. Use contrasting colors, clear labels, and intuitive placement to make fairness metrics distinct.<br>3. Implement a traffic light system (green/yellow/red) to represent fairness levels.<br>4. Place fairness metrics alongside performance indicators for informed decision-making. |
| Interactive Visualizations and Tutorials | 1. Introduce a bias comparison grid showing metrics like TPR, FPR, PRP, DPD, ΔEO.<br>2. Allow benchmarking against legal thresholds (e.g., 80% rule) and custom tolerances.<br>3. Include sliders with real-time updates on fairness-performance trade-offs.<br>4. Offer pre-set fairness optimization modes (e.g., “Maximize Precision while ΔEOpp < 0.1”).<br>5. Provide group-wise confusion matrices and visual fairness reports (e.g., bar charts, slope graphs).<br>6. Create engaging, step-by-step tutorials using real-life scenarios and interactive elements. |
| Bias Detection Tools and Customization | 1. Integrate bias detection tools and corrective prompts within AutoML workflows.<br>2. Trigger alerts for key violations (e.g., “Equal Opportunity gap > 0.3”, “TPR = 0 for a group”).<br>3. Allow users to customize fairness settings and re-evaluate models accordingly. |
| User Feedback and Transparent Disclosures | 1. Incorporate feedback loops from diverse user groups to improve fairness features. |
| | 2. Provide easy-to-use feedback forms or rating systems within the interface.<br>3. Enable active feedback for recalibrating based on group-specific fairness concerns.<br>4. Present nudges that explain fairness trade-offs.<br>5. Clearly disclose limitations of fairness evaluations.<br>6. Transparently communicate fairness criteria and reporting methods. |
| Deployment Guardrails | 1. Implement fairness threshold triggers to pause or block deployment (e.g., "PRP > 0.2").<br>2. Enable human-in-the-loop (HITL) fairness checklists before export or predictions. |

## B3. Information Architecture

The recommendation prioritizes accessible navigation, clear and intuitive visualizations, transparent communication of uncertainty, consistent auditing, and explainability features to enhance user understanding, trust, and equitable decision-making in AutoML systems.

| Fairness Strategy Area | Actionable Recommendations |
|---|---|
| Accessible Settings and Navigation | 1. Place fairness settings prominently in the main navigation.<br>2. Enable easy comparison of fairness metrics across sensitive attributes with filters and sorting.<br>3. Label control options clearly to enhance understanding of fairness-related features. |
| Clear Presentation of Fairness Information | 1. Make fairness warnings highly visible and prioritize them in layout.<br>2. Use layered presentations with key fairness metrics first, followed by detailed group-level analysis. |
| Intuitive and Informative Visualizations | 1. Use intuitive designs and clear labels for better user comprehension.<br>2. Incorporate interactive charts showing bias across user groups. |
| | 3. Enable exploration of fairness impacts by different groups.<br>4. Visualize the effects of bias with contrasting colors, clear language, and practical examples.<br>5. Ensure visualizations are accessible, non-technical, and inclusive. |
| Transparency in Uncertainty and Mitigation | 1. Display uncertainty metrics, especially for underrepresented subgroups.<br>2. Visualize potential biases and confidence levels of fairness assessments.<br>3. Clearly explain bias mitigation methods to build user trust. |
| Consistency and Auditing | 1. Apply consistent fairness definitions and criteria across the AutoML pipeline.<br>2. Enable users to conduct regular audits of training data to detect and correct bias. |
| Explainability for Fairness | 1. Surface group-wise SHAP values or feature contributions (e.g., “specific skill may +12% to male acceptance”).<br>2. Include explainability modules tailored to fairness to show which features affect different groups. |

## B4. Human Augmentation Factors

The recommendations focusses on user-defined objectives, bias-reducing data practices, continuous feedback integration, transparent documentation with user controls, and real-time fairness monitoring through reports and dashboards.

| Fairness Strategy Area | Actionable Recommendations |
|---|---|
| Defining Fairness Objectives | 1. Incorporate user input to define fairness objectives, supporting ethical alignment of AutoML systems. |
| Data Preprocessing and Sampling for Fairness | 1. Facilitate to apply preprocessing techniques to reduce data bias:<br>a. Identify and analyze data sources for bias.<br>b. Clean data (remove duplicates, fix errors, handle missing values).<br>c. Normalize data for consistency.<br>d. Balance data via oversampling/ undersampling/ synthetic generation.<br>e. Select features that minimize bias.<br>f. Conduct and document regular bias audits.<br>2. Use diverse sampling for balanced train/test datasets.<br>3. Implement automated multicollinearity detection and resolution.<br>4. Flag proxy variables strongly correlated with protected attributes.<br>5. Generate pre-model fairness reports highlighting sensitive interaction effects. |
| Establishing and Integrating User Feedback Loops | 1. Create structured processes to collect and review user feedback regularly.<br>2. Conduct routine human feedback sessions to enhance ethical decision-making.<br>3. Integrate user feedback into AutoML systems to improve fairness explanations and model adjustments.<br>4. Provide accessible feedback channels (forms, responses, reviews, resources).<br>5. Implement AutoML reports showing how user feedback impacted fairness metrics and model updates. |
| Documentation, Explanations, and User Controls | 1. Clearly document bias mitigation strategies and fairness criteria with supporting examples.<br>2. Offer diverse explanation types to improve fairness understanding.<br>3. Let users evaluate and adjust fairness metrics at different stages.<br>4. Provide manual override options and user-friendly controls for fairness-related decisions. |
| Downloadable Reports and Dashboards | 1. Enable downloadable fairness reports and bias evaluations.<br>2. Include comprehensive fairness metrics in downloadable formats for model comparison.<br>3. Provide real-time dashboards to monitor fairness and adjust bias. |

## B5. Care and Responsibility

The recommendation emphasizes transparent reporting, regular audits, user-defined fairness targets, clear communication of limitations, and integrated monitoring to ensure accountability, transparency, and continuous bias mitigation in AutoML systems

| Fairness Strategy Area | Actionable Recommendations |
|---|---|
| Transparent Reporting and Accountability | 1. Implement transparent reporting for bias mitigation strategies and fairness metrics in AutoML.<br>2. Create accountability guidelines to track fairness decisions.<br>3. Maintain detailed audit trails of fairness interventions.<br>4. Transparently report bias detection methods and compliance with fairness standards. |
| Auditing and Testing Protocols | 1. Conduct regular audits to verify fairness and alignment with human values.<br>2. Develop testing protocols to evaluate effectiveness of fairness guardrails.<br>3. Use audits, metrics, diverse datasets, feedback, and adversarial testing for robust bias detection. |
| | 4. Implement automated fairness checks before deployment.<br>5. Track and document fairness metrics after each system update. |
| Setting Fairness Targets and Mitigation | 1. Allow users to define fairness targets and apply bias mitigation techniques accordingly. |
| Communication of Limitations and Metrics | 1. Clearly communicate AutoML limitations in fairness and bias detection.<br>2. Include fairness metrics and protected attribute analysis in Data and Model Cards.<br>3. Display uncertainty levels in fairness assessments using confidence intervals or visual indicators.<br>4. Educate users on limitations to manage expectations appropriately. |
| Comprehensive Reporting and Integration | 1. Ensure fairness reporting in TEVV results using standardized formats, tools, metrics, audits, and training.<br>2. Integrate fairness metrics into AIRS to detect and address AutoML biases. |

## C. Trends of industry alignment to recommendation

Industry leaders like IBM, Google, Microsoft, and Apple have converged on best practices that align closely with the five-dimensional HCI evaluation framework proposed in this thesis. By validating each element ranging from contracts and user development to incident reporting and responsibility, this thesis confirms that HCI best practices are foundational to trustworthy AutoML design and in turn for fair machine learning using AutoML. Moreover, independently benchmarked industry tools and academic surveys reinforce that integrating HITL workflows, visual fairness interfaces, layered transparency, and post-deployment oversight are no longer optional—they are a product and regulatory imperative (Refer to Chapter 2).

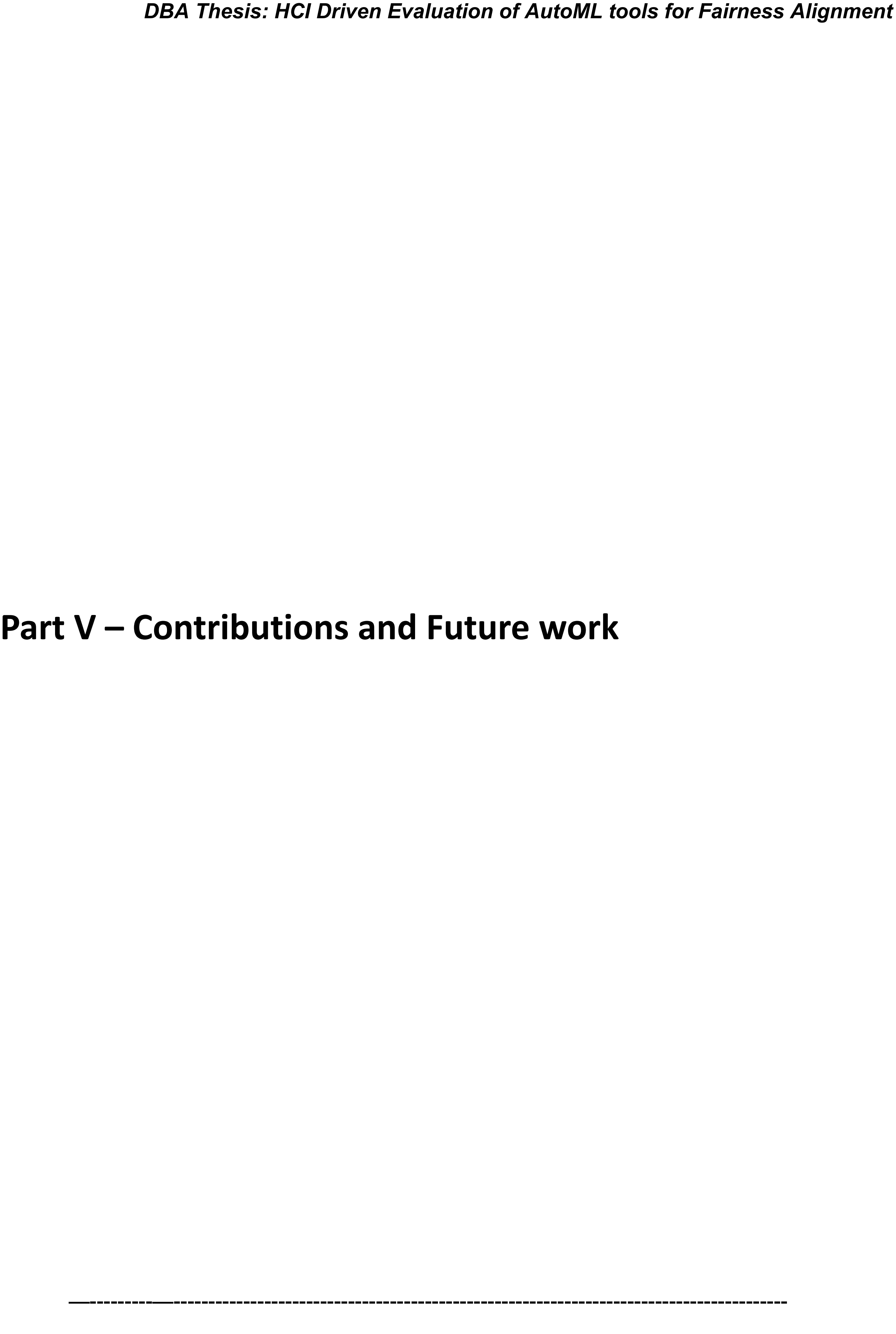

# Part V – Contributions and Future work

# Chapter 10: Contributions, Limitations and Future work

### Chapter Overview

**Contributions:** This chapter consolidates the study's theoretical, practical, and design-oriented contributions, emphasizing fairness-centered HCI as a foundation for responsible and adoptable AutoML systems.
Theoretically, it reframes fairness as a strategic product feature that enhances trust, usability, and adoption—integrating perspectives from HCI, TAM, and UX strategy. Practically, it provides a comprehensive HCI-based framework, benchmarking eight AutoML tools, and offering actionable design recommendations for fairness visibility, feedback, and explainability. The findings highlight fairness-aware HCI design as a driver of enterprise value, thereby improving compliance readiness, buyer confidence, and product differentiation in regulated domains.

**Limitations:** The chapter expresses the study's limitations including scope constraints (eight tools, HR datasets) and constraints between automation, transparency, and user control.

**Future research:** The chapter expresses that future research should validate proposed features through user studies, extend fairness analysis to new domains, and refine governance UX for continuous learning and oversight. Also, it expresses that design priorities include integrating intuitive fairness dashboards, proactive bias mitigation, and contextual feedback loops for non-expert users.

**Conclusion:** The chapter concludes that fairness must evolve from a technical safeguard to a core design value, thereby enabling AutoML systems that are transparent, inclusive, and human-centered.

## A. Theoretical Contributions

This research provides a business-oriented theoretical contribution by reframing fairness in AutoML not solely as a technical challenge, but as a strategic product feature that supports adoption, usability, and differentiation. Through an extensive literature review and original empirical audit, the study identifies key HCI-informed features critical for bias management, particularly in high-stakes domains like hiring. These features—such as transparency tools, fairness visualizations, feedback loops, and human-in-the-loop workflows—are transferable across domains, expanding the utility of the proposed framework beyond HR.

Unlike prevailing AI research that centers on optimizing predictive accuracy or model benchmarking, this study emphasizes the design and interface affordances that support bias mitigation for non-expert users. It shifts the discourse from model performance to user empowerment, offering a fresh theoretical lens that integrates fairness design with human augmentation principles. This approach is grounded in and contributes to established theories such as the Technology Acceptance Model (TAM), Innovation Diffusion Theory, and UX strategy, demonstrating how fairness-aware features influence trust, perceived usefulness, and product adoption.

This thesis advances a theory of fairness-centered HCI design in AutoML tools as not only an ethical imperative but a strategic differentiator in product development. By framing fairness features as usability and adoption enablers, the research expands traditional views of algorithmic risk into a product design paradigm that informs enterprise-focused AutoML strategy.

The research contributes to a comprehensive benchmark of AutoML tools through an HCI and fairness lens, offering a replicable foundation for future inquiry. It illustrates how fairness capabilities—when embedded into the user experience—can shape market positioning and product strategy. The thesis advances theory by showing that fairness design is not merely ethical or regulatory—it is foundational to scalable, enterprise-ready automation platforms.

Overall, the findings provide theoretical clarity on how fairness-centered HCI features intersect with enterprise value, end-user adoption, and system transparency. This lays the groundwork for future research in designing AutoML and other AI systems that are both inclusive and strategically aligned with product success.

## B. Practical contributions

The thesis showcases how bias management features may be integrated into AutoML tools, provides detailed guidelines for tool developers, and offers practical human augmentation ways to tackle algorithmic bias in recruiting practices.

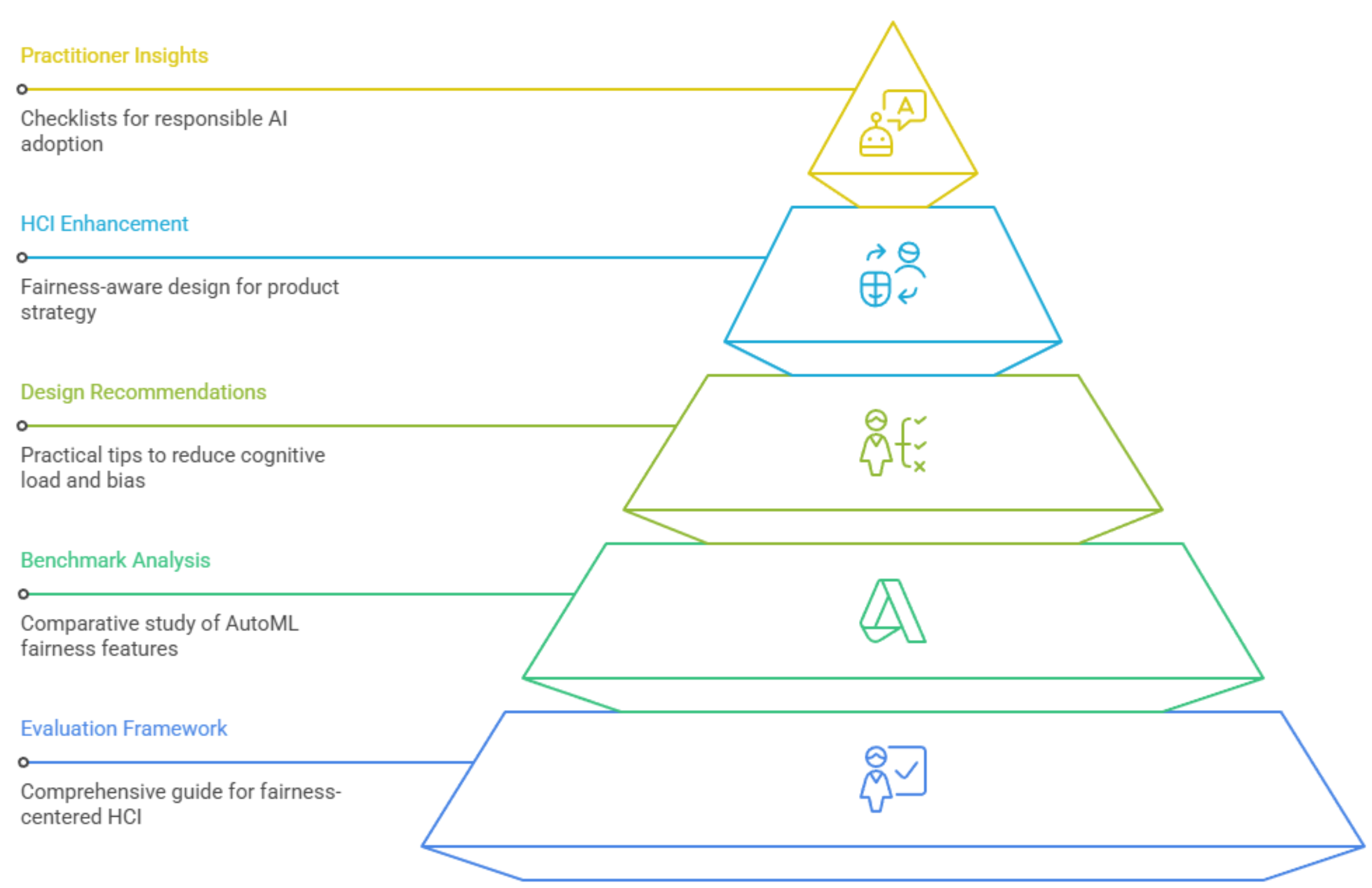

The thesis provides in-depth documentation of benchmarked AutoML tools, analyzing their performance, features, and possible applications. These contributions offer practical guidance, valuable resources, and real-world relevance to individuals engaged in the creation and implementation of AutoML tools.

- Comprehensive evaluation framework: it provides a **comprehensive evaluation framework for fairness-centered HCI in AutoML tools**, offering concrete guidance on designing interfaces, user experiences, and governance mechanisms that align with

——------——-----------------------------------------------------------------------------------------

non-expert user needs. This framework helps translate complex fairness concepts into practical, actionable design choices—supporting usability, transparency, and accountability in AutoML deployment.

- Detailed benchmark: the study generates **a detailed benchmark of eight prominent AutoML tools**, comparing their fairness affordances, visualizations, governance support, and feedback mechanisms. This comparative analysis not only surfaces systemic gaps but also identifies leading practices, offering a baseline reference for both product development teams and organizations evaluating tool capabilities.
- Practical design recommendations: it offers **practical design recommendations** that AutoML vendors can implement to reduce user cognitive load, mitigate automation bias, and improve the interpretability and configurability of fairness interventions. This includes design elements such as intuitive fairness metric visualizations, contextual nudges, automated bias detection in preprocessing, trade-off dashboards, and Human-in-the-Loop (HITL) controls, thereby reducing opportunity of buyer friction, increase trust, and improve adoption in regulated domains like HR
- Fairness-aware HCI as tool enhancement: the research outlines how **fairness-aware HCI design can directly support product and market strategy**. Fairness features not only enable compliance with emerging regulations but also reduce buyer friction, support costs, and time-to-market. By enhancing trust, usability, and auditability, these design choices help AutoML providers differentiate their products and improve adoption in sensitive domains like HR.
- Practitioner oriented insights: the thesis delivers **practitioner-oriented insights and checklists** for enterprise buyers, auditors, and tool developers. These outputs can be integrated into internal procurement criteria, model governance practices, or regulatory assessments, supporting the realistic and responsible adoption of AI technologies.

Collectively, these contributions bridge the gap between fairness research, HCI principles, and real-world implementation challenges—advancing the practical utility and ethical robustness of AutoML systems in complex, high-stakes domains. This resource also offers a comprehensive examination and assessment of different AutoML systems, analyzing their capabilities, constraints, and efficacy in bias management. It allows academics and practitioners to make well-informed judgments when choosing and using AutoML solutions for bias management.

## C. Limitations of the current research

While the thesis proposes actionable interventions for fairness in AutoML, several important limitations and tensions merit reflection:

- **Tool Scope Limitation**: The analysis was restricted to 8 AutoML tools—4 GUI-based and 4 Python libraries—selected from a broader set of 20 tools. The findings may not generalize to all AutoML systems, especially those that differ significantly in architecture, maturity, or target users.

- **Domain-Specific Generalizability**: The framework and evaluation are grounded in hiring and recruitment datasets. While principles of fairness are transferable, domain-specific considerations may limit direct applicability to fields such as healthcare, finance, or criminal justice.

- **Fairness vs. Model Performance**: Fairness-enhancing strategies, such as reweighting or post-processing, may reduce traditional metrics like accuracy or precision. Tools optimized solely for performance may underrepresent hyperparameter configurations that support fairness.

- **User Control vs. Automation**: Designing AutoML for non-experts often entails more automation, but this can suppress user agency and transparency. Excess abstraction may prevent users from identifying, questioning, or mitigating biased outcomes.

- **Transparency vs. Complexity**: Increasing explainability and fairness controls may create cognitive or UI complexity, potentially deterring users or resulting in misunderstanding of fairness interventions.

- **Assessment and Feature Limitations**: The feature selection process itself may introduce or miss bias due to the nature of data or tooling limitations. The study does not eliminate the possibility that certain AutoML design decisions—including default optimizations—may embed or propagate structural bias.

These tensions underscore that fairness is not a one-size-fits-all goal but an iterative design and governance challenge that must align with context, regulation, and user needs.

—------—-----------------------------------------------------------------------------------------

# D. Future research direction

Building upon the findings and contributions of this thesis, several future research directions are proposed to enhance the understanding and design of fairness-aware AutoML systems, particularly for non-expert users and enterprise environments.

Firstly, future research shall conduct empirical user studies (e.g., A/B testing of design prototypes) to validate the efficiency and adoption rates of the proposed HCI fairness features and build demonstrable approaches for benchmarking them. Such research should examine how fairness-related design features impact various phases of the AI lifecycle, including model monitoring, retraining, and user governance post-deployment. Research into the functioning of fairness interventions under conditions of drift, retraining triggers, or continuous learning settings would provide valuable insights for AutoML providers.

Secondly, further investigation is required to explore bias mitigation in AutoML beyond the recruitment domain. Although this thesis concentrated on HR due to its regulatory sensitivity and ethical implications, similar concerns are emerging in domains such as healthcare, lending, education, and criminal justice. Understanding domain-specific fairness trade-offs and stakeholder expectations is crucial for responsibly adapting AutoML tools.

Thirdly, more empirical evaluations of bias mitigation functionality in real-world conditions are necessary. Observational studies, field experiments, or participatory audits involving actual HR or compliance teams could help validate whether fairness features enhance user trust, compliance readiness, or hiring outcomes.

Fourthly, researchers should further investigate the challenges and trade-offs involved in integrating fairness mechanisms into AutoML tools. This includes balancing fairness with performance, automation with user control, and addressing the tension between transparency and usability, which necessitates a thoughtful application of Human-Computer Interaction (HCI), user experience (UX) strategy, and principles of behavioral economics.

Furthermore, research should be expanded to examine the societal and organizational impacts of algorithmic bias introduced by AutoML platforms. This includes understanding how bias may be influenced by architectural or optimization defaults, the implications of platform-level decisions (e.g., objective function selection), and the social consequences of unequal system performance across demographic groups.

---

In addition to these broader research directions, the following technical and design-oriented priorities are recommended for the future development of AutoML tools:

- Balancing Automation and User Control: Researchers should investigate the challenges involved in balancing fairness with performance and automation with user control. Such approach requires applying Human-Computer Interaction (HCI) and User Experience (UX) strategy principles to address these trade-offs

- Address tension between transparency and usability: Future work must address the inherent tension between maximizing transparency and maintaining system usability. Increasing explainability and fairness controls may inadvertently create cognitive or UI complexity, potentially deterring users or leading to the misunderstanding of fairness interventions.

- Enhanced Interpretability of Fairness Metrics: Interfaces should render fairness metrics (e.g., TPR, PRP, DPD) intuitively comprehensible and actionable. Statistical outputs must be translated into context-aware explanations that directly inform user interventions and business decisions.

- Proactive Fairness Interventions: Rather than merely highlighting disparities, AutoML platforms should provide integrated, automated mitigation strategies—such as preprocessing adjustments, fairness-constrained optimization, and post-processing options—tailored to the needs of non-expert users.

- Developing Standardized Fairness Governance UX: This involves creating interfaces that render fairness metrics (e.g., True Positive Rate (TPR), Predictive Rate Parity (PRP), Demographic Parity Difference (DPD)) intuitively comprehensible and actionable for non-expert users. Statistical outputs need to be translated into context-aware explanations that directly inform user interventions and business decisions.

- Intuitive Feedback Loops: Fairness-related user feedback must visibly influence model behavior, supported by visual explanations, traceability of changes, and structured updates to fairness logic based on user input.

- Standardized Fairness Governance UX: Developing coherent user experiences for fairness monitoring, documentation, and audit trails is essential. Human-in-the-loop (HITL) workflows must enable non-expert users to intervene effectively while maintaining transparency.

- Contextualized Guidance and Augmented Judgment: Fairness support must be embedded in the problem context—adapting to data types, task structure, and user goals. Tools should offer contextual alerts, fairness “nudges,” and visual aids that reduce cognitive load and bias blind spots.

- Accessible Trade-Off Analysis: AutoML tools should empower users to explore fairness–performance trade-offs visually and interactively. This includes real-time adjustment sliders, toggle comparisons, and explanations of business implications.

- By prioritizing these directions, future research and product development can facilitate more responsible, inclusive, and usable AutoML, thereby bridging the gap between algorithmic governance and everyday decision-making in regulated enterprise settings.

## E. Thesis conclusion

In conclusion, bias has been a persistent challenge throughout human history, shaping systems of inequality that continue to affect people and societies today. Even as the world strives to move beyond discrimination based on race, gender, or identity, new forms of bias are emerging through technology—especially in areas like employment, where people’s lives and ambitions often hinge on fair decisions. Automated Machine Learning (AutoML) tools have incredible potential to make advanced AI accessible to everyone, but if they are designed without careful attention to fairness and human-centered principles, they risk reinforcing or even amplifying existing biases.

This research highlights the importance of building AutoML systems that actively help users recognize and address bias. The true promise of AutoML—the democratization of AI—depends on putting people at the center of its design. Fairness must go beyond being a technical fix or an ethical requirement; it should be seen as a core design value and a driver of trust and adoption. Addressing the “black box” nature of many AutoML systems requires embedding Human-Centered AI (HCAI) principles and Human-in-the-Loop (HITL) mechanisms that give users transparency, control, and understanding.

When fairness features are treated as essential elements of good design rather than afterthoughts, AI systems become more transparent, inclusive, and trustworthy. By doing so, we can transform automation from something that risks perpetuating inequality into a powerful partner for building a fairer, more equitable future.

—------—-------------------------------------------------------------------------------------

# Appendix

## Appendix 1: Preliminary data analysis of the datasets

### C6.1 Employee Dataset (All in One)

| Category | Details |
|---|---|
| Basic Info | The dataset contains 1204 rows and 18 columns with integer, float, and object data types. |
| Missingness | There are no missing values in any of the columns (0% missingness). |
| Data Quality | - No duplicate rows were found.<br>- Outliers are present in 'Zip Code', 'Years of Experience', and notably in 'Desired Salary'.<br>- The 'Phone Number' column contains invalid placeholder entries ('################...'). - Date columns ('Application Date', 'Date of Birth') have inconsistent formats. |
| Target Variable ('Status') | - The target variable 'Status' is categorical with two well-balanced classes: 'Offered' (~50.7%) and 'Rejected' (~49.3%).<br>- Fairness analysis showed variation in 'Offered' rates across 'Education Level', with 'Master's Degree' having a higher proportion (~54.5%). |
| Feature Distributions | - Numerical features show varying distributions, some with evident outliers.<br>- Categorical features have varying counts per category.<br>- Other object-type columns ('Phone Number', 'Email', 'Address', etc.) have high cardinality. |
| Correlation | Correlation between selected numerical features ('Applicant ID', 'Zip Code', 'Years of Experience', 'Desired Salary') is very low. VIF values are below 7, indicating no significant multicollinearity. |

### C6.2 Detailed Analysis on campus recruitment & Campus Placement Prediction: Binary Classification

| Category | Details |
|---|---|
| Basic Info | - The dataset contains 215 rows and 15 columns.<br>- Data types include integers, floats, and objects (strings).<br>- The 'salary' column has missing values, while other columns are complete. |
| Missingness | - The 'salary' column has approximately 31.16% missing values. - All other columns have no missing values. |
| Data Quality | - No duplicate rows were found.<br>- Outliers are present in numerical features like 'ssc_p', 'hsc_p', 'degree_p', 'etest_p', and 'mba_p', and notably in 'salary'.<br>- Categorical columns appear to have valid entries based on value counts.<br>- 'sl_no' is a unique identifier as expected. |
| Target Variable ('Status') | - The target variable 'status' is categorical with two classes: 'Placed' and 'Not Placed'.<br>- The class distribution is imbalanced, with 'Placed' (~68.84%) being more frequent than 'Not Placed' (~31.16%). |
| Feature Distributions | - Numerical features show varying distributions, some with evident outliers (as noted in Data Quality).<br>- The 'salary' distribution is based on the non-missing values. - Categorical features show varying counts per category. |
| Correlation | - Correlation between numerical features (excluding 'sl_no' and considering non-missing 'salary') is low, though VIF values indicate some multicollinearity (e.g., high VIF for 'mba_p', 'degree_p', 'ssc_p', 'hsc_p', 'etest_p', and moderate VIF for 'salary').<br>- Fairness analysis shows variation in 'Placed' rates across demographic groups. For example:<br>o 'workex' = Yes (~86.5%) vs. No (~59.6%)<br>o Specialisation 'Mkt&Fin' (~79.2%) vs. 'Mkt&HR' (~55.8%)<br>o 'gender', 'ssc_b', 'hsc_b', 'hsc_s', and 'degree_t' also show differences in placement rates, requiring further analysis. |

## C6.3 Employees Evaluation for Promotion

| Category | Details |
| --- | --- |
| Basic Info | - The dataset contains 54,808 rows and 13 columns.<br>- Data types include integers, floats, and objects (strings). |
| Missingness | - Missing values are present in:<br>o 'education' (~4.4%)<br>o 'previous_year_rating' (~7.5%)<br>o 'avg_training_score' (~4.7%)<br>- All other columns are complete. |
| Data Quality | - No duplicate rows were found.<br>- Outliers exist in several numerical features: 'no_of_trainings', 'age', 'length_of_service', 'previous_year_rating', 'awards_won', and 'avg_training_score'.<br>- Categorical columns appear valid based on value counts.<br>- 'employee_id' is a unique identifier. |
| Target Variable ('Status') | - The target is binary: '0' (Not Promoted), '1' (Promoted).<br>- Class distribution is highly imbalanced, with only ~8.5% of employees promoted. |
| Feature Distributions | - Numerical features show varying distributions, with several outliers.<br>- Categorical features vary significantly in category counts. |
| Correlation | - Correlation among numerical features is generally low.<br>- VIF[18] indicates potential multicollinearity among: 'age', 'previous_year_rating', 'avg_training_score'.<br>- Analysis reveals disparities in promotion rates across: 'gender', 'department', 'region', 'education', 'recruitment_channel'. |

[18] VIF (Variance Inflation Factor) is a statistical measure used to detect multicollinearity among the independent variables (features) in a regression model. Multicollinearity occurs when two or more features are highly linearly related, which can distort model interpretation and reduce generalization.

## C6.4 Employee Satisfaction Index Dataset

| Category | Details |
|---|---|
| Basic Info | - The dataset contains 500 rows and 14 columns.<br>- Data types include integers, floats, and objects (strings). |
| Missingness | - There are no missing values in any of the columns (0% missingness). |
| Data Quality | - No duplicate rows were found.<br>- Outliers are present in numerical features such as 'age', 'awards', and 'salary'.<br>- Categorical columns appear to have valid entries based on value counts.<br>- 'Unnamed: 0' and 'emp_id' are unique identifiers. |
| Target Variable ('Status') | - The target variable is binary: '0' (Not Satisfied), '1' (Satisfied).<br>- Class distribution is relatively balanced: 'Satisfied' (~52.6%) & 'Not Satisfied' (~47.4%). |
| Feature Distributions | - Numerical features show varying distributions, some with evident outliers.<br>- Categorical features show diverse category counts. |
| Correlation | - High multicollinearity identified between 'job_level' and 'salary' (VIF> 100).<br>- Analysis showed varying satisfaction rates across demographic groups, such as: 'Dept', 'location', 'education', and 'recruitment_type'. |

## C6.5 IBM HR Analytics Employee Attrition & Performance

| Category | Details |
|---|---|
| Basic Info | - The dataset contains 1,470 rows and 35 columns.<br>- Data types include integers and objects (strings). |
| Missingness | - There are no missing values in any of the columns (0% missingness). |
| Data Quality | - No duplicate rows were found. - Outliers are present in several numerical features: 'Age', 'DailyRate', 'DistanceFromHome', |
| | 'HourlyRate', 'MonthlyIncome', 'MonthlyRate', 'NumCompaniesWorked', 'PercentSalaryHike', 'TotalWorkingYears', 'TrainingTimesLastYear', 'YearsAtCompany', 'YearsInCurrentRole', 'YearsSinceLastPromotion', and 'YearsWithCurrManager'.<br>- Categorical columns appear valid based on value counts.<br>- 'Over18' contains only a single value ('Y').<br>- 'EmployeeCount', 'StandardHours', and 'EmployeeNumber' are identifiers or constant columns. |
| Target Variable ('Status') | - The target variable is binary: 'No' (No Attrition), 'Yes' (Attrition).<br>- The class distribution is imbalanced: 'No' (~83.9%) 'Yes' (~16.1%). |
| Feature Distributions | - Numerical features show varying distributions, with many outliers.<br>- Categorical features show varied category counts. |
| Correlation | - Correlation among numerical features is generally low.<br>- VIF analysis indicates potential multicollinearity among features such as: 'Age', 'Education', 'HourlyRate', 'JobInvolvement', 'JobLevel', 'MonthlyIncome', 'PercentSalaryHike', 'PerformanceRating', 'RelationshipSatisfaction', 'TotalWorkingYears', 'WorkLifeBalance', 'YearsAtCompany', 'YearsInCurrentRole', 'YearsSinceLastPromotion', and 'YearsWithCurrManager'.<br>- Analysis reveals attrition rate disparities across demographic groups: Higher attrition rates among Males, Sales and HR departments, Human Resources/Technical Degree holders, Sales Representatives, Single employees, and employees working OverTime. |

### C6.6 Utrecht Fairness Recruitment dataset

| Category | Details |
|---|---|
| Basic Info | - The dataset contains 4,000 rows and 15 columns.<br>- Data types include integers, objects (strings), and booleans. |
| Missingness | - There are no missing values in any of the columns (0% missingness). |
| Data Quality | - No duplicate rows were found.<br>- Potential outliers detected in 'age' and 'ind-university_grade' based on box plots. |

—-------—--------------------------------------------------------------------------------------

| | |
|---|---|
| | - Categorical and boolean columns appear to have valid entries based on value counts.<br>- 'Id' is a unique identifier. |
| Target Variable ('Status') | - Binary target variable: 'False' (Not Accepted), 'True' (Accepted).<br>- Imbalanced distribution: 'False' (~68.3%) 'True' (~31.7%). |
| Feature Distributions | - Numerical features show varied distributions, with some potential outliers.<br>- Categorical and boolean features vary in frequency across their categories. |
| Correlation | - High multicollinearity found between 'age' and 'ind-university_grade' (VIF > 40). - 'ind-languages' shows low VIF.<br>- Analysis reveals acceptance rate disparities across demographic groups: Higher acceptance rates observed among:<br>o Males<br>o 'Rugby' sport participants<br>o Individuals with 'ind-debateclub', 'ind-entrepeneur_exp', or 'ind-international_exp'<br>o Those with a 'phd' degree<br>o Applicants from company 'A'. |

# Appendix 2: Regulatory expectations regarding bias in hiring

## A. The European Union AI Act

The EU AI Act outlines a series of requirements for data quality, traceability, transparency and human oversight. It also sets out high standards in terms of accuracy, robustness and cybersecurity when developing and deploying AI systems. The EU AI Act (EU AI Act) is an EU-developed regulatory framework for artificial Intelligence (AI). It is designed to ensure that AI systems can be trusted and are respectful of fundamental values and rights, while encouraging innovation and growth in this sector.

The EU AI act categorizes certain artificial intelligence (AI) systems as High-risk in Annex III of the regulation. The High-risk AI systems include AI systems that are deployed in the areas of Employment, workers management, and access to self-employment including (a) AI systems designed for the purpose of recruiting or selecting individuals, such as those used to target job advertisements, analyze and filter job applications, and evaluate candidates; and (b) AI systems intended for making decisions that impact the terms of work-related relationships, including the promotion or termination of contractual agreements, task allocation based on individual behavior or personal characteristics, and the monitoring and evaluation of performance and behavior within these relationships[19]

[19] https://www.europarl.europa.eu/topics/en/article/20230601STO93804/eu-ai-act-first-regulation-on-artificial-intelligence

——---------——-----------------------------------------------------------------------------------

## A1. Key expectations for High-risk AI systems

| High Risk Article | Key expectations |
| --- | --- |
| Article 8<br>Compliance with the requirements | - High-risk AI systems must comply with the requirements outlined in this section, considering their intended purposes and the current state of AI technology.<br>- Compliance with the risk management system in Article 9 is essential for meeting these requirements.<br>- Providers of products containing AI systems must ensure full compliance with both this Regulation and Union harmonisation legislation listed in Section A of Annex I.<br>- Providers have the option to integrate testing, reporting processes, and documentation related to their product with existing documentation and procedures required by Union harmonisation legislation to streamline processes and minimize additional burdens. |
| Article 9<br>Risk management system | - A risk management system must be established, implemented, documented, and maintained for high-risk AI systems.<br>- The risk management system is a continuous iterative process throughout the entire lifecycle of a high-risk AI system.<br>- Steps of the risk management system include identification and analysis of risks, estimation and evaluation of risks, evaluation of other risks, and adoption of risk management measures.<br>- Risk management measures should consider the combined application of requirements to minimize risks effectively.<br>- Measures should ensure that residual risks associated with hazards are acceptable.<br>- Risk management measures should focus on elimination or reduction of risks, implementation of mitigation measures, and provision of necessary information and training.<br>- High-risk AI systems should be tested to identify appropriate risk management measures and ensure compliance with requirements.<br>- Testing may include real-world conditions and should be done prior to placing the system on the market or putting it into service. |

—-------—--------------------------------------------------------------------------------------

| | |
|---|---|
| | - Consideration should be given to potential adverse impacts on persons under 18 and other vulnerable groups.<br>- Providers of high-risk AI systems subject to other Union law requirements may integrate risk management procedures as outlined in the text. |
| Article 10<br>Data and data governance | - High-risk AI systems using training techniques must have data sets that meet quality criteria.<br>- Data governance and management practices must be appropriate for the intended purpose.<br>- Data sets must be relevant, representative, error-free, and complete.<br>- Data sets should consider specific geographical, contextual, and functional settings.<br>- Providers may process special categories of personal data for bias detection and correction under strict conditions.<br>- Special categories of personal data must be protected with technical limitations and privacy measures.<br>- Access to special categories of personal data must be strictly controlled and documented.<br>- Special categories of personal data must be deleted once bias is corrected or retention period ends.<br>- Reasons for processing special categories of personal data must be documented in processing records. |
| Article 11<br>Technical documentation | - Technical documentation for high-risk AI systems must be prepared before the system is placed on the market or put into service.<br>- The documentation must be kept up-to-date and demonstrate compliance with regulatory requirements.<br>- It should provide clear and comprehensive information for national authorities and notified bodies to assess compliance.<br>- Minimum elements for technical documentation are specified in Annex IV, with simplified options for SMEs and startups.<br>- A single set of technical documentation is required for high-risk AI systems related to products covered by Union harmonisation legislation.<br>- The Commission can amend Annex IV through delegated acts to ensure technical documentation reflects technical progress and compliance requirements. |

| Article 12<br>Record-keeping | - High-risk AI systems must have logging capabilities to automatically record events over their lifetime.<br>- Logging capabilities should enable the recording of events relevant for identifying potential risks, facilitating post-market monitoring, and monitoring system operation.<br>- For high-risk AI systems listed in Annex III, logging capabilities must include recording the period of each system use, reference database for input data checks, input data leading to matches, and identification of natural persons involved in result verification. |
|---|---|
| Article 13<br>Transparency and provision of information to deployers | - High-risk AI systems must be transparent to enable deployers to interpret the system's output and use it appropriately.<br>- Instructions for use must be provided in a digital format or otherwise, containing concise, complete, correct, and clear information accessible to deployers.<br>- Information in the instructions for use should include:<br>o Identity and contact details of the provider and authorized representative.<br>o Characteristics, capabilities, and limitations of the AI system's performance, including intended purpose, accuracy metrics, robustness, and cybersecurity.<br>o Circumstances impacting accuracy, robustness, and cybersecurity.<br>o Risks related to health, safety, and fundamental rights.<br>o Technical capabilities to explain output and performance regarding specific persons or groups.<br>o Specifications for input data and training/validation/testing data sets.<br>o Information to interpret the output and use it appropriately.<br>o Predetermined changes to the system and human oversight measures.<br>o Computational and hardware resources, expected lifetime, maintenance measures, and software updates.<br>o Mechanisms for collecting, storing, and interpreting logs. |
| Article 14<br>Human oversight | - High-risk AI systems must be designed for effective oversight by natural persons during use.<br>- Human oversight aims to prevent or minimize risks to health, safety, or fundamental rights.<br>- Oversight measures must match the risks, autonomy level, and context of use of the AI system.<br>- Oversight can be achieved through built-in measures or measures identified by the provider and implemented by the deployer.<br>- Users must be enabled to understand the AI system's capabilities and limitations, monitor its operation, and address anomalies.<br>- Users should be aware of automation bias and be able to interpret and override the AI system's output.<br>- Users should have the ability to intervene or stop the AI system in case of emergencies.<br>- For certain high-risk AI systems, identification must be separately verified by at least two competent natural persons.<br>- The verification requirement does not apply to AI systems used for law enforcement, migration, border control, or asylum if deemed disproportionate by law. |
| Article 15<br>Accuracy, robustness and cybersecurity | - High-risk AI systems should be designed for appropriate levels of accuracy, robustness, and cybersecurity throughout their lifecycle.<br>- The Commission, in collaboration with stakeholders, should promote the development of benchmarks and measurement methodologies to assess accuracy and robustness.<br>- Accuracy levels and metrics of high-risk AI systems should be declared in accompanying instructions.<br>- Measures should be taken to make high-risk AI systems resilient against errors, faults, and inconsistencies, including technical redundancy solutions.<br>- High-risk AI systems that continue to learn should minimize the risk of biased outputs influencing future operations and address feedback loops.<br>- Systems should be resilient against unauthorized attempts to alter their use, outputs, or performance by exploiting vulnerabilities.<br>- Cybersecurity measures for high-risk AI systems should be appropriate to the circumstances and risks, including addressing AI-specific vulnerabilities like data poisoning, |
| | model poisoning, adversarial examples, and confidentiality attacks. |
| Article 16<br>Obligations of providers of high-risk AI systems | - Providers of high-risk AI systems must ensure compliance with specified requirements.<br>- They must clearly label their high-risk AI systems with their name, contact information, and other details.<br>- A quality management system must be in place, complying with regulations.<br>- Documentation and logs related to the AI system must be maintained.<br>- The AI system must undergo a conformity assessment procedure before being marketed or used.<br>- An EU declaration of conformity must be drawn up and the CE marking affixed to indicate compliance.<br>- Registration obligations must be met.<br>- Corrective actions must be taken when necessary, and information provided as required.<br>- Conformity with requirements must be demonstrated upon request by authorities.<br>- Accessibility requirements must be met as per specified directives. |
| Article 17<br>Quality management system | - Providers of high-risk AI systems must establish a quality management system to ensure compliance with regulations.<br>- The quality management system must be documented through written policies, procedures, and instructions.<br>- Key aspects of the quality management system include regulatory compliance strategy, design and development techniques, examination and validation procedures, technical specifications, data management, risk management, post-market monitoring, incident reporting, communication protocols, record-keeping, resource management, and accountability framework.<br>- Implementation of these aspects should be proportionate to the size of the provider's organization.<br>- Providers may integrate these aspects into existing quality management systems if subject to relevant sectorial Union law obligations.<br>- Financial institutions subject to Union financial services law may fulfill quality management system requirements through compliance with internal governance arrangements, with exceptions for specific points outlined in the regulation. |
| | - Harmonized standards referred to in Article 40 should be considered for compliance purposes. |
| Article 18<br>Documentation keeping | - The provider must keep technical documentation, quality management system documentation, approved changes documentation, notified bodies' decisions, and EU declaration of conformity for high-risk AI systems for 10 years after being placed on the market.<br>- Member States determine conditions for documentation retention in cases of bankruptcy or cessation of activity.<br>- Financial institutions subject to Union financial services law must include technical documentation in their documentation under the relevant law. |
| Article 19<br>Automatically generated logs | - Providers of high-risk AI systems must keep logs generated by their systems under their control.<br>- Logs must be kept for a period of at least six months, unless specified otherwise in applicable Union or national law.<br>- The duration of log retention should be appropriate to the intended purpose of the high-risk AI system.<br>- Financial institutions subject to Union financial services law must include logs from high-risk AI systems in their documentation requirements.<br>- Logs should be maintained as part of the documentation under relevant financial services law. |
| Article 20<br>Corrective actions and duty of information | - Providers of high-risk AI systems must take immediate corrective actions if they believe the system is not in conformity with regulations.<br>- Corrective actions may include bringing the system into conformity, withdrawing it, disabling it, or recalling it.<br>- Providers must inform distributors, deployers, authorized representatives, and importers of any necessary actions taken.<br>- If a high-risk AI system presents a risk and the provider becomes aware of it, they must investigate the causes in collaboration with the deployer and inform market surveillance authorities.<br>- Information must also be provided to the notified body that issued a certificate for the high-risk AI system, detailing the nature of non-compliance and any corrective actions taken. |
| Article 21<br>Cooperation with competent authorities | - Providers of high-risk AI systems must provide information and documentation to competent authorities upon request to demonstrate conformity with set requirements.<br>- Providers must give national competent authorities access to automatically generated logs of the high-risk AI system upon request.<br>- Information obtained by national competent authorities must be treated in compliance with confidentiality obligations. |
| Article 22<br>Authorised representatives of providers of high-risk AI systems | - Providers established in third countries must appoint an authorized representative in the EU before making their high-risk AI systems available on the market.<br>- The authorized representative must be enabled by the provider to perform specific tasks outlined in the mandate received from the provider.<br>- The authorized representative must provide a copy of the mandate to market surveillance authorities upon request and carry out tasks such as verifying conformity, keeping contact details and documentation for 10 years, providing information to national competent authorities, cooperating with authorities to mitigate risks, and ensuring compliance with registration obligations.<br>- The mandate empowers the authorized representative to address issues related to compliance with the regulation on behalf of the provider.<br>- The authorized representative must terminate the mandate if it believes the provider is not fulfilling its obligations under the regulation and inform the relevant authorities promptly. |
| Article 23<br>Obligations of importers | - Importers must ensure that high-risk AI systems comply with regulations before placing them on the market.<br>- Verification includes checking conformity assessment procedures, technical documentation, CE marking, and appointment of authorized representatives.<br>- If an importer suspects non-conformity or falsification, the system should not be placed on the market until rectified.<br>Importers must provide their contact information on packaging or documentation.<br>- Storage and transport conditions should not compromise compliance with requirements.<br>- Importers must retain relevant documentation for 10 years and provide it to authorities upon request.<br>- Importers must cooperate with authorities to address risks associated with high-risk AI systems. |
| Article 24<br>Obligations of distributors | - Distributors must verify that high-risk AI systems have the required CE marking, EU declaration of conformity, and instructions for use before making them available on the market.<br>- If a distributor suspects a high-risk AI system is not in conformity with requirements, they should not make it available until it meets the standards. If the system poses a risk, the distributor must inform the provider or importer.<br>- Distributors are responsible for ensuring that storage or transport conditions do not compromise the compliance of high-risk AI systems.<br>- If a distributor finds a high-risk AI system they made available is not in conformity, they must take corrective actions, withdraw or recall the system, or ensure the provider or importer does so. They must also inform relevant authorities.<br>- Distributors must provide information and documentation to national competent authorities upon request to demonstrate the conformity of high-risk AI systems.<br>- Distributors must cooperate with national competent authorities to reduce or mitigate risks posed by high-risk AI systems they made available on the market. |
| Article 25 Responsibilities along the AI value chain | - Distributors, importers, deployers, or other third parties can be considered providers of high-risk AI systems under certain circumstances.<br>- Circumstances include putting their name or trademark on a high-risk AI system, making substantial modifications to a high-risk AI system, or modifying the intended purpose of an AI system to become high-risk.<br>- The initial provider of the AI system is no longer considered the provider if the system undergoes changes, but they must cooperate with new providers and provide necessary information and technical assistance.<br>- Product manufacturers are considered providers of high-risk AI systems if the AI system is a safety component of products covered by Union harmonization legislation.<br>- Providers and third parties supplying components for high-risk AI systems must specify necessary information and assistance in a written agreement.<br>- The AI Office may develop voluntary model terms for contracts between providers and third parties supplying tools, services, or components for high-risk AI systems.<br>- Observance and protection of intellectual property rights, confidential business information, and trade secrets must be in accordance with Union and national law. |
| Article 26<br>Obligations of deployers of high-risk AI systems | - Deployers of high-risk AI systems must take appropriate technical and organizational measures to ensure compliance with instructions for use.<br>- Human oversight must be assigned to individuals with necessary competence, training, and authority, along with adequate support.<br>- Deployers must ensure input data is relevant and representative for the intended purpose of the AI system.<br>- Monitoring of the AI system's operation based on instructions for use is required, with prompt reporting of any risks or incidents to providers and authorities.<br>- Logs generated by high-risk AI systems must be kept for a specified period, unless otherwise regulated by applicable laws.<br>- Employers deploying high-risk AI systems must inform workers and worker representatives about the system's use.<br>- Public authorities and institutions must comply with registration obligations for high-risk AI systems.<br>- Data protection impact assessments may be required for high-risk AI systems under certain regulations.<br>- Authorization from judicial or administrative authorities is necessary for the use of post-remote biometric identification systems in criminal investigations.<br>- Documentation and reporting requirements exist for the use of high-risk AI systems, especially in law enforcement contexts.<br>- Cooperation with national competent authorities is mandatory for deployers to implement the regulations effectively. |
| Article 27<br>Fundamental rights impact assessment for high-risk AI systems | - Prior to deploying high-risk AI systems, certain deployers must assess the impact on fundamental rights.<br>- The assessment should include descriptions of processes, timeframes, affected individuals, risks of harm, human oversight measures, and response measures.<br>- The obligation applies to the first use of the AI system, and deployers can rely on existing impact assessments if similar.<br>- Deployers must update the assessment if any elements change during use.<br>- After the assessment, deployers must notify the market surveillance authority and may be exempt in certain cases.<br>- If data protection impact assessments have already been conducted, the fundamental rights impact assessment should complement it. |

## B. EEOC anti-discrimination guidance

In the U.S., employers shall monitor their traditional decision-making procedures to determine if they have disproportionately negative effects on protected groups under Title VII of the Civil Rights Act of 1964. The Equal Employment Opportunity Commission (EEOC) also has provided guidance on how existing Title VII requirements apply to the assessment of adverse impact in employment selection tools that use artificial intelligence (AI).

Title VII applies to all employment practices of covered employers, including recruitment, monitoring, transfer, and evaluation of employees. It requires the assessment of whether an employer's selection procedures have a disproportionately large negative effect on a protected basis, known as "disparate impact" or "adverse impact" under Title VII. Discrimination caused by use of neutral tests or selection procedures that are not job-related and are not necessary for business purposes is commonly referred to as "disparate impact" or "adverse impact". Adverse impact is measured based on selection rate in employment decisions.

The "selection rate" refers to the percentage of applicants or candidates who are hired, promoted, or selected for a position. It is calculated by dividing the number of individuals selected from a group by the total number of candidates in that group. For example, if 80 White individuals and 40 Black individuals take a personality test as part of a job application, and 48 White applicants and 12 Black applicants advance to the next round, the selection rate for Whites is 60% (48/80) and for Blacks is 30% (12/40).

The "four-fifths rule" is a general guideline used to determine if the selection rate for one group significantly differs from another. According to this rule, a selection rate is considered substantially different if the ratio of the two rates is less than four-fifths (or 80%). Using the previous example, the selection rate for Black applicants is 30% and for White applicants is 60%, resulting in a ratio of 50% (30/60). Since 50% is lower than 80%, the four-fifths rule indicates a substantial difference between the selection rates, potentially indicating discrimination against Black applicants.

However, compliance with the four-fifths rule does not guarantee that an employment procedure is free from adverse impact under Title VII. The rule is a practical and easy-to-administer test that serves as an initial indicator of potential differences in selection rates between groups. Other factors, such as the number of selections made or employer actions discouraging certain groups from applying, may also be relevant. Courts have recognized that the four-fifths rule may not always be appropriate and that additional statistical tests may be necessary to determine the lawfulness of a selection procedure under Title VII. Therefore, employers considering the use of algorithmic decision-making tools should inquire about the vendor's approach to assessing adverse impact, including whether they relied solely on the four-fifths rule or used other standards like statistical significance.

Employers may be held responsible for their use of algorithmic decision-making tools, even if the tools are designed or administered by another entity, such as a software vendor.

## C. New York City Bias Audit

Local Law 144 of 2021 prohibits employers and employment agencies from using an automated employment decision tool unless the tool has been subject to a bias audit within one year of the use of the tool, information about the bias audit is publicly available, and certain notices have been provided to employees or job candidates. In addition, the Department of Consumer and Worker Protection ("DCWP" or "Department") also added rules to implement new legislation regarding automated employment decision tools.

Compliance with the law involves not only commissioning an audit, but also meeting notification and transparency requirements for employers and employment agencies. The law applies to any company worldwide that has employees or prospective employees residing in New York City and uses Automated Employment Decision Technology (AEDT).

The law aims to prevent discrimination against protected characteristics caused using automated tools and increase transparency around their use. The Department of Consumer and Worker Protection (DCWP) enforces the law and has published rules to clarify the qualifications of independent auditors and the process for conducting bias audits. AEDTs are computation processes derived from machine learning, statistical modeling, data analytics, or AI that issue simplified outputs to assist or replace discretionary decision-making.

Bias audits assess whether an AEDT results in disparate impact against individuals based on race/ethnicity and/or sex/gender. The key metrics include:

- Impact ratio: This metric is used to calculate the selection rate for a category compared to the selection rate of the highest scoring category. It is used in the analysis of classification systems.
- Selection rate: This metric is used to calculate which individuals in a category are either selected to move forward in the hiring process or assigned a classification by an AEDT.
- Scoring rate: This metric is used to calculate the proportion of people in each group designated to the positive condition. It is used in the analysis of regression systems.

The regulation required the disparate impact is assessed for standalone groups and intersectional groups.

- Standalone groups: These are individual groups, such as male or female, that are analyzed separately for disparate impact.
- Intersectional groups: These are groups that represent a combination of protected categories, such as black male or black female, and are analyzed to assess potential biases that may arise from multiple intersecting characteristics.

Audits must be conducted using specified metrics depending on the type of AEDT, such as regression systems or classification systems. Employers and employment agencies must provide a bias audit summary on their website before using an AEDT, including information on the audit, data sources, and impact ratios for different groups. Non-compliance with the law can result in penalties starting at $500 for the first violation and increasing for subsequent defaults.

## D. Non-discrimination law in EU

In accordance with the Charter of Fundamental Rights, any form of discrimination based on sex, race, color, ethnic or social origin, genetic features, language, religion or belief, political or any other opinion, membership of a national minority, property, birth, disability, age, or sexual orientation is strictly prohibited. This prohibition extends to all areas covered by the Treaty establishing the European Community and the Treaty on European Union, except for specific provisions outlined in these treaties. Furthermore, discrimination on the grounds of nationality is also prohibited. If your rights have been violated, it is important to note that the authorities of EU countries are obligated to uphold the Charter of Fundamental Rights only when implementing EU law. However, it is worth noting that fundamental rights are safeguarded by the constitution of the respective country.

There are 2 key directives in this regard from the European Commission. They are:

- Directive 2000/78/EC against discrimination at work on grounds of religion or belief, disability, age or sexual orientation.
- Directive 2006/54/EC equal treatment for men and women in matters of employment and occupation.

—------—-----------------------------------------------------------------------------------------

## D1. Directive 2000/78/EC

The directive established a general framework for equal treatment in employment and occupation, with the aim of eliminating discrimination based on religion or belief, disability, age, or sexual orientation. The directive applied to all persons, both in the public and private sectors, and covered various aspects of employment and occupation, including access to employment, vocational training, working conditions, and membership of organizations. It was important to respect fundamental rights and freedoms, including the right to equality before the law and protection against discrimination. This directive did not prejudice freedom of association. The directive represented the importance of combating discrimination and promoting the social and economic integration of elderly and disabled people. The directive provided the need to coordinate employment policies among Member States and emphasized the importance of fostering a labor market favorable to social integration and supporting older workers. The directive provided for measures to accommodate the needs of disabled people in the workplace and allowed for certain exceptions, such as for the armed forces or in cases where a characteristic related to religion or belief, disability, age, or sexual orientation constituted a genuine and determining occupational requirement (*Directive - 2000/78 - EN - EUR-Lex*, n.d.).

## D2. Directive 2006/54/EC

Directive 2006/54/EC aimed to implement the principle of equal opportunities and equal treatment of men and women in matters of employment and occupation. It consolidated and amended several previous directives to bring together the main provisions in the field and reflect developments in case law. The directive highlighted the fundamental principle of equality between men and women in Community law, as well as the positive obligation to promote it. It emphasized that the principle of equal treatment for men and women should not be limited to discrimination based on sex, but should also include discrimination arising from gender reassignment. The directive provided a legal basis for measures to ensure equal opportunities and equal treatment in employment and occupation, including the principle of equal pay for equal work or work of equal value. It also prohibited harassment and sexual harassment and encouraged employers and those responsible for vocational training to take measures to combat all forms of discrimination. The directive emphasized the need for effective procedures for the enforcement of obligations under the directive, including the burden of proof shifting to the respondent when there was a prima facie case of discrimination. It also called for the provision of adequate compensation or reparation for victims of discrimination, and the establishment of bodies for the promotion, analysis, monitoring, and support of equal treatment (*Directive - 2006/54 - EN - EUR-Lex*, n.d.).

# Appendix 3: Compiling a systemic framework for evaluation of Human Computer Interaction considerations

## C1. Contracts and user development

This dimension focuses on establishing a clear understanding and managing the expectations between the user and the AI system. It encompasses the design of transparent disclosures regarding AI system capabilities and limitations, fostering responsible data use through mechanisms such as models and data cards, and providing comprehensive user training and guidance. Furthermore, it defines user responsibilities, including their roles in system oversight and providing constructive feedback, which are vital for safety and continuous improvement. For AutoML, this dimension ensures that users understand what an automated system can or cannot do, promoting appropriate usage and managing the inherent complexities of machine learning.

### C1.1 Acceptable Uses and Limitations

HCI plays a role in clarifying what AI systems are capable of doing and, importantly, what they are *not* capable of doing, to set appropriate expectations for users (Amershi et al., 2019). Disclosures of system limitations are part of the important design principles. Understanding these limitations is also crucial for safety, especially in safety-critical domains (Raulf et al., 2023) (Retzlaff et al., 2024).

| Evaluation criteria | AutoML specific description | Fairness context | Regulatory reference |
|---|---|---|---|
| Acceptable Uses Disclosure | Confirm that AutoML explicitly outlines the types of machine learning problems and deployment scenarios for which it is designed and recommended. | Validate that AutoML clearly communicates its appropriate use cases, particularly where fairness implications are critical, and discourages its use in contexts where fairness cannot be assured. | EU AI Act: Article 9 (Risk management system - evaluation of other risks, adoption of risk management measures to minimize risks), Article 13 (Transparency and provision of information to deployers - characteristics, capabilities, and limitations) |
| Misuse Prevention | Confirm that AutoML includes mechanisms to prevent users from | Ascertain that the system has express contractual constraints | EU AI Act: Article 9 (Risk management system - adoption of risk management |
| | applying automated models inappropriately or in contexts beyond their validated scope. | or built-in safeguards to prevent misuse of the tool for unfair or discriminatory activities. | measures to minimize risks), Article 15 (Accuracy, robustness and cybersecurity - resilience against unauthorized attempts to alter use or outputs), Article 20 (Corrective actions and duty of information - withdraw, disable, or recall non-conforming systems) |

### C1.2 Providing information regarding the use of user data for training

Ensuring visibility in ML models and datasets is a challenge that has received increasing attention (Pushkarna et al., 2022). While not always framed as a 'contract,' transparent design and the provision of documentation are key to clarifying information on data use and provenance. Tools such as Model Cards (Raulf et al., 2023) and Data Cards (Pushkarna et al., 2022) are emerging as non-technical measures to enhance the explainability and responsible deployment of intelligent systems by specifying relevant details regarding model training, intended usage, and documenting datasets.

| Evaluation criteria | AutoML specific description | Fairness context | Regulatory reference |
|---|---|---|---|
| Transparency & Disclosure | Supports user understand how their data will be used in training the autoML tool for performance | Clarify if the transparency and disclosure highlight fairness implications of the data use for model training | EU AI Act: Article 10 (Data and data governance - data quality criteria, appropriate data governance, processing special categories for bias detection/correction); NYC Bias Audit: Publicly available information about bias audit, data sources |

### C1.3 User training, guidance, and instructions for use

Responsible AI implementation requires a strong focus on user training, guidance, and clear instructions. (Li et al., 2024) emphasized that ‘communication, education, and training for users’ are pivotal for building trustworthy AI. Tailoring explanations to different users (experts vs.

non-experts) based on their information needs, context, and domain knowledge is crucial (Raulf et al., 2023). Some studies have also explored teaching user strategies to interact effectively with systems that have limited capabilities (Chromik & Butz, 2021).

| Evaluation criteria | AutoML specific description | Fairness context | Regulatory reference |
|---|---|---|---|
| User training | Evaluate how effectively users are trained to understand AutoML workflows, interpret automated model choices, and make informed decisions regarding model selection and deployment. | Examine whether training materials adequately educate users about potential algorithmic biases in AutoML outputs and provide methods for bias detection and mitigation. | EU AI Act: Article 9 (Risk management system - provision of necessary information and training for risk management measures), Article 14 (Human oversight - users enabled to understand capabilities/limitations, awareness of automation bias); EEOC: Employers should inquire about vendor's approach to assessing adverse impact. |
| Guidance and support | Assess the clarity of in-system guidance for navigating complex AutoML features, understanding system recommendations, and troubleshooting issues related to automated model building. | Determine whether guidance materials offer clear instructions on how to identify and address unfair outcomes or discriminatory impacts that might arise from AutoML-generated models. | EU AI Act: Article 9 (Risk management system - provision of necessary information and training), Article 13 (Transparency and provision of information to deployers - information to interpret output and use appropriately, human oversight measures) |
| Instructions of use | Consider how well the instructions explain the implications of different AutoML configurations, the meaning of various metrics, and the steps | Examine whether instructions explicitly highlight scenarios where fairness considerations are paramount and guide users on steps to ensure equitable | EU AI Act: Article 13 (Transparency and provision of information to deployers - instructions for use including characteristics, capabilities, limitations, risks, technical capabilities to explain |
| | for safely deploying automated models. | treatment across different demographic groups. | output/performance regarding specific persons/groups) |
| Tailoring Explanations | Assess how AutoML explanations are tailored for both expert data scientists and non-expert domain users, considering their varying understanding of machine learning concepts. | Evaluate if explanations about fairness issues are presented in a way that is understandable and actionable for all relevant user groups, regardless of their technical expertise | EU AI Act: Article 13 (Transparency and provision of information to deployers - clear information accessible to deployers, technical capabilities to explain output/performance); NYC Bias Audit: Publicly available bias audit information. |

### C1.4. User responsibilities when using the tool

While explicit user contracts are rarely mentioned, some sources discuss human responsibilities in interactions, particularly in overseeing autonomous systems and ensuring safety (Khuat et al., 2022). GDPR implies user rights to contest outcomes, which requires a system design that supports such opportunities. Design choices that allow users to provide feedback also address implicit responsibilities in contributing to system improvement (Retzlaff et al., 2024) (Ashtana et al., 2018).

| Evaluation criteria | AutoML specific description | Fairness context | Regulatory reference |
|---|---|---|---|
| Oversight of autonomous systems | Evaluate how users are guided to oversee automated model training, validation, and deployment processes to identify potential errors or unintended behaviors. | Examine whether users are informed about their responsibility to monitor AutoML outputs for signs of unfairness or bias and provide tools to do so. | EU AI Act: Article 14 (Human oversight - effective oversight by natural persons, monitor operation, address anomalies), Article 26 (Obligations of deployers - monitoring of AI system's operation, human oversight assigned, prompt reporting of risks/incidents); EEOC: Employers held responsible for use of algorithmic tools. |
| Right to contest outcomes | Assesses how AutoML tools provide information or guidance enabling users to inspect automated model decisions, understand their rationale, and contest results that appear illogical or incorrect. | Determine if users are provided with information or guidance regarding clear mechanisms to contest AutoML decisions that might lead to discriminatory outcomes, and if the system supports investigation into such claims. | EU AI Act: Article 14 (Human oversight - users able to interpret and override output), Article 20 (Corrective actions and duty of information - investigate causes of risk, inform authorities); EU Non-Discrimination Law: Right to protection against discrimination, effective procedures for enforcement. |

## C2. User Interface, Interaction and Experience Design

This dimension addresses the core sensory and interactive elements of user engagement with the AutoML system. It prioritizes the creation of intuitive Graphical User Interfaces (GUIs) that facilitate seamless interaction and ensure high usability for diverse user groups, including non-experts. Key considerations include visual explanation interfaces, hierarchical information presentation, and the integration of 'nudges' and clear metrics/visualizations to guide users through complex processes. The goal is to design an experience that is efficient, satisfying, and minimizes cognitive load, making sophisticated AutoML accessible and manageable.

### C2.1 UI/UX Design

HCI is fundamentally the design, evaluation, and implementation of interactive computing systems (Lakkshmanan et al., 2024). In AI/ML, this means focusing on the UI, interaction, and user experience (UX) (Ashtana et al., 2018) (Khuat et al., 2022) (Pop & Raţiu, 2024). For AutoML specifically, GUIs are common in industry, supporting interactive functions such as selection, exploration, reconfiguration, and value entry (Khuat et al., 2022). Designing for user experience is pivotal for the adoption and acceptance of ML technologies (Ashtana et al., 2018), specifically by non-experts. In addition, to translate qualitative interface evaluations into product development priorities, UX KPIs can offer valuable insight. For instance, measuring fairness feature usage rates or support ticket volume related to fairness confusion helps tool providers understand where usability gaps exist and how they impact non-expert user adoption.

| Evaluation criteria | AutoML specific description | Fairness context | Regulatory reference |
|---|---|---|---|
| User Interface (UI) Design | Evaluate how the AutoML GUI supports interactive functions such as selection, exploration, reconfiguration, and value entry for various model parameters and datasets. | Examine whether the UI design visually highlights potential fairness metrics or biases in model outputs, making them easily discernible to the user. | EU AI Act: Article 13 (Transparency and provision of information to deployers - information on accuracy metrics, robustness, and technical capabilities to explain output/performance); NYC Bias Audit: Metrics for bias audit, bias audit summary on website. |
| User Experience (UX) Design | Consider how the UX design in AutoML facilitates the adoption and acceptance of ML technologies by non-experts, making complex processes feel approachable | Assess whether the UX design intuitively guides users to identify and address fairness concerns, ensuring a smooth and ethical model development process. | EU AI Act: Article 14 (Human oversight - users enabled to understand capabilities/limitations, awareness of automation bias); NYC Bias Audit: Aims to prevent discrimination and increase transparency. |
| Interaction Design | Determine if the interaction design in AutoML enables seamless exploration of different model architectures, parameter settings, and performance evaluations | Investigate whether interaction patterns allow users to easily compare fairness outcomes across different model versions or data subsets | EU AI Act: Article 13 (Transparency and provision of information to deployers - information to interpret output and use appropriately); NYC Bias Audit: Disparate impact assessed for standalone and intersectional groups, key metrics. |
| Learnability | Evaluate whether the AutoML interface is designed for rapid learning, allowing users to quickly grasp automated processes and customize them effectively. | Consider whether the learnability of fairness-related features is high, enabling users to quickly understand how to assess and improve model fairness. | EU AI Act: Article 13 (Transparency and provision of information to deployers - clear information accessible to deployers); Article 14 (Human oversight - users enabled to understand capabilities/limitations). |

——-------——------------------------------------------------------------------------------

## C2.2 Usability

Usability is a key factor in the user experience of ML technologies. As AI applications become integral to our daily routines, the usability of these systems has become more important. From virtual assistants to autonomous vehicles, the ease with which users can interact and effectively utilize AI technologies is crucial. Usability in AI systems encompasses intuitive interfaces, clear instructions, and efficient task completion, ensuring that users can leverage these technologies without unnecessary complexity or confusion (Ashtana et al., 2018). Problems in system usability can arise from user interfaces (Mishra et al., 2023). Usability evaluation methods, such as formal usability evaluation complementing heuristic evaluation, are part of the iterative user-centered design process, including domain-specific environments (Rundo et al., 2020) (Acemyan & Kortum, 2012). This shall include how data are analyzed, including its diversity (Trewin et al., 2019), imbalanced data (Zhang & Zhou, 2019), used and transformed (Xanthopoulos et al., 2020) (Hong et al., 2020).

| Evaluation criteria | AutoML specific description | Fairness context | Regulatory reference |
|---|---|---|---|
| Intuitive Interfaces | Assess whether the AutoML interface allows users to intuitively select algorithms, configure parameters, and interpret automated results. | Examine whether the interface clearly and intuitively presents fairness metrics and potential biases, making them easy to grasp. | EU AI Act: Article 13 (Transparency and provision of information to deployers - clear information accessible to deployers, technical capabilities to explain output/performance); NYC Bias Audit: Key metrics for bias audit, bias audit summary. |
| Error Prevention and Recovery | Implement mechanisms in AutoML to prevent misconfigurations or data input errors and offer clear steps for correction. | Provide safeguards within AutoML to prevent the creation of highly biased models and offer clear guidance on how to rectify fairness-related errors. | EU AI Act: Article 9 (Risk management system - elimination/reduction of risks, implementation of mitigation measures), Article 15 (Accuracy, robustness, cybersecurity - resilient against errors, minimize biased outputs); Article 20 (Corrective actions). |
| User Control and Freedom | Allow users to easily pause, modify, or revert automated model-building processes and explore alternative configurations in AutoML. | Enable users to control and adjust fairness constraints or re-evaluate models based on different fairness criteria within the AutoML system. | EU AI Act: Article 14 (Human oversight - ability to interpret and override output, intervene or stop system), Article 26 (Obligations of deployers - take appropriate measures to ensure compliance with instructions for use); NYC Bias Audit: Mandatory bias audit, public disclosure. |

## C2.3 Transparency

Transparency is a consistently reported key design principle for AI tools, and it remains a key element in enhancing human–computer interaction. Transparency is vital for building trust and is often linked to explainability and understandability (Hoque et al., 2024). Transparency can involve clear agent self-identification, disclosure of system limitations, and explanations. GDPR requirements also highlight the need for transparency in algorithmic systems (Veale et al., 2018). However, many current AutoML tools obscure their internals, acting as black-box systems that hinder transparency (Khuat et al., 2022).

| Evaluation criteria | AutoML specific description | Fairness context | Regulatory reference |
|---|---|---|---|
| Internal process visibility | Provide visibility into the automated search space, the evaluation criteria used, and the intermediate steps taken by AutoML to arrive at a solution. | Examine if the interface clearly and intuitively presents fairness metrics and potential biases, making them easy to grasp. | EU AI Act: Article 13 (Transparency and provision of information to deployers - characteristics, capabilities, limitations, accuracy metrics, technical capabilities to explain output/performance); NYC Bias Audit: Bias audit summary includes data sources, impact ratios. |
| Self-Identification | Explicitly state when AutoML is making automated decisions regarding model selection, | Examine if there is a clear communication as to how AutoML considers and reports on fairness during automated processes, | EU AI Act: Article 13 (Transparency and provision of information to deployers - characteristics, capabilities, and limitations |

| | hyperparameter tuning, or data transformations. | rather than presenting it as a human decision. | of the AI system's performance, human oversight measures). |
|---|---|---|---|

Human Factors Engineering: Human factors engineering is a multidisciplinary field contributing to HCI. It considers human capabilities and limitations in system design to enhance performance, safety, and reliability (Pop & Raţiu, 2024). In the context of AI and HCI, human factors and cognitive science insights are crucial for designing effective human-AI interactions, particularly in safety-critical contexts (Raulf et al., 2023) (Chromik & Butz, 2021).

| Evaluation criteria | AutoML specific description | Fairness context | Regulatory reference |
|---|---|---|---|
| Human Capabilities and Limitations | Verify that AutoML tools accurately accommodate the cognitive load of understanding automated choices and limitations in interpreting complex ML metrics. | Confirm that AutoML interfaces present fairness information in a way that minimizes cognitive bias and supports accurate human assessment of equity. | EU AI Act: Article 14 (Human oversight - awareness of automation bias, understanding capabilities and limitations); Article 13 (Transparency and provision of information to deployers - clear and comprehensive information). |
| Performance Enhancement | Confirm that AutoML effectively enhances user performance by automating repetitive tasks, accelerating model development, and suggesting optimal solutions. | Validate that users can efficiently analyze fairness trade-offs, enabling easier selection of models that balance performance with equitable outcomes. | EU AI Act: Article 9 (Risk management system - adoption of risk management measures to minimize risks); Article 15 (Accuracy, robustness - appropriate levels of accuracy). |
| Situational Awareness Support | Confirm that AutoML provides clear dashboards and visualizations that accurately convey the progress of processes, model performance, | Verify that fairness considerations are presented in a way that helps users maintain high situational awareness regarding the model's equitable | EU AI Act: Article 13 (Transparency and provision of information to deployers - characteristics, capabilities, accuracy metrics, information to interpret output); NYC Bias |

| | and resource utilization. | performance across subgroups. | Audit: Bias audit metrics, intersectional groups. |
|---|---|---|---|

## C2.4 User Engagement

User engagement is a critical element in designing human-centric ML systems (Ashtana et al., 2018). Continuous user engagement and feedback loops are part of a user-centric ethical design paradigm that considers the need for ethics considerations. Iterative engagement, when supported by interfaces allowing feedback loops, domain knowledge injection, and model refinement, amplifies the human augmentation principle in AI-driven tools, and thereby, its quality, productivity, and trust.

| Evaluation criteria | AutoML specific description | Fairness context | Regulatory reference |
|---|---|---|---|
| Continuous Feedback Loops | Confirm that AutoML interfaces effectively support continuous feedback loops for users to comment on automated model suggestions, usability, or overall workflow. | Validate that users can easily provide feedback on perceived unfairness or bias in AutoML outputs, thus contributing to ethical system improvements. | EU AI Act: Article 9 (Risk management system - continuous iterative process), Article 20 (Corrective actions - investigate causes of risk in collaboration with deployer); EU Non-Discrimination Law: Mechanisms for enforcement. |
| Domain Knowledge Injection | Verify that AutoML tools facilitate the injection of domain knowledge by users to guide automated search processes or refine the generated model pipelines. | Confirm that users can effectively provide domain knowledge to highlight or address potential fairness issues specific to their application context or user groups. | EU AI Act: Article 10 (Data and data governance - data sets should consider specific contextual settings); Article 14 (Human oversight - understanding context of use). |

## C2.5 Nudges

'Nudges' guides users through structured processes to help them understand AI behavior (Buçinca et al., 2021) or suggests new configurations for model refinement (Khuat et al., 2022).

Consideration should be given to the cognitive effort and processes involved in users' interpretation of explanations. This includes the following:

<u>On-screen Disclosures</u>: Disclosures are mentioned as part of transparency, such as disclosing system capabilities and limitations (Amershi et al., 2019), and may include references to outcomes that the model/tool is not certain about.

<u>Warnings, Notifications/Alerts</u>: AI systems can have unpredictable behaviors. These sources imply the need for interfaces that handle such situations, although specific warnings or alerts are not detailed. In the context of autonomous driving, external HMI designs to convey messages and warnings to road users have been discussed (Chen, 2022). Further warning users for potential issues in data or model in AutoML context is supportive in decision making process (Weerts et al., 2024) including alerts about high-risk sensitive features (Li et al., 2022) (Veale et al., 2018)

<u>Metrics/Visualization/Reports</u>: Visualization via GUI components is a natural way to foster human understanding and interactivity in AutoML (Khuat et al., 2022). Interactive visualization is a key enabling technology for Human-Centered AI (HCAI) tools (Hoque et al., 2024), as it enhances comprehension, diagnosis, and iterative improvement of ML models. Furthermore, features to generate or download reports/models enhance the value of user engagement (Khuat et al., 2022).

| Evaluation criteria | AutoML specific description | Fairness context | Regulatory reference |
|---|---|---|---|
| Nudges | Verify that nudges in AutoML effectively guide users towards optimal configurations, explain automated decisions, or suggest pathways for model refinement. | Confirm that nudges effectively highlight potential fairness trade-offs or recommend adjustments to improve equitable outcomes across groups. | EU AI Act: Article 9 (Risk management system - adoption of risk management measures, provision of necessary information), Article 14 (Human oversight - awareness of automation bias). |
| On-screen Disclosures | Verify that AutoML on-screen disclosures explicitly state automated system limitations or | Validate that disclosures clearly present any known limitations or uncertainties in | EU AI Act: Article 13 (Transparency and provision of information to deployers - characteristics, capabilities, and |
| | uncertainties regarding model performance, data quality, or search completeness. | AutoML's fairness evaluation or mitigation capabilities. | limitations); Article 15 (Accuracy, robustness - accuracy levels and metrics declared). |
| Warnings, Notifications/Alerts | Check that AutoML effectively alerts users to potential issues such as data shifts, convergence failures, or unusual model behaviors during automated training. | Verify that the system generates explicit warnings or alerts when an AutoML-generated model or the loaded data exhibit statistically significant unfairness or bias. | EU AI Act: Article 9 (Risk management system - identification/analysis/evaluation of risks), Article 15 (Accuracy, robustness - minimize risk of biased outputs), Article 26 (Obligations of deployers - prompt reporting of risks/incidents); EEOC: Four-fifths rule (indicator of potential discrimination). |
| Metrics/ Visualization/ Reports | Confirm that AutoML provides intuitive metrics, interactive visualizations (e.g., performance curves, feature importance), and downloadable reports to enhance understanding and iterative model improvement. | Validate that visualizations and reports clearly present fairness metrics (e.g., disparate impact and equalized odds) and allow for easy comparison across different demographic groups or model versions. | EU AI Act: Article 13 (Transparency and provision of information to deployers - accuracy metrics, technical capabilities to explain output/performance); NYC Bias Audit: Key metrics (impact ratio, selection rate, scoring rate), intersectional groups, bias audit summary on website; EEOC: Measurement of adverse impact, selection rate. |

## C3. Information Architecture

This dimension focuses on the structural organization and presentation of information within an AutoML system to enhance user comprehension and navigation. It dictates how complex data and processes are categorized, linked, and displayed. Key elements include well-structured navigation systems, logical content hierarchies (e.g., progressive disclosure to prevent information overload), and effective information visualization techniques. A well-designed information architecture ensures that users can easily find, understand, and interact with the relevant details of their AutoML tasks, thereby building mental models of the system's operation.

### C3.1 Navigation System

Effective navigation is implicit in discussions on UI design, exploration, and content hierarchy (Khuat et al., 2022). Designing intelligent UIs is critical for supporting human-guided AutoML (Khuat et al., 2022). Successful UIs are associated with a well-defined HCI structure based on principles and guidelines. The UI design should factor in human understanding and control. User interfaces must facilitate user selection, exploration, and reconfiguration actions.

| Evaluation criteria | AutoML specific description | Fairness context | Regulatory reference |
|---|---|---|---|
| Navigational Clarity | Confirm that the AutoML navigation system clearly guides users through various stages of automated model development, from data ingestion to deployment. | Validate that the navigation system allows users to easily find and access fairness-related settings, metrics, or bias reports within the AutoML tool. | EU AI Act: Article 13 (Transparency and provision of information to deployers - clear information accessible to deployers). |
| Support for Exploration | Check that AutoML navigation facilitates the exploration of different algorithms, hyperparameter spaces, or alternative automated pipeline suggestions. | Examine if the navigation system supports users in exploring fairness metrics across different sensitive attributes or model configurations to understand impacts. | EU AI Act: Article 13 (Transparency and provision of information to deployers - information to interpret output and use appropriately). |
| Human Understanding and Control | Check that AutoML's navigation intuitively maps to the underlying automated machine-learning process, giving users a sense of control and progress. | Confirm that the navigation system design helps users understand where and how they can exert control over the fairness aspects of automated model development. | EU AI Act: Article 14 (Human oversight - understanding capabilities and limitations, ability to interpret and override output). |

### C3.2 Content Hierarchy

Content hierarchy is the organizational structure of information that can help users navigate complex AI systems more effectively. It involves prioritizing information based on importance and presenting it in layers, allowing users to focus on high-level concepts before delving into details and avoiding overwhelming the user with too much information at once (Yu, 2023).Progressive disclosure complements the content hierarchy by enabling users to access information gradually, as needed. Progressive disclosure is particularly valuable in the development of AI systems that require user trust and transparency, as it can alleviate concerns about information overload and enhance understanding through step-by-step information delivery (El Ali et al., 2024). Content hierarchy enables AI systems to be made more approachable and easier for users to engage with, thereby promoting transparency and trust (Xu et al., 2022) (Nazar et al., 2021) Content hierarchy and progressive disclosures are vital for fostering user trust and ensuring that the systems are adaptable and user-friendly (Bach et al., 2022).

| Evaluation criteria | AutoML specific description | Fairness context | Regulatory reference |
|---|---|---|---|
| Prioritization of Information | Confirm that AutoML clearly prioritizes essential information about automated model performance, key metrics, and recommended next steps for users. | Validate that fairness-related warnings, critical biases, or major fairness trade-offs are prioritized and displayed prominently to the user. | EU AI Act: Article 9 (Risk management system - focus on elimination or reduction of risks), Article 13 (Transparency and provision of information to deployers - risks related to fundamental rights). |
| Layered Information Presentation | Check that AutoML presents model details, data transformations, and automated search processes in digestible layers, thereby allowing users to delve progressively into complexity. | Examine whether fairness analyses are presented in layers, starting with high-level summaries and allowing users to drill down into group-specific metrics or bias explanations. | EU AI Act: Article 13 (Transparency and provision of information to deployers - information to interpret output, technical capabilities to explain output/performance regarding specific persons/groups). |

## C3.3 Information Visualization

Information visualization is fundamental to making complex AI processes understandable to humans (Le et al., 2020). Visualizations are crucial for understanding, diagnosing, and improving ML models (Retzlaff et al., 2024) (Khuat et al., 2022). Simple, familiar, and understandable visualizations are often preferred, especially for domain experts who are not visualization experts . Visualizations can bridge the gap between human knowledge and AI insights (Pop & Raţiu, 2024) (Hoque et al., 2024) and are particularly suitable for users with limited technical background (Melville et al., 2022) (Xanthopoulos et al., 2020).

| Evaluation criteria | AutoML specific description | Fairness context | Regulatory reference |
|---|---|---|---|
| Understandability and Simplicity | Confirm that AutoML provides simple and familiar visualizations for understanding automated model performance, hyperparameter impact, or data transformations. | Validate that fairness-related visualizations are presented simply and clearly, even for users with a limited statistical background. | EU AI Act: Article 13 (Transparency and provision of information to deployers - clear information accessible to deployers); NYC Bias Audit: Publicly available bias audit information with key metrics. |
| Diagnostic Capability | Check that AutoML visualizations allow users to effectively diagnose issues in automated model training, identify convergence problems, or pinpoint data quality concerns. | Examine whether visualizations help users diagnose specific sources of bias or unfairness in the AutoML-generated model's behavior across different groups. | EU AI Act: Article 9 (Risk management system - identification and analysis of risks), Article 12 (Record-keeping - logging capabilities for identifying potential risks); NYC Bias Audit: Bias audit assesses disparate impact. |
| Interactivity and Exploration | Validate that AutoML visualizations allow users to interactively explore different model candidates, evaluate feature importance, and compare various | Check whether fairness visualizations enable interactive exploration of group-specific performance, bias | EU AI Act: Article 13 (Transparency and provision of information to deployers - technical capabilities to explain output and |
| | automated pipeline stages. | metrics, or the impact of different fairness algorithms. | performance regarding specific persons or groups). |
| Bridging Knowledge Gaps | Verify that AutoML visualizations bridge the gap between automated ML insights and the user's domain knowledge, making complex model decisions accessible. | Confirm that fairness visualizations effectively communicate the nature of bias and disparate impacts in a way that resonates with human ethical understanding. | EU AI Act: Article 14 (Human oversight - users enabled to understand capabilities and limitations, aware of automation bias). |
| Suitability for Non-Experts | Verify that AutoML visualizations are designed to be highly suitable for non-expert users, abstracting technical complexity while retaining critical information. | Validate that fairness visualizations are tailored for non-experts, allowing them to grasp ethical implications without needing deep machine learning expertise. | EU AI Act: Article 13 (Transparency and provision of information to deployers - clear information accessible to deployers). |
| Avoiding Cognitive Biases | Verify that AutoML's presentation of rules, explanations, or automated recommendations is carefully designed to prevent misinterpretations owing to user cognitive biases. | Validate that fairness information is presented in a way that minimizes the risk of users misinterpreting biased data or making biased decisions themselves. | EU AI Act: Article 14 (Human oversight - awareness of automation bias). |

## C3.4 Content Reliability

Content reliability, or trustworthiness, is closely linked to transparency, explainability, and the perceived accuracy of the AI system's outputs (Khuat et al., 2022) (Pop & Raţiu, 2024) . A careful design of how information (such as rules or explanations) is represented is needed to avoid misunderstandings due to human cognitive biases (Pop & Raţiu, 2024). Highlighting ambiguous predictions can help users assess the trustworthiness of models (Hoque et al., 2024).

| Evaluation criteria | AutoML specific description | Fairness context | Regulatory reference |
|---|---|---|---|
| Highlighting Ambiguity/Uncertainty | Confirm that AutoML highlights cases where its automated predictions are uncertain or where the model's confidence is low, allowing users to assess reliability. | Verify that AutoML explicitly highlights any ambiguity or uncertainty in its fairness assessments, such as when data for certain subgroups are scarce. | EU AI Act: Article 13 (Transparency and provision of information to deployers - circumstances impacting accuracy/robustness); Article 10 (Data and data governance - data sets must be relevant, representative, complete). |
| Perceived Accuracy of Outputs | Confirm that users perceive AutoML's generated models and predictions as accurate and trustworthy for their specific use cases. | Validate that users perceive the fairness assessments and bias mitigation recommendations provided by AutoML as accurate and dependable. | EU AI Act: Article 15 (Accuracy, robustness - appropriate levels of accuracy declared); NYC Bias Audit: Bias audits assess disparate impact. |
| Consistency of Information | Confirm that AutoML consistently applies and presents its automated rules, explanations for model choices, and performance metrics across different iterations or projects. | Validate that fairness-related information, definitions of protected attributes, and bias mitigation strategies are consistently | EU AI Act: Article 10 (Data and data governance - data governance and management practices appropriate), Article 17 (Quality management system). |
| | | applied and presented throughout the AutoML workflow. | |
| Data Quality | Verify that AutoML provides clear information about the quality of the data used for automated model training, enhancing content reliability. | Confirm that AutoML highlights any quality issues or biases within the training data that could impact the fairness and reliability of the generated models. | EU AI Act: Article 10 (Data and data governance - data sets must meet quality criteria, relevant, representative, error-free, complete, consider geographical/contextual/functional settings, processing special categories for bias detection/correction). |

## C4. Human Augmentation Features

This dimension integrates principles from human factor engineering to optimize the collaborative relationship between humans and AutoML systems, thereby augmenting human capabilities rather than merely automating tasks. It encompasses designing for iterative engagement, allowing users to inject domain knowledge, refine models, and provide continuous feedback throughout the machine learning pipeline. This includes the development of iterative interpretability and explainability (XAI) features to make AI decisions transparent. Critical aspects also involve providing users with the ability to download reports, models, and benchmarks, empowering them with control, and enabling human oversight and override capabilities, especially in safety-critical contexts.

### C4.1 Iterative Engagement with Tool, Data, and Model/Automation

Iterative engagement is one of the core human augmentation principles in AI-driven tools (Khuat et al., 2022). This involves designing systems to support continuous feedback loops and allowing users to inject domain knowledge or refine models iteratively. Human involvement can occur throughout the entire ML pipeline(Theis et al., 2023). In addition, the tool should enable users to apply consistent preprocessing in line with the nature of the dataset (Kamiran & Calders, 2011) (Mehrabi et al., 2021) (Haluska et al., 2022). It also essentially handles fair training-testing split (Mehrabi et al., 2021), removes sensitive features (Fu et al., 2020) (Le Quy et al., 2022), avoids hyperparameter bias (Cruz et al., 2020), prevents overfitting (Zhang et al., 2022), tests on diverse data, avoids underfitting (Cunningham & Delany, 2021), evaluates using fairness-sensitive metrics (Baumann et al., 2023) (Sun et al., 2020) (Saleiro et al., 2018) (Trewin, 2018) and compares to a benchmark ((Le Quy et al., 2022), ensures model robustness to data variations , and handles missing or incomplete, noisy or corrupted, rare or unusual data points (Kallus et al., 2021) (Chen et al., 2019), or imbalanced datasets in a fair and unbiased manner (Chen et al., 2022). In addition, it shall facilitate understanding data sensitivity, handling monotonic constraints, developing fairness-aware models (Weerts et al., 2024) (Wu & Wang, 2021), explaining the prediction results, handling model documentation, allowing users to download the model/code (Doris Xin, 2022), and highlighting the limitations of the model as additional considerations for evaluation (Weerts et al., 2024).

Stakeholder engagement is iterative because ML models are produced/examined, AutoML settings are reconfigured, and search/production is rerun. It is also nonlinear, allowing stakeholders to skip steps or revisit previous phases (Khuat et al., 2022). Continuous refinement and human-in-the-loop paradigms offer advantages over fully automated approaches, enabling users to inject domain knowledge, provide feedback, and iteratively refine models (Chromik & Butz, 2021). HITL involves integrating human input during an agent's learning process, allowing iterative updates and fine-tuning based on human feedback (Retzlaff et al., 2024). AutoML systems should promote iterative, non-linear workflows to encourage the exploration and refinement of sub-components (Weerts et al., 2024).

| Evaluation criteria | AutoML specific description | Fairness context | Regulatory reference |
|---|---|---|---|
| Comprehensive Data Handling | Confirm that the AutoML tool provides automated and configurable options for consistent data preprocessing and cleaning to handle imperfections, variations, and outliers effectively. | Verify that the system handles data imperfections (e.g., missing values and noise) in a manner that does not introduce or exacerbate biases against specific subgroups. | EU AI Act: Article 10 (Data and data governance - data sets must meet quality criteria, relevant, representative, error-free, complete, consider geographical/contextual/functional settings, processing special categories for bias detection/correction). |
| Fair Data Management and Representation | Check whether the AutoML system offers functionalities to automatically or manually remove sensitive features, | Ensure that the system actively supports the creation of fair training-testing splits, tests on diverse data, handles | EU AI Act: Article 10 (Data and data governance - data sets must meet quality criteria, relevant, representative, error-free, complete, consider |
| | perform stratified or fair-aware data splitting, and apply techniques to address class or subgroup imbalances. | imbalanced datasets to prevent representation bias, and manages sensitive attributes to mitigate direct discrimination. | geographical/contextual/functional settings, processing special categories for bias detection/correction). EEOC Anti-Discrimination Guidance (U.S.) on "disparate impact" or "adverse impact" and New York City Bias Audit (Local Law 144 of 2021) |
| Iterative Model Refinement | Confirm that AutoML allows users to iteratively refine automated models by adjusting parameters, re-running searches, or selecting different pipeline components based on observed performance. | Verify that the system enables iterative refinement of models specifically targeting fairness, allowing users to improve equitable outcomes based on continuous evaluation. | EU AI Act: Article 9 (Risk management system - continuous iterative process), Article 15 (Accuracy, robustness - address feedback loops for learning systems). |
| Human-in-the-Loop (HITL) Integration | Confirm that AutoML integrates human input throughout the ML pipeline, allowing iterative updates and fine-tuning of automated model building based on user feedback. | Validate that HITL mechanisms within AutoML allow human input to directly inform and iteratively improve the fairness and ethical alignment of the models. | EU AI Act: Article 14 (Human oversight - effective oversight, users enabled to interpret and override output), Article 26 (Obligations of deployers - human oversight assigned); EU Non-Discrimination Law: Promotion of equality. |
| Non-Linear Workflow Support | Verify that AutoML supports nonlinear exploration, allowing users to skip or revisit automated steps (e.g., data preprocessing and model selection) based on evolving needs. | Confirm that users can nonlinearly engage with fairness evaluation, allowing them to revisit data, model choices, or mitigation techniques as new biases are identified. | EU AI Act: Article 9 (Risk management system - continuous iterative process); Article 14 (Human oversight - ability to interpret and override output). |
| Human Augmentation Principle | Confirm that AutoML's iterative engagement principles enhance user productivity and decision-making quality by providing flexible human control over automated processes. | Verify that iterative engagement within AutoML amplifies the human ability to identify, analyze, and resolve fairness issues, leading to more robust and trustworthy ethical outcomes. | EU AI Act: Article 14 (Human oversight - effective oversight, users enabled to understand, monitor, interpret, override); EU Non-Discrimination Law: Protection against discrimination, equal treatment. |

## C4.2 Iterative Interpretability and Explainability

Explainability (XAI) is a widely discussed concept at the nexus of AI and HCI, as it enables people to understand AI decisions (Chromik & Butz, 2021). Different types of explanations can be integrated into automated learning systems, such as visual and textual explanations (Khuat et al., 2022). Explainable AI must be designed to express helpful explanations while avoiding misunderstandings due to typical human cognitive biases. Explainability has additional feedback effects in enhancing the performance and reliability of ML solutions when human-in-the-loop collaboration is integral (Khuat et al., 2022). Explainable AI methods can make an agent's decision-making process transparent and interpretable (Retzlaff et al., 2024). This refers to providing explanations for a model's decisions, predictions, and actions. Explainability must be understood from the perspectives of human cognition and emergent human-machine cognitive systems (Khuat et al., 2022). The iterative improvement of explanations based on human feedback is a promising approach (Khuat et al., 2022).

| Evaluation criteria | AutoML specific description | Fairness context | Regulatory reference |
|---|---|---|---|
| Understandability of AI Decisions | Confirm that AutoML provides explanations that allow users to accurately understand why specific models were chosen, hyperparameter sets, or features were transformed. | Validate that explanations clarify fairness-related decisions made by AutoML, such as why a particular bias mitigation technique was applied. | EU AI Act: Article 13 (Transparency and provision of information to deployers - technical capabilities to explain output/performance regarding specific persons/groups). |

—-------—--------------------------------------------------------------------------------------

| | | | |
|---|---|---|---|
| Variety of Explanation Types | Check whether AutoML integrates various explanation types, such as visual representations of the model architecture or textual summaries of feature importance, to enhance understanding. | Examine whether AutoML provides diverse explanations for fairness issues, including visual comparisons of group performance or textual descriptions of bias sources. | EU AI Act: Article 13 (Transparency and provision of information to deployers - information to interpret the output). |
| Transparency of Decision-Making Process | This verifies that AutoML's XAI methods clearly reveal the rationale and inner workings behind the automated selection and optimization of machine learning models. | Ensure that explanations make the fairness-related decision-making process within AutoML transparent, showing how equitable considerations influence model selection. | EU AI Act: Article 13 (Transparency and provision of information to deployers - technical capabilities to explain output and performance regarding specific persons or groups). |
| Iterative Explanation Improvement | Validate that AutoML allows for the iterative refinement of its explanations of automated processes or model choices based on user queries and feedback. | Verify that AutoML actively incorporates human feedback to iteratively improve the clarity and effectiveness of its explanations of fairness and bias. | EU AI Act: Article 9 (Risk management system - continuous iterative process), Article 20 (Corrective actions - investigate causes in collaboration with deployer). |

## C4.3 Feedback Exchange on Functionality or Performance

Feedback exchange is a key principle that enables users to provide inputs on system functionality and performance. Feedback can be used during model training to push the model to align its knowledge with human decisions or by learning through imitation (Theis et al., 2023). Human feedback is crucial for refining AI models and can lead to improved model quality, productivity, and trust (Khuat et al., 2022). Designing AI systems as collaborators means considering feedback in both directions (AI understanding human intention and human understanding AI state). Effective feedback exchange mechanisms, facilitated by transparent and communicative AI, play a pivotal role in achieving successful AI-HCI integration, enhancing user perceptions, and ensuring ethical AI adoption ((Sundar & Lee, 2022) (Guzman & Lewis, 2019) .

| Evaluation criteria | AutoML specific description | Fairness context | Regulatory reference |
|---|---|---|---|
| User Input on Functionality/ Performance | Confirm that AutoML allows users to provide input on the effectiveness of its automated model building, search efficiency, or generated model performance. | Validate that users can easily submit feedback regarding the fairness-related functionality of AutoML or the perceived performance of bias-mitigation features. | EU AI Act: Article 22 (Authorized representatives - cooperate with authorities to mitigate risks). |
| Alignment with Human Decisions | Check that AutoML leverages human feedback during automated training or refinement to align its generated models more closely with the desired outcomes or human expert judgments. | Examine whether user feedback, particularly on fairness, is effectively incorporated to align AutoML-generated models with human ethical standards and equitable decision-making. | EU AI Act: Article 14 (Human oversight - interpret and override output), Article 26 (Obligations of deployers - ensure compliance with instructions for use); EU Non-Discrimination Law: Principle of equal treatment. |
| Bidirectional Feedback Mechanisms | This verifies that AutoML not only receives human feedback but also provides clear communication back to the user regarding how their input influences automated model choices or search results. | Confirm that AutoML communicates how user feedback on fairness impacts model adjustments and provides transparent insights into AI's understanding of fairness objectives. | EU AI Act: Article 13 (Transparency and provision of information to deployers - predetermined changes, human oversight measures). |

| Transparency of Feedback Impact | Verify that AutoML transparently indicates how user inputs, such as preference for certain model types or metrics, influence subsequent automated model recommendations. | Examine whether AutoML clearly demonstrates how user feedback regarding bias or fairness concerns has led to specific changes in the model or automated pipeline. | EU AI Act: Article 13 (Transparency and provision of information to deployers - predetermined changes to the system and human oversight measures). |
|---|---|---|---|

## C4.4 Download Reports, Models and Benchmarks

The feature to download reports, models, and benchmarks from AI tools serves as a significant human-computer interaction component by empowering users with control over outputs. This allows users to examine a specific ML pipeline, compare it with challenger models, and provide performance metrics. Such features to download models and benchmarks support processes for trust-building. It allows technical stakeholders to inspect, control, and manage the learning process (Khuat et al., 2022). The ability to download models and benchmarks also aligns with the principles of transparency and accountability, thereby fostering a sense of control and understanding, which are crucial for building trust in digital interactions (Zhang et al., 2024). It also allows users to independently validate the outcomes, assuring the reliability and consistency of outputs empowering users to perceive it to be a partner in decision-making rather than a black box technology (Balcombe & De Leo, 2022).

| Evaluation criteria | AutoML specific description | Fairness context | Regulatory reference |
|---|---|---|---|
| Feature Availability & Functionality | Confirm that AutoML offers robust and user-friendly features to download generated models, comprehensive reports, and relevant benchmarks. | Validate that AutoML includes direct and functional features for downloading fairness reports, bias assessments, and fairness-adjusted models. | EU AI Act: Article 13 (Transparency and provision of information to deployers - accuracy metrics, robustness, cybersecurity, specifications for input data/training/validation/testing data sets); NYC Bias Audit: Bias audit summary on website, including data sources and impact ratios. |
| User Control over Outputs | Confirm that AutoML allows users to download generated models, detailed reports, and performance benchmarks to manage their own ML assets. | Validate that users can download reports or models containing specific fairness metrics or bias analysis results for an independent assessment. | EU AI Act: Article 16 (Obligations of providers - maintain documentation and logs), Article 18 (Documentation keeping); NYC Bias Audit: Bias audit summary, key metrics. |
| Examination and Comparison | Check that AutoML's downloadable reports allow users to comprehensively examine automated ML pipelines and compare them with manually built or challenger models. | Examine whether downloadable outputs facilitate detailed comparison of fairness metrics across different AutoML-generated models or against predefined fairness benchmarks. | EU AI Act: Article 13 (Transparency and provision of information to deployers - accuracy metrics, robustness, and cybersecurity); NYC Bias Audit: Disparate impact assessed for standalone and intersectional groups, key metrics for bias audit. |

### C4.5 Human Oversight and Override

Maintaining human oversight and the ability to control or override autonomous functions is essential, particularly in safety-critical domains. Human-centered AI (HCAI) and collaborative paradigms emphasize the importance of human involvement in AI systems. Human-in-the-loop (HITL) approaches incorporate human oversight mechanisms into AI models, ensuring that humans maintain control and the ability to intervene. Human controllers play a crucial role in collaborating with and overseeing system performance and safety. The design of trustworthy autonomous systems should support high levels of human oversight and situational awareness (Khuat et al., 2022). In addition, it provides an opportunity for domain experts to collaborate with AutoML (Hanussek et al., 2020). This is achieved through interfaces that facilitate oversight and override capabilities. In safety-critical applications, prioritizing the safety of human operators and enabling on-the-fly guidance from humans can mitigate dangerous actions (Theis et al., 2023).

| Evaluation criteria | AutoML specific description | Fairness context | Regulatory reference |
|---|---|---|---|
| Level of Oversight Supported | Confirm that AutoML interfaces provide sufficient visibility and control points for users to oversee the automated model selection, training, and deployment processes. | Validate that AutoML allows users to effectively oversee fairness metrics and bias mitigation efforts, ensuring ethical alignment. | EU AI Act: Article 14 (Human oversight - effective oversight, understand capabilities/limitations, monitor operation, address anomalies), Article 26 (Obligations of deployers - human oversight assigned, monitoring of operation). |
| Override Capability | Check that AutoML offers explicit override functions allowing users to halt, modify, or reject automated model suggestions or pipeline construction. | Examine whether users can easily override AutoML's decisions that might compromise fairness, for instance, by forcing a specific bias mitigation technique. | EU AI Act: Article 14 (Human oversight - ability to interpret and override output, intervene or stop system in emergencies). |
| Collaboration and Control | Confirm that AutoML fosters a collaborative environment in which users can guide and refine automated processes while retaining ultimate control over model outcomes. | Validate that AutoML facilitates collaboration where users can guide fairness objectives, ensuring that the automated system aligns with ethical human priorities and avoids unintended bias. | EU AI Act: Article 14 (Human oversight - effective oversight to prevent/minimize risks to fundamental rights), Article 25 (Responsibilities along AI value chain - cooperation between providers and third parties). |

## C5. Care and Responsibility

This dimension addresses the ethical, safety, and accountability aspects of AutoML system deployment, ensuring responsible human-AI collaboration. It involves designing features that support decision governance, promote accountability through transparent and predictable AI behavior, and establish robust safety "guardrails." Key HCI elements include the implementation of various disclosures (e.g., data/model cards, TEVV results, and failure mode histories) to build trust and enable informed user decisions. Furthermore, this dimension necessitates mechanisms for continuous improvement through regular updates and robust Adverse Incident Reporting Systems, ensuring that AutoML systems are not only effective but also reliable, safe, and ethically managed throughout their lifecycle.

### C5.1 Decision Governance

Decision governance in the context of human-computer interaction (HCI) in AI involves enabling features that support overseeing the decision-making processes of AI systems to ensure that these interactions are transparent, trustworthy, and aligned with human values. Further, decision governance in AI must consider the balance between automation and human accountability, particularly in high-stakes environments such as healthcare, finance, and autonomous systems (Çakır, 2024) (Liu, 2021). Notably, model performance metrics and visualizations are crucial for data scientists to establish trust in autoML tools (Drozdal et al., 2020).

| Evaluation criteria | AutoML specific description | Fairness context | Regulatory reference |
|---|---|---|---|
| Oversight of AI Decision-Making | Confirm that AutoML offers mechanisms to oversee automated choices regarding model architecture, feature selection, or hyperparameter optimization throughout the pipeline. | Validate that AutoML provides features to oversee decisions related to fairness, such as how bias mitigation strategies are applied or which fairness metrics are prioritized. | EU AI Act: Article 9 (Risk management system - identification, analysis, evaluation of risks), Article 26 (Obligations of deployers - monitoring of AI system's operation). |
| Trustworthiness Alignment with Values | Verify that AutoML's automated decisions are demonstrably aligned with user-defined objectives and ethical guidelines, fostering trust in its outcomes. | Validate that AutoML's decision governance features ensure alignment with human values regarding fairness, promoting equitable outcomes, and preventing discriminatory decisions. | EU AI Act: EU AI Act: Designed to ensure AI systems can be trusted and are respectful of fundamental values and rights; Article 27 (Fundamental rights impact assessment). |

### C5.2 Accountability

Designing for accountability is part of incorporating human concerns into HCAI tools. Accountability is a desired mechanism for trustworthy autoML systems (Khuat et al., 2022). AI practitioners are obligated to take responsibility for public interaction (Mathewson, 2019). Transparency, understandability, and predictability are required for operators to hold autonomous systems accountable. Ensuring accountability is crucial for building trust in AI applications (Retzlaff et al., 2024).

| Evaluation criteria | AutoML specific description | Fairness context | Regulatory reference |
|---|---|---|---|
| Balance of Automation and Accountability | Confirm that AutoML's design maintains a clear allocation of responsibility, allowing users to understand when automation occurs and where human accountability lies for model deployment. | Verify that AutoML clearly delineates responsibility for fairness outcomes, ensuring that users can trace accountability for biased decisions to design or intervention points. | EU AI Act: Article 16 (Obligations of providers - ensure compliance), Article 25 (Responsibilities along the AI value chain), Article 26 (Obligations of deployers - compliance with instructions). |
| Auditability and Traceability | Check whether AutoML generates detailed audit trails for every automated decision point, including data transformations, model choices, and evaluation results. | Examine if AutoML's audit trails specifically document fairness-related interventions, such as bias detection reports and mitigation technique applications, for traceability. | EU AI Act: Article 12 (Record-keeping - logging capabilities to record events relevant for identifying potential risks, monitoring system operation), Article 18 (Documentation keeping). |

### C5.3 Guardrails

While "guardrails" is a primary Human Computer Interaction feature, it is critical to designing reliable, safe, and trustworthy systems (Mathewson, 2019), implementing verification processes and control mechanisms, mitigating risks in safety-critical contexts, and enabling human oversight; the ability to avert undesirable actions is fundamental to Human Computer Interaction (Khuat et al., 2022) (Theis et al., 2023).

| Evaluation criteria | AutoML specific description | Fairness context | Regulatory reference |
|---|---|---|---|
| Design for Reliability | Confirm that AutoML incorporates guardrails to ensure the reliability of automated model generation, preventing the creation of unstable or poorly performing models. | Validate that guardrails within AutoML are designed to ensure the reliable application of fairness constraints, preventing unintended biases. | EU AI Act: Article 15 (Accuracy, robustness - resilient against errors, minimize risk of biased outputs). |
| Safety Critical Contexts | Check whether AutoML implements robust guardrails to prevent the deployment of potentially unsafe or unvetted automated models in safety-critical domains. | Examine whether guardrails in AutoML specifically prevent the generation or deployment of models that could lead to discriminatory harm in sensitive or critical applications. | EU AI Act: Article 9 (Risk management system - consideration for adverse impacts on vulnerable groups), Article 14 (Human oversight - prevent/minimize risks to fundamental rights). |
| Verification Processes | Verify that AutoML's guardrails incorporate automated verification processes to confirm model quality, adherence to performance thresholds, or data integrity before deployment. | Validate that the guardrails in AutoML include verification processes to confirm that fairness metrics meet predefined thresholds before a model is considered viable. | EU AI Act: Article 9 (Risk management system - testing to ensure compliance), Article 15 (Accuracy, robustness - accuracy levels and metrics declared). |
| Control Mechanisms | Confirm that AutoML provides users with clear control mechanisms within the guardrails, | Verify that AutoML offers explicit control mechanisms within its guardrails, enabling users to set strict fairness | EU AI Act: Article 14 (Human oversight - ability to interpret and override output); Article 16 (Obligations of providers - |

| | allowing them to set bounds on search space, resource usage, or model complexity. | targets or mandating the application of specific bias-mitigation techniques. | ensure compliance with requirements). |
|---|---|---|---|

## C5.4 Disclosures

Disclosures of system capabilities, limitations, and potential errors are discussed as crucial aspects of transparency. It is also necessary for the tools to clearly disclose the assumptions and limitations, in addition to emphasizing that fairness metrics cannot be guaranteed (Weerts et al., 2024). Setting user expectations includes the following:

- Data/Model Cards: Model card (Raulf et al., 2023) and data card (Pushkarna et al., 2022) disclosure are practical mechanisms for enhancing transparency and supporting responsible AI development by providing structured documentation on models and datasets, including details relevant to training, evaluation, and intended use.
- TEVV Results: Testing, Evaluation, Validation, and Verification (TEVV) results as disclosure enable downstream adoption of such systems, including gaining adequate visibility around usability evaluation, performance metrics, safety and robustness, and through user studies (Khuat et al., 2022). Such an effort supports users in determining the adoption approaches.
- Failure modes/adverse incident history: Addressing potential failures and unpredictable behaviors is crucial for designing safe AI systems. Providing information regarding why a system fails or might fail or be unable to perform a task (e.g., a chatbot being unable to respond) is a part of disclosures (Shneiderman, 2020). Designing for human understanding and control is important so that stakeholders can interrupt anything incomprehensible and potentially dangerous. Disclosures are needed in case of errors or anticipated errors (Chromik & Butz, 2021). Highlighting and textually explaining ambiguous predictions helps users appropriately reassess their level of trust. Discrepancies between human and machine predictions indicate that an error exists, justifying the need for explanation and verification (Khuat et al., 2022).

| Evaluation criteria | AutoML specific description | Fairness context | Regulatory reference |
|---|---|---|---|
| Disclosure | Confirm that AutoML explicitly | Validate that AutoML discloses its capabilities | EU AI Act: Article 13 (Transparency and provision |

—---------—-----------------------------------------------------------------------------------------

| | | | |
|---|---|---|---|
| | communicates the types of ML tasks it can automate, the algorithms it supports, and its computational limits. | in detecting and mitigating various types of biases or ensuring specific fairness criteria. | of information to deployers - characteristics, capabilities, and limitations of performance). |
| Disclosure | Check that AutoML clearly informs users about its limitations regarding data size, model complexity, or the quality of solutions it can guarantee in specific contexts. | Examine if AutoML transparently discloses its limitations in achieving perfect fairness or its inability to detect certain subtle biases. | EU AI Act: Article 13 (Transparency and provision of information to deployers - characteristics, capabilities, and limitations of performance, circumstances impacting accuracy/robustness). |
| Disclosure | Verify that AutoML generates and presents comprehensive Model Cards for generated models and Data Cards for datasets used, detailing training, evaluation, and intended use. | Confirm that the Data and Model Cards generated by AutoML explicitly include sections on fairness considerations, protected attributes, and bias evaluation results. | EU AI Act: Article 10 (Data and data governance - data quality criteria), Article 11 (Technical documentation - demonstrate compliance, provide information); NYC Bias Audit: Bias audit summary including data sources and impact ratios, assessed for standalone and intersectional groups. |
| Disclosure | Validate that AutoML provides access to comprehensive TEVV results, including usability evaluations, performance metrics, and robustness assessments, to support adoption decisions. | Verify that the TEVV results disclosed by AutoML include detailed assessments of fairness metrics, bias detection results, and robustness of fairness-aware models. | EU AI Act: Article 9 (Risk management system - testing), Article 15 (Accuracy, robustness - accuracy levels and metrics declared); NYC Bias Audit: Bias audit summary with key metrics. |
| Disclosure | Confirm that AutoML logs and discloses past failure modes, instances where automation failed, or explanations for the inability to find an optimal model. | Validate that AutoML records and discloses any incidents where automated processes lead to unfair or biased outcomes, including the identified reasons and context. | EU AI Act: Article 12 (Record-keeping - logging capabilities for identifying potential risks), Article 20 (Corrective actions - inform market surveillance authorities of risks); NYC Bias |
| | | | Audit: Non-compliance can result in penalties. |
| Disclosure | Check whether AutoML explicitly highlights any ambiguity or low confidence in its automated predictions or model recommendations. | Confirm that AutoML clearly highlights any uncertainty in its fairness assessments, for instance, due to limited data for certain subgroups or complex bias interactions. | EU AI Act: Article 13 (Transparency and provision of information to deployers - circumstances impacting accuracy/robustness); Article 39 (Content Reliability - Highlighting Ambiguity/Uncertainty) - this seems to be a circular reference, but within the provided text, it links back to itself. |

## C5.5 Patches and Updates

Regular updates and patches to AI systems play a crucial role in enhancing human-computer interaction (HCI) by ensuring that the tools remain relevant, accurate, and aligned with user expectations. A critical aspect of regular updates is their ability to improve the predictive performance of AI systems (Bansal et al., 2019). This is particularly crucial in high-stakes domains such as healthcare and criminal justice (domains where human and AI collaboration is essential for decision-making), wherein updates may actually harm overall team performance if they are not compatible with the user's past experiences (Bansal et al., 2019). Furthermore, AI systems should strive for a balance between performance and compatibility, ensuring that enhancements in AI performance are aligned with user expectations and do not disrupt established workflows (Bansal et al., 2019). For instance, in conversational AI, regular updates driven by deep learning and data from human interactions enable systems to become more adept at understanding and responding to natural language (Yan, 2018).

| Evaluation criteria | AutoML specific description | Fairness context | Regulatory reference |
|---|---|---|---|
| Level of Oversight Supported | Confirm that AutoML's interfaces provide sufficient visibility and control points for users to oversee automated model selection, training, and deployment processes. | Validate that updates to AutoML specifically address newly identified fairness challenges or respond to user feedback on ethical considerations. | EU AI Act: Article 14 (Human oversight - effective oversight); Article 9 (Risk management system - continuous iterative process). |

——------——-------------------------------------------------------------------------------------

| | | | |
|---|---|---|---|
| Predictive Performance Improvement | Check that AutoML updates measurably enhance the predictive performance of the generated models and improve the efficiency of the automated search process. | Examine whether updates to AutoML lead to demonstrable improvements in the fairness and equitable performance of the generated models. The updates shall not create inconsistent fairness outcomes compared to its previous version | EU AI Act: Article 15 (Accuracy, robustness - appropriate levels of accuracy, minimize risk of biased outputs influencing future operations); Article 9 (Risk management system - continuous iterative process). |

## C5.6 Adverse Incident Reporting System

An Adverse Incident Reporting System (AIRS) serves as a critical component in the intersection of Human-Computer Interaction (HCI) and Artificial Intelligence (AI), playing an essential role in documenting and analyzing incidents that arise from AI systems. As AI systems become more prevalent, adverse events provide invaluable learning opportunities to refine AI algorithms and improve their interactions with humans (Lupo, 2023). A robust incident-reporting strategy facilitates the development of flexible regulatory frameworks that evolve alongside new AI technologies (Lupo, 2023). Documentation of incidents, particularly those highlighting system biases, is crucial for learning from past failures and advancing policy recommendations aimed at mitigating these risks (Turri & Dzombak, 2023). Proper incident reporting assists teams in understanding and contextualizing AI actions, fostering trust, and improving the effectiveness of human-AI collaborations (Zhang et al., 2024). Moreover, as AI systems increasingly handle complex tasks, an understanding augmented by clear incident reporting can bridge the gap between human expectations and AI functionalities (Zhang et al., 2024). Designing human-AI interactions is complicated by AI's output complexity and the uncertainty surrounding its capabilities. AIRS can provide insights into these design challenges, aiding designers and researchers in effectively addressing them (Yang et al., 2020).

| **Evaluation criteria** | **AutoML specific description** | **Fairness context** | **Regulatory reference** |
|---|---|---|---|
| Documentation & Analysis of Incidents | Confirm that AutoML's AIRS effectively documents and analyzes incidents related to automated model failures, unexpected performance, or resource issues during its operation. | Validate that AIRS specifically captures and analyzes incidents, highlighting biases in AutoML-generated models or unfair outcomes across user groups. | EU AI Act: Article 12 (Record-keeping - logging capabilities for identifying potential risks, facilitating post-market monitoring), Article 20 (Corrective actions - inform market surveillance authorities). |
| Bridging Expectation Gaps | Check that AutoML's AIRS provides insights into its operational complexities, helping to bridge user expectations with the actual functionalities of automated ML. | Examine whether the AIRS helps bridge the gap between human expectations of fair AI and AutoML's actual performance regarding equity, guiding users on realistic capabilities. | EU AI Act: Article 14 (Human oversight - awareness of automation bias); Article 13 (Transparency and provision of information to deployers - characteristics, capabilities, and limitations). |

# Appendix 4: Feature wise tabulation of the evaluation

| | Dataiku | DataRobot | H2O AutoML Studio | Altair Rapidminer | FLAML | AutoGluon | PyCaret | H2O AutoML Library |
|---|---|---|---|---|---|---|---|---|
| Acceptable Uses Disclosure | Y | Y | L | L | L | L | L | L |
| Misuse Prevention | L | L | L | L | N | N | N | L |
| Transparency & Disclosure | Y | Y | L | L | L | L | L | L |
| User Training | L | Y | L | L | N | N | N | L |
| Guidance and Support | Y | Y | L | L | N | N | N | L |
| Instructions of Use | Y | Y | L | L | L | L | L | L |
| Tailoring Explanations | Y | Y | L | L | N | N | N | L |
| Oversight of Autonomous Systems | Y | Y | L | L | N | N | N | L |
| Right to Contest Outcomes | L | Y | L | L | N | N | N | L |
| User Interface (UI) Design | Y | Y | Y | Y | N | N | N | N |
| User Experience (UX) Design | Y | Y | N | Y | N | N | N | N |
| Interaction Design | Y | Y | N | N | N | N | N | N |
| Learnability | Y | Y | Y | Y | N | N | N | N |
| Intuitive Interfaces | Y | Y | Y | Y | N | N | N | L |
| Error Prevention and Recovery | L | Y | L | L | N | N | N | L |
| User Control and Freedom | Y | Y | Y | Y | N | N | N | L |
| Internal Process Visibility | Y | Y | Y | Y | N | N | N | L |
| Understandability of AI Decisions | Y | Y | L | L | N | N | N | L |
| Variety of Explanation Types | Y | Y | L | L | N | N | N | L |
| Transparency of Decision-Making | Y | Y | L | L | N | N | N | L |
| Iterative Explanation Improvement | L | Y | L | L | N | N | N | L |
| User Input on Functionality/ Performance | L | Y | L | L | N | N | N | L |
| Alignment with Human Decisions | L | Y | L | L | N | N | N | L |
| Bidirectional Feedback Mechanisms | L | Y | L | L | N | N | N | L |
| Transparency of Feedback Impact | L | Y | L | L | N | N | N | L |
| Feature Availability & Functionality | Y | Y | Y | Y | Y | Y | Y | L |
| User Control over Outputs | Y | Y | Y | Y | N | N | N | L |
| Examination and Comparison | Y | Y | L | L | N | N | N | L |
| Level of Oversight Supported | Y | Y | L | L | N | N | N | L |
| Override Capability | Y | Y | L | L | N | N | N | L |
| Collaboration and Control | Y | Y | L | L | N | N | N | L |
| Oversight of AI Decision-Making | Y | Y | L | L | N | N | N | L |
| Trustworthiness Alignment with Values | Y | Y | L | L | N | N | N | L |
| Balance of Automation and Accountability | Y | Y | L | L | N | N | N | L |
| Auditability and Traceability | Y | Y | L | L | N | N | N | L |
| Data Linkage and Source Mapping | Y | Y | N | L | N | N | N | L |
| Data Pre-processing Flexibility | Y | L | L | L | Y | Y | Y | Y |
| Model Versioning and Rollback | Y | Y | L | L | N | N | N | L |
| Explainability Workflow Integration | Y | Y | Y | Y | L | L | L | L |
| Adversarial Robustness Testing | Y | Y | N | N | L | L | L | L |

**Legend**

Y (Yes): Full Support | L (Limited): Partial/Community Support | N (No): No Direct Support

| | Dataiku | DataRobot | H2O AutoML Studio | Altair Rapidminer | FLAML | AutoGluon | PyCaret | H2O AutoML Library |
|---|---|---|---|---|---|---|---|---|
| Auto-Feature Discovery/Generation | Y | Y | L | Y | N | N | L | L |
| Transfer Learning API Support | L | L | N | N | Y | Y | L | N |
| Low-Latency Real-Time Inference | Y | Y | Y | Y | L | L | L | L |
| Real-Time Model Monitoring | Y | Y | L | L | N | N | N | L |
| Concept and Data Drift Detection | Y | Y | L | L | N | N | N | L |
| Auto-Hyperparameter Selection | Y | Y | Y | Y | Y | Y | Y | Y |
| Cost-Sensitive Learning Configuration | L | Y | L | L | Y | L | L | L |
| Multi-label Classification Support | Y | Y | L | L | L | Y | L | L |
| Geographical/Spatial Data Support | Y | Y | N | Y | N | N | N | N |
| Private/Secure Data Handling | Y | Y | L | L | N | N | N | L |
| Auto-MLOps Integration | Y | Y | L | L | N | N | L | L |
| Model Retraining/Deployment Triggers | Y | Y | N | L | N | N | N | L |
| Pipeline Visualization | Y | Y | Y | Y | N | N | N | N |
| Collaborative Environment/Sharing | Y | Y | L | L | N | N | N | N |
| Role-Based Access Control (RBAC) | Y | Y | L | L | N | N | N | N |
| Infrastructure Auto-scaling | Y | Y | Y | Y | L | L | L | L |
| GPU/TPU Acceleration Support | Y | Y | Y | Y | Y | Y | Y | Y |
| Data Source Connectors (Native) | Y | Y | L | Y | N | N | N | L |
| Custom Code Integration (e.g., Python/R) | Y | Y | Y | Y | Y | Y | Y | Y |
| Compliance and Regulatory Reporting | Y | Y | L | L | N | N | N | L |
| Ethical AI Frameworks Integration | Y | Y | L | L | N | N | N | L |
| Performance Benchmarking Tools | Y | Y | Y | Y | L | L | L | L |
| Model Inventory Management | Y | Y | L | L | N | N | N | L |
| Data Pipeline Automation | Y | Y | L | L | N | N | N | L |
| Data Governance Tools | Y | Y | L | L | N | N | N | N |
| Automated API Endpoint Generation | Y | Y | Y | Y | L | L | L | L |
| Batch and Stream Processing Support | Y | Y | Y | Y | Y | Y | Y | Y |
| Edge/Embedded Device Deployment | L | Y | L | L | N | N | N | N |
| Git/Version Control Integration | Y | Y | L | L | N | N | N | L |
| Automatic Model Feedback Loop | L | Y | L | L | N | N | N | L |
| Scalable Storage Integration (Cloud) | Y | Y | Y | Y | L | L | L | L |
| Automated Documentation Generation | Y | Y | L | L | N | N | L | L |
| Cross-Cloud/Hybrid Compatibility | Y | Y | L | L | Y | Y | Y | Y |
| Time-to-Prediction Metric Tracking | Y | Y | Y | Y | L | L | L | L |
| Model-Agnostic Explanations (LIME/SHAP) | Y | Y | L | L | Y | Y | Y | Y |
| Fairness Metric Integration | Y | Y | L | L | L | L | L | L |
| Data Security and Encryption Standards | Y | Y | Y | Y | N | N | N | N |
| Compliance Auditing Trail | Y | Y | L | L | N | N | N | L |
| Automated Model Documentation | Y | Y | L | L | N | N | L | L |
| Interoperability Standard (ONNX/PMML) | Y | Y | L | L | Y | Y | Y | Y |

**Legend**

Y (Yes): Full Support | L (Limited): Partial/Community Support | N (No): No Direct Support

# Appendix 5: Business implications of bias in recruitment process

## A. Business implications for enterprise non-expert users

In one of the most published instances of gender bias, (Danish entrepreneur David Heinemeier Hansson's case) in the Apple Card's algorithm raised concerns about fairness and equality in financial services[20]. The case exemplifies how algorithms can perpetuate existing biases and inequalities, impacting access to credit and financial opportunities based on gender. In one of the prominent cases, Amazon developed an AI tool for recruitment purposes but abandoned it after discovering that the system was discriminating against women. The AI tool downgraded résumés containing the word "women's" and filtered out candidates from women-only colleges, perpetuating biases and preferences for male applicants[21]. The system's reliance on past résumés, which were predominantly submitted by men, led to a bias against female hires. This raised concerns about the fairness and interpretability of AI algorithms in making hiring decisions.

Further, biases inherent in human language are replicated and perpetuated by the application of standard machine learning algorithms to text data including Curriculum Vitae's of candidates. One of the studies (Caliskan et al., 2016) showed that widely used machine learning models capture and reproduce various biases found in human society, including those related to race, gender, and societal norms. This suggests that language itself contains imprints of historical biases, which can be learned and reinforced by AI systems.

In another instance, the University of Texas at Austin's computer science department used a machine-learning system called GRADE to assist with admissions decisions for its Ph.D. program but recently abandoned it due to concerns about diversity, equity, and fairness. Critics argue that the system perpetuated existing biases and inequalities in the field, as human biases can be encoded into algorithms[22]. GRADE predicted the likelihood of admissions committee approval for applicants and assigned them numerical scores based on patterns from past admissions decisions. Factors such as the presence of certain words in recommendation letters or the prestige of an applicant's undergraduate institution influenced the scores. Critics also claimed that the system disregarded the value of personal statements and recommendation letters in providing opportunities for marginalized individuals.

---

[20] *Apple Card algorithm sparks gender bias inquiry - The Washington Post*. (n.d.). Retrieved May 25, 2024, from https://www.washingtonpost.com/business/2019/11/11/apple-card-algorithm-sparks-gender-bias-allegations-against-goldman-sachs/

[21] Amazon Shuts Down AI Hiring Tool for Being Sexist https://www.reuters.com/article/world/insight-amazon-scraps-secret-ai-recruiting-tool-that-showed-bias-against-women-idUSKCN1MK0AG/

[22] *U of Texas will stop using controversial algorithm to evaluate Ph.D. applicants*. (n.d.). Retrieved May 25, 2024, from https://www.insidehighered.com/admissions/article/2020/12/14/u-texas-will-stop-using-controversial-algorithm-evaluate-phd

In yet another instance, Gates Foundation's program, Intensive Partnerships for Effective Teaching, which aimed to improve education for low-income minority students by using data and algorithms to assess teacher performance, did not show any noticeable gains for students[23]. The program focused on measures such as test scores, principal observations, and evaluations from students and parents, but these metrics may not capture the full complexity of teaching and learning. This also raised questions about the limitations of relying solely on quantitative data to assess teacher performance and improve education or making automated employment (performance) decisions using algorithms. These instances indicate that such discrimination results in losing of a capable candidate, failing on legal requirements, dent in the reputation and regulatory penalties. This paper deals with such implications of algorithmic discrimination in hiring, covering the following: ineffective hiring, legal and ethical concerns and implications, reputation damage, and fines and penalties.

The identified gaps in fairness, explainability and transparency do not impact only end-users — they affect critical enterprise stakeholders across the procurement and compliance pipeline. For example, HR technology buyers seek intuitive tools that enhance diversity reporting and fairness compliance. Legal and compliance influencers assess whether AutoML tools meet regulatory and anti-discrimination obligations. Meanwhile, IT gatekeepers evaluate platform security, interoperability, and auditability. When fairness features are opaque or unsupported, it leads to friction across all these roles, often resulting in delayed purchase decisions, pilot abandonment, or the requirement for custom compliance documentation.

## B. Methodology

The methodology employed for the paper on the business implications of algorithmic discrimination in hiring involved several key steps. Firstly, a comprehensive literature review was conducted by searching Scopus using specific keywords such as "business impact," "discrimination OR bias," and "hiring." Relevant research papers were selected and analyzed to gather insights on the business implications of algorithmic discrimination. Additionally, reported incidents of bias and discrimination in hiring the research collated by utilizing the AI Incident database. A list of these incidents, including details about the companies involved and the nature of the bias, was compiled. The research included 64 articles, of which 26 were about bias and discrimination in domains other than hiring. The rest of the articles collectively pointed towards 7 broad incidents relating to hiring and discrimination.

[23] *"An Expensive Experiment": Gates Teacher-Effectiveness Program Shows No Gains for Students*. (n.d.). Retrieved May 26, 2024, from https://www.edweek.org/teaching-learning/an-expensive-experiment-gates-teacher-effectiveness-program-shows-no-gains-for-students/2018/06

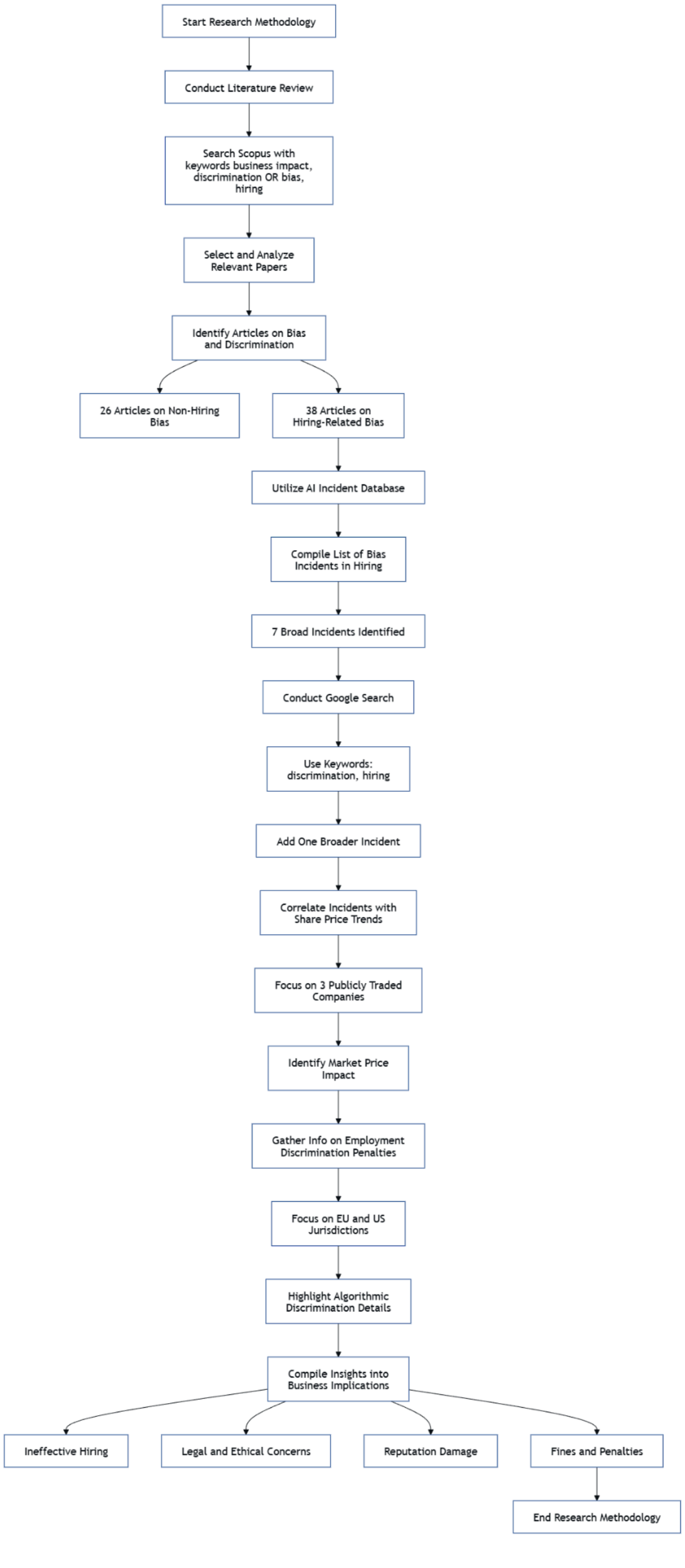

To supplement this information, a Google search was conducted using keywords "discrimination" and "hiring", adding one additional broader incident to the list of incidents referred. The gathered incidents (relating to three of the publicly traded companies) were then correlated with the trends in share prices identified from the listing, highlighting potential impact on their market price performance. Furthermore, information on potential and penalties related to employment discrimination in the EU and US jurisdictions was gathered, with an emphasis on any available details specific to algorithmic discrimination. Finally, the insights gathered from these steps were compiled into four broad categories of business implications: ineffective hiring, legal and ethical concerns and implications, reputation damage, and fines and penalties. This methodology provided a comprehensive analysis of the business implications of algorithmic discrimination in hiring, incorporating insights from academic research, reported incidents, share price trends, and fines and penalties associated with employment discrimination.

## C. Business implications - an overview

The lack of interpret-ability (Mollas et al., 2023), non-generalisability, and the black box nature of AI systems (Sikorski, 2021), which can lead to challenges in understanding and explaining how these technologies work. The failure of AI projects (Schlegel et al., 2023), the impact of the AI system's limitations, unreliable results (Alexander et al., 2023) (Miller, 2019) and potential negative effects on deployment performance (Muthusamy et al., 2018) can all contribute to and/ or result in such ineffectiveness (Sikorski, 2021). Further, the presence of bias or unfairness in AI algorithms (Baumann et al., 2023) raises legal and ethical concerns (I. Alves et al., 2024). Additionally, privacy exposure (Alexander et al., 2023) (Kikuchi et al., 2018) and security impacts (Siddiqui et al., 2021) (Barta & Görcsi, 2021) can have ethical issues and make the business vulnerable to liabilities. With regulations emerging governing artificial intelligence, it also emphasizes the effects of a lack of accountability (Yildiz & Beloff, 2020), transparency (Muthusamy et al., 2018) (Yildiz & Beloff, 2020) and legal compliance (Barta & Görcsi, 2021) In addition, the negative consequences associated with the black box nature of AI and the potential adverse effects (Muthusamy et al., 2018) of AI deployment can harm a business's reputation (Grove et al., 2020). Stakeholders may also be negatively impacted (Güngör, 2020), as their trust in the business is eroded. Lastly, failure to comply with legal and compliance requirements can result in financial penalties for the business. The value destruction (Canhoto & Clear, 2020) for the business caused by ineffective AI systems and the negative impact on customer due to privacy (Kikuchi et al., 2018) (Luo et al., 2019) or other legal exposures can further lead to financial losses.

This thesis compiles the implications of algorithmic discrimination in hiring, encompassing the following dimensions: inefficacy in hiring practices, legal and ethical considerations and ramifications, reputational harm, and financial penalties.

### C1. Key incidents

Several studies highlight concerns about the effectiveness and scientific basis of AI-driven recruitment tools in reducing bias and promoting diversity in the hiring process. Some of the major reported incidents of discrimination in hiring are listed below:

| Reported incident | Details | References |
|---|---|---|
| Amazon AI hiring tool | Amazon has shut down an artificial intelligence (AI) tool it was developing for recruiting after it was found to be discriminating against women. The tool, which was intended to help with recruitment by searching | *Amazon Shuts Down AI Hiring Tool for Being Sexist* |
| | for candidates online, downgraded résumés containing the word "women's" and filtered out potential hires who had attended women-only colleges. The tool was built using past résumés submitted to Amazon over a 10-year period, which were predominantly submitted by male applicants, perpetuating a bias against female hires. The case study highlights the challenges of developing fair and unbiased AI systems. | |
| Apple card algorithm | This article discusses the gender bias allegations against Goldman Sachs regarding its credit card practices for the Apple Card. The allegations arose when software developer David Heinemeier Hansson highlighted the differences in credit lines between male and female customers in a viral Twitter thread. Hansson revealed that his wife, Jamie Hansson, was denied a credit line increase despite having a higher credit score than him. The article highlights the regulatory investigation sparked by these allegations and emphasizes the need for transparency and fairness in credit card practices. | *Apple Card Algorithm Sparks Gender Bias Inquiry - The Washington Post* |
| Fired by Bot/ algorithm | This article discusses how algorithms used by Amazon to manage its contract drivers, including those in the Flex program, are making decisions about hiring and firing with little human oversight. The algorithms track drivers' performance and determine their eligibility for routes, often leading to terminations without clear explanations. Flex drivers are monitored for factors such as punctuality and delivery quality, but the algorithms do not always account for real-world challenges faced by drivers, such as traffic or access issues. Drivers who are wrongly terminated have limited recourse, and the appeal process is often ineffective. Amazon's use of algorithms in its human-resources operations raises concerns about transparency and fairness, as drivers face the consequences of automated decisions. | *Payout for Estée Lauder Women 'Sacked by Algorithm,'*<br><br>*Fired by Bot: Amazon Turns to Machine Managers And Workers Are Losing Out - Bloomberg* |
| | Estée Lauder has reached an out-of-court settlement with three make-up artists who lost their jobs after being assessed by an algorithm in a video interview. The women, who worked for Estée Lauder subsidiary MAC, were initially required to reapply for their positions before being informed that they were being made redundant based partly on the algorithm's judgment. The software, created by recruiting platform HireVue, analyzed the content of their answers, expressions, and other data about their job performance. The women began legal proceedings against Estée Lauder, claiming they were not informed about the nature of the assessment. The case highlights the potential issues and biases associated with automated hiring software. In an interview, one of the women expressed the importance of speaking out about the issue and stopping such practices. The women received an out-of-court settlement. | |
| Hirevue's discrimination | This article discusses the use of an artificial intelligence (AI) hiring system developed by HireVue, which analyzes facial movements, word choice, and speaking voice of job candidates to generate an "employability" score. The system has gained widespread use among prominent employers, but some experts argue that it lacks scientific reasoning and could result in unfair discrimination. Critics argue that the system may penalize nonnative speakers, nervous interviewees, or those who do not fit the model for look and speech. The article highlights concerns about the system's objectivity and the lack of transparency in its decision-making process. An official complaint has been filed against HireVue, urging the Federal Trade Commission to investigate its practices. In an update, HireVue has stated that it no longer uses visual analysis in its software since 2020, | *HireVue's AI Face-Scanning Algorithm Increasingly Decides Whether You Deserve the Job* |
| | as advances in natural language processing have made it more effective in assessments. | |
| Algorithms to evaluate Phd's | The University of Texas at Austin has announced that it will no longer use a machine-learning system called GRADE to evaluate applicants for its Ph.D. program in computer science. The system, which predicted the likelihood of an applicant's approval and assigned a numerical score, has faced criticism for potentially exacerbating existing inequalities in the field. Critics argue that the system encoded biases into its algorithms, which could have a negative impact on underrepresented groups. The creators of GRADE maintain that the system was designed to replicate the admissions committee's decision-making process and was not programmed to use race or gender in its predictions. However, detractors argue that biases can still be present in other factors used by the system. The university cited difficulties in maintaining the system as the reason for its discontinuation. | *U of Texas Will Stop Using Controversial Algorithm to Evaluate Ph.D. Applicants* |
| Flawed AI interview tools | The article discusses the testing of AI interview tools by MIT Technology Review. The tools, MyInterview and Curious Thing, use AI algorithms to evaluate candidates for job positions. The article highlights concerns about the accuracy and reliability of these algorithms. One specific issue raised is that MyInterview gave a candidate a high score for English proficiency even though she spoke only in German. The article also mentions challenges in assessing personality traits through AI interviews and the potential for bias in the hiring process. The companies behind these tools are often reluctant to share details about their algorithms, making it difficult to assess their accuracy. | We Tested AI Interview Tools. Here's What we Found. \| MIT Technology Review |
| False representation that tools are 'Bias Free' | This academic paper highlights a specific complaint against Aon, a major hiring technology vendor, regarding their online hiring tests. The complaint, filed by the ACLU, alleges that Aon deceptively markets | The Long History of Discrimination in Job Hiring |
| | these tests as "bias-free" while discriminating against job seekers based on their race or disability. The complaint also includes charges filed with the Equal Employment Opportunity Commission against both Aon and an employer using Aon's assessments. The paper discusses how Aon's personality assessment test and automated video interviewing tool, which integrate AI, assess general personality traits that are not directly related to job performance and can unfairly screen out people with disabilities. The paper also raises concerns about cognitive ability assessments, which can disadvantage Black job candidates and individuals with disabilities. The paper emphasizes the need for employers to thoroughly vet assessments for compliance with anti-discrimination laws and hold vendors accountable for designing inclusive and non-discriminatory products. It calls for a future where skills and potential, rather than bias, determine job opportunities. | Assessments \| ACLU of Florida \| we Defend the Civil Rights and Civil Liberties of All People in Florida, by Working through the Legislature, the Courts and in the Streets |

## C2. Analysis

In another instance, contract/ flex drivers for Amazon reported being terminated by algorithms through automated emails (The case of Stephen Normandin) due to reliance being placed on automated systems for employment decisions[24]. This instance also raised questions about fairness, accountability, and the impact of algorithms on employees' livelihoods.

In addition to the above instances of failures, claims made by recruitment AI companies that their tools can objectively assess candidates by removing gender and race appear to be misleading because they fail to understand the broader systems of power related to gender and race (Drage & Mackereth, 2022).

## C3. Case in point - Hireview

HireVue, a company specializing in recruitment technology, has created an assessment system powered by artificial intelligence. This system evaluates candidates based on their facial expressions, vocabulary, and speaking voice to compare them with other applicants and produce a score indicating their suitability for employment. The system has garnered significant attention from prominent employers, However, it has encountered opposition from AI academics who contend that it has a solid scientific foundation and may potentially exhibit bias against specific individuals (e.g. HireView).

[24] *Fired by Bot: Amazon Turns to Machine Managers And Workers Are Losing Out - Bloomberg*. (n.d.). Retrieved May 25, 2024, https://www.bloomberg.com/news/features/2021-06-28/fired-by-bot-amazon-turns-to-machine-managers-and-workers-are-losing-out

The AI approach employed by HireVue has gained widespread use in sectors such as hospitality and finance, prompting colleges to educate students on how to enhance their performance during the evaluations. Nevertheless, critics express doubts regarding the system's comprehension of the ideal employee and express worries regarding potential prejudices against individuals who are not native speakers and those who may be anxious during interviews[25].

HireVue, in 2022, made the decision to cease the analysis of facial expressions of applicants in its algorithmic assessment. This decision is a result of public concerns and a complaint lodged against the company regarding possible bias and privacy concerns[26] (Nightingale & Farid, 2022)

HireVue's software utilizes advanced algorithms to evaluate multiple aspects of job applicants, such as their linguistic proficiency, verbal communication skills, and now even their non-verbal cues, excluding facial expressions. The company asserts that it conducts screening to detect bias related to gender, color, and age. However, it acknowledges that it may face difficulties in identifying bias associated with factors such as income or education level.

## C4. Business implications in detail

The business implications of bias in hiring include ineffective hiring, missed opportunities caused by non-hiring of qualified candidates, legal and ethics concerns, reputation damage, and lack of diversity hindering innovation and impacting employee morale.

1. Ineffective hiring: Algorithmic recruiting methods can lead to biased candidate selection based on discriminating variables like gender, color, or handicap. This may result in the omission of competent workers and restrict diversity in the workforce. This reinforces inequity and causes organizations to miss out on benefiting from a varied array of skills and viewpoints.

[25] *HireVue's AI face-scanning algorithm increasingly decides whether you deserve the job - The Washington Post*. (n.d.). Retrieved May 25, 2024, from https://www.washingtonpost.com/technology/2019/10/22/ai-hiring-face-scanning-algorithm-increasingly-decides-whether-you-deserve-job/

[26] *Job Screening Service Halts Facial Analysis of Applicants | WIRED*. (n.d.). Retrieved May 25, 2024, from https://www.wired.com/story/job-screening-service-halts-facial-analysis-applicants/

2. Legal and ethical concerns and implications: Algorithmic biases in recruitment processes can lead to legal repercussions because of possible discriminatory results. Furthermore, ethical issues emerge about privacy, freedom of speech, and fair employment chances.
3. Reputation Damage: Biased hiring practices being revealed can harm the organization's reputation. The public's view of fairness and diversity in the recruiting process can greatly influence an organization's brand image and its capacity to attract high-caliber talent.
4. Fines and penalties: When algorithms used in hiring processes discriminate against certain protected groups, such as individuals based on race, gender, or disability, it can lead to regulatory violations and in turn may result in legal costs, fines and penalties.

### C4.1. Ineffective Hiring

Algorithmic recruiting strategies can result in biased candidate selection based on discriminatory factors resulting in exclusion of skilled employees and limit the variety of individuals in the labor force. Bias in recruiting leads to a lack of diversity in the workforce, restricting access to various viewpoints, experiences, and abilities, which in turn hinders innovation. This impedes innovation and creativity within the firm, limiting its capacity to adjust and prosper in a competitive market. Some of the key such indicators that lead to ineffective hiring are provided below:

1. Harvard Business School published a paper titled 'Hidden workers: Untapped talent', which clarified the issue of algorithmic bias in hiring and its impact on job seekers. Automated hiring systems, including CV scanners, are preventing an estimated 27 million people from finding full-time work. These systems, relied upon by as many as 75% of employers, often focus on what candidates lack rather than the value they can bring, leading to the exclusion of certain groups such as caregivers, veterans, immigrants, people with disabilities, and more. The report highlights the widening skills gap and the failure of employers to recognize the benefits of widening their candidate pool. It categorizes the affected individuals as working part-time but seeking full-time roles, unemployed job seekers, and those who are not actively seeking employment (Fuller et al., n.d.).
2. While tools like HireVue's system have been acknowledged for its ability to reduce time and expenses for organizations, simplify the hiring process, and enhance the recruitment of diverse candidates. Nonetheless, the absence of clarity regarding the algorithms and the incapacity for candidates to obtain feedback or dispute their assessment results has led to fear and exasperation among job searchers. Certain candidates perceive the process as degrading and raise concerns about the fairness of being evaluated[27]. Also, AI tools that analyze speech and bodily movements to assess a candidate's fit for a job are criticized as having no scientific basis. This approach as "modern phrenology," comparing it to the false theory that skull shape can reveal character and mental faculties[28] .

3. The use of AI in analyzing candidate videos or applications is seen as "pseudoscience" and lacks a scientific basis. The utilization of AI-powered interview software by organizations such as MyInterview and Curious Thing gives rise to worries over its correctness and trustworthiness. During a test, a candidate achieved a high score in English proficiency despite speaking exclusively in German, thereby exposing potential weaknesses in the algorithms' assessment methodology[29].
4. AI hiring tools that claim to provide neutral assessments of candidates' traits reflect the power relationship between the observer and the observed and are influenced by historical biases and current market demands (Drage & Mackereth, 2022).
5. The algorithms employed by these software companies scrutinize candidates' responses to ascertain their personality traits and suitability for the job. Experts argue that depending on tone and open-ended questions as indications of personality traits can be incorrect and unjust. The absence of regular and systematic techniques for collecting data presents difficulties in accurately evaluating candidates.
6. In addition, the researchers at the University of Cambridge ran an independent trial where they developed a simplified AI recruitment tool to evaluate candidates' pictures for personality attributes. The technique, known as the "Personality Machine," identified the "big five" personality traits: extroversion, agreeableness, openness, conscientiousness, and neuroticism. The research concluded that the software's predictions were influenced by variations in people's facial expressions, lighting conditions, backgrounds, and clothing preferences (Drage & Mackereth, 2022).

### C4.2 Legal and ethical concerns

Legal and ethical concerns arise from instances of (1) unfair and deceptive practices in hiring assessments, (2) the use of AI-generated images in fake profiles on LinkedIn undermining the reliability of job applications, and the proliferation of AI-generated content raises concerns and, (3) the lack of transparency and independent testing in AI hiring tools calls for caution in their use. Some of such key indicators are provided below:

[27] *HireVue's AI face-scanning algorithm increasingly decides whether you deserve the job - The Washington Post*. (n.d.). Retrieved May 25, 2024, from https://www.washingtonpost.com/technology/2019/10/22/ai-hiring-face-scanning-algorithm-increasingly-decides-whether-you-deserve-job/

[28] AI tools fail to reduce recruitment bias - study. (n.d.). Retrieved May 25, 2024, from https://www.bbc.com/news/technology-63228466

[29] *We tested AI interview tools. Here's what we found. | MIT Technology Review*. (n.d.). Retrieved May 25, 2024, from https://www.technologyreview.com/2021/07/07/1027916/we-tested-ai-interview-tools/

—------—------------------------------------------------------------------------------------

1. The American Civil Liberties Union (ACLU) has filed a complaint with the Federal Trade Commission (FTC) against Aon Consulting, Inc. The complaint alleges that Aon engages in unfair and deceptive practices by designing and selling online assessments to employers that discriminate against individuals based on disability and race. Specifically, the complaint states that Aon's Adaptive Employee Personality Test (ADEPT-15), which is used in hiring processes, adversely impacts individuals with disabilities and mental health conditions. The complaint also claims that Aon's video interviewing tool, vidAssess-AI, exacerbates discrimination through its use of artificial intelligence elements. Additionally, Aon's gamified cognitive assessment tool, gridChallenge, is accused of having disparities based on race and potentially screening out individuals with disabilities. The ACLU argues that Aon's marketing claims of being "bias-free" and "fair" are deceptive and violate FTC regulations. The complaint calls for an investigation into Aon's practices, an injunction, and appropriate relief[30].
2. Research (Nightingale & Farid, 2022) and regulatory concerns have been raised regarding the utilization of AI in hiring. There are suggestions to mandate employers to notify candidates when AI is employed for assessment purposes and to regularly examine algorithms through audits. Facial analysis has been prohibited in certain areas, such as Maryland, and Amazon has purportedly discontinued their resume screening technology because it produced biased outcomes. The outsourcing of diversity work to AI-powered hiring tools may unintentionally reinforce inequality and discrimination by not addressing systemic problems within organizations (Drage & Mackereth, 2022).
3. Computer-generated fake profiles using AI-generated images are being used on LinkedIn for various purposes. The use of AI-generated images in fake profiles is becoming increasingly convincing[31]. This affects the reliability of the applications submitted using Linkedin.
4. A study found that AI-generated faces are "indistinguishable" from real faces, and people have only a 50% chance of correctly guessing whether a face was created by a computer or not. The average person on the internet is essentially at chance in determining the authenticity of these images (Nightingale & Farid, 2022).
5. The proliferation of AI-generated content, including images, audio, and video, could lead to a new era of online deception. AI-generated faces tend to look trustworthy because they stick to the most average features, making them familiar to people. This raises concerns about the potential for widespread use of deepfakes and other forms of online deception using AI technology.

[30] *The Long History of Discrimination in Job Hiring Assessments | ACLU of Florida | We defend the civil rights and civil liberties of all people in Florida, by working through the legislature, the courts and in the streets.* (n.d.). Retrieved June 9, 2024, from https://www.aclufl.org/en/news/long-history-discrimination-job-hiring-assessments
[31] *The latest marketing tactic on LinkedIn: AI-generated faces : NPR*. (n.d.). Retrieved May 25, 2024, from https://www.npr.org/2022/03/27/1088140809/fake-linkedin-profiles

---

6. Numerous AI tools used in hiring lack independent testing, and the corporations responsible for their development are typically hesitant to disclose information regarding their algorithms[32]
7. The absence of transparency poses challenges for candidates and employers in assessing the veracity of the algorithms or the magnitude of their impact on recruiting choices. There are concerns regarding bias and the necessity for prudence while utilizing these tools, as they have the potential to significantly affect the economic survival of job searchers.
8. While AI technology shows promise, it is not yet ready to replace traditional recruiters and human decision-making processes in hiring. Testing a new version of the automated employment screening focused on diversity is one step towards addressing the issue (*Amazon Shuts Down AI Hiring Tool for Being Sexist*, n.d.).

### C4.3 Reputational impact

From the gathered incidents (relating to three of the publicly traded companies), an attempt was made to correlate the incidents reported with the trends in share prices to assess if there were any potential impact on market price performance of these identified companies. The results indicated that in 2 out of 3 cases there appeared to be a potential implication of the incidents of discrimination being reported publicly. The details are provided below:

| Company | Duration | Remarks | Price before news/ date | Price after news/ Date |
|---|---|---|---|---|
| Amazon (AMZN) | October 3, 2018 to October 12, 2018 | The 8% fall in share price on October 10th and 11th 2018, appears to closely correlate with the dates on which the news on Amazon's recruitment algorithm being scrapped for sexist behaviour, was published. | USD 93.52 on October 9, 2018 | USD 85.97 on October 11, 2018 |

[32] *We tested AI interview tools. Here's what we found.* | *MIT Technology Review*. (n.d.). Retrieved May 25, 2024, from https://www.technologyreview.com/2021/07/07/1027916/we-tested-ai-interview-tools/

| Apple (AAPL) | November 01, 2019 to November 12, 2019 | The prices for Apple did not get negatively affected when the news regarding Apple’s credit card being discriminatory. This could also be attributed to a dividend notified by Apple on November 7, 2019. | USD 65.04 on November 8, 2019 | USD 65.49 on November 12, 2019 |
|---|---|---|---|---|
| ESTEE LAUDER (EL) | March 2, 2022 to March 18, 2022 | The prices of Estee Lauder dropped 11% from USD 291.75 on March 2nd to USD 258.90 on March 15th, closely correlating with the dates during which the BBC documentary ‘Computer Says No’ was announced and promoted. Estee settled with the identified ex-employees in an off-the-court settlement, possibly resulting in the prices rebounding on March 16th. The price rebound also is attributed to mentioning about the settlement in the documentary when it was aired on March 16, 2022. | USD 291.75 on March 2, 2022 | USD 258.90 on March 15, 2022 and USD 270.88 on March 16, 2022 |

### C4.4 Fines and penalities

Discrimination in hiring is prohibited by various employment discrimination laws and regulations, and organizations that engage in discriminatory practices may face legal action. As a result of discriminatory hiring practices, organizations may incur legal costs associated with defending themselves against lawsuits filed by aggrieved individuals or advocacy groups. They may also face fines and penalties imposed by regulatory bodies for violating anti-discrimination laws. These legal consequences can be significant and have a negative impact on an organization's reputation, finances, and overall business operations.

In the case of Filcam VGIL Bologna and others v Deliveroo Italia SRL in the Bologna Labour Court, there was an instance of indirect discrimination leading to penalty. Indirect discrimination occurs when rules or arrangements that appear neutral in practice disadvantage employees with protected characteristics. The case involves Deliveroo's algorithm, which determined workers' priority to access delivery time slots, and was found to be discriminatory. The algorithm scored riders based on various criteria, but it failed to consider the reasons for cancellation or non-participation in a shift, potentially leading to indirect discrimination. The court ruled that Deliveroo failed to objectively justify the system and lacked transparency by not disclosing the algorithm's workings. The case highlights the importance of understanding and questioning the output of algorithms to avoid unintended discrimination and emphasizes the need for human oversight in AI systems.

In addition, on analyzing the list of AI related litigations as maintained by George Washington University lists 6 cases when filtered for "Employment" and "Hiring" in the application area of the page. A snapshot of the same is presented below:

| Caption | Brief Description | Algorithm | Jurisdiction | Application Areas | Cause of Action | Issues | Date Action Filed | New Activity |
|---|---|---|---|---|---|---|---|---|
| Aerotek, Inc. v. Boyd | Texas Supreme Court rules that e-signatures on new hire paperwork can compel arbitration for a company that was sued for racial discrimination | | State: Texas Circuit Court | Employment, Hiring | Contracts, Arbitration | Use of Race | 1/1/2018 | 7/1/2021 |
| Baker v. CVS Health Corporation | A candidate interviewing for a position with CVS Health has alleged that he and other interviewees were subjected to an artificial intelligence-powered lie detector test without the appropriate notification, according to a class action complaint recently removed to the U.S. District Court for the District of Massachusetts. | | U.S. District Court for the District of Massachusetts | Civil Rights, Employment, Hiring | Civil Rights, Permanent Injunction | Failure to Disclosure, Misuse of AI, Unaware of Use of Algorithm | 6/30/2023 | 1/26/2024 |
| Equal Employment Opportunity Commission v. iTutorGroup, Inc. | In the lawsuit, EEOC v. iTutorGroup, Inc., the EEOC alleged that iTutorGroup's hiring software automatically rejected older job applicants in violation of the Age Discrimination in Employment Act ("ADEA"). This settlement comes amongst the EEOC's stated intent to enforce anti-discrimination laws in connection with the use of AI in workplace decisions, and is likely the first of many more litigations and settlements in this area. | | U.S. District Court for the Eastern District of New York | Employment, Hiring | Civil Rights, Permanent Injunction | Programmer Bias, Socioeconomics Bias | 5/5/2022 | 9/8/2023 |
| Houston Federation of Teachers v. Houston Independent School District | Texas Supreme Court rules in favor of teachers' procedural due process objections to the school district not turning over the code of how the teacher VAM (value added model) ratings were calculated that resulting in employment terminations, pay raises, and tenure | EVAAS | Federal: US Dist. Ct. S.D. Texas | Constitutional Law, Employment, Performance Assessment, Termination | 42 USC 1983, Procedural Due Process | Accountability, Transparency/Trade Secrecy | 4/30/2014 | 8/13/2017 |
| In the Matter of HireVue, Inc. | EPIC filed a complaint and request for investigation, injunction, and other relief with the FTC, arguing HireVue's opaque and proprietary tools were unproven, biased, and invasive as it claimed to measure "cognitive ability," "psychological traits," "emotional intelligence," and "social aptitudes" of job applicants without using facial recognition technology. | HireVue | Government: FTC | Employment, Facial Recognition, Hiring | FTC Act, Public Policy, Unfair and Deceptive Trade Practices | Lack of Scholarly Review, Programmer Bias, Transparency/Trade Secrecy, Use of Race, User of Gender | 11/9/2019 | 1/21/2021 |
| Mobley v. Workday, Inc. | A class action lawsuit was brought to challenge to Workday's artificial intelligence systems and screening tools, arguing that these tools are biased against applicants who are either over the age of 40, Black, or disabled. The lawsuit argues that Workday violated the Age Discrimination in Employment Act, the Americans with Disabilities Act, and the Civil Rights Act. | ATS | N.D. Cal | Civil Rights, Employment, Hiring | Civil Rights | Programmer Bias, Transparency/Trade Secrecy, Use of Race | 2/21/2023 | 1/11/2024 |

Records 1-6 of 6

Of the above, (1) Equal Employment Opportunity Commission v. iTutorGroup, Inc., (2) In the Matter of HireVue, Inc., and (3) Mobley v. Workday, Inc. were relating to programmer bias. The details and outcome of these cases are presented below:

| Case ref | Excerpts from the case | Fine or penalty |
|---|---|---|
| Equal Employment Opportunity Commission v. iTutorGroup, Inc. | On May 5, 2022, the EEOC filed a complaint against iTutorGroup, an organization that hires remote English tutors for students in China, in the Eastern District of New York. The EEOC alleged that the company violated the ADEA by implementing a software hiring program that "intentionally discriminated against older applicants because of their age" by "automatically reject[ing] female applicants age 55 or older and male applicants age 60 or older," effectively screening out over 200 applicants. The purported discriminatory software was discovered when an applicant submitted two applications identical in all but birth date. According to the EEOC, the applicant used one application with their real date of birth and filed a second application with a more recent date of birth. The candidate allegedly received an interview only when using the more recent date of birth. On August 9, 2023, the Equal Employment Opportunity Commission ("EEOC") announced the settlement of the agency's first lawsuit involving the alleged discriminatory use of artificial intelligence ("AI") in the workplace. On September 8, 2023, federal court approved a consent decree from the Equal Employment Opportunity Commission (EEOC) with iTutorGroup Inc. and its affiliates ("iTutor") over alleged age discrimination in hiring, stemming from automated systems in recruiting software. | According to the consent decree, iTutor will pay $365,000 to over 200 job candidates who were automatically screened out by iTutor's recruiting software to resolve the EEOC's claims. |
| In the Matter of HireVue, Inc. | EPIC filed a complaint with the FTC on November 19, 2019 that HireVue has violated Section 5 of the FTC Act engaging in unfair and deceptive | The case is ongoing. |
| | practices by making claims to consumers without a reasonable basis for the claims, and causes or is likely to cause substantial injury to the consumer, and misrepresented the amount of control consumers had over their privacy. HireVue uses both an applicant-recorded video interview and online games to evaluate "thousands of data points" from a candidate, then using predictive algorithms to calculate the job candidate's employability. EPIC notes that job candidates do not have access to the "training data, factors, logic, or techniques used to generate each algorithmic assessment. In some cases, even HireVue is unaware of the basis for an algorithmic assessment." EPIC claims that HireVue is incorrect to say it does not engage facial recognition technology. Next, EPIC claims that HireVue's recruiting tools are biased and discriminatory by gender, those with neurological disorders, by race, and attempting to identify sexual orientation. Along with the FTC Act, EPIC brings this action citing to public policy AI Principles from the Organization for Economic Cooperation and Development ("OECD") and the Universal Guidelines for Artificial Intelligence ("UGAI") The complaint states HireVue violated the policies of Fairness, Security/Safety, Transparency and Explainability, Accountability, Accuracy/REliability and Validity. | Hirevue is under investigation and announced that it will stop relying on "facial analysis" to assess job candidates. |
| Mobley v. Workday, Inc. | Derek Mobley, the Plaintiff, filed suit against Workday alleging violation of the Civil Rights Act of 1964, Age Discrimination in Employment Act of 1967, and the ADA Amendments Act of 2008 because Workday provides potential employers an algorithm which they can use to screen applicants. The plaintiff alleges that the program disproportionately discriminates against African | The case is ongoing. |
| | Americans, people over the age of 40, and the disabled. | |

## D. Business implications for AutoML provider

Fairness opacity not only raises ethical concerns but also creates friction for enterprise buyers. Procurement teams often flag tools lacking interpretability as compliance risks. Moreover, the absence of intuitive fairness workflows increases support burden, with end-users relying on external consultation or vendor tickets to understand bias metrics. This reduces scalability and inflates the total cost of ownership (TCO) for the AutoML provider.

For AutoML providers, integrating fairness-by-design features is not only an ethical imperative but also a strategic enabler. Tools that proactively support fairness reduce enterprise buyer friction by shortening procurement cycles, lowering the need for custom compliance documentation, and improving buyer trust. In turn, this can reduce post-sale support costs and accelerate time to market in highly regulated domains like HR.

## E. Conclusion

The analysis presented demonstrates that algorithmic discrimination in hiring has substantial commercial ramifications. Biased algorithms in hiring can lead to unproductive procedures that exclude qualified workers and restrict diversity in the workforce, impeding innovation and creativity within a company. This can ultimately have a significant influence on a company's capacity to compete in the market and adapt to evolving demands. The article showcases various cases in which algorithmic hiring tools have been discovered to propagate biases and engage in discriminatory practices against specific populations.

The use of unfair and deceptive methods in recruiting assessments, the utilization of AI-generated photos in fraudulent profiles, and the absence of transparency and independent evaluation in AI hiring tools give rise to ethical problems and have the potential to infringe upon anti-discrimination laws. Companies that participate in discriminatory practices may be subject to legal consequences, such as lawsuits and fines imposed by regulatory authorities. The legal ramifications can adversely affect a company's reputation, profits, and overall business operations.

The paper emphasizes the correlation between instances of employment discrimination and the potential negative effect on a company's stock prices. Instances of discrimination that are made known to the public can result in adverse publicity and harm a company's standing, potentially impacting its associations with consumers, investors, and other interested parties.

In essence, the analysis highlights the importance for firms to recognize the possible biases and discriminatory effects of algorithmic hiring tools and the need for transparency, autonomous testing, and human supervision in the creation and utilization of these tools. Failure to address algorithmic discrimination in hiring exposes companies to both legal ramifications and impedes their capacity to attract and retain a broad pool of personnel, stifle innovation, and uphold a favorable reputation.

# Acronyms

A list of acronyms and abbreviations used in the thesis are structured as follows:

| Acronym | Expansion | Chapter with prominent references |
|---|---|---|
| AEDT | Automated Employment Decision Technology | Chapter 4 |
| AI | Artificial Intelligence | Chapter 1 |
| AIA | AI Act | Chapter 4 |
| AIML | Artificial intelligence/Machine learning | Chapter 1 |
| AIRS | Adverse Incident Reporting System | Chapter 5 & Appendix |
| AutoML | Automated Machine Learning | Chapter 2 |
| BO | Bayesian optimization | Chapter 2 |
| CAGR | Compound annual growth rate | Chapter 2 |
| CHI | Computer-human interaction | Chapter 2 |
| CLI | Command Line Interface | Chapter 3 |
| CLT | Cognitive Load Theory | Chapter 3 |
| DBA | Doctorate in Business Administration | Chapter 1 |
| DCWP | Department of Consumer and Worker Protection | Chapter 4 |
| DEI | Diversity, equity, inclusion | Chapter 2 |
| DETR | Detection Transformer | Chapter 2 |
| DI | Disparate Impact | Chapter 2 |
| DPD | Demographic Parity Difference | Chapter 3 |
| DSA | Digital Service Act | Chapter 4 |
| EEOC | Equal Employment Opportunity Commission | Chapter 4 |
| EMOO | Evolutionary Many-Objective Optimization | Chapter 2 |
| EOpp | Equality of Opportunity | Chapter 3 |
| ERM | Enterprise Risk Management | Chapter 1 |
| ESG | Environmental, Social, and Governance | Chapter 9 |
| EU AI Act | European Union AI Act | Chapter 4 |
| FaaS | Fairness as a Service | Chapter 9 && Chapter 2 |
| FL | Federated learning | Chapter 2 |
| FN | False Negatives | Chapter 3 |
| FP | False Positives | Chapter 3 |
| FPR | False Positive Rate | Chapter 3 |
| FS-DD | Fair Selection with the Differentiable Distribution Difference | Chapter 2 |
| GDPR | General Data Protection Regulation | Chapter 4 |
| GRADE | Graduate Recommendation Algorithm for Departmental Admissions | Chapter 9 & Chapter 2 |
| GUI | Graphical User Interface | Chapter 3 |
| HAI | Human-AI collaboration | Chapter 5 |
| HCAI | Human-Centered AI | Chapter 3 |
| HCI | Human-Computer Interaction | Chapter 5 |
| HCOMP | Human Computation | Chapter 5 & Appendix |
| HITL | Human-in-the-loop | Chapter 5 |
| HMI | Human-machine interface | Chapter 5 |
| HPO | Hyperparameter optimization | Chapter 3 |
| HR | Human Resources | Chapter 2 |
| HRI | Human-robot interaction | Chapter 5 |
| IDT | Innovation Diffusion Theory | Chapter 3 |
| ILP | Inductive logic programming | Chapter 2 |
| iML | Interactive machine learning | Chapter 2 |
| IoT | Internet of Things | Chapter 2 |
| IP | Intellectual property | Chapter 4 |
| KPI | Key Performance Indicator | Chapter 5 |
| KDE | Kernel density estimation | Chapter 2 |
| LFIT | Learning from interpretation transition | Chapter 2 |
| ML | Machine Learning | Chapter 2 |
| NAS | Neural Architecture Search | Chapter 2 |

| NNI | Neural Network Intelligence | Chapter 3 |
|---|---|---|
| NLP | Natural language processing | Chapter 2 |
| PE | Proportional Equality | Chapter 2 |
| PRP | Predictive Rate Parity | Chapter 3 |
| ROI | Return on Investment | Chapter 9 & Chapter 2 |
| R&D | Research and Development | Chapter 1 |
| SBFT | Search-based fairness testing | Chapter 2 |
| SHR | Smart Human Resource | Chapter 1 |
| SMEs | Small and medium-sized enterprises | Chapter 2 |
| TAM | Technology Acceptance Model | Chapter 3 |
| TEVV | Testing, Evaluation, Validation, and Verification | Chapter 5 |
| TN | True Negatives | Chapter 3 |
| TP | True Positives | Chapter 3 |
| TPR | True Positive Rate | Chapter 3 |
| UI | User Interface | Chapter 5 |
| UX | User Experience | Chapter 5 |
| XAI | Explainable AI | Chapter 5 |
| YOLOv5 | You Only Look Once version 5 | Chapter 1 |
| ΔEO | Equalized Odds Difference | Chapter 3 |
| ΔEOpp | Equality of Opportunity | Chapter 3 |

# Bibliography

1. Achchab, S., & Temsamani, Y. K. (2022). Use of Artificial Intelligence in Human Resource Management: "Application of Machine Learning Algorithms to an Intelligent Recruitment System." In Lecture Notes in Networks and Systems (Vol. 249). https://doi.org/10.1007/978-3-030-85365-5_20
2. Acemyan, C. Z., & Kortum, P. (2012). The Relationship Between Trust and Usability in Systems. Proceedings of the Human Factors and Ergonomics Society Annual Meeting, 56(1), 1842–1846. https://doi.org/10.1177/1071181312561371
3. Alamelu, M., Kumar, D. S., Sanjana, R., Sree, J. S., Devi, A. S., & Kavitha, D. (2021). Resume Validation and Filtration using Natural Language Processing. IEMECON 2021 - 10th International Conference on Internet of Everything, Microwave Engineering, Communication and Networks. https://doi.org/10.1109/IEMECON53809.2021.9689075
4. Albaroudi, E., Mansouri, T., & Alameer, A. (2024). A Comprehensive Review of AI Techniques for Addressing Algorithmic Bias in Job Hiring. AI (Switzerland), 5(1), 383–404. https://doi.org/10.3390/ai5010019
5. Alelyani, S. (2021). Detection and evaluation of machine learning bias. Applied Sciences (Switzerland), 11(14). https://doi.org/10.3390/app11146271
6. Alexander, C. S. (2022). Text Mining for Bias: A Recommendation Letter Experiment. American Business Law Journal, 59(1), 5–59. https://doi.org/10.1111/ablj.12198
7. Alexander, C. S., Smith, A., & Ivanek, R. (2023). Safer not to know? Shaping liability law and policy to incentivize adoption of predictive AI technologies in the food system. Frontiers in Artificial Intelligence, 6. https://doi.org/10.3389/frai.2023.1298604
8. Alves, G., Bernier, F., Couceiro, M., Makhlouf, K., Palamidessi, C., & Zhioua, S. (2023). Survey on fairness notions and related tensions. EURO Journal on Decision Processes, 11, 100033. https://doi.org/10.1016/J.EJDP.2023.100033
9. Alves, I., Leite, L. A. F., Meirelles, P., Kon, F., & Aguiar, C. S. R. (2024). Practices for Managing Machine Learning Products: A Multivocal Literature Review. IEEE Transactions on Engineering Management, 71, 7425–7455. https://doi.org/10.1109/TEM.2023.3287759
10. Amershi, S., Vorvoreanu, M., Teevan, J., Horvitz, E., Fourney, A., Suh, J., Kikin-Gil, R., Bennett, P. N., Nushi, B., Collisson, P., Iqbal, S., Weld, D., & Inkpen, K. (2019, May 2). Guidelines for Human-AI Interaction. https://doi.org/10.1145/3290605.3300233
11. Ashtana, R., Vikash, V., & Omprakash, O. (2018). Human-centric machine learning: Addressing user experience and ethical considerations. International Journal of Applied Research, 4(12), 65–69. https://doi.org/10.22271/allresearch.2018.v4.i12a.11467
12. Atabek, A., Eralp, E., & Gursoy, M. E. (2023). Trust, Privacy and Security Aspects of Bias and Fairness in Machine Learning. Proceedings - 2023 5th IEEE International Conference

—------—-------------------------------------------------------------------------------------

on Trust, Privacy and Security in Intelligent Systems and Applications, TPS-ISA 2023, 111–121. https://doi.org/10.1109/TPS-ISA58951.2023.00023

13. Azevedo, K., Quaranta, L., Calefato, F., & Kalinowski, M. (2024). A multivocal literature review on the benefits and limitations of Automated Machine Learning tools. https://doi.org/10.48550/arXiv.2401.11366
14. Bach, T. A., Khan, A., Hallock, H., Beltrão, G., & Sousa, S. (2022). A Systematic Literature Review of User Trust in AI-Enabled Systems: An HCI Perspective. International Journal of Human–Computer Interaction, 40(5), 1251–1266. https://doi.org/10.1080/10447318.2022.2138826
15. Balcombe, L., & De Leo, D. (2022). Human-Computer Interaction in Digital Mental Health. Informatics, 9(1), 14. https://doi.org/10.3390/informatics9010014
16. Bansal, G., Nushi, B., Kamar, E., Horvitz, E., Weld, D. S., & Lasecki, W. S. (2019). Updates in Human-AI Teams: Understanding and Addressing the Performance/Compatibility Tradeoff. Proceedings of the AAAI Conference on Artificial Intelligence, 33(01), 2429–2437. https://doi.org/10.1609/aaai.v33i01.33012429
17. Barbierato, E., Pozzi, A., & Tessera, D. (2023). Controlling Bias Between Categorical Attributes in Datasets: A Two-Step Optimization Algorithm Leveraging Structural Equation Modeling. IEEE Access, 11, 115493–115510. https://doi.org/10.1109/ACCESS.2023.3325235
18. Barta, G., & Görcsi, G. (2021). Risk management considerations for artificial intelligence business applications. International Journal of Economics and Business Research, 21(1), 87–106. https://doi.org/10.1504/IJEBR.2021.112012
19. Barocas, S., Hardt, M., & Narayanan, A. (2019). Fairness and Machine Learning. fairmlbook.org
20. Bathen, L. A. D., & Jadav, D. (2022). Trustless AutoML for the Age of Internet of Things. IEEE International Conference on Blockchain and Cryptocurrency, ICBC 2022. https://doi.org/10.1109/ICBC54727.2022.9805535
21. Baudel, T., Verbockhaven, M., Cousergue, V., Roy, G., & Laarach, R. (2021). ObjectivAIze: Measuring Performance and Biases in Augmented Business Decision Systems. In Lecture Notes in Computer Science (including subseries Lecture Notes in Artificial Intelligence and Lecture Notes in Bioinformatics): Vol. 12934 LNCS. https://doi.org/10.1007/978-3-030-85613-7_22
22. Baumann, J., Castelnovo, A., Crupi, R., Inverardi, N., & Regoli, D. (2023). Bias on Demand: A Modelling Framework That Generates Synthetic Data With Bias. ACM International Conference Proceeding Series, 1002–1013. https://doi.org/10.1145/3593013.3594058
23. Bellamy, R. K. E., Mojsilovic, A., Nagar, S., Ramamurthy, K. N., Richards, J., Saha, D., Sattigeri, P., Singh, M., Varshney, K. R., Zhang, Y., Martino, J., & Mehta, S. (2019). AI Fairness 360: An extensible toolkit for detecting and mitigating algorithmic bias. IBM

Journal of Research and Development, 63(4–5). https://doi.org/10.1147/JRD.2019.2942287

24. Bender, J., Trat, M., & Ovtcharova, J. (2022). Benchmarking AutoML-Supported Lead Time Prediction. Procedia Computer Science, 200, 482–494. https://doi.org/10.1016/j.procs.2022.01.246
25. Bied, G., Gaillac, C., Hoffmann, M., Caillou, P., Crépon, B., Nathan, S., & Sebag, M. (2023). Fairness in job recommendations: estimating, explaining, and reducing gender gaps. CEUR Workshop Proceedings, 3523.
26. Biswas, A., & Mukherjee, S. (2019). Fairness through the lens of proportional equality. Proceedings of the International Joint Conference on Autonomous Agents and Multiagent Systems, AAMAS, 3, 1832–1834.
27. Biswas, S., & Rajan, H. (2020). Do the machine learning models on a crowd sourced platform exhibit bias? an empirical study on model fairness. Proceedings of the 28th ACM Joint Meeting on European Software Engineering Conference and Symposium on the Foundations of Software Engineering, null, null. https://doi.org/10.1145/3368089.3409704
28. Blohm, M., Hanussek, M., & Kintz, M. (2021). Leveraging automated machine learning for text classification: Evaluation of AutoML tools and comparison with human performance. Proceedings of the 13th International Conference on Agents and Artificial Intelligence, 1131–1136. https://doi.org/10.5220/0010331411311136
29. Booth, B. M., Hickman, L., Subburaj, S. K., Tay, L., Woo, S. E., & D'Mello, S. K. (2021). Bias and Fairness in Multimodal Machine Learning: A Case Study of Automated Video Interviews. ICMI 2021 - Proceedings of the 2021 International Conference on Multimodal Interaction, 268–277. https://doi.org/10.1145/3462244.3479897
30. Breck, E., Polyzotis, N., Roy, S., Whang, S. E., & Zinkevich, M. (2019). Data validation for machine learning. Proceedings.Mlsys.OrgN Polyzotis, M Zinkevich, S Roy, E Breck, S WhangProceedings of Machine Learning and Systems, 2019•proceedings.Mlsys.Org. https://proceedings.mlsys.org/paper_files/paper/2019/hash/928f1160e52192e3e0017fb63ab65391-Abstract.html
31. Buçinca, Z., Malaya, M., & Gajos, K. (2021). To Trust or to Think: Cognitive Forcing Functions Can Reduce Overreliance on AI in AI-assisted Decision-making. https://doi.org/10.48550/arxiv.2102.09692
32. Cai, C. J., Reif, E., Hegde, N., Hipp, J., Kim, B., Smilkov, D., Wattenberg, M., Viegas, F., Corrado, G. S., Stumpe, M. C., & Terry, M. (2019). Human-centered tools for coping with imperfect algorithms during medical decision-making. Conference on Human Factors in Computing Systems - Proceedings, 14. https://doi.org/10.1145/3290605.3300234
33. Çakır, A. M. (2024). Human AI Collaboration in Decision-Making Auto Systems With AI. Human Computer Interaction, 8(1), 123. https://doi.org/10.62802/b4z5p105

—------—-------------------------------------------------------------------------------------

34. Caldera, A., Hettiarachchi, S., Bandara, H. M. R. M., Abeywickrama, Y. S., Fernando, B. D. R., & Wijesuriya, I. M. (2023). Interview Bot Using Natural Language Processing and Machine Learning. ICAC 2023 - 5th International Conference on Advancements in Computing: Technological Innovation for a Sustainable Economy, Proceedings, 161–166. https://doi.org/10.1109/ICAC60630.2023.10417234
35. Caliskan, A., Bryson, J. J., & Narayanan, A. (2016). Semantics derived automatically from language corpora contain human-like biases. Science, 356(6334), 183–186. https://doi.org/10.1126/science.aal4230
36. Canhoto, A. I., & Clear, F. (2020). Artificial intelligence and machine learning as business tools: A framework for diagnosing value destruction potential. Business Horizons, 63(2), 183–193. https://doi.org/10.1016/j.bushor.2019.11.003
37. Canetti, R., et al. (2018). From Soft Classifiers to Hard Decisions: How fair can we be? FAT. https://doi.org/10.48550/arXiv.1810.02003
38. Carroll, J. M. (1997). HUMAN-COMPUTER INTERACTION: Psychology as a Science of Design. Annual Review of Psychology, 48(1), 61–83. https://doi.org/10.1146/annurev.psych.48.1.61
39. Castelnovo, A., Crupi, R., Greco, G., Regoli, D., Penco, I. G., & Cosentini, A. C. (2022). A clarification of the nuances in the fairness metrics landscape. Scientific Reports 2022 12:1, 12(1), 1–21. https://doi.org/10.1038/s41598-022-07939-1
40. Castelnovo, A., Sanpaolo, I., Crupi, R., Greco, G., Regoli, D., Cosentini, A. C., & Penco, I. G. (2021). The Zoo of Fairness Metrics in Machine Learning. https://doi.org/10.21203/rs.3.rs-1162350/v1
41. Chakraborty, J., Peng, K., & Menzies, T. (2020). Making Fair ML Software using Trustworthy Explanation. Proceedings - 2020 35th IEEE/ACM International Conference on Automated Software Engineering, ASE 2020, 1229–1233. https://doi.org/10.1145/3324884.3418932
42. Chen, J. Y. C. (2022). Transparent Human–Agent Communications. International Journal of Human–Computer Interaction, 38(18–20), 1737–1738. https://doi.org/10.1080/10447318.2022.2120173
43. Chen, J., Kallus, N., Mao, X., Svacha, G., & Udell, M. (2019). Fairness under unawareness: Assessing disparity when protected class is unobserved. FAT* 2019 - Proceedings of the 2019 Conference on Fairness, Accountability, and Transparency, 339–348. https://doi.org/10.1145/3287560.3287594
44. Chen, R. (2024). A Study Applying Rogers' Innovation Diffusion Theory on the Adoption Process of New Teaching Methods in Secondary Education. Research and Advances in Education, 3(2), 6–10. https://doi.org/10.56397/rae.2024.02.02

45. Chen, Y.-W., Song, Q., & Hu, X. (2019). Techniques for Automated Machine Learning. ACM SIGKDD Explorations Newsletter, 22(2), 35–50. https://doi.org/10.48550/arxiv.1907.08908
46. Chen, Z., Zhang, J. M., Sarro, F., & Harman, M. (2022). A Comprehensive Empirical Study of Bias Mitigation Methods for Machine Learning Classifiers. https://doi.org/10.48550/arxiv.2207.03277
47. Cheng, M., De-Arteaga, M., Mackey, L., & Kalai, A. T. (2023). Social norm bias: residual harms of fairness-aware algorithms. Data Mining and Knowledge Discovery, 37(5), 1858–1884. https://doi.org/10.1007/s10618-022-00910-8
48. Cho, J., Hwang, G., & Suh, C. (2020). A fair classifier using kernel density estimation. Advances in Neural Information Processing Systems, 2020-Decem. https://dl.acm.org/doi/10.5555/3495724.3496989
49. Choi, W., Choi, T., & Heo, S. (2023). A comparative study of automated machine learning platforms for exercise anthropometry-based typology analysis: Performance evaluation of AWS SageMaker, GCP VertexAI, and MS Azure. Bioengineering (Basel), 10(8). https://doi.org/10.3390/bioengineering10080891
50. Chouldechova, A. (2017). Fair prediction with disparate impact: A study of bias in recidivism prediction instruments. Big Data, 5(2), 153-163. arXiv:1703.00056
51. Chromik, M., & Butz, A. (2021). Human-XAI Interaction: A Review and Design Principles for Explanation User Interfaces (pp. 619–640). springer. https://doi.org/10.1007/978-3-030-85616-8_36
52. Cohen, M. C., Pan, A., Salehi, P., Ba, Y., Blasch, E., Sung, J., Chiou, E. K., Bhatti, S., Mancenido, M. V., & Kim, N. (2025). PADTHAI-MM: Principles-based approach for designing trustworthy, human-centered AI using the MAST methodology. AI Magazine, 46(1). https://doi.org/10.1002/aaai.70000
53. Crisan, A., & Fiore-Gartland, B. (2021). Fits and Starts: Enterprise Use of AutoML and the Role of Humans in the Loop. 1–15. https://doi.org/10.1145/3411764.3445775
54. Cruz, A. F., Saleiro, P., Belém, C., Soares, C., & Bizarro, P. (2020). A Bandit-Based Algorithm for Fairness-Aware Hyperparameter Optimization. https://doi.org/10.48550/arxiv.2010.03665
55. Cunningham, P., & Delany, S. J. (2021). Underestimation Bias and Underfitting in Machine Learning. Lecture Notes in Computer Science (Including Subseries Lecture Notes in Artificial Intelligence and Lecture Notes in Bioinformatics), 12641 LNAI, 20–31. https://doi.org/10.1007/978-3-030-73959-1_2/FIGURES/7
56. Datta, A., Fredrikson, M., Ko, G., Mardziel, P., & Sen, S. (2017). Use Privacy in Data-Driven Systems Theory and Experiments with Machine Learnt Programs. https://doi.org/10.1145/3133956.3134097

---

57. Delecraz, S., Eltarr, L., Becuwe, M., Bouxin, H., Boutin, N., & Oullier, O. (2022). Making Recruitment More Inclusive: Unfairness Monitoring With A Job Matching Machine-Learning Algorithm. Proceedings - International Workshop on Equitable Data and Technology, FairWare 2022, 34–41. https://doi.org/10.1145/3524491.3527309
58. Drage, E., & Mackereth, K. (2022). Does AI Debias Recruitment? Race, Gender, and AI's "Eradication of Difference." Philosophy and Technology, 35(4), 1–25. https://doi.org/10.1007/S13347-022-00543-1/METRICS
59. Drozdal, J., Weisz, J., Wang, D., Dass, G., Yao, B., Zhao, C., Muller, M., Ju, L., & Su, H. (2020). Trust in AutoML. International Conference on Intelligent User Interfaces, Proceedings IUI, 20, 297–307. https://doi.org/10.1145/3377325.3377501
60. Doris Xin. (2022). *AutoML: the Promise vs. Reality According to Practitioners*. InfoQ.
61. D'Souza, J., Kadam, V., Shinde, P., & Saxena, K. (2023). The Quest for Fairness: A Comparative Study of Accuracy in AI Hiring Systems. 2023 3rd Asian Conference on Innovation in Technology, ASIANCON 2023. https://doi.org/10.1109/ASIANCON58793.2023.10269895
62. Dutta, S., Venkatesh, P., Mardziel, P., Datta, A., & Grover, P. (2021). Fairness under Feature Exemptions: Counterfactual and Observational Measures. IEEE Transactions on Information Theory, 67(10), 6675–6710. https://doi.org/10.1109/TIT.2021.3103206
63. Ehsan, U., Muller, M., Liao, Q. V., Riedl, M. O., & Weisz, J. D. (2021). Expanding Explainability: Towards Social Transparency in AI systems. 1–19. https://doi.org/10.1145/3411764.3445188
64. El Ali, A., Venkatraj, K. P., Morosoli, S., Cesar, P., Naudts, L., & Helberger, N. (2024). Transparent AI Disclosure Obligations: Who, What, When, Where, Why, How. 43, 1–11. https://doi.org/10.1145/3613905.3650750
65. Eldeeb, H., Maher, M., Elshawi, R., & Sakr, S. (2024). AutoMLBench: A comprehensive experimental evaluation of automated machine learning frameworks. Expert Syst. Appl., 243(122877), 122877. https://doi.org/10.1016/j.eswa.2023.122877
66. Elisa, B., Andrea, B., Elena, M., Claudia, C., & Simone, S. (2023). Toward a cost-effective fairness-aware ML lifecycle. CEUR Workshop Proceedings, 3486, 312–317.
67. Escalante, H. J., Guyon, I., Escalera, S., Jacques, J., Madadi, M., Baro, X., Ayache, S., Viegas, E., Gucluturk, Y., Guclu, U., Van Gerven, M. A. J., & Van Lier, R. (2017). Design of an explainable machine learning challenge for video interviews. Proceedings of the International Joint Conference on Neural Networks, 2017-May, 3688–3695. https://doi.org/10.1109/IJCNN.2017.7966320
68. Esmaeili, S. A., Duppala, S., Dickerson, J. P., & Brubach, B. (2022). Fair Labeled Clustering. Proceedings of the ACM SIGKDD International Conference on Knowledge Discovery and Data Mining, 327–335. https://doi.org/10.1145/3534678.3539451

69. Evans, B. P., Xue, B., & Zhang, M. (2020). Improving generalisation of AutoML systems with dynamic fitness evaluations. GECCO 2020 - Proceedings of the 2020 Genetic and Evolutionary Computation Conference, 324–332. https://doi.org/10.1145/3377930.3389805
70. Ezzeldin, Y. H., Yan, S., He, C., Ferrara, E., & Avestimehr, S. (2023). FairFed: Enabling Group Fairness in Federated Learning. Proceedings of the 37th AAAI Conference on Artificial Intelligence, AAAI 2023, 37, 7494–7502. https://doi.org/10.48550/arXiv.2110.00857
71. Fabris Advisor, A., Antonio Susto Co-advisor, G., Silvello Coordinator, G., & Neviani, A. (2023). Algorithmic Fairness Datasets: Curation, Selection, and Applications. https://www.research.unipd.it/handle/11577/3478853
72. Fabris, A., Baranowska, N., Dennis, M. J., Hacker, P., Saldivar, J., Fabra, P., Frederik, S., Borgesius, Z., Biega, A. J., Graus, D., & Borgesius, F. Z. (2023). Fairness and Bias in Algorithmic Hiring: a Multidisciplinary Survey. 1. https://doi.org/10.1145/3696457
73. Fabris, A., Dennis, M. J., Saldivar, J., Hacker, P., Baranowska, N., Graus, D., Zuiderveen Borgesius, F., & Biega, A. J. (2025). Fairness and Bias in Algorithmic Hiring: A Multidisciplinary Survey. ACM Transactions on Intelligent Systems and Technology, 16(1), 1–54. https://doi.org/10.1145/3696457
74. Fabris, A., Messina, S., Silvello, G., & Susto, G. A. (2022). Algorithmic fairness datasets: the story so far. Data Mining and Knowledge Discovery, 36, 2074–2152. https://doi.org/10.1007/s10618-022-00854-z
75. Fallman, D. (2007). Why Research-Oriented Design Isn't Design-Oriented Research: On the Tensions Between Design and Research in an Implicit Design Discipline. Knowledge, Technology & Policy, 20(3), 193–200. https://doi.org/10.1007/s12130-007-9022-8
76. Faudzi, M. A., Sharudin, S. A., Cob, Z. C., Omar, R., & Ghazali, M. (2022). The Evaluation of Cognitive Load Significance for Mobile Learning Application via User Interface Design Violations. 392–397. https://doi.org/10.1109/icoco56118.2022.10031943
77. Feng, K. J. K., & Mcdonald, D. W. (2023). Addressing UX Practitioners' Challenges in Designing ML Applications: an Interactive Machine Learning Approach. 337–352. https://doi.org/10.1145/3581641.3584064
78. Ferreira, L., Pilastri, A., Martins, C., Pires, P. M., & Cortez, P. (2021). A Comparison of AutoML Tools for Machine Learning, Deep Learning and XGBoost. https://doi.org/10.1109/IJCNN52387.2021.9534091
79. Ferreira, L., Pilastri, A., Martins, C., Santos, P., & Cortez, P. (2020). An automated and distributed machine learning framework for telecommunications risk management. ICAART 2020 - Proceedings of the 12th International Conference on Agents and Artificial Intelligence, 2, 99–107. https://doi.org/10.5220/0008952800990107

—-------—-----------------------------------------------------------------------------------

80. Feurer, M., Klein, A., Eggensperger, K., Springenberg, J., Blum, M., & Hutter, F. (2015). Efficient and Robust Automated Machine Learning. Advances in Neural Information Processing Systems, 28. https://dl.acm.org/doi/10.5555/2969442.2969547
81. Frank Hutter, R. C. R. B. M. B. I. G. B. K. and H. L. (2014). *AutoML workshop @ ICML'14*. Https://Sites.Google.Com/Site/Automlwsicml14/.
82. Foody, G. M. (2023). Challenges in the real world use of classification accuracy metrics: From recall and precision to the Matthews correlation coefficient. PLoS One, 18(10), e0291908. https://doi.org/10.1371/journal.pone.0291908
83. Fox, S., & Rey, V. F. (2024). A Cognitive Load Theory (CLT) Analysis of Machine Learning Explainability, Transparency, Interpretability, and Shared Interpretability. Machine Learning and Knowledge Extraction, 6(3), 1494–1509. https://doi.org/10.3390/make6030071
84. Fu, R., Huang, Y., & Singh, P. V. (2020). Artificial Intelligence and Algorithmic Bias: Source, Detection, Mitigation, and Implications. Pushing the Boundaries: Frontiers in Impactful OR/OM Research, 39–63. https://doi.org/10.1287/EDUC.2020.0215
85. Fuller, J. B., Raman, M., Sage-Gavin, E., Hines, K., Thomas, J., Papoutsis, S., Flynn, S., King, D., Robertson, T., Yiannakis, C., Malinska, J., Ganchinho, M., Bonanno, C., Saverino, K., Ryan, M., Maruca, R., & Hintermann, F. (n.d.). How leaders can improve hiring practices to uncover missed talent pools, close skills gaps, and improve diversity.
86. Gade, K., Geyik, S. C., Kenthapadi, K., Mithal, V., & Taly, A. (2019). Explainable AI in industry. Proceedings of the ACM SIGKDD International Conference on Knowledge Discovery and Data Mining, 3203–3204. https://doi.org/10.1145/3292500.3332281
87. Gade, K., Geyik, S., Kenthapadi, K., Mithal, V., & Taly, A. (2020). Explainable AI in Industry: Practical Challenges and Lessons Learned. The Web Conference 2020 - Companion of the World Wide Web Conference, WWW 2020, 303–304. https://doi.org/10.1145/3366424.3383110
88. Geden, M., & Andrews, J. (2021). Fair and Interpretable Algorithmic Hiring using Evolutionary Many Objective Optimization. Proceedings of the AAAI Conference on Artificial Intelligence, 35(17), 14795–14803. https://doi.org/10.1609/AAAI.V35I17.17737
89. Ghai, B. (2023). Towards Fair and Explainable AI using a Human-Centered AI Approach. Cornell university. https://doi.org/10.48550/arxiv.2306.07427
90. Ghadekar, P., Kabra, A., Gangwal, K., Kinage, A., Agarwal, K., & Chaudhari, K. (2023). A Semantic Approach for Automated Hiring using Artificial Intelligence & Computer Vision. 2023 IEEE 8th International Conference for Convergence in Technology, I2CT 2023. https://doi.org/10.1109/I2CT57861.2023.10126463
91. Gilbert, J. E. (2021). Equitable AI. Conference on Human Factors in Computing Systems - Proceedings. https://doi.org/10.1145/3411763.3457780

---

92. Gijsbers, P., LeDell, E., Thomas, J., Poirier, S., Bischl, B., & Vanschoren, J. (2019). An Open Source AutoML Benchmark. ArXiv, abs/1907.00909, null. https://doi.org/10.48550/arXiv.1907.00909
93. Goretzko, D., & Israel, L. S. F. (2022). Pitfalls of Machine Learning-Based Personnel Selection: Fairness, Transparency, and Data Quality. Journal of Personnel Psychology, 21(1), 37–47. https://doi.org/10.1027/1866-5888/a000287
94. Gössl, S. L. (2023). Recommender Systems and Discrimination. In International Library of Ethics, Law and Technology (Vol. 40). https://doi.org/10.1007/978-3-031-34804-4_2
95. Greif, E., & Grosz, T. (2023). To see, or not to see: Online job advertisement and EU non-discrimination law. European Labour Law Journal, 14(3), 376–390. https://doi.org/10.1177/20319525231172089
96. Griep, K., Stone, L., Edwards, S., Hill, M., Cui, J., Liu, C., Jia, Q., & Song, Y. (2023). Ensuring Ethical, Transparent, and Auditable Use of Education Data and Algorithms on AutoML. ACM International Conference Proceeding Series, 66–72. https://doi.org/10.1145/3639479.3639492
97. Grove, H., Clouse, M., & Xu, T. (2020). New risks related to emerging technologies and reputation for corporate governance. Journal of Governance and Regulation, 9(2), 64–74. https://doi.org/10.22495/jgrv9i2art4
98. Güngör, H. (2020). Creating Value with Artificial Intelligence: A Multi-stakeholder Perspective. Journal of Creating Value, 6(1), 72–85. https://doi.org/10.1177/2394964320921071
99. Guo, J., Chen, Z., Ji, Y., Zhang, L., Luo, D., Li, Z., & Shen, Y. (2024). UniAutoML: A Human-Centered Framework for Unified Discriminative and Generative AutoML with Large Language Models. https://doi.org/10.48550/arxiv.2410.12841
100. Guzman, A. L., & Lewis, S. C. (2019). Artificial intelligence and communication: A Human–Machine Communication research agenda. New Media & Society, 22(1), 70–86. https://doi.org/10.1177/1461444819858691
101. Haluška, R., Brabec, J., & Komárek, T. (2023). Benchmark of Data Preprocessing Methods for Imbalanced Classification. Proceedings - 2022 IEEE International Conference on Big Data, Big Data 2022, 2970–2979. https://doi.org/10.1109/BigData55660.2022.10021118
102. Halvari, T., Nurminen, J. K., & Mikkonen, T. (2021). Robustness of AutoML for Time Series Forecasting in Sensor Networks. 2021 IFIP Networking Conference (IFIP Networking), 1–3. https://doi.org/10.23919/IFIPNetworking52078.2021.9472199
103. Hanussek, M., Blohm, M., & Kintz, M. (2020). Can AutoML outperform humans? An evaluation on popular OpenML datasets using AutoML Benchmark. https://doi.org/10.48550/arXiv.2009.01564
104. Han, X., et al. (2023). Retiring ΔDP: New Distribution-Level Metrics for Demographic Parity. arXiv. https://doi.org/10.48550/arXiv.2301.13443

—------—------------------------------------------------------------------------------------------

105. Hardt, M., Price, E., & Srebro, N. (2016). Equality of Opportunity in Supervised Learning. NeurIPS. Link https://doi.org/10.48550/arXiv.1610.02413
106. Harris, C. G. (2024). Combining Human-in-the-Loop Systems and AI Fairness Toolkits to Reduce Age Bias in AI Job Hiring Algorithms. Proceedings - 2024 IEEE International Conference on Big Data and Smart Computing, BigComp 2024, 60–66. https://doi.org/10.1109/BIGCOMP60711.2024.00019
107. Heidrich, L., Slany, E., Scheele, S., & Schmid, U. (2023). FairCaipi: A Combination of Explanatory Interactive and Fair Machine Learning for Human and Machine Bias Reduction. Machine Learning and Knowledge Extraction, 5(4), 1519–1538. https://doi.org/10.3390/make5040076
108. Hegarty, S. E., et al. (2025). Assessing Algorithm Fairness Requires Adjustment for Risk Distribution Differences: Re-Considering the Equal Opportunity Criterion. https://doi.org/10.1101/2025.01.31.25321489
109. He, Y., Burghardt, K., & Lerman, K. (2020). A geometric solution to fair representations. AIES 2020 - Proceedings of the AAAI/ACM Conference on AI, Ethics, and Society, 7(20), 279–285. https://doi.org/10.1145/3375627.3375864
110. Hiles, A. (2015). Enterprise Risk Management. In The Definitive Handbook of Business Continuity Management (pp. 1–21). John Wiley & Sons, Inc. https://doi.org/10.1002/9781119205883.ch1
111. Hinduja, S., et al. (2024). Bias-Based Cyberbullying Among Early Adolescents: Associations With Cognitive and Affective Empathy ScienceDirect - https://doi.org/10.1177/02724316221088757
112. Hilliard, A., Kazim, E., Bitsakis, T., & Leutner, F. (2022). Measuring Personality through Images: Validating a Forced-Choice Image-Based Assessment of the Big Five Personality Traits. Journal of Intelligence, 10(1). https://doi.org/10.3390/jintelligence10010012
113. Hinnefeld, J. H., Cooman, P., Mammo, N., & Deese, R. (2018). Evaluating Fairness Metrics in the Presence of Dataset Bias. https://arxiv.org/abs/1809.09245v1
114. Holzinger, A., Plass, M., Holzinger, K., Crişan, G. C., Pintea, C. M., & Palade, V. (2016). Towards interactive machine learning (iML): Applying ant colony algorithms to solve the traveling salesman problem with the human-in-the-loop approach. Lecture Notes in Computer Science (Including Subseries Lecture Notes in Artificial Intelligence and Lecture Notes in Bioinformatics), 9817 LNCS, 81–95. https://doi.org/10.1007/978-3-319-45507-5_6/FIGURES/4
115. Holzinger, A., Zatloukal, K., & Müller, H. (2025). Is human oversight to AI systems still possible? New Biotechnology, 85, 59–62. https://doi.org/10.1016/j.nbt.2024.12.003
116. Hong, S. R., Castelo, S., D’Orazio, V., Benthune, C., Santos, A., Langevin Scott and Jonker, D., Bertini, E., & Freire, J. (2020). Towards evaluating exploratory model building process with AutoML systems. https://doi.org/10.48550/arXiv.2009.00449

———————————————————————————————

117. Hoque, M., Shin, S., & Elmqvist, N. (2024). Visualization for Human-Centered AI Tools. https://doi.org/10.48550/arxiv.2404.02147
118. Human-Computer Interaction. (2003). crc. https://doi.org/10.1201/9780367804787
119. Hullermeier, E., Mohr, F., Tornede, A., & Wever, M. (2021). Automated Machine Learning, Bounded Rationality, and Rational Metareasoning. ArXiv, abs/2109.04744, null. https://www.semanticscholar.org/paper/b30cd10ccb786966289cd44baa7aeeff2ef5f14f
120. Hutchinson, B., & Mitchell, M. (2019). 50 Years of Test (Un)fairness: Lessons for machine learning. FAT* 2019 - Proceedings of the 2019 Conference on Fairness, Accountability, and Transparency, 49–58. https://doi.org/10.1145/3287560.3287600
121. Jain, B., Huber, M., Elmasri, R., & Fegaras, L. (2020). Using Bias Parity Score to Find Feature-Rich Models with Least Relative Bias. Technologies, 8(4). https://doi.org/10.3390/technologies8040068
122. Jayaratne, M., & Jayatilleke, B. (2020). Predicting Personality Using Answers to Open-Ended Interview Questions. IEEE Access, 8, 115345–115355. https://doi.org/10.1109/ACCESS.2020.3004002
123. Jiang, X., Dai, Y., & Wu, Y. (2023). Fair Selection through Kernel Density Estimation. Proceedings of the International Joint Conference on Neural Networks, 2023-June. https://doi.org/10.1109/IJCNN54540.2023.10191616
124. Kallus, N., Mao, X., & Zhou, A. (2021). Assessing Algorithmic Fairness with Unobserved Protected Class Using Data Combination. Https://Doi.Org/10.1287/Mnsc.2020.3850, 68(3), 1959–1981. https://doi.org/10.1287/MNSC.2020.3850
125. Kamiran, F., & Calders, T. (2012). Data preprocessing techniques for classification without discrimination. Knowledge and Information Systems, 33(1), 1–33. https://doi.org/10.1007/S10115-011-0463-8/METRICS
126. Karmaker (“Santu”), S. K., Hassan, M. M., Zhai, C., Smith, M. J., Xu, L., & Veeramachaneni, K. (2021). AutoML to Date and Beyond: Challenges and Opportunities. ACM Computing Surveys, 54(8), 1–36. https://doi.org/10.1145/3470918
127. Kheder, H. A. (2023). HUMAN-COMPUTER INTERACTION: ENHANCING USER EXPERIENCE IN INTERACTIVE SYSTEMS. Kufa Journal of Engineering, 14(4), 23–41. https://doi.org/10.30572/2018/kje/140403
128. Khuat, T., Kedziora, D., & Gabrys, B. (2022). The Roles and Modes of Human Interactions with Automated Machine Learning Systems. cornell university. https://doi.org/10.48550/arxiv.2205.04139
129. Kikuchi, H., Yamaguchi, T., Hamada, K., Yamaoka, Y., Oguri, H., & Sakuma, J. (2018). Study on record linkage of anonymizied data. IEICE Transactions on Fundamentals of Electronics, Communications and Computer Sciences, E101A(1), 19–28. https://doi.org/10.1587/transfun.E101.A.19

—------—------------------------------------------------------------------------------------

130. Kilickaya, M., & Vanschoren, J. (2023). What can AutoML do for continual learning? https://doi.org/10.48550/arXiv.2311.11963
131. Kinger, S., Kinger, D., Thakkar, S., & Bhake, D. (2024). Towards smarter hiring: resume parsing and ranking with YOLOv5 and DistilBERT. Multimedia Tools and Applications. https://doi.org/10.1007/s11042-024-18778-9
132. Kosch, T., Zagermann, J., Schmidt, A., Reiterer, H., Karolus, J., & Woźniak, P. W. (2023). A Survey on Measuring Cognitive Workload in Human-Computer Interaction. ACM Computing Surveys, 55(13s), 1–39. https://doi.org/10.1145/3582272
133. Kshetri, N. (2020). Evolving uses of artificial intelligence in human resource management in emerging economies in the global South: some preliminary evidence. Management Research Review, 44(7), 970–990. https://doi.org/10.1108/MRR-03-2020-0168
134. Lakkshmanan, A., Tyagi, A. K., Sree, P. H., & Sharma, A. K. (2024). Engineering Applications of Artificial Intelligence (pp. 166–179). igi global. https://doi.org/10.4018/979-8-3693-5261-8.ch010
135. Le Quy, T., Roy, A., Iosifidis, V., Zhang, W., & Ntoutsi, E. (2022). A survey on datasets for fairness-aware machine learning. Wiley Interdisciplinary Reviews: Data Mining and Knowledge Discovery, 12(3), e1452. https://doi.org/10.1002/WIDM.1452
136. Leavy, S., Meaney, G., Wade, K., & Greene, D. (2020). Mitigating gender bias in machine learning data sets. In Communications in Computer and Information Science: Vol. 1245 CCIS. https://doi.org/10.1007/978-3-030-52485-2_2
137. Li D. J. L., & Macke, S. (2020). A Human-in-the-loop Perspective on AutoML: Milestones and the Road Ahead. IEEE Data Engineering Bulletin.
138. Li, M., Zhang, X., Thrampoulidis, C., Chen, J., & Oymak, S. (2021). AutoBalance: Optimized Loss Functions for Imbalanced Data. Advances in Neural Information Processing Systems, 34, 3163–3177.
139. Li, Y., Lasecki, W. S., Hilliges, O., & Kumar, R. (2020, April 25). Artificial Intelligence for HCI: A Modern Approach. https://doi.org/10.1145/3334480.3375147
140. Li, Y., Wu, B., Huang, Y., & Luan, S. (2024). Developing trustworthy artificial intelligence: insights from research on interpersonal, human-automation, and human-AI trust. Frontiers in Psychology, 15. https://doi.org/10.3389/fpsyg.2024.1382693
141. Liao, Y., & Naghizadeh, P. (2022). The Impacts of Labeling Biases on Fairness Criteria. arXiv. https://doi.org/10.48550/arXiv.2206.00137
142. Lindauer, M., Karl, F., Klier, A., Moosbauer, J., Tornede, A., Mueller, A., Hutter, F., Feurer, M., & Bischl, B. (2024). Position: A Call to Action for a Human-Centered AutoML Paradigm. https://doi.org/10.48550/arxiv.2406.03348
143. Liu, B. (2021). In AI We Trust? Effects of Agency Locus and Transparency on Uncertainty Reduction in Human–AI Interaction. Journal of Computer-Mediated Communication, 26(6), 384–402. https://doi.org/10.1093/jcmc/zmab013

——------——---------------------------------------------------------------------------------------

144. Lohaus, M., Perrot, M., & Luxburg, U. V. (2020). Too Relaxed to Be Fair. https://www.semanticscholar.org/paper/dce01c01fa0c46c12ad59672b2d717b1c2bf0197
145. Lukaszewski, K. M., & Stone, D. L. (2024). Will the use of AI in human resources create a digital Frankenstein? Organizational Dynamics, 53(1). https://doi.org/10.1016/j.orgdyn.2024.101033
146. Lupo, G. (2023). Risky Artificial Intelligence: The Role of Incidents in the Path to AI Regulation. Law, Technology and Humans, 5(1), 133–152. https://doi.org/10.5204/lthj.2682
147. Luo, X., Tong, S., Fang, Z., & Qu, Z. (2019). Frontiers: Machines vs. humans: The impact of artificial intelligence chatbot disclosure on customer purchases. Marketing Science, 38(6), 937–947. https://doi.org/10.1287/mksc.2019.1192
148. Lütz, F. (2022). Gender equality and artificial intelligence in Europe. Addressing direct and indirect impacts of algorithms on gender-based discrimination. ERA Forum, 23(1), 33–52. https://doi.org/10.1007/s12027-022-00709-6
149. Majidi, F., Openja, M., Khomh, F., & Li, H. (2022). An Empirical Study on the Usage of Automated Machine Learning Tools. 12, 59–70. https://doi.org/10.1109/icsme55016.2022.00014
150. Mamman, H., Basri, S., Balogun, A. O., Imam, A. A., Kumar, G., & Capretz, L. F. (2024). BiasTrap: Runtime Detection of Biased Prediction in Machine Learning Systems. Journal of Advanced Research in Applied Sciences and Engineering Technology, 40(2), 127–139. https://doi.org/10.37934/araset.40.2.127139
151. Mathewson, K. (2019). A Human-Centered Approach to Interactive Machine Learning. https://doi.org/10.48550/arxiv.1905.06289
152. Mckinsey. (2019). The risk-based approach to cybersecurity. https://www.mckinsey.com/capabilities/risk-and-resilience/our-insights/the-risk-based-approach-to-cybersecurity
153. Mehrabi, N., Morstatter, F., Saxena, N., Lerman, K., & Galstyan, A. (2021). A Survey on Bias and Fairness in Machine Learning. ACM Computing Surveys (CSUR), 54(6). https://doi.org/10.1145/3457607
154. Melville, N. P., Xiao, X., & Robert, L. (2022). Putting humans back in the loop: An affordance conceptualization of the 4th industrial revolution. Information Systems Journal, 33(4), 733–757. https://doi.org/10.1111/isj.12422
155. Miller, D. D. (2019). The medical AI insurgency: what physicians must know about data to practice with intelligent machines. Npj Digital Medicine, 2(1). https://doi.org/10.1038/s41746-019-0138-5
156. Miller, R. L. (2015). Rogers' Innovation Diffusion Theory (1962, 1995) (pp. 261–274). igi global. https://doi.org/10.4018/978-1-4666-8156-9.ch016

—------—------------------------------------------------------------------------------

157. Mishra, R., Pati, B., & Satpathy, R. (2023). Human Computer Interaction Applications in Healthcare: An Integrative Review. EAI Endorsed Transactions on Pervasive Health and Technology, 9. https://doi.org/10.4108/eetpht.9.4186
158. Mollas, I., Bassiliades, N., & Tsoumakas, G. (2023). Truthful meta-explanations for local interpretability of machine learning models. Applied Intelligence, 53(22), 26927–26948. https://doi.org/10.1007/s10489-023-04944-3
159. Mueller, F. 'Floyd,' Andres, J., Mehta, Y., Semertzidis, N., Li, X., Benford, S., Marshall, J., & Matjeka, L. (2023). Toward Understanding the Design of Intertwined Human–Computer Integrations. ACM Transactions on Computer-Human Interaction, 30(5), 1–45. https://doi.org/10.1145/3590766
160. Mujtaba, D. F., & Mahapatra, N. R. (2019). Ethical Considerations in AI-Based Recruitment. International Symposium on Technology and Society, Proceedings, 2019-Novem. https://doi.org/10.1109/ISTAS48451.2019.8937920
161. Mujtaba, D. F., & Mahapatra, N. R. (2024). Fairness in AI-Driven Recruitment: Challenges, Metrics, Methods, and Future Directions. https://arxiv.org/abs/2405.19699v2
162. Mukherjee, D., Yurochkin, M., Banerjee, M., & Sun, Y. (2020). Two Simple Ways to Learn Individual Fairness Metrics from Data (pp. 7097–7107). PMLR. https://doi.org/10.48550/arXiv.2006.11439
163. Muthusamy, V., Slominski, A., & Ishakian, V. (2018). Towards enterprise-ready AI deployments minimizing the risk of consuming AI models in business applications. Proceedings - 2018 1st IEEE International Conference on Artificial Intelligence for Industries, AI4I 2018, 108–109. https://doi.org/10.1109/AI4I.2018.8665685
164. Nadeem, A., Marjanovic, O., & Abedin, B. (2021). Gender Bias in AI: Implications for Managerial Practices. In Lecture Notes in Computer Science (including subseries Lecture Notes in Artificial Intelligence and Lecture Notes in Bioinformatics): Vol. 12896 LNCS. https://doi.org/10.1007/978-3-030-85447-8_23
165. Nakao, Y., Strappelli, L., Stumpf, S., Naseer, A., Regoli, D., & Gamba, G. D. (2022). Towards Responsible AI: A Design Space Exploration of Human-Centered Artificial Intelligence User Interfaces to Investigate Fairness. International Journal of Human–Computer Interaction, ahead-of-print(ahead-of-print), 1762–1788. https://doi.org/10.1080/10447318.2022.2067936
166. Narkar, S., Zhang, Y., Liao, Q. V., Wang, D., & Weisz, J. D. (2021). Model LineUpper: Supporting Interactive Model Comparison at Multiple Levels for AutoML. International Conference on Intelligent User Interfaces, Proceedings IUI, 170–174. https://doi.org/10.1145/3397481.3450658
167. Narayanan, S. (2023). Democratize with Care: The need for fairness specific features in user-interface based open source AutoML tools. https://doi.org/10.48550/arxiv.2312.12460

—-------—-------------------------------------------------------------------------------------

168. Nazar, M., Yafi, E., Su'Ud, M. M., & Alam, M. M. (2021). A Systematic Review of Human–Computer Interaction and Explainable Artificial Intelligence in Healthcare With Artificial Intelligence Techniques. IEEE Access, 9, 153316–153348. https://doi.org/10.1109/access.2021.3127881
169. Nguyen, G., Biswas, S., & Rajan, H. (2023). Fix Fairness, Don't Ruin Accuracy: Performance Aware Fairness Repair using AutoML. ESEC/FSE 2023 - Proceedings of the 31st ACM Joint Meeting European Software Engineering Conference and Symposium on the Foundations of Software Engineering, 502–514. https://doi.org/10.1145/3611643.3616257
170. Nightingale, S. J., & Farid, H. (2022). AI-synthesized faces are indistinguishable from real faces and more trustworthy. Proceedings of the National Academy of Sciences of the United States of America, 119(8). https://doi.org/10.1073/PNAS.2120481119
171. Nikolinakos, N. T. (2023). The European Parliament's 2020 Resolution: Proposal for a Regulation on Ethical Principles for the Development, Deployment and Use of Artificial Intelligence, Robotics and Related Technologies. In Law, Governance and Technology Series (Vol. 53). https://doi.org/10.1007/978-3-031-27953-9_6
172. Nirmala, R., & Chitte, N. (2021). Industry 4.0: The Human Resource Perspective. In Fourth Industrial Revolution and Business Dynamics: Issues and Implications. https://doi.org/10.1007/978-981-16-3250-1_14
173. Ntoutsi, E., Fafalios, P., Gadiraju, U., Iosifidis, V., Nejdl, W., Vidal, M. E., Ruggieri, S., Turini, F., Papadopoulos, S., Krasanakis, E., Kompatsiaris, I., Kinder-Kurlanda, K., Wagner, C., Karimi, F., Fernandez, M., Alani, H., Berendt, B., Kruegel, T., Heinze, C., … Staab, S. (2020). Bias in data-driven artificial intelligence systems—An introductory survey. Wiley Interdisciplinary Reviews: Data Mining and Knowledge Discovery, 10(3). https://doi.org/10.1002/WIDM.1356
174. Olson, G. M., & Olson, J. S. (2002). Human-Computer Interaction: Psychological Aspects of the Human Use of Computing. Annual Review of Psychology, 54(1), 491–516. https://doi.org/10.1146/annurev.psych.54.101601.145044
175. Ortega, A., Fierrez, J., Morales, A., Wang, Z., de la Cruz, M., Alonso, C. L., & Ribeiro, T. (2021). Symbolic AI for XAI: Evaluating LFIT inductive programming for explaining biases in machine learning. *Computers, 10*(11), 154. https://doi.org/10.3390/computers10110154
176. Páez, A. (2021). Negligent Algorithmic Discrimination. Law and Contemporary Problems, 84(3), 19–33. https://dx.doi.org/10.2139/ssrn.3765778
177. Pagano, T. P., Araujo, M. M., Loureiro, R. B., Santos, L. L., Oliveira, E. L. S., Cruz, G. O. R., Winkler, I., Nascimento, E. G. S., Peixoto, R. M., Guimarães, G. A. S., Cruz, M. A. S., & Lisboa, F. V. N. (2023). Bias and Unfairness in Machine Learning Models: A Systematic

Review on Datasets, Tools, Fairness Metrics, and Identification and Mitigation Methods. Big Data and Cognitive Computing, 7(1), 15. https://doi.org/10.3390/bdcc7010015

178. Paladino, L. M., Hughes, A., Perera Alexander and Topsakal, O., & Akinci, T. C. (2023). Evaluating the performance of automated machine learning (AutoML) tools for heart disease diagnosis and prediction. AI (Basel), 4(4), 1036–1058.
179. Pang, R., Xi, Z., Ji, S., Luo, X., & Wang, T. (2021). On the Security Risks of AutoML. Proceedings of the 31st USENIX Security Symposium, Security 2022, 3953–3970. https://arxiv.org/abs/2110.06018v1
180. Parameswara, M. A., Mahendra, F. S. P., Mardika, M. R. C., Puspasari, I., & Utama, N. P. (2023). Big Five Personality Prediction Based on Indonesian Tweets and Personality Test. Proceedings of the International Conference on Electrical Engineering and Informatics. https://doi.org/10.1109/ICEEI59426.2023.10346812
181. Paullada, A., Raji, I. D., Bender, E. M., Denton, E., & Hanna, A. (2020). Data and its (dis)contents: A survey of dataset development and use in machine learning research. Patterns, 2(11). https://doi.org/10.1016/j.patter.2021.100336
182. Peisl, T., & Edlmann, R. (2020). Exploring Technology Acceptance and Planned Behaviour by the Adoption of Predictive HR Analytics During Recruitment. In Communications in Computer and Information Science: Vol. 1251 CCIS. https://doi.org/10.1007/978-3-030-56441-4_13
183. Pendyala, V. S., Atrey, N., Aggarwal, T., & Goyal, S. (2022). Enhanced Algorithmic Job Matching based on a Comprehensive Candidate Profile using NLP and Machine Learning. Proceedings - IEEE 8th International Conference on Big Data Computing Service and Applications, BigDataService 2022, 183–184. https://doi.org/10.1109/BigDataService55688.2022.00040
184. Perera, A., Aleti, A., Tantithamthavorn, C., Jiarpakdee, J., Turhan, B., Kuhn, L., & Walker, K. (2022). Search-based fairness testing for regression-based machine learning systems. Empirical Software Engineering, 27(3). https://doi.org/10.1007/s10664-022-10116-7
185. Perrone, V., Donini, M., Zafar, M. B., Schmucker, R., Kenthapadi, K., & Archambeau, C. (2021). Fair Bayesian Optimization. AIES 2021 - Proceedings of the 2021 AAAI/ACM Conference on AI, Ethics, and Society, 854–863. https://doi.org/10.1145/3461702.3462629
186. Pfisterer, F., Coors, S., Thomas, J., & Bischl, B. (2019). Multi-Objective Automatic Machine Learning with AutoxgboostMC. ArXiv, abs/1908.10796, https://doi.org/10.48550/arXiv.1908.10796
187. Pop, E.-L., & Raţiu, A. (2024). Human-Computer Interaction in Artificial Intelligence with Applications in Healthcare: A Review. institut f r ost und s dosteuropaforschung. https://doi.org/10.3233/faia241213

—-------—-------------------------------------------------------------------------------------

188. Pushkarna, M., Kjartansson, O., & Zaldivar, A. (2022). Data Cards: Purposeful and Transparent Dataset Documentation for Responsible AI. 1776–1826. https://doi.org/10.1145/3531146.3533231
189. Rahman, M. M., Pan, S., & Foulds, J. R. (2024). Towards A Unifying Human-Centered AI Fairness Framework. 88–92. https://doi.org/10.1145/3677525.3678645
190. Ramezanzadehmoghadam, M., Chi, H., Jones, E. L., & Chi, Z. (2021). Inherent Discriminability of BERT Towards Racial Minority Associated Data. In Lecture Notes in Computer Science (including subseries Lecture Notes in Artificial Intelligence and Lecture Notes in Bioinformatics): Vol. 12951 LNCS. https://doi.org/10.1007/978-3-030-86970-0_19
191. Raulf, A., Bruder, C., Berro, C., Deligiannaki, F., Theis, S., & Jentzsch, S. (2023). Requirements for Explainability and Acceptance of Artificial Intelligence in Collaborative Work. https://doi.org/10.48550/arxiv.2306.15394
192. Research Methods for Human-Computer Interaction. (2008). cambridge university. https://doi.org/10.1017/cbo9780511814570
193. Retzlaff, C. O., Afshari, M., Mousavi, P., Wayllace, C., Das, S., Holzinger, A., Saranti, A., Yang, T., Angerschmid, A., & Taylor, M. E. (2024). Human-in-the-Loop Reinforcement Learning: A Survey and Position on Requirements, Challenges, and Opportunities. Journal of Artificial Intelligence Research, 79, 359–415. https://doi.org/10.1613/jair.1.15348
194. Roundtree, A. (2025). AI Tool Compliance Reporting: A Heuristic Analysis of Survey Data Using Natural Language Processing. 163. https://doi.org/10.54941/ahfe1006053
195. Rundo, L., Sala, E., Vitabile, S., Gambino, O., & Pirrone, R. (2020). Recent advances of HCI in decision-making tasks for optimized clinical workflows and precision medicine. Journal of Biomedical Informatics, 108, 103479. https://doi.org/10.1016/j.jbi.2020.103479
196. Rus, C., de Rijke, M., & Yates, A. (2023). Counterfactual Representations for Intersectional Fair Ranking in Recruitment. CEUR Workshop Proceedings, 3490.
197. Sadgrove, K. (2016). The Complete Guide to Business Risk Management. The Complete Guide to Business Risk Management. https://doi.org/10.4324/9781315614915
198. Saha, D., Schumann, C., McElfresh, D. C., Dickerson, J. P., Mazurek, M. L., & Tschantz, M. C. (2020). Measuring non-expert comprehension of machine learning fairness metrics. 37th International Conference on Machine Learning, ICML 2020, PartF16814, 8346–8356. https://doi.org/10.48550/arXiv.2001.00089
199. Saleiro, P., Kuester, B., Hinkson, L., London, J., Stevens, A., Anisfeld, A., Rodolfa, K. T., & Ghani, R. (2018). Aequitas: A Bias and Fairness Audit Toolkit. https://doi.org/10.48550/arxiv.1811.05577
200. Saxena, A., Buhukya, S., Sumalatha, I., Dutt, A., Shaaker, A. M., & Asha, V. (2023). Machine Learning and Human Resource Management: A Path to Efficient Workforce

Management. 2023 10th IEEE Uttar Pradesh Section International Conference on Electrical, Electronics and Computer Engineering, UPCON 2023, 1709–1714. https://doi.org/10.1109/UPCON59197.2023.10434761

201. Schlegel, D., Schuler, K., & Westenberger, J. (2023). Failure factors of AI projects: results from expert interviews. International Journal of Information Systems and Project Management, 11(3), 25–40. https://doi.org/10.12821/ijispm110302
202. Shah, V., Lacanlale, J., Kumar, P., Yang, K., & Kumar, A. (2021). Towards Benchmarking Feature Type Inference for AutoML Platforms. Proceedings of the ACM SIGMOD International Conference on Management of Data, 1584–1596. https://doi.org/10.1145/3448016.3457274
203. Shneiderman, B. (2020). Human-Centered Artificial Intelligence: Three Fresh Ideas. AIS Transactions on Human-Computer Interaction, 12(3), 109–124. https://doi.org/10.17705/1thci.00131
204. Siddiqui, F., Khan, R., & Sezer, S. (2021). Bird's-eye view on the Automotive Cybersecurity Landscape Challenges in adopting AI/ML. 2021 6th International Conference on Fog and Mobile Edge Computing, FMEC 2021. https://doi.org/10.1109/FMEC54266.2021.9732568
205. Sikorski, M. (2021). Digital Innovations and Smart Solutions for Society and Economy: Pros and Cons. Foundations of Management, 13(1), 103–116. https://doi.org/10.2478/fman-2021-0008
206. Singh, P., & Vanschoren, J. (n.d.). Automated Imbalanced Learning. https://doi.org/10.48550/arXiv.2211.00376
207. Soleimani, M., Intezari, A., & Pauleen, D. J. (2022). Mitigating cognitive biases in developing ai-assisted recruitment systems: A knowledge-sharing approach. International Journal of Knowledge Management, 18(1). https://doi.org/10.4018/IJKM.290022
208. Souverain, T., et al. (2024). Implementing Fairness in AI Classification: The Role of Explainability. arXiv. https://doi.org/10.48550/arXiv.2407.14766
209. Stone, P., Jessup, S. A., Ganapathy, S., & Harel, A. (2022). Design Thinking Framework for Integration of Transparency Measures in Time-Critical Decision Support. International Journal of Human–Computer Interaction, 38(18–20), 1874–1890. https://doi.org/10.1080/10447318.2022.2068745
210. Sudha, G., Sasipriya, K. K., Sri Janani, S., Nivethitha, D., Saranya, S., & Karthick Thyagesh, G. (2021). Personality Prediction Through CV Analysis using Machine Learning Algorithms for Automated E-Recruitment Process. Proceedings of the 2021 4th International Conference on Computing and Communications Technologies, ICCCT 2021, 617–622. https://doi.org/10.1109/ICCCT53315.2021.9711787

—------—-------------------------------------------------------------------------------------

211. Sun, Y., Song, Q., Gui, X., Ma, F., & Wang, T. (2023). AutoML in The Wild: Obstacles, Workarounds, and Expectations. Proceedings of the 2023 CHI Conference on Human Factors in Computing Systems, null, null. https://doi.org/10.1145/3544548.3581082
212. Sundar, S. S., & Lee, E.-J. (2022). Rethinking Communication in the Era of Artificial Intelligence. Human Communication Research, 48(3), 379–385. https://doi.org/10.1093/hcr/hqac014
213. Suresh, H., & Guttag, J. V. (2021). A Framework for Understanding Unintended Consequences of Machine Learning. ACM FAccT. doi:10.1145/3457607
214. Tellez, N., Serra, J., Ebreso, U., Opara, K., Kumar, Y., Li, J. J., & Morreale, P. (2022). An Assure AI Bot (AAAI bot). 2022 International Symposium on Networks, Computers and Communications, ISNCC 2022. https://doi.org/10.1109/ISNCC55209.2022.9851759
215. Tilmes, N. (2022). Disability, fairness, and algorithmic bias in AI recruitment. Ethics and Information Technology, 24(2). https://doi.org/10.1007/s10676-022-09633-2
216. Trewin, S. (2018). AI Fairness for People with Disabilities: Point of View. https://doi.org/10.48550/arxiv.1811.10670
217. Trewin, S., Muller, M., Manser, E., Lyckowski, N., Gruen, D., Hebert, D., Branham, S., Basson, S., & Treviranus, J. (2019). Considerations for AI fairness for people with disabilities. AI Matters, 5(3), 40–63. https://doi.org/10.1145/3362077.3362086
218. Truong, A., Walters, A., Goodsitt, J., Hines, K., Bruss, C. B., & Farivar, R. (2019). Towards automated machine learning: Evaluation and comparison of AutoML approaches and tools. Proceedings - International Conference on Tools with Artificial Intelligence, ICTAI, 2019-Novem, 1471–1479. https://doi.org/10.1109/ICTAI.2019.00209
219. Turri, V., & Dzombak, R. (2023). Why We Need to Know More: Exploring the State of AI Incident Documentation Practices. 576–583. https://doi.org/10.1145/3600211.3604700
220. Veale, M., Binns, R., & Van Kleek, M. (2018). Some HCI Priorities for GDPR-Compliant Machine Learning. center for open science. https://doi.org/10.31228/osf.io/wm6yk
221. Veale, M., Van Kleek, M., & Binns, R. (2018). Fairness and accountability design needs for algorithmic support in high-stakes public sector decision-making. Conference on Human Factors in Computing Systems - Proceedings, 2018-April. https://doi.org/10.1145/3173574.3174014
222. Veitch, E., & Alsos, O. A. (2021). Human-Centered Explainable Artificial Intelligence for Marine Autonomous Surface Vehicles. Journal of Marine Science and Engineering, 9(11), 1227. https://doi.org/10.3390/jmse9111227
223. Vorm, E. S., & Combs, D. J. Y. (2022). Integrating Transparency, Trust, and Acceptance: The Intelligent Systems Technology Acceptance Model (ISTAM). International Journal of Human–Computer Interaction, 38(18–20), 1828–1845. https://doi.org/10.1080/10447318.2022.2070107

—------—--------------------------------------------------------------------------------

224. Wang, D., Weisz, J. D., Muller, M., Ram, P., Geyer, W., Dugan, C., Tausczik, Y., Samulowitz, H., & Gray, A. (2019). Human-AI Collaboration in Data Science. Proceedings of the ACM on Human-Computer Interaction, 3(CSCW), 24. https://doi.org/10.1145/3359313
225. Weerts, H. J. P., Pfisterer, F., Feurer, M., Eggensperger, K., Bergman, E., Awad, N. H., Vanschoren, J., Pechenizkiy, M., Bischl, B., & Hutter, F. (2023). Can Fairness be Automated? Guidelines and Opportunities for Fairness-aware AutoML. ArXiv, abs/2303.08485, null. https://doi.org/10.48550/arXiv.2303.08485
226. Weidele, D. K. I., Weisz, J. D., Oduor, E., Muller, M., Andres, J., Gray, A., & Wang, D. (2020). AutoAIViz: Opening the blackbox of automated artificial intelligence with conditional parallel coordinates. International Conference on Intelligent User Interfaces, Proceedings IUI, 308–312. https://doi.org/10.1145/3377325.3377538
227. Whaiduzzaman, M., Shahrier, L., Barros, A., Jan, T., Rahman, M. S., Sakib, A., Fidge, C., Khan, N. J., Thompson-Whiteside, S., Mahi, M. J. N., Ghosh, S., & Chaki, S. (2023). Concept to Reality: An Integrated Approach to Testing Software User Interfaces. Applied Sciences, 13(21), 11997. https://doi.org/10.3390/app132111997
228. Wicaksono, A., & Maharani, A. (2020). The Effect of Perceived Usefulness and Perceived Ease of Use on the Technology Acceptance Model to Use Online Travel Agency. Journal of Business Management Review, 1(5), 313–328. https://doi.org/10.47153/jbmr15.502020
229. Winter, M., & Jackson, P. (2020). Flatpack ML: How to Support Designers in Creating a New Generation of Customizable Machine Learning Applications. Lecture Notes in Computer Science (Including Subseries Lecture Notes in Artificial Intelligence and Lecture Notes in Bioinformatics), 12201 LNCS, 175–193. https://doi.org/10.1007/978-3-030-49760-6_12/FIGURES/4
230. Wu, Q., & Wang, C. (2021). FairAutoML: Embracing Unfairness Mitigation in AutoML. https://arxiv.org/abs/2111.06495v2
231. Xanthopoulos, I., Tsamardinos, I., Christophides, V., Simon, E., & Salinger, A. (2020). Putting the human back in the AutoML loop. CEUR Workshop Proceedings, 2578.
232. Xin, D., Wu, E. Y., Lee, D. J. L., Salehi, N., & Parameswaran, A. (2021). Whither automl? understanding the role of automation in machine learningworkflows. Conference on Human Factors in Computing Systems - Proceedings. https://doi.org/10.1145/3411764.3445306
233. Xu, W., Dainoff, M. J., Ge, L., & Gao, Z. (2022). Transitioning to Human Interaction with AI Systems: New Challenges and Opportunities for HCI Professionals to Enable Human-Centered AI. International Journal of Human–Computer Interaction, ahead-of-print(ahead-of-print), 494–518. https://doi.org/10.1080/10447318.2022.2041900

—-------—------------------------------------------------------------------------------------

234. Yamani, A. Z., Baslyman, M., & Al-Shammare, H. A. (2024). Establishing Heuristics for Improving the Usability of GUI Machine Learning Tools for Novice Users. 1–19. https://doi.org/10.1145/3613904.3642087

235. Yam, J., & Skorburg, J. A. (2021). From human resources to human rights: Impact assessments for hiring algorithms. Ethics and Information Technology, 23(4), 611–623. https://doi.org/10.1007/s10676-021-09599-7

236. Yan, R. (2018). “Chitty-Chitty-Chat Bot”: Deep Learning for Conversational AI. 5520–5526. https://doi.org/10.24963/ijcai.2018/778

237. Yang, F., Cisse, M., & Koyejo, S. (2020). Fairness with Overlapping Groups. Advances in Neural Information Processing Systems, 2020-December. https://arxiv.org/abs/2006.13485v1

238. Yang, Q., Steinfeld, A., Zimmerman, J., & Rosé, C. (2020). Re-examining Whether, Why, and How Human-AI Interaction Is Uniquely Difficult to Design. 1–13. https://doi.org/10.1145/3313831.3376301

239. Yao, Q., Wang, M., Chen, Y., Dai, W., Li, Y., Tu, W., Yang, Q., & Yu, Y. (2018). Taking human out of learning applications: A survey on automated machine learning. ArXiv, 1810.13306.

240. Yildiz, Z. O., & Beloff, N. (2020). The Emerging AI Policy for e-commerce Industry. ACM International Conference Proceeding Series, 66–70. https://doi.org/10.1145/3385209.3385210

241. Yu, S. (2023). Towards Trustworthy and Understandable AI: Unraveling Explainability Strategies on Simplifying Algorithms, Appropriate Information Disclosure, and High-level Collaboration. 133–143. https://doi.org/10.1145/3616961.3616965

242. Zender, A., Holzheuser, A., & Humm, B. G. (2024). Successfully Improving the User Experience of an Artificial Intelligence System. 39, 253–258. https://doi.org/10.15439/2024f2707

243. Zhang, K., Khosravi, B., Vahdati, S., Faghani, S., Nugen, F., Rassoulinejad-Mousavi, S. M., Moassefi, M., Jagtap, J. M. M., Singh, Y., Rouzrokh, P., & Erickson, B. J. (2022). Mitigating Bias in Radiology Machine Learning: 2. Model Development. Https://Doi.Org/10.1148/Ryai.220010, 4(5). https://doi.org/10.1148/RYAI.220010

244. Zhang, R., Duan, W., Knijnenburg, B., Flathmann, C., Mcneese, N. J., Schelble, B., & Musick, G. (2024). I Know This Looks Bad, But I Can Explain: Understanding When AI Should Explain Actions In Human-AI Teams. ACM Transactions on Interactive Intelligent Systems, 14(1), 1–23. https://doi.org/10.1145/3635474

245. Zhang, Y., & Zhou, L. (2019). Fairness Assessment for Artificial Intelligence in Financial Industry. https://doi.org/10.48550/arxiv.1912.07211

246. Zhang, Y., Bellamy, R., & Varshney, K. (2020). Joint Optimization of AI Fairness and Utility. 400–406. https://doi.org/10.1145/3375627.3375862

—---------—-----------------------------------------------------------------------------------

247. Zhao, J., Wang, T., Yatskar, M., Ordonez, V., & Chang, K. W. (2018). Gender Bias in Coreference Resolution: Evaluation and Debiasing Methods. NAACL HLT 2018 - 2018 Conference of the North American Chapter of the Association for Computational Linguistics: Human Language Technologies - Proceedings of the Conference, 2, 15–20. https://doi.org/10.18653/V1/N18-2003
248. Zöller, M.-A., & Huber, M. F. (2021). Benchmark and survey of automated machine learning frameworks. J. Artif. Intell. Res., 70, 409–472. https://doi.org/10.48550/arXiv.1904.12054